\documentclass[a4paper,12pt]{memoir}

\ExplSyntaxOn
\iow_new:N \g_book_bbl_iow
\ior_new:N \g_book_bbl_ior
\iow_open:Nn \g_book_bbl_iow {\jobname.bbl}
\iow_now:Nx \g_book_bbl_iow {\c_percent_str\c_space_tl $~biblatex~auxiliary~file~$}
\iow_now:Nx \g_book_bbl_iow {\c_percent_str\c_space_tl $~biblatex~bbl~format~version~3.2~$}
\ior_open:Nn \g_book_bbl_ior {\jobname-bibdata.tmp}
\ior_str_map_inline:Nn \g_book_bbl_ior {\iow_now:Nn \g_book_bbl_iow {#1}}
\ior_close:N \g_book_bbl_ior
\iow_close:N \g_book_bbl_iow
\ExplSyntaxOff

\usepackage[left=3cm,top=3cm,bottom=3.4cm,right=3cm]{geometry}
\usepackage[british]{babel}
\usepackage{csquotes}
\setsecnumdepth{subsection}
\usepackage{amsfonts,amsmath,amssymb,amsthm,mathtools,mathrsfs,bbm}

\usepackage{xcolor}
\definecolor{green(munsell)}{rgb}{0.0, 0.66, 0.47}
\definecolor{BlueGreenn}{rgb}{0.3,0.5,0.8}
\definecolor{DB}{rgb}{0.3,0.3,0.3}
\definecolor{DOr}{rgb}{0.7,0.3,0.3}
\definecolor{DGr}{rgb}{0.3,0.7,0.3}
\definecolor{DBl}{rgb}{0.1,0.3,0.5}
\definecolor{arylideyellow}{rgb}{0.91, 0.84, 0.42}
\definecolor{burntorange}{rgb}{0.8, 0.33, 0.0}
\definecolor{chromeyellow}{rgb}{1.0, 0.65, 0.0}
\definecolor{niceyellow}{RGB}{255,204,0}
\definecolor{lightgray}{rgb}{0.96,0.96,0.96}
\definecolor{darkgreen}{rgb}{0.1,0.45,0.06}
\definecolor{darkred}{rgb}{0.65,0.1,0.1}
\definecolor{lightred}{rgb}{1,0.9,0.9}
\definecolor{darkblue}{rgb}{0.1,0.1,0.5}
\definecolor{verydarkblue}{rgb}{0.1,0.1,0.3}
\definecolor{lightblue}{rgb}{0.88,0.92,0.99}
\definecolor{lightorange}{rgb}{1,0.95,0.89}
\definecolor{lightyellow}{rgb}{1,1,0.87}
\definecolor{burntlightorange}{RGB}{245,235,211}
\definecolor{skde}{RGB}{0,80,158}
\definecolor{skdelight}{RGB}{198,215,231}
\definecolor{skdeverylight}{RGB}{228,245,251}
\definecolor{skdelys}{RGB}{122,178,220}
\definecolor{pink}{RGB}{232,88,208}

\usepackage[hypertexnames=true,citecolor=burntorange,colorlinks=true,linkcolor=skde,
  pdffitwindow=false,urlbordercolor=white,pdfborder={0 0 0}]{hyperref}

\usepackage{tikz-cd}

\usepackage{tikz}
\usetikzlibrary{shapes.geometric}
\usepackage[shortlabels]{enumitem}
\setlist{noitemsep}
\setlist{nolistsep}
\usepackage{float}

\usepackage{xspace,comment,mdframed,environ,bbding,pifont,makeidx,ifthen,datetime,nicefrac,euscript,etoc,needspace}
\AddToHook{cmd/section/before}{\Needspace{4\baselineskip}}
\AddToHook{cmd/subsection/before}{\Needspace{3\baselineskip}}

\newtheoremstyle{bookstyle}{6pt}{6pt}{\itshape}{}{\bfseries}{.}{0.5em}{\thmname{#1}\ \thmnumber{#2}\thmnote{ \textbf{(#3)}}}

\theoremstyle{bookstyle}
\newtheorem{theorem}{Theorem}[section]
\newtheorem{proposition}[theorem]{Proposition}
\newtheorem{lemma}[theorem]{Lemma}

\newtheorem{corollary}[theorem]{Corollary}

\newcounter{problem}

\newtheoremstyle{bookdefinitionstyle}{6pt}{6pt}{\upshape}{}{\bfseries}{.}{0.5em}{\thmname{#1}\ \thmnumber{#2}\thmnote{ \textbf{(#3)}}}

\theoremstyle{bookdefinitionstyle}
\newtheorem{definition}[theorem]{Definition}
\newtheorem{remark}[theorem]{Remark}
\newtheorem{example}[theorem]{Example}

\numberwithin{equation}{section}

\usepackage[style=alphabetic, labelnumber, defernumbers=true, firstinits=true, maxbibnames=99, url=true, doi=true, isbn=false, eprint=false, backend=biber]{biblatex}
\renewbibmacro{in:}{}
\DeclareFieldFormat{pages}{#1}

\AtEveryBibitem{\iffieldundef{doi}{}{\clearfield{url}}}

\copypagestyle{fnsizeheadings}{headings}
\makeevenhead{fnsizeheadings}{\footnotesize\thepage}{}{\footnotesize\leftmark}
\makeoddhead{fnsizeheadings}{\footnotesize\rightmark}{}{\footnotesize\thepage}
\nouppercaseheads

  \usepackage{calc}
  
    \usepackage{fontspec}
    \usepackage[math-style=TeX,bold-style=ISO]{unicode-math}
    \newfontfamily\libertinustitleface[LetterSpace=-2]{LibertinusSans-Bold.otf}
    \newfontfamily\libertinussubtitleface{LibertinusSerif-Semibold.otf}
    \setboldmathrm{LibertinusSerif-Bold.otf}[ItalicFont=LibertinusSerif-BoldItalic.otf]
    \AtBeginDocument{\RenewDocumentCommand{\Im}{}{\operatorname{Im}}\RenewDocumentCommand{\Re}{}{\operatorname{Re}}}

\DeclareMathOperator{\sech}{sech}


  \excludecomment{secondrelease}

\newcommand{\booktheoremindex}[2]{\index{theorem!#2@#2 (\protect\ref*{#1})}}

\renewcommand{\Re}{\operatorname{Re}}
\renewcommand{\Im}{\operatorname{Im}}
\newcommand{\Std}{\operatorname{Std}}
\newcommand{\bStd}{\boldsymbol{\operatorname{Std}}}

\newcommand{\Aut}{\operatorname{Aut}}
\newcommand{\Con}{\operatorname{Con}}
\providecommand{\Tr}{\operatorname{Tr}}
\newcommand{\AU}{\operatorname{AU}}
\newcommand{\U}{\operatorname{U}}
\newcommand{\Fix}{\operatorname{Fix}}
\newcommand{\Inv}{\operatorname{Inv}}
\newcommand{\dom}{\operatorname{dom}}
\newcommand{\St}{\operatorname{St}}
\newcommand{\Ran}{\operatorname{Ran}}
\newcommand{\spn}{\operatorname{span}}
\newcommand{\rank}{\operatorname{rank}}
\newcommand{\id}{\operatorname{id}}
\newcommand{\rlspn}{\Rl\text{-}\operatorname{span}}

\DeclareDocumentCommand{\innerproduct}{s m g}{\IfBooleanTF{#1}{\langle{#2},{\IfNoValueTF{#3}{#2}{#3}}\rangle}{\left\langle{#2},{\IfNoValueTF{#3}{#2}{#3}}\right\rangle}}
\let\ip\innerproduct

\newcommand{\scpr}{\langle\,\cdot\,,\,\cdot\,\rangle}
\newcommand{\Rescpr}{\Re\langle\,\cdot\,,\,\cdot\,\rangle}

\ifdefined\symbfit
  \let\bookboldvector\symbfit
\else
  \let\bookboldvector\boldsymbol
\fi
\newcommand{\bp}{\bookboldvector{p}}

\newcommand{\Sob}{\mathrm{H}}

\newcommand{\Hil}{{\mathcal H}}
\newcommand{\Kil}{{\mathcal K}}
\newcommand{\CL}{{\mathcal L}}
\newcommand{\CR}{{\mathcal R}}
\newcommand{\CA}{{\mathcal A}}
\newcommand{\CD}{{\mathcal D}}
\newcommand{\CJ}{{\mathcal J}}
\newcommand{\CM}{{\mathcal M}}

\newcommand{\CI}{{\mathcal I}}
\newcommand{\CK}{{\mathcal K}}

\newcommand{\CU}{{\mathcal U}}
\newcommand{\CV}{{\mathcal V}}

\newcommand{\EuF}{\EuScript{F}}

\newcommand{\Sch}{{\mathscr S}}

\newcommand{\TripNorm}[1]{|\!|\!|#1|\!|\!|}

\newcommand{\te}{\theta}

\def\A{{\mathcal A}}

\def\E{{\mathcal  E}}

\def\H{{\mathcal  H}}
\def\CP{{\mathcal  P}}
\def\M{{\mathcal  M}}

\def\H{{\mathcal H}}
\def\K{{\mathcal  K}}

\def\V{{\mathcal  V}}

\def\a{{\alpha}}
\def\l{{\lambda}}

\newcommand{\rperp}{{\perp_{\Rl}}}

\newcommand{\col}{\,:\,}
\newcommand{\sa}{{\mathrm{sa}}}

\newcommand{\Rl}{{\mathbb R}}
\newcommand{\RR}{{\mathbb R}}

\newcommand{\Cl}{{\mathbb C}}
\newcommand{\Nl}{{\mathbb N}}
\newcommand{\Zl}{{\mathbb Z}}
\newcommand{\Kl}{{\mathbb K}}
\newcommand{\Hardy}{\mathbb{H}}
\newcommand{\Hinfty}{{\mathbb H}^\infty}
\newcommand{\HinftyC}{{\mathbb H}^\infty_C}

\newcommand{\HinftyCs}{{\mathbb H}^\infty_{C,\mathrm{s}}}
\newcommand{\Hbv}{\mathbb{H}^\infty_{\text{c.bv}}}

\newcommand{\Strip}{{\mathbb S}}
\newcommand{\Stripb}{{\mathbb S}_\beta}
\newcommand{\Disc}{{\mathbb D}}

\newcommand{\Om}{\Omega}
\newcommand{\om}{\omega}
\newcommand{\la}{\lambda}
\newcommand{\eps}{\varepsilon}

\newcommand{\Bor}{{\mathfrak B}}

\newcommand\restr[1]{\raisebox{-0.5ex}{\ensuremath{|_{#1}}}}
\newcommand{\Span}{\operatorname{span}}
\newcommand{\spec}{\operatorname{sp}}
\newcommand{\specp}{\operatorname{sp}_{\mathrm{p}}}
\newcommand{\aff}{{\mathrm{aff}}}
\newcommand{\rad}{{\operatorname{rad}}}
\newcommand{\supp}{{\operatorname{supp}}}

\DeclareMathOperator{\sign}{sign} \DeclareMathOperator{\arctanh}{arctanh}

\newcommand{\FM}{\mathfrak M}

\DeclareMathOperator{\GL}{GL} \DeclareMathOperator{\Sub}{Sub} \DeclareMathOperator{\SubR}{Sub_{\Rl}} \DeclareMathOperator{\Gr}{Gr} \DeclareMathOperator{\Real}{\Gr_{\Rl}}
\DeclareMathOperator{\Aff}{Aff}

\makeindex
\AddToHook{cmd/theindex/after}{\label{back:SubjectIndex}}
\usepackage[style=super,acronym,nogroupskip,toc]{glossaries}
\makenoidxglossaries
\glssanitizesorttrue
\glsnoexpandfields

\newcommand{\symboltableheading}{\textbf{Symbol} & \textbf{Meaning} & \textbf{Page}\\\hline}
\newglossarystyle{booksymbols}{\setglossarystyle{long3col}}

\setglossarypreamble{\label{back:ListOfSymbols}}

\newglossaryentry{SH}
{name={$S_H$},
  description={Tomita operator of standard subspace $H$},
  sort=SH,
}
\newglossaryentry{StdCategory}
{name={$\bStd$},
  description={tensor category of standard subspaces},
  sort=StdCategory,
}
\newglossaryentry{StdH}
{name={$\Std(\Hil)$},
  description={Set of standard subspaces in Hilbert space $\Hil$},
  sort=StdH,
}
\newglossaryentry{AUH}
{name={$\AU(\Hil)$},
  description={group of unitary and antiunitary operators in Hilbert space $\Hil$},
  sort=AUH,
}
\newglossaryentry{precMeasures}
{name={$\nu\prec\mu$},
  description={$\nu$ is absolutely continuous with respect to $\mu$},
  sort=precMeasures,
}
\newglossaryentry{precStandardSubspaces}
{name={$K\prec H$},
  description={$K$ is unitarily equivalent to a direct summand of $H$},
  sort=precStandardSubspaces,
}
\newglossaryentry{Hrperp}
{name={$H^\rperp$},
  description={real orthogonal complement of $H$},
  sort=Hrperp,
}
\newglossaryentry{Hperp}
{name={$H^\perp$},
  description={complex orthogonal complement of $H$},
  sort=Hperp,
}
\newglossaryentry{Hprime}
{name={$H'$},
  description={symplectic complement of $H$},
  sort=H00,
}
\newglossaryentry{FixA}
{name={$\Fix(A)$},
  description={fixed-point space of (real) linear map $A$},
  sort=FixA,
}
\newglossaryentry{RealH}
{name={$\Real(\Hil)$},
  description={lattice of closed real-linear subspaces in $\Hil$},
  sort=RealH,
}
\newglossaryentry{JH}
{name={$J_H$},
  description={modular conjugation of $H$},
  sort=JH,
}
\newglossaryentry{DeltaH}
{name={$\Delta_H$},
  description={modular operator of $H$},
  sort=DeltaH,
}
\newglossaryentry{HlowC}
{name={$H_\Cl$},
  description={$H\cap iH$},
  sort=HlowC,
}
\newglossaryentry{HupC}
{name={$H^\Cl$},
  description={$\overline{H+iH}$},
  sort=HlowC,
}
\newglossaryentry{CH}
{name={$\Cl[H]$},
  description={closed complex span of $H$},
  sort=CH,
}
\newglossaryentry{EA}
{name={$E^A$},
  description={spectral measure of $A$},
  sort=EA,
}
\newglossaryentry{Bor}
{name={$\Bor$},
  description={set of finite Borel functions $f:\Rl\to\Cl$},
  sort=Bor,
}
\newglossaryentry{sA}
{name={$\spec(A)$},
  description={spectrum of $A$},
  sort=sA,
}
\newglossaryentry{spA}
{name={$\specp(A)$},
  description={point spectrum of $A$},
  sort=spA,
}
\newglossaryentry{I1H}
{name={$\CI_1(\Hil)$},
  description={trace class in $B(\Hil)$},
  sort=I1H,
}
\newglossaryentry{I2H}
{name={$\CI_2(\Hil)$},
  description={Hilbert-Schmidt class in $B(\Hil)$},
  sort=I2H,
}
\newglossaryentry{BH}
{name={$B(\Hil)$},
  description={bounded complex-linear operators on $\Hil$},
  sort=BH,
}
\newglossaryentry{BRH}
{name={$B_\Rl(\Hil)$},
  description={bounded real-linear operators on $\Hil$},
  sort=BH,
}
\newglossaryentry{Hil}
{name={$\Hil$},
  description={complex Hilbert space},
  sort=Hil,
}
\newglossaryentry{InvH}
{name={$\Inv(\Hil)$},
  description={closed antilinear involutions in $\Hil$},
  sort=InvH,
}
\newglossaryentry{EH}
{name={$E_H$},
  description={(real) orthogonal projection onto closed (real) subspace $H$},
  sort=EH,
}
\newglossaryentry{Stripb}
{name={$\Stripb$},
  description={strip region $\Rl+i(\min\{0,\beta\},\max\{0,\beta\})$},
  sort=Stripb,
}
\newglossaryentry{Stripab}
{name={$\Strip_{a,b}$},
  description={strip region $\Rl+i(a,b)$},
  sort=Stripab,
}
\newglossaryentry{Hinfty}
{name={$\Hinfty(\Strip_{a,b})$},
  description={bounded analytic functions on $\Strip_{a,b}$},
  sort=Hinfty,
}
\newglossaryentry{Hbv}
{name={$\Hbv(\Strip_{a,b})$},
  description={bounded analytic functions with continuous extension to $\overline{\Strip_{a,b}}$},
  sort=Hbv,
}
\newglossaryentry{leftmult}
{name={$\ell$},
  description={left multiplication},
  sort=l,
}
\newglossaryentry{rightmult}
{name={$r$},
  description={right multiplication},
  sort=r,
}
\newglossaryentry{EAxi}
{name={$E^A_\xi$},
  description={expectation value of spectral measure, $\langle\xi, E^A(\cdot)\xi\rangle$},
  sort=EAxi,
}
\newglossaryentry{Mpredual}
{name={$\M_{*}$},
  description={predual of the von Neumann algebra $\M$},
  sort=Mstar,
}
\newglossaryentry{Dens}
{name={$\CD(\Hil)$},
  description={space of density matrices on $\Hil$},
  sort=Dense,
}
\newglossaryentry{rho}
{name={$\rho$},
  description={density matrix},
  sort=rho,
}

\newglossaryentry{Tr}
{name={$\Tr$},
  description={Hilbert space trace},
  sort=Tr,
}
\newglossaryentry{scpr}
{name={$\scpr$},
  description={scalar product on $\Hil$, typically complex},
  sort=0scpr,
}
\newglossaryentry{scprR}
{name={$\scpr_\Rl$},
  description={real part of complex scalar product},
  sort=0scprR,
}
\newglossaryentry{scprHS}
{name={$\scpr_2$},
  description={Hilbert-Schmidt scalar product on $\CI_2(\Hil)$},
  sort=0scprHS,
}
\newglossaryentry{HSnorm}
{name={$\|\cdot\|_2$},
  description={Hilbert-Schmidt norm on $\CI_2(\Hil)$},
  sort=0norm2,
}
\newglossaryentry{Asa}
{name={$\CA_\sa$},
  description={real subspace of self-adjoint elements of ${}^*$-algebra $\CA$},
  sort=Asa,
}
\newglossaryentry{omega}
{name={$\om$},
  description={state on a unital ${}^*$-algebra},
  sort=omega,
}
\newglossaryentry{IinftyH}
{name={$\CI_\infty(\Hil)$},
  description={compact operators on $\Hil$},
  sort=IinftyH,
}
\newglossaryentry{KH}
{name={$\CK(\Hil)$},
  description={compact operators on $\Hil$},
  sort=KH,
}
\newglossaryentry{IpH}
{name={$\CI_p(\Hil)$},
  description={$p$-th Schatten ideal in $B(\Hil)$},
  sort=IpH,
}
\newglossaryentry{TrClNorm}
{name={$\|\cdot\|_1$},
  description={trace norm on $\CI_1(\Hil)$},
  sort=0norm1,
}

\newglossaryentry{cutting}
{name={$P_H$},
  description={cutting projection of factorial subspace $H$},
  sort=PH,
}
\newglossaryentry{polarizer}
{name={$D_H$},
  description={polarizer of $H\in\Real(\Hil)$},
  sort=DH,
}
\newglossaryentry{FH}
{name={$F(H)$},
  description={factorial part of $H$},
  sort=FH,
}
\newglossaryentry{ZH}
{name={$Z(H)$},
  description={$Z(H)=H\cap H'$, centre of $H\in\Real(\Hil)$},
  sort=ZH,
}
\newglossaryentry{Real}
{name={$\Re$},
  description={real part of element of ${}^*$-algebra},
  sort=Re,
}
\newglossaryentry{Imaginary}
{name={$\Im$},
  description={imaginary part of element of ${}^*$-algebra},
  sort=Im,
}
\newglossaryentry{HardyStrip}
{name={$\Hardy^2(\Strip_\beta)$},
  description={Hardy space on the strip $\Strip_\beta$},
  sort=H2S,
}
\newglossaryentry{compactTestFunctions}
{name={$C_c^\infty(O)$},
  description={smooth complex-valued functions with compact support in the open set $O\subset\Rl^d$. The real-valued subspace is $C_{c,\Rl}^\infty(O)$},
  sort=Cc,
}
\newglossaryentry{Sobolev}
{name={$\Sob^s_{\Rl}(\Rl^d)$},
  description={real Sobolev space of order $s$, consisting of real tempered distributions $f$ with $(1+|\bp|^2)^{s/2}\tilde f\in L^2(\Rl^d)$},
  sort=HSobolev,
}
\newglossaryentry{Schwartz}
{name={$\Sch(\Rl)$},
  description={Schwartz space on $\Rl$},
  sort=Schwartz,
}
\newglossaryentry{Affiliated}
{name={$\M_\aff$},
  description={Affiliated operators to a von Neumann algebra $\M$},
  sort=Maff,
}
\newglossaryentry{Rplus}
{name={$\Rl_+$},
  description={open positive half-line $(0,\infty)$},
  sort=Rplus,
}
\newglossaryentry{Cplus}
{name={$\Cl_+$},
  description={open upper half-plane in $\Cl$},
  sort=Cplus,
}
\newglossaryentry{H2radC+}
{name={$\Hardy_\rad^2(\Cl_+)$},
  description={radial Hardy space on the upper half-plane},
  sort=H2radC+,
}
\newglossaryentry{H2C+}
{name={$\Hardy^2(\Cl_+)$},
  description={Hardy space on the upper half-plane},
  sort=H2C+,
}
\newglossaryentry{PRH}
{name={$\CP_{\Rl}(\Hil)$},
  description={lattice of all real-linear real orthogonal projections on $\Hil$},
  sort=PRH,
}
\newglossaryentry{entropyvector}
{name={$\E_H(\psi)$},
  description={entropy of a vector $\psi\in\Hil$ w.r.t. $H\in \Std(\Hil)$},
  sort=EntropyHvector,
}
\newglossaryentry{HbC}
{name={$\HinftyC(\Strip_{a,b})$},
description={bounded analytic functions with continuous extension to $\overline{\Strip_{a,b}}$},
sort=HC,
}
\newglossaryentry{ConH}
{name={$\Con(\Hil)$},
description={set of all conjugations on complex Hilbert space $\Hil$},
sort=ConH,
}
\newglossaryentry{sH}
{name={$s_H$},
description={(pre-)symplectic form of $H\in\Real(\Hil)$},
sort=sH,
}
\newglossaryentry{PH}
{name={$\CP(\Hil)$},
description={lattice of all orthogonal everywhere defined projections on $\Hil$},
sort=ProjH,
}
\newglossaryentry{UH}
{name={$\U(\Hil)$},
description={group of unitary operators on $\Hil$ (orthogonal in the real case)},
sort=UH,
}
\newglossaryentry{SubH}
{name={$\Sub(\Hil)$},
description={family of all subspaces of $\Hil$},
sort=SubH,
}
\newglossaryentry{GrH}
{name={$\Gr(\Hil)$},
description={Grassmannian of all closed subspaces of $\Hil$},
sort=GrassH,
}
\newglossaryentry{SubR}
{name={$\Sub_\Rl(\Hil)$},
description={family of all real subspaces of $\Hil$ (orthogonal in the real case)},
sort=SubR,
}
\newglossaryentry{GrR}
{name={$\Gr_\Rl(\Hil)$},
description={Real Grassmannian of all closed real subspaces of $\Hil$},
sort=GrassR,
}
\newglossaryentry{vee}
{name={$\vee$},
description={supremum in projection lattice or Grassmannian},
sort=0vee,
}
\newglossaryentry{closure}
{name={$\overline{\,\cdot\,}$},
description={norm closure},
sort=0closure,
}
\newglossaryentry{equivalence}
{name={$\cong$},
description={unitary equivalence},
sort=0unitaryequivalence,
}
\newglossaryentry{PHK}
{name={$P_{H,K}$},
description={projection with range $H$ and kernel $K$ \eqref{eq:def2subspaceprojection}},
sort=PHK,
}
\newglossaryentry{SHK}
{name={$S_{H,K}$},
description={involution with $(\pm1)$-eigenspaces $H$ and $K$},
sort=SHK,
}
\newglossaryentry{EplusH}
{name={$\E_+(\Hil)$},
description={positive non-singular contractions on $\Hil$ \eqref{eq:EplusH}},
sort=EplusH,
}
\newglossaryentry{EDeltaSmall}
{name={$E^{\Delta_H}_<$},
description={spectral projection of $\Delta_H$ corresponding to $(0,1)$\eqref{eq:SpectralProjectionsDelta}},
sort=EDeltaSmall,
}
\newglossaryentry{EDeltaSmallE}
{name={$E^{\Delta_H}_\leq$},
description={spectral projection of $\Delta_H$ corresponding to $(0,1]$\eqref{eq:SpectralProjectionsDelta}},
sort=EDeltaSmallE,
}
\newglossaryentry{EDeltaBig}
{name={$E^{\Delta_H}_>$},
description={spectral projection of $\Delta_H$ corresponding to $(1,\infty)$\eqref{eq:SpectralProjectionsDelta}},
sort=EDeltaBig,
}
\newglossaryentry{EDeltaBigE}
{name={$E^{\Delta_H}_\geq$},
description={spectral projection of $\Delta_H$ corresponding to $[1,\infty)$\eqref{eq:SpectralProjectionsDelta}},
sort=EDeltaBigE,
}
\newglossaryentry{EDelta1}
{name={$E^{\Delta_H}_1$},
description={spectral projection of $\Delta_H$ corresponding to $\{1\}$\eqref{eq:SpectralProjectionsDelta}},
sort=EDelta1,
}
\newglossaryentry{IH}
{name={$I_H$},
description={phase of the polarizer $D_H$ of $H$ (Lemma~\ref{lemma:Polarizer})},
sort=IH,
}
\newglossaryentry{BJH}
{name={$B_H$},
description={real orthogonal isomorphism $\Fix J_H\to H$  (Prop.~\ref{proposition:ModularDataProperties}~\ref{item:FixJvsFixS})},
sort=BJH,
}

\newglossaryentry{modularGroup}
{name={$\Delta_H^{it}$},
  description={modular unitary at time $t$ for the standard subspace $H$},
  sort=DeltaHt,
}

\newglossaryentry{modularHamiltonian}
{name={$L_H$},
  description={modular Hamiltonian $\log\Delta_H$ of $H$},
  sort=LH,
}

\newglossaryentry{stripStar}
{name={$f^*$},
  description={strip reflection $f^*(\zeta)=\overline{f(i\beta+\overline\zeta)}$},
  sort=fstar,
}

\newglossaryentry{stripSymmetric}
{name={$\HinftyCs(\Strip_\beta)$},
  description={real subspace of $\HinftyC(\Strip_\beta)$ fixed by strip reflection},
  sort=HCs,
}

\newglossaryentry{localCurrent}
{name={$H(I)$},
  description={local real subspace of the $U(1)$ current associated with $I\subset\Rl$},
  sort=HI,
}

\newglossaryentry{currentHalfLines}
{name={$H_\pm$},
  description={half-line standard subspaces $H(\Rl_\pm)$ of the $U(1)$ current},
  sort=Hpm,
}

\newglossaryentry{fullFourier}
{name={$\EuF$},
  description={Fourier transform with kernel $(2\pi)^{-1/2}e^{ipx}$},
  sort=Fourier,
}

\newglossaryentry{positiveFourier}
{name={$\EuF_+$},
  description={positive-frequency Fourier map on real Schwartz functions},
  sort=FourierPlus,
}

\newglossaryentry{affineGroup}
{name={$\Aff(\Rl)$},
  description={affine group of the real line},
  sort=Aff,
}

\newglossaryentry{positiveAffine}
{name={$\Aff_+(\Rl)$},
  description={orientation-preserving affine group of the real line},
  sort=AffPlus,
}

\newglossaryentry{affineReflection}
{name={$j$},
  description={affine reflection $x\mapsto-x$},
  sort=j,
}

\newglossaryentry{standardTensor}
{name={$H\otimes K$},
  description={closure of the real span of simple tensors in the ambient complex Hilbert tensor product},
  sort=TensorH,
}

\newglossaryentry{realTensor}
{name={$H\otimes_\Rl K$},
  description={Hilbert tensor product of the real Hilbert spaces $H$ and $K$},
  sort=TensorR,
}

\newglossaryentry{restrictedModular}
{name={$\Delta_{H,\leq}$},
  description={restriction of $\Delta_H$ to its spectral subspace for $(0,1]$},
  sort=DeltaHleq,
}

\newglossaryentry{entropyOperator}
{name={$\E_H$},
  description={entropy operator of the standard subspace $H$ (Definition~\ref{def:entropyoperator})},
  sort=EntropyH,
}

\newglossaryentry{boundaryTransfer}
{name={$D_\beta$},
  description={boundary-to-boundary operator $F_0\mapsto F_\beta$},
  sort=Dbeta,
}

\newglossaryentry{boundaryDomain}
{name={$\CD_\beta$},
  description={space of lower boundary values, the domain of $D_\beta$ in $L^2(\Rl)$},
  sort=DbetaDomain,
}

\newglossaryentry{fourierBoundaryDomain}
{name={$\widetilde{\CD}_\beta$},
  description={Fourier-side boundary data $g\in L^2(\Rl)$ with $e^{\beta\xi}g(\xi)\in L^2(\Rl)$},
  sort=DbetaFourier,
}

\newglossaryentry{realAdjoint}
{name={$T^\dagger$},
  description={adjoint of $T$ with respect to the real Hilbert space scalar products},
  sort=Tadj,
}

\newglossaryentry{externalComplexification}
{name={$K_\Cl$},
  description={external complexification of the real Hilbert space $K$},
  sort=KC,
}

\newglossaryentry{complexifiedOperator}
{name={$T_\Cl$},
  description={complexification of $T$, with domain $\dom T+i\dom T$},
  sort=TC,
}

\newglossaryentry{complexifiedSpectrum}
{name={$\sigma_\Cl(T)$},
  description={spectrum of the external complexification $T_\Cl$},
  sort=SpectrumC,
}

\newglossaryentry{pencilSpectrum}
{name={$\sigma_{\mathrm{pencil}}(T)$},
  description={spectrum of the real-linear scalar pencil for the given complex structure},
  sort=SpectrumPencil,
}

\newglossaryentry{realSingularValue}
{name={$s_n(T)$},
  description={$n$th singular value of a compact operator between real Hilbert spaces},
  sort=snT,
}

\newcommand{\version}{1}
\newcommand{\bookabstract}{This is a preview of an upcoming monograph on standard subspaces and modular theory. At the moment, it consists of the first chapter. This starts with a gentle introduction to the subject, and then describes the modular theory of standard subspaces, the characterization of their modular data by the KMS condition and reflection positivity, and various constructions of standard subspaces. Further topics in Chapter~1 include the connection of standard subspaces to complexifications and symplectic structures, including their polarizers, and operations with standard subspaces, namely direct sums, direct integrals, and tensor products. Many examples and classification results are included as well. This first chapter also contains a description of the free $U(1)$-current and some other applications to quantum field theory, such as KMS states of type I factors, the local subspaces of the free Klein--Gordon field, and the entropy operator defined by a standard subspace.
}
\newsavebox{\booktitlelines}
\newcommand{\booksubtitle}{General theory and applications to QFT}
\renewcommand{\maketitle}{\thispagestyle{empty}\par
  \moveright.05\textwidth\vbox to \textheight{\hsize=.95\textwidth
    \linewidth=\hsize
    \parindent=0pt
    \parskip=0pt
    \setbox\booktitlelines=\vbox{\libertinustitleface
      \fontsize{40}{50}\selectfont
      \hbox{Standard Subspaces}\hbox{and Modular Theory}}\copy\booktitlelines
    \vskip 18pt
    {
       \libertinussubtitleface\fontsize{20.25}{27}\selectfont
       \resizebox{\wd\booktitlelines}{!}{\booksubtitle}\par}
    \vskip 12pt plus 1fil
    {\Large
      Ricardo Correa da Silva\textsuperscript{a}\\
      Gandalf Lechner\textsuperscript{b}\\
      Roberto Longo\textsuperscript{c}\par}
    \vskip 12pt plus 1fil
    {\normalsize\textbf{Abstract:} \bookabstract\par}
    \vskip 8pt
    {\small
      Version: \version. 5 November 2026\par}
    \vskip 12pt plus 1.25fil
    {\small
      \rule{2cm}{0.5pt}\par
      \textsuperscript{a} ricardo.correa.silva@usp.br, Instituto de Física, Universidade de São Paulo, Rua do Matão 1371, 05508-090 São Paulo, Brazil\par
      \textsuperscript{b} gandalf.lechner@fau.de, Department Mathematik, Friedrich-Alexander Universität Erlangen-Nürnberg, Cauerstra{\ss}e 11, 91058 Erlangen, Germany\par
      \textsuperscript{c} longo@mat.uniroma2.it, Dipartimento di Matematica, Università di Roma ``Tor Vergata'', Via della Ricerca Scientifica 1, 00133 Roma, Italy\par}\kern0pt
  }}

\begin{document}
\maketitle
\frontmatter
\clearpage
\chapter*{Preface}

\vspace*{3.25cm}

\subsubsection*{Modular theory and standard subspaces}

Modular theory is one of the most remarkable developments in the theory of operator algebras\footnote{The name ``modular'' derives from the modular function of a locally compact group, see Example~\ref{example:HaarMeasure}.}. The origins of this theory go back to Tomita's work in the 1960s. Tomita's results, initially circulated in manuscript form \cite{Tomita:1967}, were developed and made accessible by Takesaki \cite{Takesaki:1970} and soon became a central tool in the theory of von Neumann algebras. As witnessed by landmark results on type III von Neumann algebras by Connes and Takesaki \cite{Connes:1973,Takesaki:1973b,ConnesTakesaki:1977}, much of the modern theory of von Neumann algebras would be difficult to imagine without modular theory. Furthermore, modular theory has become a crucial tool in the applications of von Neumann algebras to quantum physics, be it in the description of thermal equilibrium states \cite{HaagHugenholtzWinnink:1967,BratteliRobinson:1997}, the connection between geometric and algebraic structures in quantum field theory \cite{Borchers:2000}, or the analysis of relative entropies \cite{Araki:1976,Araki:1977}.

Various excellent expositions of modular theory exist in textbook form \cite{BratteliRobinson:1987,Takesaki:2003,Stratila:2020}. While these books focus on the original von Neumann algebraic formulation of modular theory and the purely mathematical theory, or its applications in quantum statistical mechanics, the purpose of this book is to focus on the spatial formulation of modular theory through standard subspaces, and to present the theory alongside its applications in quantum field theory.

Starting from a von Neumann algebra $\M$ acting on a Hilbert space $\Hil$ and a cyclic separating vector $\Omega$, many results of modular theory are based on the closure of the involution
\begin{align}
    S_{\M,\Om}^0 : \M\Om\to\M\Om,\qquad A\Om\mapsto A^*\Om,
\end{align}
the \emph{Tomita operator} $S_{\M,\Om}$ of $(\M,\Om)$. The spatial approach consists in realizing that the closed real subspace $H:=\{A\Om\col A=A^*\in\M\}^{\|\cdot\|}\subset\Hil$ satisfies
\begin{align}\label{eq:StandardInPreface}
    H\cap iH=\{0\},\qquad \overline{H+iH}=\Hil,
\end{align}
and therefore allows for the definition of the \emph{Tomita operator of $H$}, the densely defined closed antilinear involution
\begin{align}
    S_H:H+iH\to H+iH,\qquad h_1+ih_2\mapsto h_1-ih_2.
\end{align}
As the two properties $H\cap iH=\{0\}$ and $\overline{H+iH}=\Hil$ encode that $\Om$ is a standard (cyclic and separating) vector for $\M$, closed real spaces $H\subset\Hil$ with these properties are called \emph{standard subspaces}. This name first appeared in the paper of Leyland, Roberts, and Testard \cite{LeylandsRobertsTestard:1978} with the aim to study the modular structure in second quantization, but the object had already appeared before in the work of Rieffel and Van Daele on a bounded operator approach to the Tomita-Takesaki Theorem \cite{RieffelvanDaele:1977}.

It turns out that $S_H$ coincides with $S_{\M,\Om}$. Hence the standard subspace $H$ determines the Tomita operator of the pair $(\M,\Om)$, and it becomes apparent that modular theory can be formulated entirely in terms of standard subspaces of a Hilbert space, without mentioning von Neumann algebras. For example, one may also choose a formulation in terms of a pair of closed, real-linear subspaces in ``generic position'', as done by Eckmann and Osterwalder \cite{EckmannOsterwalder:1973}. We will however stick to the formulation of standard subspaces explained above. While this perspective is more general than the von Neumann algebraic setup because not every standard subspace derives from a von Neumann algebra with standard vector, interesting constructions of von Neumann algebras can be based on standard subspaces \cite{Araki:1963,LeylandsRobertsTestard:1978,CorreaDaSilvaLechner:2023}.

Whereas those results of modular theory that are concerned with the action of the modular data on a von Neumann algebra are not entirely encoded in the corresponding standard subspace (like for example the Tomita-Takesaki Theorem), they do have their counterparts for standard subspaces. Results that are concerned with the modular data themselves (like for example Borchers' Theorem) are fully encoded in standard subspace language, and this perspective often leads to simplified proofs.

In this book, we will systematically develop the theory of standard subspaces and their modular theory from the basics to contemporary applications. Despite the simple definition of the main object of interest, a rich theory unfolds, in particular when configurations of several standard subspaces (inclusions, or nets) are concerned. Many results of this nature can already be found in Longo's lecture notes \cite{Longo:2008_2}. Neeb's lecture notes \cite{Neeb:2026b} present a substantial amount of standard subspace theory from the point of view of representation theory and new developments on nets of standard subspaces. However, there are many things that are not contained in these two sets of lecture notes, and with this book, we plan to present a fairly complete perspective on the subject.

\subsubsection*{Standard subspaces and quantum field theory}

Our motivation to study modular theory and standard subspaces derives mostly from the operator-algebraic formulation of quantum field theory \cite{Haag:1996,Araki:1999,Advances-AQFT:2015}. For a quantum field theory on Minkowski space, the Reeh-Schlieder Theorem \cite{ReehSchlieder:1961} states that the vacuum is standard for all local observable algebras under suitable assumptions, hence modular theory applies. It has long been known that these modular data are sometimes connected to geometric symmetries \cite{BisognanoWichmann:1975,BisognanoWichmann:1976,HislopLongo:1982,BuchholzDreyerFlorigSummers:2000}, and are related to questions of statistical independence via the split property \cite{DoplicherLongo:1984,BuchholzDAntoniLongo:1990,BuchholzDAntoniLongo:1990-1}. Other applications include the formulation of thermal equilibrium states (KMS states) \cite{HaagHugenholtzWinnink:1967,BratteliRobinson:1997}, constructions of quantum field theoretic models \cite{LongoWitten:2010,Lechner:AQFT-book:2015}, and the formulation and analysis of relative entropies of states in quantum field theory \cite{Araki:1976,Araki:1977}, including entanglement entropy \cite{HollandsSanders:2017}.

The role of standard subspaces is often to represent localization regions, and standard subspaces can be used as a one-particle input to formulate examples of quantum field theories. This opens the road to define free quantum field theories entirely in terms of a net of standard subspaces derived from a positive-energy representation of the Poincaré group \cite{BrunettiGuidoLongo:2002}, a perspective that has been generalized to other Lie groups and spacetimes \cite{MorinelliNeeb:2023}. An interpretation of these subspaces in terms of localization observables can be found in \cite{LechnerdeOliveira:2026}. In chiral conformal theories, real subspaces associated with intervals on a line or circle connect localization with positive-energy M\"obius symmetry \cite{Longo:2008_2,LongoWitten:2010}.

In the Lecture Notes \cite{Longo:2008_2}, the results by Borchers \cite{Borchers:1992} and Wiesbrock \cite{Wiesbrock:1993-1} and Araki-Zsid\'o \cite{ArakiZsido:2005} on half-sided modular inclusions of von Neumann algebras have been translated into the standard subspace setting. Although the proofs are essentially the same, this gives a new perspective relating standard subspaces and symmetries. For instance, a positive-energy, unitary representation of PSL$(2,\Rl)$ can be equivalently given by three commuting standard subspaces, with suitable half-sided modular conditions.

A relevant piece of the theory of standard subspaces concerns entropy. The notion of an entropy operator $\E_H$ of a standard subspace $H$, and a corresponding quadratic form $\psi\mapsto\E_H(\psi)$ on $\Hil$ was defined and analysed in \cite{Longo:2019,CiolliLongoRuzzi:2019}. This entropy quantity is on the one hand equal to the relative entropy between the vacuum state and a coherent state in a free quantum field theory on the Bose-Fock space over~$\Hil$ \cite{BostelmannCadamuroVecchio:2021}, and on the other hand, is meaningful in itself.

This overview hints at the many applications of standard subspaces in quantum field theory, some of which will be presented in this book.

\subsubsection*{Organization of this book}

This book addresses both students who want to learn about standard subspaces, and more experienced researchers who can use it as a reference that collects many results scattered in the literature. While most of the things presented here are known in some form, also several new results and strengthenings of known results are included. Apart from this preface, we do not give any references in the main text, but explain the origin of the results recorded here and the corresponding references in notes and remarks sections at the end of each chapter. These sections also contain links to further articles.

For the first arXiv publication, this manuscript contains Chapter 1 on fundamentals of standard subspaces. It can be read by anybody with a good background in functional analysis, and uses operator algebras only in a few isolated examples. As we have separated the mathematical theory from its applications in physics (some of which are already contained in Chapter~\ref{chapter:Fundamentals}), it is also possible to read this chapter without a background in quantum field theory. See also the \hyperref[guide:ChapterOneReadingPaths]{guide for reading Chapter~\ref*{chapter:Fundamentals}} on page~\pageref{guide:ChapterOneReadingPaths}.

\newpage

At the end of the book, there is also an \hyperref[back:SubjectIndex]{index} and a \hyperref[back:ListOfSymbols]{list of symbols} to make it easier to navigate the book. Any standard subspace is accompanied with various other spaces and operators. The most common ones are collected in the table \hyperref[back:StandardSubspaceNotation]{Notation for a standard subspace}.

\bigskip

The rough plan for future editions of this book is to include the following chapters and corresponding background material in appendices.

\medskip

\begin{itemize}
    \item[$\bullet$] {\bfseries Chapter 2:} Standard subspaces and operator algebras, including in particular second quantization von Neumann algebras on Fock spaces and the type of a standard subspace.
    \item[$\bullet$] {\bfseries Chapter 3:} Inclusions and endomorphisms of standard subspaces, including in particular half-sided modular inclusions and split inclusions.
    \item[$\bullet$] {\bfseries Chapter 4:} Nets of standard subspaces and quantum field theory.
\end{itemize}

\subsubsection*{Acknowledgements}

We have benefitted from discussing standard subspaces and modular theory with many colleagues over the years, and wish to thank all of them. Financial support by the Deutsche Forschungsgemeinschaft (DFG, German Research Foundation) --- project no. 463380247 within the Heisenberg grant LE 2222/3-1 (``Quantenfelder und Operatoralgebren'') is gratefully acknowledged (GL). AI has been used to proofread this manuscript.

\vfill

\noindent
S\~ao Paulo, Brazil\hfill Ricardo Correa da Silva
\\
\noindent
Erlangen, Germany\hfill Gandalf Lechner\\
\noindent
Rome, Italy\hfill Roberto Longo\\
\vspace*{3mm}
\\
\noindent
06.11.2026

\clearpage

\newpage
\begin{KeepFromToc}
  \setcounter{tocdepth}{1}
  \tableofcontents
  \thispagestyle{empty}
\end{KeepFromToc}
\mainmatter

\chapter{Fundamentals of standard subspaces}\label{chapter:Fundamentals}

In this chapter we introduce standard subspaces and their modular theory. The idea is to give a thorough introduction to the subject that is fairly complete as far as questions about a single standard subspace or the family $\Std(\Hil)$ of all standard subspaces in a complex Hilbert space~$\Hil$ is concerned. The more intricate properties of families of standard subspaces, such as inclusions and nets, will be the subject of later chapters. Also the important connection to von Neumann algebras will only be considered in Chapter 2. The content of this chapter is therefore entirely based in Hilbert space theory and accessible with a basic knowledge in functional analysis and in particular Hilbert spaces, including unbounded Hilbert space operators.

\medskip

Section~\ref{section:RealSubspaces} prepares the ground with some material about subspaces of Hilbert spaces, including real subspaces of complex Hilbert spaces, and various associated projection and involution operators. Standard subspaces are defined in Section~\ref{section:StandardSubspacesBasics}. Their basic properties along with some elementary examples are presented. Furthermore, the class of abelian standard subspaces is discussed, and the decomposition of a closed real subspace into an abelian space and a factorial space is explained.

While these investigations can be carried out with elementary techniques, the theory begins properly in Section~\ref{section:ModularData}, where the modular objects of a standard subspace are introduced. A main theorem in this section is the Tomita-Takesaki Theorem for standard subspaces (Theorem~\ref{theorem:Tomita}), and we give some canonical examples of standard subspaces coming from measure spaces and locally compact groups. We also describe how standard subspaces can be classified by positive contractions or graphs.

Intrinsic characterizations of the modular group and modular conjugation by the KMS condition and reflection positivity, respectively, are given in Section~\ref{section:Characterization}. As an application, Section~\ref{section:U1current} discusses the important example of the $U(1)$-current standard subspace in full detail, drawing on Appendix~\ref{section:HardySpaces} for information about Hardy spaces.

The interplay of standard subspaces with complex structures and complexifications of real Hilbert spaces is discussed in Section~\ref{section:ComplexStructures}. This includes in particular a study of the polarizer of a standard subspace, and the intrinsic characterization of a standard subspace as a real polarized Hilbert space or a real Hilbert space with an orthogonal dynamics. Furthermore, it is explained how standard subspaces can be realized via dual pairs of real spaces, a technique that is applied in the context of the local spaces of a free Klein-Gordon field later.

Section~\ref{section:Operations} is devoted to operations on standard subspaces living in possibly different Hilbert spaces, such as direct sums, direct integrals, and tensor products. This results in the family $\bStd$ of all standard subspaces having the structure of a tensor category. As an application of tensor products, we give a characterization of the modular Hamiltonian in terms of complete passivity and as an application of direct integrals, we describe the decomposition of multiplicity-free standard subspaces.

Although this first chapter is introductory, its contents already allow for interesting applications in quantum physics, which are collected in Section~\ref{section:Chapter1Applications}. These applications include the local standard subspaces of a free quantum Klein-Gordon field, and the standard subspace entropy operator defined via the cutting projection. We also mention the KMS states of a type I factor, which is partially a precursor of the operator algebraic structure of Chapter~2.

\vfill

\subsection*{Contents of Chapter \ref{chapter:Fundamentals}}
\etocsettocstyle{}{}
\etocsetnexttocdepth{subsection} \localtableofcontents

\clearpage

\subsection*{Guide for reading Chapter~\ref{chapter:Fundamentals}}
\phantomsection\label{guide:ChapterOneReadingPaths}

Here we give a guide for readers that are particularly interested in some aspects of standard subspaces.

\medskip

\noindent\textbf{Main theory.} If you are interested to learn about the main theoretical aspects, start with real subspaces and their symplectic complements in Sections~\ref{section:RealVsComplex} and~\ref{section:CyclicSeparating}. Read Section~\ref{section:StandardSubspacesBasics} for the definition of standardness, abelian and factorial subspaces. The central result is the Tomita--Takesaki theorem in
Section~\ref{section:TomitaOperators}. Follow it with the relation between the modular spectrum and the centre in Section~\ref{section:SpectrumDelta}.
The details on unbounded projections in Section~\ref{section:SubspaceLattices} can be consulted as they enter the proofs. Section~\ref{section:KMS} on the KMS condition is also fundamental.

\medskip

\noindent\textbf{Examples and classification.} There are various families of examples and constructions of standard subspaces this chapter, some of which can be turned into classification results: Example~\ref{example:2dStandardSubspaces} is about two-dimensional standard subspaces, and Examples~\ref{example:MatrixStandardSubspace} and \ref{example:HilbertSchmidt} are canonical examples involving matrices and Hilbert-Schmidt operators. Example~\ref{example:StandardSubspacesFromMeasures} builds standard subspaces from symmetric measures, and Example~\ref{example:HaarMeasure} from Haar measures on locally compact groups. Example~\ref{example:PositiveContractions} realizes standard subspaces in terms of positive contractions, and Theorem~\ref{theorem:StdAndContractions} shows that up to unitary equivalence, every standard subspace is of this form. Similarly, Example~\ref{example:StandardSubspacesFromGraphs} builds examples of standard subspaces as graphs of positive operators, and Theorem~\ref{theorem:GraphStandardSubspaces} shows that most standard subspaces are of this form. All of Section~\ref{section:U1current} is dedicated to a discussion of an example that is important in quantum field theory, the local standard subspaces of the free $U(1)$-current. In Section~\ref{section:complexifications}, standard subspaces are constructed by complexification procedures, involving in particular the dual pair and complex structure defined by a positive self-adjoint operator. Theorem~\ref{theorem:PositiveOperatorStandardSubspaces} shows that every factorial standard subspace is of this form. As an application of this method, the local standard subspaces of a free Klein-Gordon field are discussed in Section~\ref{section:KleinGordonSpaces}.

\medskip

\noindent\textbf{Further theory.} The further structural elements of the theory involve the orthogonal and cutting projections associated with a standard subspace (Section~\ref{section:SpectrumDelta}) and the connection to complexifications and (pre-)symplectic forms and their polarizers (Section~\ref{section:Polarizers}). Basic operations like direct sum, direct integrals, and tensor products are presented in Section~\ref{section:Operations}.

\medskip

\noindent\textbf{Applications to QFT.} We have included some applications of standard subspaces already in this chapter, although much more will appear in later chapters. KMS states of type I factors are presented in Section~\ref{section:KMSApplication} from the perspective of standard subspaces. The dual pair construction in Section~\ref{section:complexifications} supplies the Cauchy data description of the standard subspaces of the Klein-Gordon field (Section~\ref{section:KleinGordonSpaces}). The cutting projection lies at the basis of defining the entropy operator of a standard subspace (\ref{section:Entropy}), and the Klein-Gordon spaces and entropy are connected in Section~\ref{section:FieldMomentumEntropy}.

\clearpage

\section{Subspaces of Hilbert spaces}\label{section:RealSubspaces}

Standard subspaces will be defined as certain {\em real} subspaces $H$ of a {\em complex} Hilbert space\footnote{We do not assume that $\Hil$ is separable unless explicitly mentioned otherwise.}~$\Hil$\glsadd{Hil}. To prepare the ground for their definition and analysis that starts in Section~\ref{section:StandardSubspacesBasics}, we collect in this section various basic facts about subspaces of Hilbert spaces that will be used throughout the book.

In Section~\ref{section:SubspaceLattices}, we consider subspaces and projections in Hilbert spaces. We recall the structure of the set (lattice) of all closed subspaces, and discuss the most general family of projections in a Hilbert space, indexed by ordered pairs of linearly independent subspaces. Here all structures are understood over the same number field, either the real numbers $\Kl=\Rl$ or the complex numbers $\Kl=\Cl$.

 In Section~\ref{section:RealVsComplex}, we consider real objects in complex Hilbert spaces. This includes in particular real subspaces of complex Hilbert spaces, but also real-linear and antilinear operators, conjugations, and real orthogonal projections.

Still in the context of real subspaces of complex Hilbert spaces, Section~\ref{section:CyclicSeparating} introduces concepts that are at the basis of defining standard subspaces, namely cyclic and separating spaces, and the symplectic complement.

\subsection{Subspaces and projections}\label{section:SubspaceLattices}

 In this section, $\Hil$\glsadd{Hil} denotes a real or complex Hilbert space. Its scalar product and norm are denoted $\scpr$\glsadd{scpr} and $\|\cdot\|$. Three basic classes of linear operators $\Hil\to\Hil$ are the real or complex $C^*$-algebra $B(\Hil)$\glsadd{BH}, respectively, of bounded linear operators, the lattice $\CP(\Hil)$\glsadd{PH} of orthogonal projections, and the group $\U(\Hil)$\glsadd{UH} of unitary operators,
 \begin{align}
    B(\Hil)
    &=
    \{A:\Hil\to\Hil\col A \text{ linear}, \|A\|<\infty\},
    \\
    \CP(\Hil)
    &=
    \{E\in B(\Hil)\col E^2=E=E^*\},
    \\
    \U(\Hil)
    &=
    \{U\in B(\Hil)\col U^*U=UU^*=1\}.
\end{align}
In the real case, we refer to unitary operators as orthogonal operators.

By a \emph{subspace}\index{subspace} $H$ of $\Hil$ we will always mean a linear subspace, i.e. a non-empty subset that is closed under linear combinations. Here ``linear'' of course refers to the field $\Kl=\Rl$ or $\Kl=\Cl$ over which $\Hil$ is defined. We write $\overline H$\glsadd{closure} for the closure of a subset $H\subset\Hil$, and\glsadd{SubH}\glsadd{GrH}\index{Grassmannian}
\begin{align}\label{eq:SubGr}
    \Sub(\Hil)
    &:=
    \{H\subset\Hil\col H \text{ is a linear subspace} \},
    \\
    \Gr(\Hil)
    &:=
    \{H\subset\Hil\col H=\overline H \text{ is a closed linear subspace} \},
\end{align}
for the family of all subspaces and all closed subspaces (the \emph{Grassmannian} of $\Hil$), respectively. Two subspaces $H,K\in\Sub(\Hil)$ will be called \emph{unitarily equivalent}, written $H\cong K$, if there exists $U\in\U(\Hil)$ such that $H=UK$.\glsadd{equivalence}

As shorthand notation, we shall write\glsadd{vee}
\begin{align}
    H+K
    &:=
    \spn\{H,K\}\in\Sub(\Hil),
    \\
    H\vee K
    &:=
    \overline{\spn\{H,K\}}\in\Gr(\Hil),
\end{align}
 for the algebraic and closed linear span of any two nonempty subsets $H,K\subset\Hil$, respectively. We emphasize that even for closed spaces $H,K\in\Gr(\Hil)$, we typically have\footnote{$H+K$ is closed if and only if $\sup\{|\langle h,k\rangle|\col h\in H_0,k\in K_0,\|h\|,\|k\|\leq1\}<1$, where $H_0=H\cap(H\cap K)^\perp$ and $K_0=K\cap(H\cap K)^\perp$.} $H+K\subsetneq H\vee K$. For finite-dimensional $K$ and closed $H$, the algebraic linear span $H+K\in\Gr(\Hil)$ is however always closed.

For any subset $\V\subset\Hil$, the \emph{orthogonal complement} \index{complement!orthogonal}
\begin{align}
    \V^\perp:=\{\xi\in\Hil\col \langle\xi,v\rangle=0\,\forall v\in \V\}
\end{align}
lies in $\Gr(\Hil)$, and
\begin{align}
    (\V^\perp)^\perp=\overline{\spn(\V)}.
\end{align}

A basic theorem about Hilbert spaces says that the Grassmannian $\Gr(\Hil)$ is a complete bounded orthocomplemented orthomodular lattice:

 \begin{theorem}[The lattice $\boldsymbol{\Gr(\Hil)}$ of closed subspaces]\label{theorem:Grassmannian}\booktheoremindex{theorem:Grassmannian}{Lattice of closed subspaces}
    Let $\Hil$ be a Hilbert space.
    \begin{enumerate}
        \item\label{item:GrassmannianComplete} $\Gr(\Hil)$ is a complete bounded lattice, namely it is partially ordered by inclusion, has the minimal element $\{0\}$, the maximal element $\Hil$, and for any family\footnote{Unless specified otherwise, a family is indexed by an index set $\CJ$ of arbitrary cardinality.} $(H_j)_{j\in\CJ}\subset\Gr(\Hil)$,
        \begin{align}
            \bigwedge_{j\in\CJ} H_j := \bigcap_{j\in\CJ} H_j,
            \qquad
            \bigvee_{j\in\CJ} H_j := \overline{\spn\{H_j\col j\in\CJ\}}
        \end{align}
        are the infimum and supremum of $(H_j)_{j\in\CJ}$ w.r.t. the partial order by inclusion, and lie in~$\Gr(\Hil)$.

        \item\label{item:OrthogonalComplementation} The \emph{orthogonal complementation}
        \begin{align*}
            \Gr(\Hil)\to\Gr(\Hil),\qquad
            H\mapsto H^\perp,
        \end{align*}
        is an order-reversing involution that is an orthocomplement, namely
        \begin{align}
            H\cap H^\perp=\{0\},\qquad
            H+H^\perp=\Hil,\qquad
            H\in\Gr(\Hil).
        \end{align}

        \item\label{item:GrassmannianOrthomodular} The orthogonal complement is {\em orthomodular}, namely for $K,H\in\Gr(\Hil)$,
        \begin{align}\label{eq:Orthomodularity}
            K\subset H\Rightarrow H=K\vee(K^\perp\cap H).
        \end{align}

        \item\label{item:GrassmannianDeMorgan} De Morgan's Laws hold, namely for any family $(H_j)_{j\in\CJ}\subset\Gr(\Hil)$,
        \begin{align}
            \left(\bigwedge_{j\in\CJ} H_j\right)^\perp = \bigvee_{j\in\CJ} H_j^\perp,
            \qquad
            \left(\bigvee_{j\in\CJ} H_j\right)^\perp = \bigwedge_{j\in\CJ} H_j^\perp.
        \end{align}

        \item\label{item:GrassmannianUnitaryAction} The unitary group $\U(\Hil)$ acts on $\Gr(\Hil)$ by automorphisms. In particular, $UH\in\Gr(\Hil)$ for $U\in\U(\Hil)$ and $H\in\Gr(\Hil)$, and $(UH)^\perp=U(H^\perp)$.
    \end{enumerate}
\end{theorem}

Just as $\Gr(\Hil)$, also the orthogonal projections $\CP(\Hil)$ form a complete bounded orthocomplemented orthomodular lattice: The partial order is operator order, the minimal and maximal projections are $0$ and $1$, and the orthogonal complement is
\begin{align}\label{eq:Eperp}
    E^\perp
    =
    1-E,\qquad
    E\in\CP(\Hil).
\end{align}
 The unitary group acts on $\CP(\Hil)$ according to $E\mapsto UEU^*$, $E\in\CP(\Hil)$, $U\in\U(\Hil)$.

These two lattices are isomorphic:

 \begin{theorem}[Closed subspaces and orthogonal projections]\label{theorem:ClosedSpacesAndOrthogonalProjections}\booktheoremindex{theorem:ClosedSpacesAndOrthogonalProjections}{Closed subspaces and orthogonal projections}
    Let $\Hil$ be a Hilbert space. The map
    \begin{align}
        \CP(\Hil)\to\Gr(\Hil),\qquad E\mapsto \Ran(E):=E(\Hil)
    \end{align}
    is a $\U(\Hil)$-equivariant isomorphism of complete bounded orthocomplemented orthomodular lattices, with inverse\glsadd{EH}
    \begin{align}\label{eq:EH}
        \Gr(\Hil)&\to\CP(\Hil),\qquad H\mapsto E_H,\\
        E_H(h+h^\perp)&=h,\qquad h\in H,\;h^\perp\in H^\perp.
    \end{align}
\end{theorem}

All of these facts are well known, so we do not give a proof here. We will however also frequently work with projections that are more general than the bounded everywhere defined orthogonal projections $E_H$, $H\in\Gr(\Hil)$ \eqref{eq:EH}. These more general projections will be introduced now.

The initial definitions are of purely algebraic nature and work in any vector space. Nonetheless, we choose a formulation suitable for Hilbert space operators from the beginning.

\begin{definition}[Projections]\label{def:Projection}\index{projection!general}
    Let $V$ be a vector space. A \emph{projection} on $V$ is a linear map $P:\dom P\to V$ that is defined on a subspace $\dom P\subset V$ and satisfies
    \begin{align}
        \Ran P:=P(\dom P)\subset\dom P,
        \qquad
        P^2\xi=P\xi,\quad \xi\in\dom P.
    \end{align}
\end{definition}

The following terminology is equivalent to linear independence in the current context, but will be helpful later in order to avoid confusion regarding real and complex linear independence.

 \begin{definition}[Independent subspaces]\label{def:IndependentSubspaces}
    Two subspaces $H,K$ of a vector space $V$ are called \emph{independent}\index{subspace!independent} if $H\cap K=\{0\}$.
\end{definition}

\begin{lemma}[Projections and pairs of independent subspaces]\label{lemma:ProjectionsAndPairsOfIndependentSpaces}
    Let $V$ be a vector space.
    \begin{enumerate}
        \item\label{item:PHK} Let $H,K\subset V$ be independent subspaces. Then
        \begin{equation}\label{eq:def2subspaceprojection}\glsadd{PHK}
            P_{H,K}:H+K\to V, \qquad h+k\mapsto h,
        \end{equation}
        is a projection with $\Ran P=H$, $\ker P=K$, $\dom P=H+K$.

        \item\label{item:PGL} Let $A:V\to V$ be an invertible linear map, and $P_{H,K}$ the projection defined in \ref{item:PHK}. Then also
        \begin{align}
            AP_{H,K}A^{-1}
            =
            P_{AH,AK},
            \qquad
            \dom(AP_{H,K}A^{-1})
            =
            AH+AK,
        \end{align}
        is a projection.

        \item\label{item:PHKplusPKH} Let $H,K\subset V$ be independent subspaces. Then
        \begin{align}
            P_{H,K}+P_{K,H}={\id_V}|_{H+K}.
        \end{align}

        \item\label{item:AllProjections} Every projection $P$ on $V$ is of the form $P=P_{H,K}$ described in \ref{item:PHK}, with $H=\Ran P$ and $K=\ker P$.
    \end{enumerate}
\end{lemma}
\begin{proof}
    \ref{item:PHK}, \ref{item:PGL}, \ref{item:PHKplusPKH} follow immediately from the definition.

    \ref{item:AllProjections} Let $P:\dom P\to V$ be a projection. Writing $v\in\dom P$ as $v=(Pv)+(1-P)v$, we see $\ker P=\Ran(1-P)$, $\dom P=\Ran P+\ker P$, and the independence of these spaces follows because
    $v=Pw\in \Ran P\cap \ker P$ satisfies $0=Pv=P^2w=Pw=v$. Given $h=Pv\in\Ran P$ and $k=(1-P)w\in\ker P$, we have $P(h+k)=h$. This shows $P=P_{\Ran P,\ker P}$.
\end{proof}

In other words, the family of ordered pairs of independent subspaces is in natural and $\GL(V)$-equivariant bijection with the family of projections on $V$. In particular, $P_{H,K}=P_{\tilde H,\tilde K}$ is equivalent to $H=\tilde H$, $K=\tilde K$.

\medskip

We now consider the case that $V=\Hil$ is a Hilbert space. Here the simplest pairs of independent subspaces $H,K$ are orthogonal ones, i.e. $H\subset K^\perp$, but also more general configurations are of interest. The functional analytic properties of projections are studied in the following proposition.

\begin{proposition}[Properties of the projections $\boldsymbol{P_{H,K}}$]\label{proposition:twosubspacesprojection}
    Let $\Hil$ be a Hilbert space with two independent subspaces $H,K\subset \Hil$.
    \begin{enumerate}
        \item\label{item:2spacesprojectionclosable} $P_{H,K}$ is closable if and only if $\overline{H}\cap \overline{K}=\{0\}$. In this case, its closure is
        \begin{align}
            \overline{P_{H,K}}=P_{\overline{H},\overline{K}}.
        \end{align}

 \item\label{item:core2spaceprojection} A linear space $D\subset H+K$ such that $P_{H,K}D\subset D$ is a core for $P_{H,K}$ if and only if $\overline{D\cap H}=\overline H$ and $\overline{D\cap K}=\overline K$. (Here ``core'' means that the graph of the restriction to $D$ is dense in the graph of $P_{H,K}$, even if $P_{H,K}$ is not closable.)

        \item\label{item:2spacesprojectionclosed} $P_{H,K}$ is closed if and only if $H$ and $K$ are closed.

        \item\label{item:2spacesprojectionbounded} $P_{H,K}$ is bounded if and only if $\overline H\cap \overline K=\{0\}$ and $\overline H+\overline K$ is closed.

        \item\label{item:adj2spaceprojection} If $\overline{H+K}=\Hil$, the adjoint of $P_{H,K}$ is well defined, and given by
        \begin{align}
            P_{H,K}^*=P_{K^\perp,H^\perp}.
        \end{align}
    \end{enumerate}
\end{proposition}
\begin{proof}
    \ref{item:2spacesprojectionclosable} We claim that the closure $\overline{\Gamma(P_{H,K})}$ of the graph of $P_{H,K}$ is
    \begin{align*}
        \overline{\{(h+k,h)\col h\in H,k\in K\}}
        =
        \{(\overline h+\overline k,\overline h)\col \overline h\in \overline H,\overline k\in \overline K\}.
    \end{align*}
    The inclusion $\supset$ is clear. For the reverse inclusion, let $(k_n+h_n,h_n)\to(\xi,\eta)\in\overline{\Gamma(P_{H,K})}$, with two sequences $(h_n)\subset H$, $(k_n)\subset K$. This implies that $(h_n)$ and $(k_n)$ converge individually, hence $\xi\in\overline{H}+\overline{K}$, $\eta\in\overline{H}$.

    The map $P_{H,K}$ is closable if and only if $\overline{\Gamma(P_{H,K})}$ is the graph of a linear operator, which is the case if and only if $\overline H\cap\overline K=\{0\}$. In this case, $\overline{\Gamma(P_{H,K})}=\Gamma(P_{\overline H,\overline K})$, hence $\overline{P_{H,K}}=P_{\overline{H},\overline{K}}$.

 \ref{item:core2spaceprojection}
    A subspace $D\subset H+K$ is a core if and only if for every $h+k\in\dom P_{H,K}$, there exists a sequence $d_n=h_n+k_n\in D$ with $h_n\in H$, $k_n\in K$, such that $d_n\to h+k$ and $P_{H,K}d_n=h_n\to P_{H,K}(h+k)=h$.
    Now, $P_{H,K}D\subset D$ yields $h_n\in H\cap D$ and $k_n\in K\cap D$. Therefore, $D$ being a core becomes equivalent to $\overline{D\cap H}=\overline H$ and $\overline{D\cap K}=\overline K$.

    \ref{item:2spacesprojectionclosed} As $P_{H,K}$ is closed if and only if it is closable and $P_{H,K}=\overline{P_{H,K}}$, the claim follows immediately from~\ref{item:2spacesprojectionclosable}.

    \ref{item:2spacesprojectionbounded} Suppose $\overline H\cap\overline K=\{0\}$ and $\overline H+\overline K$ is closed. The first assumption implies that $P_{H,K}$ is closable by \ref{item:2spacesprojectionclosable}, and the second assumption implies that the domain $\overline H+\overline K$ of $\overline{P_{H,K}}$ is closed. Hence $\overline{P_{H,K}}$ is bounded by the closed graph theorem, so $P_{H,K}$ is bounded as well.

 Suppose now conversely that $P_{H,K}$ is bounded. Then it is closable with closure $\overline{P_{H,K}}$ equal to the continuous extension of $P_{H,K}$ to $\overline{H+K}$. In particular, $\overline H\cap\overline K=\{0\}$ by \ref{item:2spacesprojectionclosable}. Consider $\eta\in\overline{H+K}$, i.e. $\eta=\lim_n(h_n+k_n)$ with sequences $(k_n)\subset K$, $(h_{n})\subset H$. By continuity, the two limits $\lim_n h_n=\overline{P_{H,K}}\eta$ and $\lim_nk_n=(1-\overline{P_{H,K}})\eta$ exist, and $\eta\in\overline H+\overline K$ follows. This shows $\overline{H+K}=\overline H+\overline K$ (the inclusion $\supset$ is obvious), so $\overline H+\overline K$ is closed.

    \ref{item:adj2spaceprojection}
 The projection $P_{H,K}$ being densely defined is equivalent to the adjoint existing. Let $\xi=h+k\in \dom(P_{H,K})$, where $h\in H$ and $k\in K$, and $\psi\in \dom(P_{H,K}^\ast)$. Then,
 $$\langle h,\psi\rangle=\langle P_{H,K}\xi,\psi\rangle=\langle h,P_{H,K}^\ast\psi\rangle+\langle k,P_{H,K}^\ast\psi\rangle$$
    requires that (i) $\langle k,P_{H,K}^\ast\psi\rangle$ is independent of $k\in K$, which is equivalent to $P_{H,K}^\ast\psi\in K^{\perp}$; and (ii) $\langle h,\psi-P_{H,K}^\ast\psi\rangle=0$ for all $h\in H$, which is equivalent to $\psi-P_{H,K}^\ast\psi\in H^{\perp}$. Hence, $\psi\in K^{\perp}+H^{\perp}=\dom(P_{H,K}^\ast)$ and $P_{H,K}^\ast=P_{K^{\perp},H^{\perp}}$.
      
 \end{proof}

 As a very special case, we note that $P_{H,K}$ is densely defined and self-adjoint if and only if $K=H^\perp$ and $H=K^\perp$. In this case, $\dom P_{K,H}=\Hil$ and $P_{H,K}=E_H$, i.e. we recover the bijection between orthogonal projections and closed subspaces of $\Hil$ (Theorem~\ref{theorem:ClosedSpacesAndOrthogonalProjections}).

\bigskip

 Instead of the projection $E_H$, a closed subspace $H\in\Gr(\Hil)$ can equivalently be described as the fixed-point space of the orthogonal reflection about it, namely the unitary involution
 \begin{align}
    H+H^\perp \ni h+h^\perp\mapsto h-h^\perp \in H+H^\perp.
\end{align}

As involutions will play an important role in modular theory, we now introduce the corresponding general concept.

\begin{definition}[Involutions]\label{def:Involutions}\index{involution (Def.~\ref{def:Involutions})}
    Let $V$ be a vector space. An \emph{involution} on $V$ is a linear map $S:\dom S\to V$ that is defined on a subspace $\dom S\subset V$ and satisfies
    \begin{align}
        \Ran S:=S(\dom S)\subset\dom S,
        \qquad
        S^2\xi=\xi,\quad \xi\in\dom S.
    \end{align}
\end{definition}

As for projections, one sees that any involution defines a pair of independent subspaces, namely its eigenspaces (here and in the following, we write \glsadd{FixA}$\Fix(A):=\ker(A-1)$ for the fixed points of a linear map~$A$)
\begin{align}
    \Fix(S)&=\{\xi\in\dom S\col S\xi=\xi\},
    \\
    \Fix(-S)&=\{\xi\in\dom S\col S\xi=-\xi\},
    \\
    \dom(S)
    &=
    \Fix(S)+\Fix(-S),\qquad S(h_++h_-)=h_+-h_-,\qquad h_\pm\in\Fix(\pm S),
\end{align}
and any ordered pair $(H,K)$ of independent spaces defines an involution
\begin{align}\label{eq:SHK}\glsadd{SHK}
    S_{H,K}:H+K\to H+K,\qquad h+k\mapsto h-k.
\end{align}

\begin{lemma}[Involutions and projections]\label{lemma:InvolutionsAndProjections}
    \leavevmode
    \begin{enumerate}
        \item\label{item:PvsS} The projections and involutions on a vector space are in bijection via
        \begin{align}
            P_{H,K}\mapsto S_{H,K}
            :=
            2P_{H,K}-1
            \iff
            P_{H,K}
            =
            \frac12(S_{H,K}+1).
        \end{align}
        \item\label{item:PvsSFA} On a Hilbert space, $S_{H,K}$ is closable (closed, bounded, densely defined) if and only if $P_{H,K}$ is closable (closed, bounded, densely defined). If $H+K$ is dense, we have
        \begin{align}
            S_{H,K}^*=S_{K^\perp,H^\perp}.
        \end{align}
    \end{enumerate}
\end{lemma}
\begin{proof}
    The bijection in \ref{item:PvsS} is clear from the definition, and the claims in \ref{item:PvsSFA} are immediate consequences of the relation $S_{H,K}=2P_{H,K}-1$ and Proposition~\ref{proposition:twosubspacesprojection}.
\end{proof}

\subsection{Real linear subspaces and operators in complex Hilbert spaces}\label{section:RealVsComplex}

Whereas so far the Hilbert space $\Hil$ could be real or complex, we now specialize to \emph{complex}~$\Hil$, and consider real subspaces and operators in it. We follow the convention that the scalar product $\scpr$ is linear in its second argument.

 \begin{lemma}[The real form of a complex Hilbert space]\label{lemma:RealFormOfComplexHilbertSpace}
 Let $\Hil$ be a complex Hilbert space. Considered as a real vector space $\Hil_\Rl$ (i.e. by forgetting the complex structure), and equipped with the real scalar product\glsadd{scprR}\glsadd{Real}
 \begin{align}\label{rw:RealScalarProduct}
        \scpr_\Rl
        :=
        \Re\scpr,
    \end{align}
    $(\Hil_\Rl,\scpr_\Rl)$ is a real Hilbert space with the same norm as $\Hil$.
\end{lemma}
\begin{proof}
    As $\langle\xi,\xi\rangle=\Re\langle\xi,\xi\rangle$ for any $\xi\in\Hil$, the inner products $\scpr_\Rl$ and $\scpr$ define the same norm, $\scpr_\Rl$ is positive definite, and $(\Hil_\Rl,\scpr_\Rl)$ is complete.
\end{proof}

 A \emph{real-linear subspace} (or \emph{real subspace}, for short) is a subspace of $\Hil_\Rl$, i.e. a non-empty subset of $\Hil$ that is closed under real-linear, but not necessarily under complex linear combinations. We write\index{Grassmannian!real}\glsadd{GrR}\glsadd{SubR}
\begin{align}\label{eq:SubGrR}
    \Sub_\Rl(\Hil)
    &:=
    \{H\subset\Hil\col H \text{ is a real-linear subspace} \}\supset\Sub(\Hil),
    \\
    \Gr_\Rl(\Hil)
    &:=
    \{H\subset\Hil\col H=\overline H \text{ is a closed real-linear subspace} \}\supset\Gr(\Hil),
\end{align}
 for the family of all real subspaces and all closed real subspaces (the \emph{real Grassmannian} of $\Hil$), respectively. For example, $\Gr(\Cl)=\{\{0\},\Cl\}$ and $\Gr_\Rl(\Cl)=\Gr(\Cl)\cup\{e^{i\alpha}\Rl\col\alpha\in[0,\pi)\}$.

As the inner products $\scpr_\Rl$ and $\scpr$ define the same norm, our symbol $\overline H$ for closures is unambiguous. All results of the previous section apply to $\Sub(\Hil_\Rl)=\Sub_\Rl(\Hil)$ and $\Gr(\Hil_\Rl)=\Gr_\Rl(\Hil)$. We will however need to distinguish between orthogonal complements w.r.t. $\scpr$ and $\scpr_\Rl$, which will be written as\glsadd{Hrperp}
\begin{align}
    \V^{\perp}
    &=
    \{\xi\in\Hil\col \langle\xi,v\rangle=0\,\forall v\in \V\},
    \\
    \V^{\rperp}
    &=
    \{\xi\in\Hil\col \Re\langle\xi,v\rangle=0\,\forall v\in \V\}.
\end{align}
In the same way, we will need to distinguish the complex span $\spn(\V)$ of a subset $\V\subset\Hil$ from its real span $\rlspn(\V)$.

 \begin{definition}[Real linear and antilinear maps]\label{def:RealLinearAndAntilinearMaps}
    Let $\Hil$ be a complex Hilbert space.
    \begin{enumerate}
 \item\label{item:RealLinearMap} A \emph{real-linear map} is a map $A:\dom A\to V$ into a real vector space $V$, defined on a real-linear subspace $\dom A\in\Sub_\Rl(\Hil)$ (which could be $\dom A=\Hil$), and respecting real linear combinations.
        \item\label{item:AntilinearMap} If $V$ is a complex vector space, a real-linear map $A:\dom A\to V$ is called \emph{antilinear} (or \emph{conjugate-linear}) if
 \begin{align}
            i\dom A=\dom A,\qquad iA\xi=-Ai\xi,\qquad \xi\in\dom A.
        \end{align}
    \end{enumerate}
\end{definition}

 We write $B_\Rl(\Hil)$\glsadd{BRH} for the real $C^*$-algebra of all bounded real-linear operators $A:\dom A=\Hil\to\Hil$. The ${}^*$-operation in $B_\Rl(\Hil)$ is the one induced by the real scalar product, namely
 \begin{align}\label{eq:RealAdjoint}
    \Re\langle\xi,A\eta\rangle
    =
    \Re\langle A^*\xi,\eta\rangle,
    \qquad
    A\in B_\Rl(\Hil),\quad
    \xi,\eta\in\Hil.
\end{align}
 In case $A$ is real-linear but only densely defined (not necessarily bounded), $A^*$ is defined on $\dom A^*=\{\xi\in\Hil\col\exists c>0\col|\Re\langle\xi,A\eta\rangle|\leq c\|\eta\|\;\forall\eta\in\dom A\}$ by the same formula \eqref{eq:RealAdjoint}. We emphasize that for closed densely defined real-linear operators, the polar decomposition theorem holds, viewing them as closed operators on the real Hilbert space $\Hil_\Rl$.

It is easy to check that in case $A$ is even linear (that is, \emph{complex} linear), its real adjoint $A^*$ coincides with the adjoint defined w.r.t. $\scpr$, and that in case $A$ is antilinear, its adjoint satisfies
\begin{align}
    \langle\xi,A\eta\rangle
    =
    \overline{\langle A^*\xi,\eta\rangle},
    \qquad
    A\;\text{antilinear},\quad
    \xi\in\dom(A^*),\;\eta\in\dom(A).
\end{align}
We may therefore denote all adjoints by the same symbol.

Some classes of operators that emerge from this and will be used frequently are the following.

 \begin{definition}[Real orthogonal projections]\label{def:RealOrthogonalProjection}
    Let $\Hil$ be a complex Hilbert space. A \emph{real-linear projection} is a real-linear operator that is a projection in the sense of Def.~\ref{def:Projection}. A \emph{real orthogonal projection} is a real-linear bounded operator $E:\Hil\to\Hil$ such that $E^2=E=E^*$. The set of all real orthogonal projections on $\Hil$ is denoted $\CP_\Rl(\Hil)$.\glsadd{PRH}\index{projection!real orthogonal (Def.~\ref{def:RealOrthogonalProjection})}
\end{definition}

Note that since $\CP_\Rl(\Hil)=\CP(\Hil_\Rl)$, the real orthogonal projections form a complete bounded orthocomplemented orthomodular lattice that contains $\CP(\Hil)$ and is isomorphic to $\Gr_\Rl(\Hil)$.

\begin{definition}[Antiunitary operators and conjugations]\label{def:AntiunitariesAndConjugations}
    Let $\Hil$ be a complex Hilbert space.
    \begin{enumerate}
 \item\label{item:AntiunitaryMap} An \emph{antiunitary} map is a bounded antilinear map $V:\Hil\to\Hil$ with $V^*V=1=VV^*$.\index{antiunitary (Def.~\ref{def:AntiunitariesAndConjugations})} The \emph{(anti-)unitary group} is\glsadd{AUH}
        \begin{align}\label{eq:AUH}
            \AU(\Hil)
            :=
            \{V:\Hil\to\Hil\col\text{unitary or antiunitary}\}
            .
        \end{align}

 \item\label{item:Conjugation} A \emph{conjugation} is an antiunitary involution, i.e. an antilinear operator $\Gamma:\Hil\to\Hil$ satisfying $\Gamma=\Gamma^*=\Gamma^{-1}$. The set of all conjugations on $\Hil$ is denoted $\Con(\Hil)$\glsadd{ConH}\index{conjugation (Def.~\ref{def:AntiunitariesAndConjugations})}.
    \end{enumerate}
\end{definition}

\begin{remark}[Antiunitary equivalence and conjugations]\label{remark:AntiunitariesAndConjugations}
    \leavevmode
    \begin{enumerate}
 \item\label{item:AntiunitaryGroup} The (anti-)unitary group is contained in the real unitary group (the orthogonal group), $\AU(\Hil)\subset\U(\Hil_\Rl)$, and therefore acts on $\CP_\Rl(\Hil)$ and $\Gr_\Rl(\Hil)$ as described before.

        We will see later that in case $K,H\in\Gr_\Rl(\Hil)$ are \emph{antiunitarily equivalent}, i.e. $K=VH$ with an {\em anti}unitary~$V$, then there also exists a unitary $U$ such that $K=UH$, i.e. $H\cong K$ (Corollary~\ref{corollary:AntiunitaryAndUnitaryImages}). We therefore do not need to distinguish between unitary and antiunitary equivalence in our terminology.

        \item\label{item:RealAndImaginaryPartofConjugation} The \emph{real and imaginary part projections} of a conjugation $\Gamma\in\Con(\Hil)$ are
        \begin{align}
            \Gamma_\pm
            :=
            \tfrac12(1\pm\Gamma).
        \end{align}
        It is easy to check that $\Gamma_\pm\in\CP_\Rl(\Hil)$ are real orthogonal projections satisfying
        \begin{align}\label{eq:GammaElementaryProperties}
            \Gamma_\pm^\perp=\Gamma_\mp,\qquad
            \Gamma_\pm\Gamma_\mp=0,\qquad
            \Gamma\Gamma_\pm=\pm\Gamma_\pm,\qquad
            i\Gamma_\pm=\Gamma_\mp i.
        \end{align}
 In particular, the fixed-point space
 \begin{align}
            \Fix\Gamma
            :=
            \{\xi\in\Hil\col\xi=\Gamma\xi\}
            =
            \Gamma_+(\Hil)
        \end{align}
        lies in $\Gr_\Rl(\Hil)$ and satisfies $\Fix\Gamma\cap i\Fix\Gamma=\{0\}$ and $\Fix\Gamma+i\Fix\Gamma=\Hil$.

        The modular theory of standard subspaces presents a far-reaching generalization of this simple situation.
    \end{enumerate}
\end{remark}

We refer to Section~\ref{section:ComplexStructures} for a description of complex structures on real Hilbert spaces.

\subsection{Cyclic and separating spaces. The symplectic complement.}\label{section:CyclicSeparating}

We now introduce some concepts that will lead us to standard subspaces. From now on,~$\Hil$ always denotes an arbitrary complex Hilbert space.

 Every closed real subspace $H\in\Sub_\Rl(\Hil)$ defines two closed {\em complex} linear subspaces, namely
 \begin{align}\label{eq:TwoComplexSpaces}
    H_\Cl:=H\cap iH,\qquad
    H^\Cl:=H\vee iH=\overline{H+iH}.
\end{align}
Note that $H_\Cl$\glsadd{HlowC} is the largest complex subspace contained in $H$, and $H^\Cl$\glsadd{HupC} is the smallest closed complex subspace containing $H$. For typographical reasons, we will sometimes also use the alternative notation $\Cl[H]:=H+iH$\glsadd{CH}. The two spaces $H_\Cl$ and $H^\Cl$ motivate the following definition.

\begin{definition}[Cyclic and separating spaces]\label{definition:CyclicSeparating}
   Let $H\in\Sub_\Rl(\Hil)$ be a real subspace.
    \begin{enumerate}
 \item\label{item:SeparatingSubspace} $H$ is called {\em separating\index{subspace!separating (Def.~\ref{definition:CyclicSeparating})}} if $H\cap iH=\{0\}$.
 \item\label{item:CyclicSubspace} $H$ is called {\em cyclic\index{subspace!cyclic (Def.~\ref{definition:CyclicSeparating})}} if $H^\Cl=\Hil$, i.e. if its complex span $H+iH$ is dense in $\Hil$.
    \end{enumerate}
\end{definition}

The terminology introduced here has its origin in the context of cyclic and separating vectors for von Neumann algebras. This connection will become visible in Chapter 2, for now we accept these terms simply as definitions.

It is clear that the properties of being separating or cyclic are preserved under the action of any (anti)unitary operator $V\in\AU(\Hil)$.

A space $H\in\Real(\Hil)$ is cyclic if and only if its complex orthogonal complement is trivial, \glsadd{Hperp}$H^\perp=\{0\}$. However, the real orthogonal complement $H^\rperp$ of a cyclic space $H$ can be non-trivial. The same is true for the symplectic complement, i.e. the ``complement'' w.r.t. the symplectic form $\Im\scpr$.

 \begin{definition}[Symplectic complement]\label{definition:SymplecticComplement}\glsadd{Imaginary}
    Let $\V\subset\Hil$ be a subset. The {\em symplectic complement of $\V$}\index{complement!symplectic (Def.~\ref{definition:SymplecticComplement})} is \glsadd{Hprime}
    \begin{align}
        \V':=i\V^\rperp = \{\xi\in\Hil\col\Im\langle\xi,v\rangle=0\,\forall v\in \V\}.
    \end{align}
\end{definition}

Clearly, for any subset $\V\subset\Hil$, the symplectic complement $\V'\in\Real(\Hil)$ is a closed real subspace. Some further straightforward properties of the symplectic complement are listed in the following lemma.

\begin{lemma}[Basic properties of the symplectic complement]\label{lemma:CyclicSeparating}
    Let $H,K\in\Sub_\Rl(\Hil)$ be real subspaces.
    \begin{enumerate}
        \item\label{item:CyclicVsSeparating} $H$ is cyclic if and only if $H'$ is separating.
        \item\label{item:InclusionsAndCyclicSeparating} Let $K\subset H$. If $K$ is cyclic, then $H$ is cyclic. If $H$ is separating, then $K$ is separating.
        \item\label{item:DoubleComplement} $H''=\overline H$.
        \item\label{item:SymplecticComplementInclusions} $H\subset K\Rightarrow K'\subset H'$.
 \item\label{item:SymplecticComplementAndLatticeStructures} $(H+K)'=H'\cap K'$ and $(\overline H\cap\overline K)'=\overline{H'+K'}$.
 \item\label{item:AUandComplements} Let $V\in\AU(\Hil)$. Then $(VH)'=V(H')=:VH'$.
    \end{enumerate}
\end{lemma}

Note that combining \ref{item:CyclicVsSeparating} and \ref{item:DoubleComplement} shows that $H'$ is cyclic if and only if $\overline H$ is separating. For general $H\in\Sub_\Rl(\Hil)$, it is not true that $H'$ cyclic is equivalent to $H$ separating.

\begin{proof}
    \ref{item:CyclicVsSeparating} From the definition of symplectic complement, we see
    \begin{align}
        H'\cap iH'=H'\cap H^\rperp=H^\perp=(H+iH)^\perp.
    \end{align}
 The space $H'\cap iH'$ on the left is $\{0\}$ if and only if $H'$ is separating, and the space $(H+iH)^\perp$ on the right is $\{0\}$ if and only if $H$ is cyclic. Part \ref{item:InclusionsAndCyclicSeparating} is immediate from Definition~\ref{definition:CyclicSeparating}.
    Parts~\ref{item:DoubleComplement}, \ref{item:SymplecticComplementInclusions}, and~\ref{item:AUandComplements} are direct consequences of $H'=iH^\rperp$ and Theorem~\ref{theorem:Grassmannian}.

    \ref{item:SymplecticComplementAndLatticeStructures} The equality $(H+K)'=H'\cap K'$ follows directly from the definition of symplectic complement. Applying it to $H'$ and $K'$ and using~\ref{item:DoubleComplement}, we obtain
    \begin{align*}
    	\overline H\cap\overline K=H''\cap K''=  (\overline{H'+K'})'.
    \end{align*}
 \end{proof}

 This lemma shows in particular that $H\mapsto H'$, just as $H\mapsto H^\rperp$, is an order-reversing involution on the lattice $\Real(\Hil)$, where order means order by inclusion. However, whereas $\rperp$ is a complementation (see Theorem~\ref{theorem:Grassmannian}~\ref{item:OrthogonalComplementation}), the symplectic complement does, despite its name, {\em not} have this property. This can be seen by trivial examples such as the real subspace $\Rl\subset\Cl$ which satisfies $\Rl'=\Rl$ and hence $\Rl'\cap\Rl\neq\{0\}$, $\Rl+\Rl'\neq\Cl$. Also the orthomodularity property $K\subset H\Rightarrow H=K\vee(K'\cap H)$ does {\em not} hold (cf. \eqref{eq:Orthomodularity}), as can be seen by choosing $K=\Rl$ and $H=\Hil=\Cl$. This is a first sign that the symplectic structure is often more involved than the real orthogonal structure.

It is also worth pointing out that the projection $E_{H'}$ onto the symplectic complement of $H\in\Real(\Hil)$ takes the form
\begin{align}
    E_{H'}=E_{iH^\rperp}=-iE_{H^\rperp}i=1+iE_H i,
\end{align}
which is slightly less obvious than its real orthogonal analogue $E_{H^\rperp}=1-E_H$ \eqref{eq:Eperp}. Nonetheless, the symplectic complement will be more natural and more important than the real orthogonal complement in the context of standard subspaces.

\section{Basic structure of standard subspaces}\label{section:StandardSubspacesBasics}

This short section provides a gentle introduction to our main subject, standard subspaces. In Section~\ref{section:DefinitionExamplesStandardSubspaces}, we start with the basic definition and some introductory, mostly finite-dimensional, examples. Afterwards we have a look at the simplest class of standard subspaces in Section~\ref{section:AbelianSpaces}, the maximal abelian standard subspaces. 

 \subsection{Definition and first examples of standard subspaces}\label{section:DefinitionExamplesStandardSubspaces}

 \begin{definition}[Standard subspace]\label{definition:standardsubspace}\index{standard subspace|see{subspace, standard}}
    A {\em standard subspace \index{subspace!standard (Def.~\ref{definition:standardsubspace})}} of $\Hil$ is a closed real subspace $H\in\Real(\Hil)$ that is cyclic and separating. The set of all standard subspaces of~$\Hil$ is denoted $\Std(\Hil)$\glsadd{StdH}.
\end{definition}

Trivial examples of standard subspaces are given by $\Rl^n\subset\Cl^n$ or the subspace of real-valued functions in a complex $L^2$-space $L^2(X,d\mu)$ over a measure space $(X,\mu)$. More interesting examples will follow below.

\begin{remark}[Standard subspaces and real structures]\label{remark:RealStructures}
 In case $H\in\Std(\Hil)$ is such that its complex span is closed, then it is nothing else than a \emph{real structure} on the complex space $\Hil$, namely a real subspace $H$ satisfying $H\cap iH=\{0\}$ and $H+iH=\Hil$. This is always the case for finite-dimensional~$\Hil$, but, as we shall see, represents a quite special situation for infinite-dimensional~$\Hil$.

    A discussion of standard subspaces and complexifications will be given in Section~\ref{section:ComplexStructures}.
\end{remark}

Standard subspaces occur naturally in the context of $\Real(\Hil)$: Any closed real subspace $H\in\Real(\Hil)$ canonically decomposes into a direct sum of a complex Hilbert space, a trivial space, and a standard subspace.

\begin{lemma}[Canonical decomposition of real closed subspaces]\label{lemma:DecomposeClosedRealSubspace}
    Let $H\in\Real(\Hil)$. Then there exists a canonical complex orthogonal decomposition  such that
    \begin{align}
        \Hil&=\Hil_1\oplus\Hil_2\oplus\Hil_3\\
        H&=\Hil_1 \oplus \{0\} \oplus H_{\mathrm{std}},\label{eq:Hdecomposition}
    \end{align}
 with $H_{\mathrm{std}}\in\Std(\Hil_3)$, namely $\Hil_1=H_\Cl$, $\Hil_2=(H^\Cl)^\perp$, $\Hil_3=H^\Cl\cap H_\Cl^\perp$, and $H_{\mathrm{std}}=H\cap\Hil_3$. If $H$ is cyclic, $\Hil_2=\{0\}$.
 \end{lemma}
\begin{proof}
    We recall the two closed complex subspaces $H_\Cl:=H\cap iH$ and $H^\Cl:=\overline{H+iH}$ \eqref{eq:TwoComplexSpaces}, and first split $\Hil$ as $\Hil=H^\Cl\oplus(H^\Cl)^\perp$. The real space $H$ is contained in $H^\Cl$, hence $H=H\oplus\{0\}$ in this decomposition. Furthermore, $H$ is cyclic in $H^\Cl$. We may therefore assume $\Hil=H^\Cl$ and $H$ cyclic, and now split the Hilbert space again, according to $\Hil=H_\Cl\oplus H_\Cl^\perp$. Let us denote the complex orthogonal projection onto $H_\Cl$ by $P$. Any $h\in H$ splits as $h=Ph+P^\perp h$, and as $H_\Cl\subset H$, both $Ph$ and $P^\perp h$ lie in $H$.

 The part $P^\perp H$ of $H$ in $H_\Cl^\perp$ is closed and $P^\perp(H+iH)$ is dense in $P^\perp\Hil$. As $P^\perp H\cap iP^\perp H=P^\perp H_\Cl=\{0\}$, we see that $P^\perp H$ is standard in $P^\perp\Hil$.
 \end{proof}

From the discussion in the previous section it is clear that the group $\AU(\Hil)$ \eqref{eq:AUH} acts on $\Std(\Hil)$, i.e. $VH$ is standard if $H\in\Std(\Hil)$ and $V\in\AU(\Hil)$.

Furthermore, Lemma~\ref{lemma:CyclicSeparating}~\ref{item:CyclicVsSeparating} implies directly
\begin{align}
    H\in\Std(\Hil)
    \Leftrightarrow
    H'\in\Std(\Hil)
    \Leftrightarrow
    H^\rperp\in\Std(\Hil),
    \qquad
    H\in\Real(\Hil).
\end{align}

At this stage, $H^\rperp$ and $H'$ appear as equally good ``mirror images'' of a standard subspace $H\in\Std(\Hil)$. Later, modular theory will show that $H'$ is more natural than~$H^\rperp$ (Section~\ref{section:ModularData}).

\medskip

We now give some elementary but instructive examples of standard subspaces that give an idea of later developments of the theory. For simplicity, these will be mostly finite-dimensional, although the structure becomes much richer in the infinite-dimen\-sional case. We begin with the following simple observation.

\begin{lemma}[Finite-dimensional standard subspaces]\label{lemma:CyclicEquivalentToSeparatingInFiniteDimension}
    Let $\Hil$ be a complex Hilbert space of finite (complex) dimension $d:=\dim\Hil$. Then any standard subspace $H\in\Std(\Hil)$ has real dimension $\dim_\Rl(H)=d$, and a real subspace $H\subset\Hil$ with $\dim_\Rl(H)=d$ is cyclic if and only if it is separating.
\end{lemma}
\begin{proof}
    Let $H\in\Std(\Hil)$. As $H$ and $iH$ intersect trivially, have the same real dimension~$d_H$, and span $\Hil$, we have $2d_H=\dim_\Rl(H+iH)=\dim_\Rl(\Hil)=2d$, namely $d_H=d$.

    If $H$ is a real subspace of dimension $\dim_\Rl(H)=d$, the dimension formula
    \begin{align}
        \dim_\Rl(H\cap iH)=\dim_\Rl(H)+\dim_\Rl(iH)-\dim_\Rl(H+iH)=2d-\dim_\Rl(H+iH)
    \end{align}
    shows that $H$ being separating, namely $\dim_\Rl(H\cap iH)=0$, is equivalent to $H$ being cyclic, namely $\dim_\Rl(H+iH)=2d$.
\end{proof}

In a one-dimensional Hilbert space $\Hil\cong\Cl$, every real one-dimensional subspace $H\cong e^{i\alpha}\Rl$, $\alpha\in\Rl$, is standard. Thus, up to unitary equivalence, there is only the example $\Rl\subset\Cl$. Already in two dimensions, the situation is more interesting: Up to unitary equivalence, there is a one-parameter family of standard subspaces.

\begin{example}[Two-dimensional standard subspaces]\label{example:2dStandardSubspaces}
    {\em For every two-dimensional standard subspace $H$, there exists a unique $0<\lambda\leq1$ such that $H$ is unitarily equivalent to the standard subspace
    \begin{align}\label{eq:Hlambda}
        H_\lambda
        :=
        \left\{
            \begin{pmatrix}
                \lambda^{1/2}z\\\overline{z}
            \end{pmatrix}
            \col z\in\Cl
        \right\}
        \in\Std(\Cl^2)
        .
    \end{align}
    }
\end{example}
\begin{proof}
    Let $H$ be a two-dimensional real subspace. We pick an orthonormal basis $\{h_1,h_2\}$ of $H$ w.r.t. $\Re\scpr$, and consider
    \begin{align}\label{eq:h1h2product}
        \alpha_H:=|\langle h_1,h_2\rangle|.
    \end{align}
 The parameter $\alpha_H$ does not depend on the choice of orthonormal basis because every orthonormal basis of $H$ arises from $\{h_1,h_2\}$ by a rotation or reflection\footnote{By a small calculation, one can check $|E_HiE_H|_H=\alpha_H\cdot 1|_H$, which makes the basis independence of $\alpha_H$ manifest.} in $\rlspn\{h_1,h_2\}\cong\Rl^2$. Furthermore, $\alpha_{VH}=\alpha_H$ for any $V\in\AU(\Hil)$, and $\alpha_H\in[0,1]$ by real orthonormality of $\{h_1,h_2\}$.

    We claim that $H$ is standard if and only if $\alpha_H<1$. To prove this, we first note that for a real two-dimensional subspace $H$ of a two-dimensional Hilbert space~$\Hil$, the complex subspace $H\cap iH$ has either real dimension $0$ (when $H$ is separating, hence standard by Lemma~\ref{lemma:CyclicEquivalentToSeparatingInFiniteDimension}) or real dimension $2$, i.e. $H=iH$. So $H$ not being standard is equivalent to $H=iH$, which is equivalent to $h_1=zh_2$ for some $z\in\Cl$, $|z|=1$. By the extremal case of the Cauchy-Schwarz Inequality, this is in turn equivalent to $\alpha_H=1$, as claimed.

    We next show that all values of $\alpha_H\in[0,1)$ are realized by standard subspaces. Fix $\Hil=\Cl^2$, a parameter $0<\lambda\leq1$, and consider the two unit vectors
    \begin{align}\label{eq:2dBasis}
        h_1
        &=
        \frac{1}{\sqrt{1+\lambda}}
        \begin{pmatrix}
         \lambda^{1/2}\\ 1
        \end{pmatrix}
        ,\qquad
        h_2
        =
        \frac{1}{\sqrt{1+\lambda}}
        \begin{pmatrix}
         -i\lambda^{1/2}\\ i
        \end{pmatrix},
    \end{align}
    which satisfy $\langle h_1,h_2\rangle=i\frac{1-\la}{1+\la}$, so are in particular real orthogonal. The parameter $\alpha_H=\frac{1-\la}{1+\la}$ varies over $[0,1)$ as $\la$ varies over $(0,1]$. The real span of $\{h_1,h_2\}$ is $H_\lambda$ \eqref{eq:Hlambda}, so we have shown that $H_\lambda$ is standard and that each two-dimensional standard subspace is (anti-) unitarily equivalent to $H_\lambda$, and hence unitarily equivalent to $H_\la$ (Corollary~\ref{corollary:AntiunitaryAndUnitaryImages}), for a unique $\la\in(0,1]$.
\end{proof}

We will revisit this example in Remark~\ref{remark:2dExampleRevisited} and Example~\ref{example:ModularDataOf2dStandardSubspace} to add more information to it. Despite their simplicity, the one- and two-dimensional standard subspaces described here are the fundamental building blocks: Any standard subspace arises as a direct integral of them (Corollary~\ref{cor:multifree}).

 Furthermore, we observe the following. When we define the positive operator $\Delta$ (depending on $\lambda\in(0,1]$) and the conjugation (antiunitary involution) $J$,
\begin{align}\label{eq:JDelta2d}
    \Delta=
    \begin{pmatrix}
        \lambda^{-1} &0\\
        0&\lambda
    \end{pmatrix},
    \qquad
    J\begin{pmatrix}
        z_1\\z_2
    \end{pmatrix}
    :=
    \begin{pmatrix}
        \overline{z_2}\\
        \overline{z_1}
    \end{pmatrix},
\end{align}
 then the standard subspace $H_\lambda$ \eqref{eq:Hlambda} coincides with the fixed-point space of the antilinear involution $J\Delta^{1/2}$,
 \begin{align}
    H_\lambda=\Fix(J\Delta^{1/2}).
\end{align}
In Section~\ref{section:ModularData}, we will see that such a description exists canonically for every standard subspace.

 \begin{example}[Standard subspaces in matrix spaces]\label{example:MatrixStandardSubspace}
    {\em Let $\Hil=M_d$ be the Hilbert space of complex $(d\times d)$-matrices with the scalar product $\langle A,B\rangle=\Tr(A^*B)$. Fix an invertible matrix $\Om\in M_d$, and consider
    \begin{align}\label{eq:HMatrixCase}
        H
        :=
        \{A\Om\in M_d \col A=A^*\}.
    \end{align}
    Then $H$ is a standard subspace, and its symplectic complement is
    \begin{align}\label{eq:MatrixExampleSymplecticComplement}
        H'
        =
        \{\Om A\in M_d \col A=A^*\}.
    \end{align}
    }
\end{example}

 For clarity, we emphasize that for $d\geq2$, not all standard subspaces of $M_d$ are unitarily equivalent to one of this form.

\begin{proof}
 We first observe that the real subspace $(M_d)_\sa$ of all self-adjoint matrices\glsadd{Asa} is standard in~$M_d$. Indeed, it is cyclic because any $A\in M_d$ is a complex linear combination of self-adjoint matrices, and it is separating because $A=iB$ for $A=A^*$, $B=B^*$, implies $A=0$. As $\Om$ is invertible, also $H$ has real dimension $d^2$, and $M_d \Om=H+iH=\Hil$, so $H$ is cyclic and hence standard by Lemma~\ref{lemma:CyclicEquivalentToSeparatingInFiniteDimension}.

    To prove \eqref{eq:MatrixExampleSymplecticComplement}, let $X\in\Hil$, which we write as $X=\Om B$ with $B=\Om^{-1}X\in M_d$, and let $A=A^*\in M_d$. Then
    \begin{align}
        \Im\langle X,A\Om\rangle
        =
        \Im\langle\Om B,A\Om\rangle
        =
        \frac{1}{2i}\Tr((B^*-B)\Om^*A\Om),
    \end{align}
    and $X\in H'$ is equivalent to this trace vanishing for all $A\in(M_d)_\sa$. By linearity, this vanishing condition extends to all $A\in M_d$, so that we may take $\Om^*A\Om=(B^*-B)^*$ and use the faithfulness of the trace to see that it is equivalent to $B=B^*$. This proves \eqref{eq:MatrixExampleSymplecticComplement}.
\end{proof}

 We next describe an infinite-dimensional generalization of the matrix example which uses the (complex) trace class and Hilbert-Schmidt ideals $\CI_1(\Hil)\subset\CI_2(\Hil)\subset B(\Hil)$\glsadd{I1H}\glsadd{I2H}. The left and right multiplication operators \glsadd{leftmult}\glsadd{rightmult}
 \begin{align}\label{eq:leftrightmultiplicationoperators}
    \ell,r:B(\Hil)\to B(\CI_2(\Hil)),\qquad \ell(A)(X):=AX,\qquad r(A)(X):=XA,
\end{align}
 are well-defined bounded operators on the (complex) Hilbert space $\CI_2(\Hil)$ of Hilbert-Schmidt operators, equipped with the Hilbert-Schmidt scalar product $\langle A,B\rangle_2=\Tr(A^*B)$\glsadd{scprHS}. We note that $\ell$ and $r$ are linear and isometric, $\ell$ a unital ${}^*$-homomorphism and $r$ a unital ${}^*$-antihomomorphism.

\begin{example}[Standard subspaces in Hilbert-Schmidt operators]\label{example:HilbertSchmidt}
 {\em Let $\Om\in\CI_2(\Hil)$ be a Hilbert-Schmidt operator with $\ker\Om=\{0\}$ and $\ker\Om^*=\{0\}$, and consider
    \begin{align}\label{eq:HHS1}
        H
        :=
        \overline{\ell(B(\Hil)_{\sa})\Om},
        \qquad
        B(\Hil)_{\sa}
        :=
        \{A\in B(\Hil) \col A=A^*\}.
    \end{align}
    Then $H$ is a standard subspace of $\CI_2(\Hil)$, and its symplectic complement is
    \begin{align}\label{eq:HilbertSchmidtExampleSymplecticComplement}
        H'
        =
        \overline{r(B(\Hil)_\sa)\Om}.
    \end{align}
    }
 \end{example}

\begin{proof}
 According to our assumptions, there exist two orthonormal bases $(e_j)_{j\in\CJ}$ and $(f_j)_{j\in\CJ}$ of~$\Hil$ and a positive square-summable sequence $\la\in l^2(\CJ,(0,\infty))$ such that $\Om=\sum_j \la_j|f_j\rangle \langle e_j|$.

    We first show that $H_0:=\ell(B(\Hil)_\sa)\Om$ is cyclic, which will imply that $H\supset H_0$ is cyclic (in general, $H_0$ is a proper non-closed subspace of $H$). Clearly $H_0+iH_0=\ell(B(\Hil))\Om$, and choosing $A\in B(\Hil)$ as $A=\la_j^{-1}|e_k\rangle\langle f_j|$, we see that $\ell(A)\Om=|e_k\rangle\langle e_j|$ lies in $H_0+iH_0$. Thus any $X\in(H_0+iH_0)^\perp$ must satisfy $0=\langle X,A\Om\rangle_2=\langle e_j,X^*e_k\rangle$ for all $j,k\in\CJ$, and hence $X=0$. Thus $H_0$ is cyclic, and nearly identical arguments show that $\tilde H_0:=r(B(\Hil)_\sa)\Om$ is cyclic as well (here the assumption $\ker\Om^*=\{0\}$ is used).

 For self-adjoint operators $A,B\in B(\Hil)_\sa$, we have
 \begin{align*}
        \Im\langle A\Om,\Om B\rangle_2
        =
        \frac{1}{2i}\left(\Tr(\Om^*A\Om B-B\Om^*A\Om)\right)
        =
        0
    \end{align*}
    by the cyclicity of the trace. Hence $H_0\subset \tilde H_0'$, which implies $H=\overline{H_0}\subset \tilde H_0'$ because $\tilde H_0'$ is closed. As $\tilde H_0$ is cyclic, $\tilde H_0'$ is separating, so its subspace $H$ is separating as well, thus standard.

 Taking symplectic complements, this shows the inclusion ``$\supset$'' in \eqref{eq:HilbertSchmidtExampleSymplecticComplement}. To prove the reverse inclusion, let $X\in H'$. For a finite subset $F\subset\CJ$, let $P_F$ and $Q_F$ be the orthogonal projections onto $\spn\{e_j\col j\in F\}$ and $\spn\{f_j\col j\in F\}$, respectively. Then $\Om_F:=Q_F\Om P_F$ is invertible from $P_F\Hil$ onto $Q_F\Hil$. Set $X_F:=Q_FXP_F$. For every self-adjoint operator $A=Q_FAQ_F$, we have
    \begin{align*}
        \Im\langle X_F,A\Om_F\rangle_2
        =\Im\langle X,A\Om\rangle_2=0.
    \end{align*}
    Identifying $P_F\Hil$ and $Q_F\Hil$ with $\Cl^{|F|}$ by their chosen bases, Example~\ref{example:MatrixStandardSubspace} gives $X_F=\Om_FB_F$ for a self-adjoint operator $B_F$ on $P_F\Hil$. Extending $B_F$ by zero on $(P_F\Hil)^\perp$ yields $X_F=\Om B_F$. In the limit $F\to\CJ$, the claim follows.
 \end{proof}

This example can naturally be viewed from an operator-algebraic perspective. We explain this point of view in the following lemma, which as an exception to the material of this chapter uses operator algebra terminology. In particular, we will use a prime on a subset $\CM\subset B(\Hil)$ of operators to denote its commutant, $\CM'=\{A\in B(\Hil)\col [A,M]=0\,\forall M\in\CM\}$.

\Needspace{10\baselineskip}
\begin{lemma}[Symplectic complement and commutant]\label{lemma:SymplecticComplementCommutant}
 In the situation of Example~\ref{example:HilbertSchmidt} with $\Om$ normalized to Hilbert-Schmidt norm $\|\Om\|_2=1$\glsadd{HSnorm}, consider the faithful state
 \begin{align}\label{eq:DensityMatrixState}
        \om:B(\Hil)\to\Cl,\qquad
        \om(A):=\langle\Om,\ell(A)\Om\rangle_2
        =
        \Tr(\Om\Om^* A).
    \end{align}
    \begin{enumerate}
        \item\label{item:GNSHS} $(\CI_2(\Hil),\ell,\Om)$ is the GNS triple of $(B(\Hil),\om)$.
        \item\label{item:CommutantHS} The commutant of $\CL:=\ell(B(\Hil))$ is $\CL'=r(B(\Hil))=:\CR$, the von Neumann algebra of all right multipliers.
        \item\label{item:HHS} The standard subspace $H$ of Example~\ref{example:HilbertSchmidt} and its symplectic complement are given by
        \begin{align}
            H
            =
            \overline{\CL_{\sa}\Om},
            \qquad
            H'
            =
            \overline{\CL'_{\sa}\Om}.
        \end{align}
    \end{enumerate}
\end{lemma}
\begin{proof}
    \ref{item:GNSHS} The map $\ell$ is a ${}^*$-representation of $B(\Hil)$, and $\Om$ is a cyclic unit vector as shown in Example~\ref{example:HilbertSchmidt}. The GNS property of $\Om,\ell$ holds by definition of $\om$.

 For \ref{item:CommutantHS}, we first note that left and right multipliers commute, i.e. $\CL\subset\CR'$. For the other implication, let $T'\in\CL'$, and $\xi\in\Hil$, $\|\xi\|=1$. The range of $\ell(|\xi\rangle\langle\xi|)$ consists of all rank-one operators of the form $|\xi\rangle\langle\eta|$, $\eta\in\Hil$. This range is preserved by $T'$, so it follows that there exists $A\in B(\Hil)$ such that $T'(|\xi\rangle\langle\eta|)=|\xi\rangle\langle A\eta|=|\xi\rangle\langle\eta|A^*=r(A^*)|\xi\rangle\langle\eta|$. Hence $T'$ and $r(A^*)$ agree on all finite-rank operators, hence on all of $\CI_2(\Hil)$ by continuity.

    \ref{item:HHS} is a reformulation of \eqref{eq:HHS1} and \eqref{eq:HilbertSchmidtExampleSymplecticComplement}.
\end{proof}

This observation motivates a closer look at operator algebras and standard subspaces, which will be done in Chapter~2. It also explains our choice of notation for symplectic complements: The dash in $H'$ refers to the dash indicating commutants $\CM'$.

\subsection{Abelian and factorial spaces}\label{section:AbelianSpaces}

 The relation between algebra and commutant observed in Lemma~\ref{lemma:SymplecticComplementCommutant} motivates the following terminology, which imports algebraic language into the language of standard subspaces. An algebra $\M$ acting on a linear space is abelian if $\M\subset\M'$, its centre is $\M\cap\M'$, and $\M$ is a factor (a term mostly used for von Neumann algebras, see Chapter 2) if it has trivial centre, namely $\M\cap\M'=\Cl1$. We formulate the corresponding subspace properties in the general setting of $\Real(\Hil)$.

\begin{definition}[Abelian spaces, centres, factorial spaces]\label{definition:HAbelianCenterEtc}
    Let $H\in\Real(\Hil)$.
    \begin{enumerate}
        \item\label{item:AbelianSubspace} $H$ is called {\em abelian}\index{subspace!abelian (Def.~\ref{definition:HAbelianCenterEtc})} if $H\subset H'$.
        \item\label{item:Center} The {\em centre of $H$}\index{centre of a subspace@centre of a subspace} is $Z(H):=H\cap H'$.
        \item\label{item:FactorialSubspace} $H$ is called {\em factorial}\index{subspace!factorial (Def.~\ref{definition:HAbelianCenterEtc})} if it has trivial centre, $H\cap H'=\{0\}$.
    \end{enumerate}
\end{definition}

Owing to the fact that the imaginary part of the scalar product, $s:=\Im\scpr$, is a symplectic form on $\Hil$ that restricts to a pre-symplectic form $s_H$ on any $H\in\Real(\Hil)$, also the terminology of symplectic geometry applies. We will not use it here, but at least give a quick dictionary. Let $H\in\Real(\Hil)$.
\begin{enumerate}[(i)]
 \item\label{item:SymplecticSubspace} $H$ is called {\em symplectic}\index{subspace!symplectic} if $s_H$ is non-degenerate. This is equivalent to $H$ being factorial (Proposition~\ref{proposition:proppolirizer}~\ref{item:DHKernel}).
 \item\label{item:IsotropicSubspace} $H$ is called {\em isotropic}\index{subspace!isotropic} if $s_H=0$. This is equivalent to $H$ being abelian.
 \item\label{item:CoisotropicSubspace} $H$ is called {\em coisotropic} if $H'\subset H$\index{subspace!coisotropic}. This is equivalent to $H'$ being abelian.
 \item\label{item:LagrangianSubspace} $H$ is called {\em Lagrangian}\index{subspace!Lagrangian} if it is isotropic and has no proper isotropic extension (maximal isotropic). This will be called maximal abelian here.
\end{enumerate}

 Whereas in symplectic geometry, the Lagrangian (maximal abelian) spaces are often at the centre of interest, in standard subspaces and modular theory the most interesting spaces are the factorial ones. In contrast, abelian standard subspaces have a very simple structure which we clarify now. This also gives us further examples of standard subspaces.

\begin{proposition}[Abelian spaces]\label{proposition:AbelianSpaces}
    Let $H\in\Real(\Hil)$. If $H$ is abelian, then it is separating. Moreover, the following properties of $H\in\Real(\Hil)$ are equivalent:
    \begin{enumerate}
        \item\label{item:HMaximalAbelian} $H$ is maximal abelian, i.e. it has no proper abelian extension\index{subspace!maximal abelian (Prop.~\ref{proposition:AbelianSpaces})}
        \item\label{item:H=HPrime} $H=H'$.
        \item\label{item:HAbelianAndCyclic} $H$ is abelian and cyclic (hence standard).
        \item\label{item:HwithONB} There exists a complex orthonormal basis $(e_j)_{j\in\CJ}$ of $\Hil$ such that
        \begin{align}\label{eq:HONB}
            H=\overline{\rlspn\{e_j\col j\in\CJ\}}.
        \end{align}
        \item\label{item:HConj} There exists a conjugation $\Gamma\in\Con(\Hil)$ that leaves $H$ pointwise invariant, $H=\Fix\Gamma$.
    \end{enumerate}
    All maximal abelian standard subspaces are unitarily equivalent.
\end{proposition}

 In Proposition~\ref{proposition:CenterAndMaximalAbelianness}, modular theory is used to derive yet another characterization of maximal abelian subspaces, and a characterization of the centre of a general standard subspace.

\begin{proof}
    Abelian spaces are separating because $H\cap iH\subset H\cap iH'=H\cap H^\rperp=\{0\}$.

    \ref{item:HMaximalAbelian}$\Rightarrow$\ref{item:H=HPrime} Suppose that $H$ is maximal abelian but $H\subsetneq H'$. Let $h'\in H'$. Then the scalar products $\langle h_1+r_1h',h_2+r_2h'\rangle$ are real for all $h_1,h_2\in H$, $r_1,r_2\in\Rl$. Hence the closed real subspace $K:=\rlspn\{H,\Rl h'\}$ is an abelian extension of $H$. Since $H$ is maximal abelian, we conclude $h'\in H$, and $H=H'$ follows.

    \ref{item:H=HPrime}$\Rightarrow$\ref{item:HMaximalAbelian} Let $K$ be an abelian extension of $H$, i.e. $H'=H\subset K\subset K'$. Taking symplectic complements shows $H\supset K$, hence $K=H$. So $H$ is maximal abelian.

    \ref{item:H=HPrime}$\Rightarrow$\ref{item:HAbelianAndCyclic} $H=H'$ clearly implies that $H$ is abelian, and since $H=H''$ (Lemma~\ref{lemma:CyclicSeparating}~\ref{item:DoubleComplement}), also that $H'\subset H$ is abelian. The first statement of this proposition now shows that $H'$ is separating, so $H$ is cyclic (Lemma~\ref{lemma:CyclicSeparating}~\ref{item:CyclicVsSeparating}).

    \ref{item:HAbelianAndCyclic}$\Rightarrow$\ref{item:HwithONB} Choose a real orthonormal basis $(e_j)_{j\in\CJ}$ of the real Hilbert space $(H,\Rescpr)$. As $H$ is abelian, we have $\langle e_j,e_k\rangle=\delta_{jk}+i\Im\langle e_j,e_k\rangle=\delta_{jk}$ for any two $j,k\in\CJ$. Thus $(e_j)_{j\in\CJ}$ is also orthonormal w.r.t. the complex scalar product, and since $H$ is cyclic, it is an orthonormal basis of $\Hil$.

    \ref{item:HwithONB}$\Rightarrow$\ref{item:HConj} With the orthonormal basis $(e_j)$ \eqref{eq:HONB}, the map
    \begin{align*}
        \Gamma:\Hil\to\Hil,\qquad \sum_j c_j\,e_j\mapsto\sum_j \overline{c_j}e_j
        ,\qquad
        \sum_j|c_j|^2<\infty,
    \end{align*}
 is a conjugation satisfying $\Fix\Gamma=H$

    \Needspace{7\baselineskip}
    \ref{item:HConj}$\Rightarrow$\ref{item:H=HPrime} We consider the conjugation $\Gamma$ as a real-linear operator, and observe that its spectral decomposition is $\Gamma=\Gamma_+-\Gamma_-$, with the two real orthogonal projections $\Gamma_\pm=\frac12(1\pm\Gamma)$ (Remark~\ref{remark:AntiunitariesAndConjugations}~\ref{item:RealAndImaginaryPartofConjugation}). This yields
 \begin{align*}
        H'=
        (\Fix\Gamma)'
        &=
        (i\Gamma_+\Hil)^\rperp
        =
        (\Gamma_-\Hil)^\rperp
        =
        \Gamma_+\Hil
        =
        \Fix\Gamma
        =
        H.
    \end{align*}

    The claim on unitary equivalence is obvious from \ref{item:HwithONB}.
\end{proof}

Note that we have shown in particular that every abelian standard subspace is maximal abelian, and that every Hilbert space has standard subspaces.

A general closed real subspace can always be decomposed into a direct sum of a maximal abelian standard subspace and a factorial subspace.

\begin{proposition}[Decomposition into maximal abelian and factorial part]\label{proposition:DecomposeHIntoMaximallyAbelianAndFactorPart}
    Let $H\in\Real(\Hil)$ and define its {\em factorial part}\index{factorial part (Prop.~\ref{proposition:DecomposeHIntoMaximallyAbelianAndFactorPart})} as \glsadd{FH}
    \begin{align}
        F(H)
        :=
        Z(H)^\rperp\cap H,
    \end{align}
 where \glsadd{ZH}$Z(H)=H\cap H'$ is the centre of $H$. Then:
 \begin{enumerate}
        \item\label{item:ZFDecomposition} $H=Z(H)\oplus_\Cl F(H)$ as a complex orthogonal direct sum.
        \item\label{item:ZMaximalAbelian} $Z(H)\in\Std(\Hil_Z)$ is a maximal abelian standard subspace of $\Hil_Z:=\overline{\Cl[H\cap H']}$.
        \item\label{item:FFactorial} $F(H)$ is a closed factorial subspace of $\Hil_Z^\perp$, and $F(H)\in\Std(\Hil_Z^\perp)$ in case $H\in\Std(\Hil)$ is standard.
    \end{enumerate}
\end{proposition}
\begin{proof}
    \ref{item:ZFDecomposition} The decomposition $H=Z(H)\oplus F(H)$ is real orthogonal by definition. Furthermore, $F(H)\subset H\subset H\vee H'=Z(H)'$, so the decomposition also holds as a complex orthogonal decomposition.

 \ref{item:ZMaximalAbelian} By definition of $\Hil_Z$, the centre $Z(H)$ is a cyclic closed subspace of $\Hil_Z$. As $Z(H)'\cap\Hil_Z=\overline{H+H'}\cap\Hil_Z\supset Z(H)$, we see that $Z(H)$ is an abelian subspace of $\Hil_Z$ and hence a maximal abelian standard subspace by Proposition~\ref{proposition:AbelianSpaces}.

    \ref{item:FFactorial} $F(H)$ is a closed subspace of $\Hil_Z^\perp$ by definition. Any vector $f\in Z(F(H))$ satisfies $0=\Im\langle f,\tilde f\rangle$ for all $\tilde f\in F(H)$ by definition of $Z(F(H))$, and in view of \ref{item:ZFDecomposition}, also $0=\Im\langle f,h\rangle$ for all $h\in Z(H)$. As $F(H)$ and $Z(H)$ span $H$, this implies $f\in H'$. Thus $f\in H\cap H'=Z(H)$, which is orthogonal to $F(H)$ and implies $f=0$.

    In case $H$ is standard, $H+iH\subset\Hil$ is dense and $F(H)+iF(H)$ is dense in $\Hil_Z^\perp$. Furthermore, $F(H)$ is separating in this case because $F(H)\cap iF(H)\subset H\cap iH=\{0\}$, so $F(H)\in\Std(\Hil_Z^\perp)$.
\end{proof}

\begin{remark}[Maximally abelian and factorial spaces in $\boldsymbol{d=1,2}$]\label{remark:2dExampleRevisited}
    Any one-dimensional standard subspace is unitarily equivalent to $\Rl\subset\Cl$, so maximal abelian. A two-dimensional standard subspace $H$ may be decomposed according to Proposition~\ref{proposition:DecomposeHIntoMaximallyAbelianAndFactorPart} as the direct sum $H=Z(H)\oplus F(H)$ of a maximal abelian standard subspace $Z(H)$ and a factorial standard subspace $F(H)$. As every one-dimensional standard subspace is maximal abelian, we either have $F(H)=\{0\}$, i.e. $H$ is maximal abelian, or $F(H)=H$, i.e. $H$ is factorial.

    In the classification given in Example~\ref{example:2dStandardSubspaces}, the real orthonormal basis \eqref{eq:2dBasis} is complex orthonormal if and only if $\alpha=0$, i.e. $\la=1$ (see \eqref{eq:h1h2product}). So $H_\la$ \eqref{eq:Hlambda} is maximal abelian for $\la=1$, and factorial for $0<\la<1$.
\end{remark}

\Needspace{8\baselineskip}
\section{Modular data of standard subspaces}\label{section:ModularData}

 In this section the theory of standard subspaces starts properly and various objects of central importance are introduced. In Section~\ref{section:TomitaOperators} we explain how standard subspaces are in bijection with closed antilinear involutions. The involution $S_H$ of a standard subspace $H\in\Std(\Hil)$ is called its {\em Tomita operator} and is the basic tool to analyse~$H$. The polar decomposition of the Tomita operator provides us with a conjugation, the {\em modular conjugation} $J_H$, and a positive operator, the {\em modular operator} $\Delta_H$. The action of the conjugation $J_H$ and the unitaries $\Delta_H^{it}$, $t\in\Rl$, is the content of the Tomita-Takesaki Theorem for standard subspaces (Theorem~\ref{theorem:Tomita}).

Once these fundamental structures are established, we consider properties of the spectrum of $\Delta_H$ and discuss some associated operators in Section~\ref{section:SpectrumDelta}. These include in particular the {\em modular Hamiltonian} $\log\Delta_H$ and the unitary one-parameter group that it generates, the {\em modular group}. We show that the action of this group on the underlying standard subspace is passive, and classify all passive one-parameter groups (Proposition~\ref{prop:ClassificationOfPassiveOneParameterGroups}). As another basic tool for studying standard subspaces we also consider the real orthogonal projection $E_H$ onto~$H$, and -- in the case of factorial spaces -- the (typically unbounded) {\em cutting projection} $P_H$. Both these projections can be expressed in terms of the modular data $J_H,\Delta_H$ in a direct manner (Proposition~\ref{proposition:orthogonalprojectionmodulardata} and Proposition~\ref{proposition:propertiescutting}).

In Section~\ref{section:ConstructionsAndExamples}, we explain how modular data can be used as a starting point for constructing standard subspaces. This includes in particular examples of standard subspaces coming from measure spaces (Example~\ref{example:StandardSubspacesFromMeasures}) and locally compact groups (Example~\ref{example:HaarMeasure}).

 Section~\ref{section:ContractionsAndGraphs} discusses a classification of standard subspaces up to unitary equivalence by certain self-adjoint operators (Theorem~\ref{theorem:StdAndContractions}). This result establishes in particular that the modular operator $\Delta_H$ alone (without the modular conjugation) already determines $H$ up to unitary equivalence. A related construction classifies standard subspaces as graphs of positive operators in a doubled Hilbert space with twisted complex structure (Theorem~\ref{theorem:GraphStandardSubspaces}), again up to unitary equivalence.

\subsection{Tomita operators and the Tomita--Takesaki Theorem}\label{section:TomitaOperators}

As observed in Remark~\ref{remark:RealStructures}, standard subspaces $H$ with closed complex span define real structures on $\Hil$, and hence an associated ``complex conjugation''. The generalization of this observation to arbitrary standard subspaces leads us to Tomita operators, which are the fundamental tool for understanding standard subspaces.

\begin{definition}[Tomita operator]\label{definition:Tomitaoperator}
    The {\em Tomita operator\index{Tomita operator (Def.~\ref{definition:Tomitaoperator})}} of a standard subspace $H\in\Std(\Hil)$ is the map~\glsadd{SH}
    \begin{align}\label{def:SH}
        S_H:H+iH\to H+iH,\qquad h_1+ih_2\mapsto h_1-ih_2.
    \end{align}
\end{definition}

Note that $S_H$ is well-defined because $H$ is separating, i.e. $H$ and $iH$ are independent (Def.~\ref{def:IndependentSubspaces}). The Tomita operator is nothing else than the involution $S_H=S_{H,iH}$ of the pair $(H,iH)$ \eqref{eq:SHK}, and can equivalently be expressed by the projection \eqref{eq:def2subspaceprojection}, namely
\begin{align}\label{eq:SHvsPHiH}
    S_H&=2P_{H,iH}-1,\qquad
    P_{H,iH}:H+iH\to \Hil,\qquad h_1+ih_2\mapsto h_1.
\end{align}

We emphasize that despite their formal similarity to complex conjugation, Tomita operators are typically not isometric and not even bounded.

\Needspace{6\baselineskip}
\begin{lemma}[Closedness and boundedness of Tomita operators]\label{lemma:SHBounded}
    Let $H\in\Std(\Hil)$ with Tomita operator $S_H$.
    \begin{enumerate}
        \item\label{item:SHclosed} $S_H$ is closed.
        \item\label{item:SHbounded} $S_H$ is bounded if and only if $H+iH=\Hil$.
    \end{enumerate}
\end{lemma}
\begin{proof}
    According to Lemma~\ref{lemma:InvolutionsAndProjections} and Proposition~\ref{proposition:twosubspacesprojection}, $S_H$ is closed because $H$ and $iH$ are closed, and bounded if and only if $H+iH$ is closed. As $H+iH$ is a dense subspace by definition, the claim follows.
\end{proof}

 As Tomita operators are typically unbounded, we will need to pay proper attention to their domains\footnote{In the following, equalities $A=B$ between (real or complex-linear) operators $A,B$ are always understood as $\dom(A)=\dom(B)$ and $A\xi=B\xi$ for all $\xi\in\dom A$, whereas extensions are denoted $A\subset B$. We refer to general functional analysis books such as \cite{Schmudgen:2012} for background material on unbounded operators.}, namely
 \begin{align}
    \dom S_H=H+iH=\Ran S_H.
\end{align}
According to Lemma~\ref{lemma:SHBounded}, Tomita operators are special types of involutions (Def.~\ref{def:Involutions}), namely elements of\glsadd{InvH}
\begin{align}\label{eq:InvH}
    \Inv(\Hil)
    :=
    \{S:\dom S\to\dom S\col S \text{ densely defined closed antilinear involution}\}.
\end{align}

 We will from now on rely on standard notions and results from operator theory, including the spectral theorem for complex-linear self-adjoint operators (for a self-adjoint operator $A$, we denote its spectral measure by $E^A$\glsadd{EA}, and $E^A_\xi$\glsadd{EAxi} is the number-valued measure $I\mapsto\langle\xi,E^A(I)\xi\rangle$), the corresponding functional calculus, and spectrum (we write $\spec(A)$\glsadd{sA} for the spectrum of $A$ and \glsadd{spA}$\specp(A)\subset\spec(A)$ for its point spectrum).

It turns out that $\Inv(\Hil)$ is in natural bijection with $\Std(\Hil)$.

\begin{proposition}[Standard subspaces and their Tomita operators]\label{proposition:StandardSubspacesVsClosedInvolutions}\index{classification of standard subspaces!by Tomita operators}
\leavevmode
    \begin{enumerate}
        \item\label{item:StdToInv} $S_H\in\Inv(\Hil)$ for every $H\in\Std(\Hil)$.

        \item\label{item:InvToStd} $\Fix(S):=\ker(S-1)\in\Std(\Hil)$ for any $S\in\Inv(\Hil)$.

        \item\label{item:StdEquivInv} The maps
        \begin{align}\label{eq:StdInv}
            \Std(\Hil) \ni H \mapsto S_H \in \Inv(\Hil),
            \qquad
            \Inv(\Hil) \ni S \mapsto \Fix(S) \in \Std(\Hil)
        \end{align}
        defined in \ref{item:StdToInv} and \ref{item:InvToStd} are inverses of each other, i.e. $\Std(\Hil)$ and $\Inv(\Hil)$ are in bijection.

       \item\label{item:TomitaAUH} The maps \eqref{eq:StdInv} are $\AU(\Hil)$-equivariant,
       \begin{align}
            S_{VH}=VS_HV^*,\qquad H\in\Std(\Hil),\quad V\in\AU(\Hil).
       \end{align}

        \item\label{item:TomitaComplement} The map \eqref{eq:StdInv} carries symplectic complements into adjoints,
        \begin{align}
            S_{H'}=S_H^*,\qquad H\in\Std(\Hil).
        \end{align}
    \end{enumerate}
\end{proposition}
\begin{proof}
    \ref{item:StdToInv} $S_H$ is a densely defined antilinear involution, and closed by Lemma~\ref{lemma:SHBounded}~\ref{item:SHclosed}.

 \ref{item:InvToStd} The space $H:=\Fix(S)$ is real-linear because $S$ is antilinear, and closed because $S$ is closed. Clearly $iH$ is the eigenspace of $S$ with eigenvalue $-1$, so $H\cap iH=\{0\}$. Given $\xi\in\dom(S)$, the vectors $\xi_+:=\frac12(\xi+S\xi)$ and $\xi_-:=\frac{1}{2i}(\xi-S\xi)$ lie in $H$. Hence $\xi=\xi_++i\xi_-$ shows that $H+iH$ is dense, i.e. $H$ is standard.

    \ref{item:StdEquivInv} We clearly have $\Fix(S_H)=H$ and $S_{\Fix(S)}=S$, which proves the claim.

    \ref{item:TomitaAUH} We have $\dom(VS_HV^*)=V\dom(S_H)=V(H+iH)=\dom S_{VH}$ and, with $h_1,h_2\in~H$,
    \begin{align*}
        VS_HV^*V(h_1+ih_2)=V(h_1-ih_2)=S_{VH}V(h_1+ih_2).
    \end{align*}
    This shows $VS_HV^*=S_{VH}$.

 \ref{item:TomitaComplement} We use Lemma~\ref{lemma:InvolutionsAndProjections}~\ref{item:PvsSFA}, viewing $S_H=S_{H,iH}$ as a real-linear involution on the real Hilbert space $\Hil$. We therefore have to use the real orthogonal complement, and find
 \begin{align*}
        S_H^*
        =
        (S_{H,iH})^*
        =
        S_{(iH)^\rperp,H^\rperp}
        =
        S_{H',iH'}
        =
        S_{H'}.
    \end{align*}
 \end{proof}

The relation $S_H^*=S_{H'}$ shows that the symplectic complement is the natural ``complement'' on $\Std(\Hil)$. The real orthogonal complement has instead the Tomita operator
\begin{align}
    S_{H^\rperp}
    &=
    S_{iH'}
    =
    -iS_{H'}i
    =
    -S_H^*.
\end{align}

 As the involutions $S_H\in\Inv(\Hil)$ are closed real-linear maps, they have a real polar decomposition (see Section~\ref{section:RealSubspaces}, following~\eqref{eq:RealAdjoint}), which is traditionally written as \begin{align}\label{eq:SPolarDecomposition}
    S_H
    =
    J_H\Delta_H^{1/2}.
\end{align}
 In this polar decomposition, $J_H$ is the real-linear partial isometry, called the {\em modular conjugation\index{modular!conjugation}\glsadd{JH} of $H$}, and \glsadd{DeltaH}$\Delta_H=S_H^*S_H$ is the square of the real-linear positive self-adjoint part of $S_H$, called the {\em modular operator\index{modular!operator} of $H$}. Collectively the pair $J_H,\Delta_H$ are also referred to as the {\em modular data} of $H$. Since~$S_H$ is an antilinear involution, these operators have special properties that we spell out now.

\begin{proposition}[Basic properties of modular data]\label{proposition:ModularDataProperties}
    \leavevmode
    Let $S_H\in\Inv(\Hil)$ be a closed antilinear involution with polar decomposition \eqref{eq:SPolarDecomposition}.
    \begin{enumerate}
        \item\label{item:JProperties} The modular conjugation is a conjugation, i.e.
        \begin{align}
            J_H = J_H^* = J_H^{-1}.
        \end{align}
 \item\label{item:DeltaProperties} The modular operator $\Delta_H=S_H^*S_H$ is a positive, non-singular (i.e. $\ker\Delta_H=\{0\}$) self-adjoint linear operator with dense range.

        \item\label{item:JDeltaCommutation} For every complex Borel function $f:\Rl_+\to\Cl$ on the positive half-line $\Rl_+:=(0,\infty)$\glsadd{Rplus}, we have
        \begin{align}\label{eq:JfDelta}
            J_Hf(\Delta_H) J_H = \bar f(\Delta_H^{-1}).
        \end{align}
        In particular, we have the \emph{modular relation}
 \begin{align}\label{eq:JDeltaCommutation}
            J_H\Delta_H J_H = \Delta_H^{-1}.
        \end{align}

        \item\label{item:JDEquivariance} For any $V\in\AU(\Hil)$,
        \begin{align}
            J_{VH}=VJ_HV^*,\qquad \Delta_{VH}=V\Delta_H V^*.
        \end{align}

        \item\label{item:JDComplement} The symplectic complement $H'$ has the modular data
        \begin{align}
            J_{H'}=J_H,\qquad \Delta_{H'}=\Delta_H^{-1}.
        \end{align}

        \item\label{item:FixJvsFixS} The map\glsadd{BJH}
        \begin{align}
            B_H:\Fix J_H\to\Fix S_H=H,\qquad \xi\mapsto \sqrt{2}(1+\Delta_H)^{-1/2}\xi
        \end{align}
        is a real orthogonal isomorphism.
    \end{enumerate}
\end{proposition}
\begin{proof}
 The modular operator $\Delta_H=S_H^*S_H$ is linear because $S_H$ is antilinear, and self-adjoint positive by definition. Hence $J_H$ is antilinear and in fact antiunitary because $S_H$ has trivial kernel and dense range.

    Any $\xi\in\ker\Delta_H$ lies also in $\ker\Delta_H^{1/2}$ and hence satisfies $S_H\xi=J_H\Delta_H^{1/2}\xi=0$. As $S_H$ is an involution, this implies $\xi=0$, so $\ker\Delta_H=\{0\}$, and $\overline{\Ran\Delta_H}=\ker(\Delta_H)^\perp=\Hil$, proving \ref{item:DeltaProperties}.

    The involutivity of $S_H$ (in particular, $\Ran S_H=\dom S_H$) implies
    \begin{align}
        J_H\cdot \Delta_H^{1/2}
        =
        S_H
        =
        S_H^{-1}
        =
        \Delta_H^{-1/2}J_H^*
        =
        J_H^*\cdot J_H\Delta_H^{-1/2}J_H^*.
    \end{align}
 As $J_H^*$ is a real-linear isometry and $J_H\Delta_H^{-1/2}J_H^*$ positive self-adjoint, the uniqueness of the polar decomposition yields $J_H^*=J_H$, proving  \ref{item:JProperties}, and $J_H\Delta_H^{-1/2}J_H=\Delta_H^{1/2}$. As a consequence, we also have \eqref{eq:JDeltaCommutation}, namely \eqref{eq:JfDelta} in the special case $f=\id$.

    To get the general statement in \ref{item:JDeltaCommutation}, we consider the spectral measure $E:=E^{\Delta_H}$ of $\Delta_H$ and define on Borel sets $A\subset(0,\infty)$ the (complex-linear complex orthogonal) projections $\tilde E(A):=J_HE(A^{-1})J_H$, where $A^{-1}=\{\la^{-1}\col\la\in A\}$. Extended by zero to $(-\infty,0]$, this is another spectral measure on $\Rl$, and we have
 \begin{align*}
        \int \la\,d\tilde E(\la)
        =
        J_H\int \la^{-1}\,dE(\la)J_H
        =
        J_H \Delta_H^{-1}J_H
        =
        \Delta_H
        =
        \int\la \,dE(\la).
    \end{align*}
    The uniqueness of the spectral measure now implies $E=\tilde E$. By integration, this yields for any Borel function $f:\Rl\to\Cl$
    \begin{align*}
        J_Hf(\Delta_H)J_H
        =
        \int \overline{f(\la)}\,J_H dE(\la)J_H
        =
        \int \overline{f(\la^{-1})}\,d\tilde E(\la)
        =
        \int \overline{f(\la^{-1})}\,dE(\la)
        =
        \overline{f}(\Delta_H^{-1}),
    \end{align*}
    as claimed.

    \ref{item:JDEquivariance} is a direct consequence of Proposition~\ref{proposition:StandardSubspacesVsClosedInvolutions}~\ref{item:TomitaAUH} and the uniqueness of the polar decomposition.

    \ref{item:JDComplement} We know already $S_{H'}=S_H^*$ (Prop.~\ref{proposition:StandardSubspacesVsClosedInvolutions}~\ref{item:TomitaComplement}), and the modular relation $J_H\Delta_H J_H=\Delta_H^{-1}$. Hence
    \begin{align*}
        J_{H'}\Delta_{H'}^{1/2}
        =
        S_{H'}=S_H^*=\Delta_H^{1/2}J_H=J_H\Delta_H^{-1/2},
    \end{align*}
    and the uniqueness of the polar decomposition finishes the proof.

    \ref{item:FixJvsFixS} The range of $B_H$ is contained in $\dom\Delta_H^{1/2}=\dom S_H$. Part~\ref{item:JDeltaCommutation} implies for $\xi\in\Fix J_H$
    \begin{align*}
        S_HB_H\xi
        &=
        \sqrt{2}\,J_H\frac{\Delta_H^{1/2}}{(1+\Delta_H)^{1/2}}\xi
        =
        \sqrt{2}\,\frac{\Delta_H^{-1/2}}{(1+\Delta_H^{-1})^{1/2}}\xi
        =
        B_H\xi
        \\
        \|B_H\xi\|^2
        &=
        \langle\xi,(1+\Delta_H)^{-1}\xi\rangle
        +
        \langle J_H\xi,(1+\Delta_H)^{-1}J_H\xi\rangle
        \\
        &=
        \big\langle\xi,
        \big((1+\Delta_H)^{-1}+(1+\Delta_H^{-1})^{-1}\big)\xi
        \big\rangle
        \\
        &=
        \|\xi\|^2.
    \end{align*}
    Hence $B_H$ maps $\Fix J_H$ into $H$ and is isometric. It is surjective because, for $h\in H$, the vector $(1+\Delta_H)^{1/2}h/\sqrt{2}$ is well defined, fixed by $J_H$ because of~\ref{item:JDeltaCommutation}, and mapped to $h$ by $B_H$ by definition of $B_H$.
\end{proof}

The following theorem is the standard subspace version of the fundamental Tomita-Takesaki theorem for von Neumann algebras, stating that the modular unitaries are automorphisms of~$H$ and the modular conjugation provides an identification between $H$ and $H'$. In contrast to its von Neumann algebraic version, to be discussed in Chapter~2, it has a much easier proof.

\begin{theorem}[Tomita-Takesaki Theorem for standard subspaces]\label{theorem:Tomita}\booktheoremindex{theorem:Tomita}{Tomita--Takesaki theorem for standard subspaces}\index{Tomita-Takesaki theorem@Tomita--Takesaki theorem!for standard subspaces}
    Let $H\in\Std(\Hil)$. Then
    \begin{align}
        \Delta_H^{it}H = H,\quad t\in\Rl,\qquad J_HH =H'.
    \end{align}
\end{theorem}
\begin{proof}
 The modular unitaries $\Delta_H^{it}$ commute with $\Delta_H$ and with $J_H$ by application of \eqref{eq:JfDelta} with $f(\la)=\la^{it}$, so $\Delta_H^{it}H=H$ follows from Proposition~\ref{proposition:ModularDataProperties}~\ref{item:JDEquivariance}.
 Concerning the second relation, notice that if $h\in H$, then
    \begin{align*}
        \langle h,J_Hh\rangle
        =
        \langle h,J_HS_H h\rangle
        =
        \langle h,\Delta_H^{1/2}h\rangle
        \in\Rl
    \end{align*}
 because $\Delta_H^{1/2}$ is self-adjoint. Thus for all $h_1,h_2\in H$
 \begin{align*}
        \langle h_1,J_Hh_2\rangle
        &=
        \frac12\left(
            \langle h_1,J_Hh_2\rangle
            +
            \langle h_2,J_Hh_1\rangle
        \right)
        \\
        &=
        \frac12\left(
            \langle h_1+h_2,J_H(h_1+h_2)\rangle
            -
            \langle h_1,J_Hh_1\rangle
            -
            \langle h_2,J_Hh_2\rangle
        \right)
        \in\Rl,
    \end{align*}
    namely $\Im \langle h_1,J_Hh_2\rangle=0$, and we conclude $J_H H\subset H'$.

    As $J_H = J_{H'}$ (Prop.~\ref{proposition:ModularDataProperties}~\ref{item:JDComplement}), we also have $J_H H'\subset H''= H$, which by involutivity of $J_H$ implies $H'\subset J_HH$. This finishes the proof of $J_HH=H'$.
\end{proof}

We illustrate the Tomita-Takesaki Theorem with some examples.

\begin{example}[Modular data of two-dimensional standard subspaces]\label{example:ModularDataOf2dStandardSubspace}
    In Example~\ref{example:2dStandardSubspaces} we have seen that any two-dimensional standard subspace is unitarily equivalent to one of the form
    \begin{align}\label{eq:Hlambda2}
        H_\lambda
        :=
        \left\{
            \begin{pmatrix}
                \lambda^{1/2}z\\\overline{z}
            \end{pmatrix}
            \col z\in\Cl
        \right\}\in\Std(\Cl^2),
    \end{align}
    with a unique parameter $0<\la\leq1$. We had also seen $H_\la=\Fix(J\Delta^{1/2})$ with the conjugation $J$ and the positive operator $\Delta$ given in \eqref{eq:JDelta2d}, namely
    \begin{align}
        \Delta=
        \begin{pmatrix}
            \lambda^{-1} &0\\
            0&\lambda
        \end{pmatrix},
        \qquad
        J\begin{pmatrix}
            z_1\\z_2
        \end{pmatrix}
        :=
        \begin{pmatrix}
            \overline{z_2}\\
            \overline{z_1}
        \end{pmatrix}.
    \end{align}
    It is now clear that $J=J_H$ and $\Delta=\Delta_H$ are the modular data of $H$ \eqref{eq:Hlambda}. The modular group acts by automorphisms of $H_\la$,
    \begin{align}
        \Delta^{it}\begin{pmatrix}
                \lambda^{1/2}z\\\overline{z}
            \end{pmatrix}
            =
            \begin{pmatrix}
                \la^{-it}\lambda^{1/2}z\\\la^{it}\overline{z}
            \end{pmatrix}
            =
            \begin{pmatrix}
                \lambda^{1/2}z_t\\\overline{z_t}
            \end{pmatrix}
            ,\qquad
            z_t=\la^{-it}z,
    \end{align}
    in agreement with Theorem~\ref{theorem:Tomita}, and the symplectic complement of $H_\la$ is
    \begin{align}
        H_\la'=JH_\la=H_{1/\la}.
    \end{align}
\end{example}

\begin{example}[Modular data of standard subspaces in matrix spaces]\label{example:ModularDataOfMatrix}
    In Example~\ref{example:MatrixStandardSubspace}, we considered an invertible $(d\times d)$-matrix $\Om\in M_d$ and the corresponding standard subspace $H=\ell(M_d)_\sa\Om$ (with $\ell$ denoting left and $r$ denoting right multiplication) in the Hilbert space $M_d$ with Hilbert-Schmidt scalar product.

    We claim that its Tomita operator is
    \begin{align}\label{eq:TomitaMatrixExample}
        S_H:M_d\to M_d,\qquad S_H(X)=(\Om^{-1})^* X^*\Om,
    \end{align}
    and that its modular conjugation and modular group are, $X\in M_d$,
     \begin{align}\label{eq:ModularDataMatrixExample}
        J_H(X)=U^*X^*U^*,
        \qquad
        \Delta_H^{it}(X)
        =
        \rho^{it}XU\rho^{-it}U^*,
    \end{align}
    where $\Om^*=U\rho^{1/2}$, $\rho^{1/2}=|\Om^*|$, is the polar decomposition\footnote{This notation is chosen because then the density matrix $\rho:=\Om\Om^*$ is denoted by its usual symbol~$\rho$.} of $\Om^*$.

    These formulae for $J_H$ and $\Delta_H^{it}$ imply that the modular conjugation and modular group act on left multipliers $\ell(A)$, $A\in M_d$, according to
    \begin{align}
        J_H\ell(A)J_H
        &=
        r(UA^*U^*),\qquad
        \Delta_H^{it}\ell(A)\Delta_H^{-it}
        =
        \ell(\rho^{it}A\rho^{-it}).
    \end{align}
    Thus $J_H$ conjugates left multipliers to right multipliers $r(UA^*U^*)$, i.e. maps $H$ onto $H'$ (cf. Example~\ref{example:MatrixStandardSubspace}), and the modular group maps left multipliers to left multipliers, i.e. preserves~$H$. These observations exemplify Theorem~\ref{theorem:Tomita}.
\end{example}
\begin{proof}
 By decomposing $A\in M_d$ as $A=\frac12(A+A^*)+i\cdot\frac{1}{2i}(A-A^*)$ as a complex linear combination of two self-adjoint matrices, one sees that the Tomita operator of $H$ acts according to $S_H(A\Om)=A^*\Om$, $A\in M_d$. One directly checks that the operator in \eqref{eq:TomitaMatrixExample} agrees with this.

    To compute the polar decomposition of $S_H$, let $\Om^*=U\rho^{1/2}$ be the polar decomposition of $\Om^*$ (with $U$ unitary because $\Om^*$ is invertible). Then we find, with $\St:M_d\to M_d$, $\St(A)=A^*$, the adjoint map and the obvious exchange relation $\ell(A)\St=\St r(A^*)$,
    \begin{align*}
        S_H
        &=
        \ell((\Om^*)^{-1})\St \ell(\Om^*)
        \\
        &=
        \ell(\rho^{-1/2})\ell(U^*)\St\ell(U)\ell(\rho^{1/2})
        \\
        &=
        r(U^*)\left(\St \cdot \ell(\rho^{1/2})r(\rho^{-1/2})\right)r(U).
    \end{align*}
    One easily checks that $r(U^*)$ is a unitary, $\St$ is a conjugation, and $\ell(\rho^{1/2})r(\rho^{-1/2})$ is positive w.r.t. the Hilbert-Schmidt inner product. Therefore uniqueness of the polar decomposition implies
    \begin{align*}
        J_H
        &=
        r(U^*)\St r(U),
        \qquad
        \Delta_H^{1/2}
        =
        r(U^*)\ell(\rho^{1/2})r(\rho^{-1/2})r(U)
        .
    \end{align*}
    Evaluation on $X\in M_d$ gives \eqref{eq:ModularDataMatrixExample}.
\end{proof}

 \subsection{Modular spectrum, dynamics, and projections}\label{section:SpectrumDelta}

Any standard subspace provides us with two data: A conjugation $J_H$ and a positive operator~$\Delta_H$. Since all conjugations are unitarily equivalent (Prop.~\ref{proposition:AbelianSpaces}), $\Delta_H$ contains more information about $H$ than $J_H$, in particular spectral information. In this section we will see how this allows us to characterize the centre of a standard subspace. Spectral properties of $\Delta_H$ also determine the type of $H$ (Chapter 2). 

We will also consider some associated operators, in particular the unitary one parameter group $(\Delta_H^{it})_{t\in\Rl}$ and study its properties. It turns out that this group acts passively on $H$, and all passive one-parameter groups can be specified. Furthermore, we define the orthogonal and cutting projections $E_H$ and $P_H$, and explain their main properties.

\begin{lemma}[Symmetry of spectrum of $\boldsymbol{\Delta_H}$]\label{lemma:SpectralSymmetryOfDelta}
    Let $H\in\Std(\Hil)$. The spectrum\footnote{Note that $0$ is never an eigenvalue of $\Delta_H$ because $S_H$ has trivial kernel, but may lie in the spectrum $\spec(\Delta_H)$.} $\spec(\Delta_H)\backslash\{0\}$ of the modular operator is invariant under $\la\leftrightarrow\la^{-1}$. In particular, $\Delta_H$ is unbounded if and only if $0\in\spec(\Delta_H)$, and isometric if and only if $\Delta_H=1$.
\end{lemma}
\begin{proof}
    This is clear from the modular relation \eqref{eq:JDeltaCommutation}.
\end{proof}

 In view of this symmetry of $\spec(\Delta_H)$, the point $\la=1$ is special, and the decomposition $(0,1)\cup\{1\}\cup(1,\infty)$ of $(0,\infty)$, and correspondingly of $\spec(\Delta_H)\setminus\{0\}$, will appear frequently. We therefore define the following shorthand notation for the corresponding spectral projections and subspaces:\glsadd{EDeltaSmall}\glsadd{EDeltaSmallE}\glsadd{EDeltaBig}\glsadd{EDeltaBigE}\glsadd{EDelta1}
 \begin{equation}\label{eq:SpectralProjectionsDelta}
\begin{aligned}
    E^{\Delta_H}_<
    &:=
    E^{\Delta_H}((0,1)),
    &\qquad
    \Hil^H_<&:=E^{\Delta_H}_<\Hil,\\
    E^{\Delta_H}_\leq
    &:=
    E^{\Delta_H}((0,1]),&\qquad
    \Hil^H_\leq&:=E^{\Delta_H}_\leq\Hil,\\
    E^{\Delta_H}_>
    &:=
    E^{\Delta_H}((1,\infty)),
    &\qquad
    \Hil^H_>
    &:=
    E^{\Delta_H}_>\Hil,\\
    E^{\Delta_H}_\geq
    &:=
    E^{\Delta_H}([1,\infty)),
    &\qquad
    \Hil^H_\geq&:=E^{\Delta_H}_\geq\Hil,
    \\
    E^{\Delta_H}_1
    &:=
    E^{\Delta_H}(\{1\}),&\qquad \Hil^H_1&:=E^{\Delta_H}_1\Hil.
\end{aligned}
\end{equation}

{
All these spectral subspaces are complex subspaces of $\Hil$; the superscript $H$ records the standard subspace determining them. Since $\ker\Delta_H=\{0\}$, its spectral projection at $0$ vanishes, and
\begin{align*}
    \Hil&=\Hil^H_<\oplus\Hil^H_1\oplus\Hil^H_>,\qquad
    \Hil^H_1=\ker(\Delta_H-1),\\
    \Hil^H_\leq&=\Hil^H_<\oplus\Hil^H_1,\qquad
    \Hil^H_\geq=\Hil^H_1\oplus\Hil^H_>.
\end{align*}
The modular relation gives $J_H\Hil^H_<=\Hil^H_>$ and $J_H\Hil^H_1=\Hil^H_1$.
}

 \begin{remark}[Spectral decomposition of modular operator in matrix example]
    The spectral decomposition of $\Delta_H^{1/2}$ in the matrix example (Example~\ref{example:ModularDataOfMatrix}) is
    \begin{align}
        \Delta_H^{1/2}
        =
        r(U^*)\sum_{\la,\mu\in\spec(\rho)}\sqrt{\frac{\la}{\mu}}\,\ell(P_\la)r(P_\mu)\,r(U),
    \end{align}
    with $P_\la$ the spectral projection of $\rho$ corresponding to eigenvalue $\la$. Thus the spectrum of $\Delta_H$ is $\spec(\Delta_H)=\{\frac{\la}{\mu} \col \la,\mu\in\spec(\rho)\}$, and the symmetry pointed out in  Lemma~\ref{lemma:SpectralSymmetryOfDelta} is apparent.
\end{remark}

This symmetry suggests that a possible spectral value $1$ plays a special role.

\begin{definition}[Spectral gap]\label{def:SpectralGap}\index{spectral gap (Def.~\ref{def:SpectralGap})}
    The modular operator $\Delta_H$ of a standard subspace $H\in\Std(\Hil)$ is said to have a \emph{spectral gap} if $1\not\in\spec(\Delta_H)$.
\end{definition}

Standard subspaces with spectral gap are in several ways simpler than those without, as we will see later in the context of cutting projections (Prop.~\ref{proposition:propertiescutting}) and polarizers (Prop.~\ref{proposition:proppolirizer}).

 If $\Delta_H$ has no spectral gap, $1$ could or could not be an eigenvalue, i.e. an element of the point spectrum $\specp(\Delta_H)$. This turns out to determine the centre of $H$, as we show next. Furthermore, we obtain yet another characterization of maximal abelian standard subspaces, which also explains why maximal abelian standard subspaces are very special.

\bigskip

 Before we study standard subspaces and spectral properties further, we pause for a technical lemma that is often a useful tool.

\begin{lemma}[Domains of functions of modular operators]\label{lemma:DeltaDomains}
    Let $H\in\Std(\Hil)$.
    \begin{enumerate}
        \item\label{item:DomDeltaHalf} $\dom\Delta_H^{1/2}=H+iH$.
 \item\label{item:DomDeltapOrder} Let $p,q\in\Rl$ with $pq\geq0$. Then $\dom\Delta_H^p\subset\dom\Delta_H^q$ for $|p|\geq|q|$.
        \item\label{item:HInvariance} Let $f:\Rl_+\to\Rl$ be a bounded real Borel function such that $f(\la^{-1})=f(\la)$ for $E^{\Delta_H}$-almost all $\la>0$. Then $f(\Delta_H)H\subset H$.
 \item\label{item:DeltaAnalyticVectors} There exists a dense real subspace of $H$ consisting of vectors $h$ that are entire analytic for~$\log\Delta_H$, i.e. $z\mapsto \Delta_H^{iz}h$ is an entire $\Hil$-valued function.
 \item\label{item:DeltaCore} A linear subspace $D\subset\dom\Delta_H^{1/2}$ with $S_HD\subset D$ is a core for $\Delta_H^{1/2}$ if and only if $\overline{D\cap H}=H$.
 \end{enumerate}
\end{lemma}

 \begin{proof}
    \ref{item:DomDeltaHalf} As the modular conjugation $J_H$ is bounded, we have $\dom\Delta_H^{1/2}=\dom S_H=H+iH$.

    \ref{item:DomDeltapOrder} By spectral theory, the domain of $\Delta_H^p$, $p\in\Rl$, is given by
	\begin{align*}
		\dom \Delta_H^p
		=
		\left\{\xi\in\Hil\col \int_0^\infty \la^{2p} dE^{\Delta_H}_\xi(\la)<\infty \right\}.
	\end{align*}
	For a positive power $p\geq0$ and arbitrary $\xi\in\Hil$, we have $\int_0^\infty \la^{2p} dE^{\Delta_H}_\xi(\la)<\infty$ if and only if $\int_1^\infty \la^{2p} dE^{\Delta_H}_\xi(\la)<\infty$. Hence for $p\geq q\geq0$, we see that $\xi\in\dom\Delta_H^p$ implies $\xi\in\dom\Delta_H^q$.

	For a negative power $p\leq0$ and arbitrary $\xi\in\Hil$, we have $\int_0^\infty \la^{2p} dE^{\Delta_H}_\xi(\la)<\infty$ if and only if $\int_0^1 \la^{2p} dE^{\Delta_H}_\xi(\la)<\infty$. Hence for $p\leq q\leq0$, we also see that $\xi\in\dom\Delta_H^p$ implies $\xi\in\dom\Delta_H^q$.

	\ref{item:HInvariance} As $f$ is bounded, $H$ is contained in the domain $\dom f(\Delta_H)=\Hil$. Clearly, $f(\Delta_H)$ commutes with $\Delta_H$, and in view of \eqref{eq:JfDelta}, also with $J_H$. So for any $h\in H$, we have $f(\Delta_H)h=f(\Delta_H)J_H\Delta_H^{1/2}h=J_H\Delta_H^{1/2}f(\Delta_H)h$, i.e. $f(\Delta_H)h\in H$.

    \ref{item:DeltaAnalyticVectors} Given $0<\eps<1$, the characteristic function $\chi_{[\eps,\eps^{-1}]}$, satisfies the requirements of~\ref{item:HInvariance}. Hence $\chi_{[\eps,\eps^{-1}]}(\Delta_H)H\subset H$, and as $\eps\to0$, we obtain a dense subspace of $H$, consisting of vectors that are analytic for $\log\Delta_H$.

 \ref{item:DeltaCore} We know that $D$ is a core for $\Delta_H^{1/2}$ if and only if it is a core for $S_H$, which in turn is equivalent to being a core for the projection $P_{H,iH}$. As $P_{H,iH}=\frac12(S_H+1)$ leaves $D$ invariant by assumption, $D$ is a core for $S_H$ if and only if $\overline{D\cap H}=\overline H$ and $\overline{D\cap iH}=\overline{iH}$ (Proposition~\ref{proposition:twosubspacesprojection}~\ref{item:core2spaceprojection}). As $H$ is closed and $D$ is linear, the claim follows.
 \end{proof}

 We now use spectral analysis to discuss standard subspaces further, and begin with analysing the centre.

\begin{proposition}[Characterization of centre and maximal abelianness by $\boldsymbol{\Delta_H}$]\label{proposition:CenterAndMaximalAbelianness}
    \leavevmode
    Let $H\in\Std(\Hil)$.
    \begin{enumerate}
        \item\label{item:centerH} The centre of $H$ is
        \begin{align}
            H\cap H'
            =
            \{h\in H\cap\dom\Delta_H \col \Delta_H h=h\}
            =
            \{h\in H \col \Delta_H^{it}h=h\;\forall t\in\Rl \}
            .
        \end{align}

        \item\label{item:factorialH} A standard subspace is factorial if and only if $1\not\in\specp(\Delta_H)$.

        \item\label{item:maximalabelianH} A standard subspace is maximal abelian ($H=H'$) if and only if $\Delta_H=1$.
    \end{enumerate}
\end{proposition}
 \begin{proof}
    \ref{item:centerH} Let $h\in H\cap H'$. Then $h=S_Hh=S_H^*h$, and therefore $h=S_H^*S_Hh=\Delta_Hh$, proving ``$\subset$'' in the first equality. Conversely, if $h\in H\cap\dom\Delta_H$ satisfies $h=\Delta_Hh$, then $h=S_H^*S_Hh=S_H^*h$, so $h\in \Fix(S_H^*)=H'$. This shows the first equality.

    The inclusion ``$\subset$'' in the second equality follows because if $\Delta_Hh=h$, then $h=E^{\Delta_H}_1h$ \eqref{eq:SpectralProjectionsDelta}, so $h$ is invariant under the modular unitaries. Conversely, $\Delta_H^{it}h=h$ implies $h\in\dom\log(\Delta_H)$ and $\log(\Delta_H)h=0$ by Stone's Theorem. Hence $h=E^{\log(\Delta_H)}(\{0\})h=E^{\Delta_H}_1h$, so $h$ is in the domain of $\Delta_H$ and $\Delta_Hh=h$.

 \ref{item:factorialH} The complex linear bounded projection $E_1^{\Delta_H}$ preserves $H$ by Lemma~\ref{lemma:DeltaDomains}~\ref{item:HInvariance}. Hence $\Hil^H_1=E_1^{\Delta_H}(\overline{H+iH})=\overline{E_1^{\Delta_H}H+iE_1^{\Delta_H}H}$. By part~\ref{item:centerH}, $E_1^{\Delta_H}H=Z(H)$. Hence $\Hil^H_1$ is trivial (i.e. $1\not\in\specp(\Delta_H)$) if and only if $H$ is factorial (i.e. $Z(H)=\{0\}$).

    \ref{item:maximalabelianH} If $H=H'$ is maximal abelian, we have $\Delta_H^{-1}=\Delta_{H'}=\Delta_H$, hence $\Delta_H=1$. If $\Delta_H=1$, the standard subspace $H$ is pointwise invariant under the conjugation $J_H$ and therefore maximal abelian by Proposition~\ref{proposition:AbelianSpaces}.
\end{proof}

 We now have a closer look at the canonical dynamics defined by a standard subspace and its generator. As $0$ is not an eigenvalue of $\Delta_H$, the following functions of $\Delta_H$ are well-defined.

\Needspace{10\baselineskip}
\begin{definition}[Modular unitaries and modular Hamiltonian]\label{def:ModularHamiltonianAndGroup}Let $H\in\Std(\Hil)$.
    \begin{enumerate}
        \item The {\em modular Hamiltonian of $H$}\index{modular!Hamiltonian}\glsadd{modularHamiltonian}
        \begin{align}
            L_H:=\log\Delta_H.
        \end{align}

        \item The {\em modular unitaries}\index{modular!unitaries}\glsadd{modularGroup} are the unitary operators $e^{it\log\Delta_H}=\Delta_H^{it}$, $t\in\Rl$. This strongly continuous unitary one-parameter group is also called the {\em modular group of $H$}\index{modular!group}.

    \end{enumerate}
\end{definition}

Note that \eqref{eq:JfDelta} implies in particular that the modular conjugation and the modular group commute,
\begin{align}\label{eq:JCommutesWithModularGroup}
    J_H\Delta_H^{it}
    =
    \Delta_H^{it}J_H,\qquad t\in\Rl.
\end{align}
We may therefore think of a standard subspace $H$ as a dynamical system with internal dynamics given by the group generated by the modular Hamiltonian $\log\Delta_H$, and $J_H$ a compatible conjugation.

The symmetry of the spectrum of $\Delta_H$ implies that the spectrum of the modular Hamiltonian $\log\Delta_H$ is symmetric around $0$. In particular, it is not positive or negative unless $H$ is maximal abelian and $\Delta_H=1$, whence $\log\Delta_H=0$.

We also observe that any vector $h\in H$ has balanced spectral support w.r.t. $\log\Delta_H$ because if $E^{\log(\Delta_H)}(I)h=0$ for some Borel set $I\subset\Rl$ and $h\in H$, then
\begin{align}
    E^{\log\Delta_H}(-I)h
    =
    E^{\log\Delta_H}(-I)S_H h
    =
    S_HE^{\log\Delta_H}(I)h
    =0
\end{align}
 vanishes as well. The modular Hamiltonian $L_H=\log\Delta_H$ has non-positive expectation values on $H\cap\dom L_H$.

Nonetheless, the generator of the modular group as an operator on $H$ itself will turn out to be negative, a property that is related to {\em passivity}. We refer to the Notes and Remarks in Section~\ref{section:NotesAndRemarksChapter1} for an explanation of this physics terminology.

For now, we introduce the concept of passivity in our current context.

\begin{definition}[Passivity]\label{def:passivity}\index{passivity}
 Let $H\in \Std(\H)$. A one-parameter unitary group $U(t):=e^{itA}$ with self-adjoint generator $A$ is said to be passive with respect to $H$ if
  \begin{enumerate}
		\item $U(t)H=H$ for every $t\in\Rl$,
		\item $\langle h,Ah\rangle\leq0$ for all $h\in \dom(A)\cap H$.
	\end{enumerate}
 \end{definition}

\index{passivity of $\Delta_H$ (Prop.~\ref{prop:PassivityOfDelta})}
\begin{proposition}[Passivity of the modular Hamiltonian]\label{prop:PassivityOfDelta}
 Let $H\in\Std(\Hil)$. Then its modular group is passive w.r.t. $H$, i.e.
 \begin{align}
        \langle h,(\log\Delta_H) h\rangle\leq0,\qquad h\in H\cap\dom(\log\Delta_H),
    \end{align}
 and $\langle h,(\log\Delta_H) h\rangle=0$ if and only if $h\in Z(H)$.
 \end{proposition}
\begin{proof}
    That $\Delta_H^{it}$ keeps $H$ globally invariant is the content of Theorem \ref{theorem:Tomita}. Let $h\in H\cap\dom(\log\Delta_H)$. Without loss of generality, we may assume $\|h\|=1$. By Jensen's Inequality applied to the convex function $-\log$, we find
    \begin{align*}
        \langle h,\log\Delta_H\,h\rangle
        &=
        \int \log(\la)\,dE^{\Delta_H}_h(\la)
        \\
        &\leq
        \log\left(\int \la\,dE^{\Delta_H}_h(\la)\right)
        \\
        &=
        \log\langle \Delta_H^{1/2}h,\Delta_H^{1/2} h\rangle
        \\
        &=
        \log\langle J_Hh,J_H h\rangle
        \\
        &=
        \log(\|h\|^2)
        \\
        &=0.
    \end{align*}
 As $-\log$ is strictly convex, the inequality in this estimate is an equality if and only if $h$ has $\Delta_H$-spectral support $\{1\}$, i.e. if and only if $h\in Z(H)$.
 \end{proof}

 Having established that the modular group is passive w.r.t. its own standard subspace, of course also the unitary one-parameter groups with generators $aL_H$, $a\geq0$, have this property. There are however further passive one-parameter groups, and we classify all of them in the following proposition.

We will use the shorthand notation \eqref{eq:SpectralProjectionsDelta} and $L_H:=\log\Delta_H$.

The negative, zero and positive spectral subspaces of $L_H$ are respectively $\Hil^H_<$, $\Hil^H_1$ and $\Hil^H_>$. Thus the subscript $1$ refers to the modular operator $\Delta_H$, even when we discuss its logarithm.

 {
\begin{proposition}[Classification of passive one-parameter groups]\label{prop:passivityAlogH}\label{lemma:passivityAlogH}\label{prop:ClassificationOfPassiveOneParameterGroups}
    Let $H\in\Std(\Hil)$, and let $U(t)=e^{itA}$ be a strongly continuous unitary one-parameter group.
 The following are equivalent.
    \begin{enumerate}
        \item\label{item:passivegroup} $U$ is passive with respect to $H$.
        \item\label{item:passivejointsign} $A$ and $L_H$ commute strongly, $J_HAJ_H=-A$, and $AL_H\geq0$, where $AL_H=f(A,L_H)$ is the self-adjoint operator defined by the joint functional calculus of $A$ and $L_H$, and $f(a,\ell)=a\ell$.
        \item\label{item:passiveblockform} Relative to the spectral decomposition $\Hil=\Hil_>^H\oplus\Hil^H_1\oplus\Hil^H_<$,
        \begin{align}\label{eq:passivegenerators}
            A=B\oplus A_0\oplus(-J_HBJ_H),\qquad
            \dom A=\dom B\oplus\dom A_0\oplus J_H\dom B,
        \end{align}
        where $B$ is any positive self-adjoint operator on $\Hil^H_>$ strongly commuting with $(L_H)|_{\Hil^H_>}$, and $A_0$ is any self-adjoint operator on $\Hil^H_1$ satisfying $J_1A_0J_1=-A_0$, where $J_1:=(J_H)|_{\Hil^H_1}$.
    \end{enumerate}
    In particular, every choice of $B$ and $A_0$ in~\ref{item:passiveblockform} defines a passive group by $U(t)=e^{itA}$.
\end{proposition}
\begin{proof}
    \ref{item:passivegroup}$\Rightarrow$\ref{item:passivejointsign} By Proposition~\ref{proposition:ModularDataProperties}~\ref{item:JDEquivariance}, the global stability of $H$ under $U(t)$ implies that $U(t)$ commutes with $J_H$ and $\Delta_H^{is}=e^{isL_H}$, $s\in\Rl$. Conversely, these commutation relations imply that $U(t)$ commutes with $S_H$ and preserves $H=\Fix S_H$. By Stone's Theorem, invariance is therefore equivalent to strong commutation of $A$ and $L_H$ together with $J_HAJ_H=-A$. By strong commutation, the spectral projections of $A$ and $L_H$ commute, and we set
    \begin{align*}
        F_{n,m}:=E^A([-n,n])E^{L_H}([-m,m]),\qquad
        \CD:=\bigcup_{n,m\geq1} F_{n,m}\Hil \subset \dom(A)\cap\dom(L_H).
    \end{align*}
    Each range lies in $\dom A\cap\dom L_H$. The projections $F_n:=F_{n,n}$ increase strongly to $1$ and commute with both generators. Thus $F_{n}\xi\to\xi$ in the graph norm of each generator for every $\xi$ in its respective domain, and we conclude that $\CD$ is a common core for $A$ and $L_H$.

    Now, since $\CD\subset \dom(A)\cap\dom(L_H)\cap \dom(S_H)$, for all $\xi\in \CD$ we have that $(1+S_H)\xi\in H$, and passivity yields
    \begin{align*}
    	0&\geq\ip{(1+S_H)\xi}{A(1+S_H)\xi} \qquad \text{and}\\
    	0&\geq\ip{(1+S_H)i\xi}{A(1+S_H)i\xi}=\ip{(1-S_H)\xi}{A(1-S_H)\xi}.
    \end{align*}
    Therefore, summing up the two equations above,
    \begin{align*}
    	0&\geq \ip{\xi}{A\xi}+\ip{S_H\xi}{AS_H\xi}\\
    	&=\ip{\xi}{A\xi}+\ip{J_H\Delta_H^{\nicefrac{1}{2}}\xi}{AJ_H\Delta_H^{\nicefrac{1}{2}}\xi}\\
    	&=\ip{\xi}{A\xi}-\ip{J_H\Delta_H^{\nicefrac{1}{2}}\xi}{J_HA\Delta_H^{\nicefrac{1}{2}}\xi}\\
    	&=\ip{\xi}{A\xi}-\ip{\xi}{A\Delta_H\xi}\\
    	&=\ip{\xi}{A(1-\Delta_H)\xi}.
    \end{align*}

    Since, $F_n\H\subset \CD$, the calculation above implies that $a(1-e^\ell)\leq0$ almost everywhere for the joint spectral measure of $(A,L)$ on $[-n,n]^2$. This implies $a\ell\geq0$ almost everywhere, namely $AL_H\geq0$.

    \ref{item:passivejointsign}$\Rightarrow$\ref{item:passivegroup} As in the previous argument, we have that $A(1-e^{L_H})$ is negative on $F_n\Hil$. For $h\in H\cap F_n\Hil$, we have $S_Hh=h$, and therefore $2\langle h,Ah\rangle=\langle h,A(1-e^{L_H})h\rangle\leq0$. For arbitrary $h\in H\cap\dom A$, the vectors $F_nh\in H\cap F_n\Hil$ converge to $h$ in the graph norm of $A$. Passing to the limit proves $\langle h,Ah\rangle\leq0$, and hence~\ref{item:passivegroup}.

    \ref{item:passivejointsign}$\Rightarrow$\ref{item:passiveblockform} The three spectral subspaces of $L_H$ reduce $A$. On $\Hil^H_>$ the condition $a\ell\geq0$ is equivalent to positivity of $B:=A|_{\Hil^H_>}$. The exchange relation $J_HL_HJ_H=-L_H$ shows that $J\Hil^H_>=\Hil^H_<$ and $J_H\Hil^H_1=\Hil^H_1$. Thus $J_HAJ_H=-A$ determines the restriction to $\Hil^H_<$ as $-J_HBJ_H$, with domain $J_H\dom B$, and gives $J_1A|_{\Hil^H_1}J_1=-A|_{\Hil^H_1}$. This proves~\ref{item:passiveblockform}.

    \ref{item:passiveblockform}$\Rightarrow$\ref{item:passivejointsign} The self-adjoint direct sum~\eqref{eq:passivegenerators} strongly commutes with $L_H$, satisfies $J_HAJ_H=-A$ and has $a\ell\geq0$ on each summand. It therefore satisfies~\ref{item:passivejointsign}.
\end{proof}

Note that the freedom in $A_0$ amounts to an arbitrary strongly continuous orthogonal one-parameter group on the real Hilbert space $Z(H)$, extending complex-linearly to $\Hil^H_1=Z(H)+iZ(H)$. Its expectation values on $Z(H)\cap\dom A_0$ vanish. In particular, if $H$ is maximal abelian, every unitary one-parameter group preserving $H$ is passive.
 }

\medskip

 This analysis shows in particular that the modular group is not characterized by passivity. In Section~\ref{section:Operations}, we will see that a stronger version called complete passivity does characterize the modular group up to rescalings of its generator when $H$ is not maximal abelian. In Section~\ref{section:Characterization}, we will see a different characterization of the modular group based on holomorphic properties.

For now, we return to the general question how the modular data $J_H,\Delta_H$ of a standard subspace $H$ inform us about properties of $H$. As a closed real subspace, $H$ is uniquely determined by the real orthogonal projection \eqref{eq:EH}
 \begin{align}
    E_H:\Hil=H+H^\rperp\to\Hil,\qquad h+h^\perp\mapsto h.
\end{align}
On the other hand, it is uniquely determined by its modular data $J_H,\Delta_H$. Hence it is clear that~$E_H$ can be expressed in terms of the modular data. We now derive such a formula for $E_H$ together with a related convenient property of the bounded operator $(1+\Delta_H)^{-1}$ (part~\ref{item:ConvenientProperty}) and an expression of the modular data in terms of projections (part~\ref{item:JviaE}).

\begin{proposition}[{Real orthogonal projection in terms of modular data}]\label{proposition:orthogonalprojectionmodulardata}
 Let $H\in \Std(\Hil)$ and $(J_H,\Delta_H)$ its modular data.
 \begin{enumerate}
        \item\label{item:ConvenientProperty} $\Ran(1+\Delta_H)^{-1}\subset \dom(S_H)$ and $(1+\Delta_H)^{-1} H'\subset H$.
        \item\label{item:EHModular} The real orthogonal projection $E_H$ onto $H$ satisfies
        \begin{equation}
            E_H =(1+S_H)(1+\Delta_H)^{-1}.
        \end{equation}
        \item\label{item:JviaE} $J_H=\sign(E_H-E_{iH})$ and $(1+\Delta_H)^{-1}=\frac12(E_H+E_{iH})$.
    \end{enumerate}
\end{proposition}
\begin{proof}
    \ref{item:ConvenientProperty} The first claimed inclusion holds because
    \begin{align*}
        \Ran(1+\Delta_H)^{-1}\subset \dom(1+\Delta_H)\subset\dom(\Delta_H^{{1/2}}) = \dom(S_H),
    \end{align*}
    and the second inclusion can easily be checked using the commutation relations stated in Proposition \ref{proposition:ModularDataProperties}~\ref{item:JDeltaCommutation} as follows: For any vector $h'\in H'$,
    \begin{equation*}
        S_H (1+\Delta_H)^{-1} h'=\frac{\Delta_H^{-{1/2}}}{1+\Delta_H^{-1}} J_Hh'=\frac{\Delta_H^{{1/2}}}{1+\Delta_H} J_Hh'=(1+\Delta_H)^{-1} S_H^\ast h'=(1+\Delta_H)^{-1} h'.
    \end{equation*}
    Hence, $(1+\Delta_H)^{-1} h'\in H$.

    \ref{item:EHModular} Recall that $H^\rperp=iH'$. Hence the orthogonal decomposition of any vector $\xi\in \Hil$ is of the form $\xi=h+ih'$ with $h=E_H\xi\in H$ and $h'\in H'$. Applying the operator $(1+\Delta_H)^{-1}\in B(\Hil)$ to $\xi$ we get with the help of \ref{item:ConvenientProperty}
    \begin{align}
    \begin{aligned}\label{eq:ConvenientProperty}
        (1+\Delta_H)^{-1}\xi&=(1+\Delta_H)^{-1}h+i(1+\Delta_H)^{-1}h'\\
        S_H(1+\Delta_H)^{-1}\xi&=S_H(1+\Delta_H)^{-1}h-i(1+\Delta_H)^{-1}h'.
    \end{aligned}
    \end{align}
    Adding these two equations gives
    \begin{align*}
        (1+S_H)(1+\Delta_H)^{-1}\xi
        &=
        (1+S_H)(1+\Delta_H)^{-1}h
        \\
        &=
        \left(\frac{1}{1+\Delta_H}+\frac{1}{1+\Delta_H^{-1}}\right)h
        =h
        =E_H\xi.
    \end{align*}

 \ref{item:JviaE} The operator $E_H-E_{iH}$ is real self-adjoint and has trivial kernel because $H$ is separating. Hence the phase of its polar decomposition is $\sign(E_H-E_{iH})$. On the other hand
 \begin{align*}
        E_H-E_{iH}
        &=
        E_H+iE_{H}i
        \\
        &=
        (1+S_H)(1+\Delta_H)^{-1}+i(1+S_H)(1+\Delta_H)^{-1}i
        \\
        &=
        J_H\cdot 2\Delta_H^{1/2}(1+\Delta_H)^{-1}.
    \end{align*}
    As $J_H$ is real unitary and $2\Delta_H^{1/2}(1+\Delta_H)^{-1}$ is positive, the uniqueness of the polar decomposition implies the result.

 The second equation follows analogously by adding instead of subtracting the two projections $E_H$ and $E_{iH}$.
 \end{proof}

In case $H$ is factorial, it is independent from its symplectic complement $H'$ in the sense $H\cap H'=\{0\}$ (Def.~\ref{def:IndependentSubspaces}), and another natural projection associated with $H$ appears.

\begin{definition}[Cutting projection]\label{definition:cuttingprojection}
    Let $H\in \Std(\Hil)$ be factorial. The {\em cutting projection}\index{projection!cutting (Def.~\ref{definition:cuttingprojection})} of $H$ is \glsadd{cutting}
    \begin{align}
        P_{H}
        :=
        P_{H,H'}:H+H'\to\Hil,\qquad h+h'\mapsto h.
    \end{align}
\end{definition}

 The cutting projection has its name because in relevant examples such as the half-line space of the $U(1)$-current, it amounts to cutting (multiplying with a characteristic function). This will be explained in Proposition~\ref{proposition:U1Cutting}.

Besides also being a projection, the properties of $P_H$ are quite different from those of $E_H$.

\begin{proposition}[Properties of the cutting projection]\label{proposition:propertiescutting}
Let $H\in \Std(\Hil)$ be factorial.
\begin{enumerate}
 \item\label{item:PHdense} $P_H$ is a densely defined closed real-linear projection,
 \item\label{item:PHformula} $P_H =\overline{(1+S_H)(1-\Delta_H)^{-1}}$  (closure),
	\item\label{item:PHstar} $P_H^\ast=-iP_H i=P_{iH}$,
	\item\label{item:PHinvariant} \label{item:cuttingprojectioncommutemodulargroup}$\Delta_H^{it}P_H=P_H\Delta_H^{it}$,
	\item\label{item:PHbounded} $P_H$ is bounded if and only if $\Delta_H$ has a spectral gap, i.e. if and only if $(1-\Delta_H)^{-1}\in B(\Hil)$.
\end{enumerate}
\end{proposition}
\begin{proof}
    \ref{item:PHdense} As $H$ is factorial, we have $(H+H')'=H'\cap H=\{0\}$, so $P_H$ is densely defined. It is closed because $H$ and $H'$ are closed (Proposition~\ref{proposition:twosubspacesprojection}~\ref{item:2spacesprojectionclosed}).

    \ref{item:PHformula} As in the proof of Proposition~\ref{proposition:orthogonalprojectionmodulardata}, we consider the general equations \eqref{eq:ConvenientProperty}, but now for $\xi=h+h'$ (no factor $i$ in front of $h'$). Subtracting these two equations results in

\begin{equation}\label{eq:calccuttingprojection}
	(1-S_H)(1+\Delta_H)^{-1}\xi=(1-S_H)(1+\Delta_H)^{-1}h=\frac{1-\Delta_H}{1+\Delta_H}h.
\end{equation}
 Using again that $H$ is factorial, Proposition~\ref{proposition:CenterAndMaximalAbelianness} ensures that $1\notin\specp(\Delta_H)$ and $(1-\Delta_H)^{-1}$ exists as a densely defined closed operator. Therefore,
 \begin{equation}\label{eq:orthogonalprojectionmodulardata0}
	P_H\xi=h=\left(\frac{1+\Delta_H}{1-\Delta_H}\right)(1-S_H)(1+\Delta_H)^{-1}\xi.
\end{equation}

Let us abbreviate $Q_H:=\left(\frac{1+\Delta_H}{1-\Delta_H}\right)(1-S_H)(1+\Delta_H)^{-1}$. One can check that $\xi \in \dom Q_H$ implies
$Q_H\xi\in H$ and $\left(1-Q_H\right)\xi\in H'$. Thus,
 \begin{equation}\label{eq:orthogonalprojectionmodulardata}
	P_H =Q_H=
	(1+S_H)\left(\frac{1+\Delta_H}{1-\Delta_H}\right)(1+\Delta_H)^{-1}.
\end{equation}

Finally, notice that $(1+S_H)\left(\frac{1+\Delta_H}{1-\Delta_H}\right)(1+\Delta_H)^{-1}\restr{\dom((1-\Delta_H)^{-1})}=(1+S_H)(1-\Delta_H)^{-1}$ and $\dom((1-\Delta_H)^{-1})$ is a core for this operator due to Proposition \ref{proposition:twosubspacesprojection}~\ref{item:core2spaceprojection}.

\ref{item:PHstar} According to Proposition \ref{proposition:twosubspacesprojection}~\ref{item:adj2spaceprojection}, we have
\begin{align*}
    P_H^\ast=P_{H,H'}^\ast=P_{(H')^\rperp,H^\rperp}=P_{iH,iH'}=P_{iH}=-iP_H i.
\end{align*}

\ref{item:PHinvariant} It is enough to recall that the Tomita-Takesaki Theorem implies $\Delta_H^{it}H\subset H$ and $\Delta_H^{it}H'\subset H'$ for all $t\in \Rl$. Therefore, $P_H\Delta_H^{it}(h+h')=\Delta_H^{it}h=\Delta_H^{it}P_H(h+h')$ for all $h\in H$ and $h'\in H'$.

\ref{item:PHbounded} {
If $1\notin\spec(\Delta_H)$, then $(1-\Delta_H)^{-1}$ and $\Delta_H^{1/2}(1-\Delta_H)^{-1}$ are bounded. Hence $(1+S_H)(1-\Delta_H)^{-1}$ is bounded and so is $P_H$ by part~\ref{item:PHformula}. Conversely, notice that for all $\xi \in\dom((1-\Delta_H)^{-1})$ we have $P_H\xi=(1+S_H)(1-\Delta_H)^{-1}\xi$. Therefore, since $(1-\Delta_H)^{-1}$ is complex-linear and $S_H(1-\Delta_H)^{-1}$ is antilinear, we have
    \begin{align*}
        (P_H-iP_Hi)\xi=2(1-\Delta_H)^{-1}\xi,\qquad \forall \xi\in\dom(1-\Delta_H)^{-1}.
    \end{align*}
    If $P_H$ is bounded, this shows that also $(1-\Delta_H)^{-1}$ is bounded. Thus $1\notin\spec(\Delta_H)$.
}
\end{proof}

An explicit example of a cutting projection can be found in the contex of the $U(1)$-current in Lemma~\ref{lemma:CutFunctions}, which also explains the name ``cutting''. The cutting projection will also appear in the context of entropy in Section~\ref{section:Entropy}.

\subsection{Constructions and examples of standard subspaces}\label{section:ConstructionsAndExamples}

The fundamental results of the previous section give us also new methods of constructing examples of standard subspaces. In this section, we will consider two classes of examples. The idea is that in view of the bijection between standard subspaces and closed antilinear involutions established in Proposition~\ref{proposition:StandardSubspacesVsClosedInvolutions}, standard subspaces may be defined by prescribing their Tomita operator (a closed antilinear involution). The following corollary to Proposition~\ref{proposition:StandardSubspacesVsClosedInvolutions} presents two simple but useful variations of this idea.

\begin{corollary}[Standard subspaces and one-parameter groups]\label{corollary:StandardSubspacesAndOneParameterGroups}
    \leavevmode
    \begin{enumerate}
 \item\label{item:JDelta}\index{classification of standard subspaces!by modular data} Let $J$ be a conjugation and $\Delta$ a positive self-adjoint non-singular operator such that the modular relation $J\Delta J=\Delta^{-1}$ holds. Then
 \begin{align}
            S:=J\Delta^{1/2}
        \end{align}
        is a closed antilinear involution, and $H:=\Fix(S)$ is the standard subspace with modular data $J_H=J$, $\Delta_H=\Delta$.
        \item\label{item:JU} Let $U(t)=e^{itP}$, $t\in\Rl$, be a strongly continuous unitary one-parameter group with generator $P$, and $J$ a conjugation commuting with $U(t)$, $t\in\Rl$. Then
        \begin{align}
            S:=Je^{P/2}
            =
            JU(-\tfrac{i}{2})
        \end{align}
        is a closed antilinear involution, and $H:=\Fix(S)$ a standard subspace with modular data $J_H=J$, $\Delta_H=e^{P}$.
    \end{enumerate}
\end{corollary}
\begin{proof}
    \ref{item:JDelta} The operator $S:=J\Delta^{1/2}$ is closed because $\Delta$ is closed and $J$ is bounded, and antilinear because $\Delta$ is linear and $J$ is antilinear. It is an involution because of $J\Delta J=\Delta^{-1}$. The claim now follows from Proposition~\ref{proposition:StandardSubspacesVsClosedInvolutions}.

 \ref{item:JU} By Stone's Theorem, $P$ is self-adjoint, so $e^{P/2}$ is positive, self-adjoint, and non-singular. Furthermore, $\dom P$ consists precisely of those vectors $\xi$ for which $t\mapsto U(t)\xi$ is strongly differentiable. Since $J$ commutes with $U(t)$ and is antiunitary, this implies $J\dom P=\dom P$. Let $\xi\in\dom P$. Then differentiating $Je^{itP}J\xi=e^{itP}\xi$ at $t=0$ yields $-iJPJ\xi=JiPJ\xi=iP\xi$, namely $JP=-PJ$. Thus $Je^{P/2}J=e^{-P/2}$, and we are in the situation of part \ref{item:JDelta}.
 \end{proof}

 We illustrate these ideas with two examples. The first example defines standard subspaces by prescribing the modular group and modular conjugation.

\begin{example}[Standard subspaces from measure spaces]\label{example:StandardSubspacesFromMeasures}
    Let $\mu$ be a measure on $\Rl$ that is symmetric in the sense
    \begin{align}
        r_*\mu=\mu,\qquad r(x)=-x.
    \end{align}
    On the $L^2$-space $\Hil:=L^2(\Rl,\mu)$, we consider the strongly continuous unitary one-parameter group $(U(t)f)(x)=e^{itx}f(x)$, $t\in\Rl$. Thanks to the symmetry assumption on $\mu$, the operator $(Jf)(x)=\overline{f(-x)}$ is a well-defined conjugation on $L^2(\Rl,d\mu)$. It clearly commutes with $U(t)$, $t\in\Rl$. Hence there exists a unique standard subspace $H\subset\Hil$ with $\Delta_H^{it}=U(t)$ and $J_H=J$, namely $H=\Fix(JU(-\frac{i}{2}))$ (Corollary~\ref{corollary:StandardSubspacesAndOneParameterGroups}~\ref{item:JU}). To describe it, we note
    \begin{align*}
        \dom U(-\tfrac{i}{2})
        =
        \left\{f\in L^2(\Rl,\mu)\col \int_\Rl e^{x}|f(x)|^2d\mu(x)<\infty\right\}.
    \end{align*}
    Thus
    \begin{align*}
        H
        =
        \left\{
            f\in L^2(\Rl,\mu) \col \int_\Rl e^{x}|f(x)|^2d\mu(x)<\infty,\quad f(-x)=e^{x/2}\overline{f(x)},\; \text{for } \mu\text{-almost all }\,x
        \right\}.
    \end{align*}
\end{example}

 In this example, the modular operator is multiplicity-free in the sense that $\{\Delta_H^{it}\col t\in\Rl\}$ has abelian commutant. If $\mu$ is $\sigma$-finite, the modular Hamiltonian $\log\Delta_H$ has a cyclic vector. A classification of all such multiplicity-free standard subspaces will be carried out in Section~\ref{section:MultiplicityFreeSpaces}.

\begin{example}[The $\boldsymbol{U(1)}$-current standard subspace in the strip picture]\label{example:U1current}
    Consider the standard subspace $H\in\Std(L^2(\Rl))$ given by choosing the measure $\mu$ to be Lebesgue measure in the previous example, and let $\EuF:L^2(\Rl)\to L^2(\Rl)$ denote the unitary Fourier transform. Then $\tilde H:=\EuF^{-1} H$ is a standard subspace in $L^2(\Rl)$, with modular group and conjugation
    \begin{align}
        (\Delta_{\tilde H}^{it}f)(\te)=f(\te-t),\qquad (J_{\tilde H}f)(\te)=\overline{f(\te)},\qquad \te\in\Rl.
    \end{align}
    This follows directly by Fourier transforming the modular data of the previous example, taking into account Proposition~\ref{proposition:ModularDataProperties}~\ref{item:JDEquivariance}.

    The description of the functions in this standard subspace requires methods from complex analysis and will be presented after we have introduced the KMS condition in Section~\ref{section:U1current}. The name ``$U(1)$-current'' derives from its role in quantum field theory, see the notes and remarks in Section~\ref{section:NotesAndRemarksChapter1}.
\end{example}

 The next example defines standard subspaces by prescribing the modular operator and modular conjugation. This example was the original motivation of Tomita for developing what is now called Tomita-Takesaki modular theory (see notes and remarks in Section~\ref{section:NotesAndRemarksChapter1}).

In preparation, we recall some facts from harmonic analysis \cite{DeitmarEchterhoff:2009}: Any locally compact group $G$ admits a non-zero left-invariant outer Radon measure, unique up to positive multiples. It is called {\em Haar measure} $\mu$. As also $\mu_g(\cdot):=\mu(\cdot g)$ is a Haar measure, there exists a function $\delta_G:G\to(0,\infty)$, called the {\em modular function of $G$}\index{modular!function}, such that $\mu_g=\delta_G(g)\cdot \mu$. The modular function is a continuous group homomorphism from $G$ into the multiplicative group $\Rl_+$. Groups $G$ with $\delta_G(g)=1$ for all $g\in G$ are called {\em unimodular}, these are groups in which every left-invariant Haar measure is also right-invariant. In particular, abelian and compact groups are unimodular. For $\mu$-integrable functions $f:G\to\Cl$ on a general locally compact group $G$, one has
\begin{align}\label{eq:HaarSubstitution}
    \int_G f(g^{-1})\delta_G(g^{-1})\,d\mu(g)
    =
    \int_G f(g)\,d\mu(g).
\end{align}

\begin{example}[Standard subspaces from locally compact groups]\label{example:HaarMeasure}
 {\em
    Let $G$ be a locally compact group with left-invariant Haar measure $\mu$ and modular function $\delta_G:G\to\Rl_+$. In the complex Hilbert space $\Hil:=L^2(G,\mu)$, let
    \begin{align}
        \CD
        &:=
        \left\{f\in\Hil\col \int_G \delta_G(g)|f(g)|^2\,d\mu(g)<\infty\right\},
        \\
        H
        &:=
        \{f\in\CD \col f(g)=\delta_G(g^{-1})\overline{f(g^{-1})} \quad \text{for } \mu\text{-almost all }\;g\in G\}
        \label{eq:HaarStandardSubspace}.
    \end{align}
    Then $H$ is a standard subspace, and its modular data are
    \begin{align}
        (J_Hf)(g)
        &=
        \sqrt{\delta_G(g^{-1})}\,\overline{f(g^{-1})},
        \\
        (\Delta_Hf)(g)
        &=
        \delta_G(g)f(g).
    \end{align}
    $H$ is maximal abelian if and only if $G$ is unimodular.
    }
 \end{example}
\begin{proof}
    Let $(Jf)(g):=\sqrt{\delta_G(g^{-1})}\,\overline{f(g^{-1})}$ and $(\Delta^{1/2} f)(g):=\sqrt{\delta_G(g)}f(g).$ Because $\delta_G$ is a homomorphism, $J$ is an antilinear involution. Let $f_1,f_2\in\Hil$. Using \eqref{eq:HaarSubstitution}, we find
    \begin{align*}
        \langle Jf_1,Jf_2\rangle
        &=
        \int_G \delta_G(g^{-1})f_1(g^{-1})\overline{f_2(g^{-1})}\,d\mu(g)
        \\
        &=
        \int_G f_1(g)\overline{f_2(g)}\,d\mu(g)
        =
        \langle f_2,f_1\rangle,
    \end{align*}
    and conclude that $J$ is a conjugation.

 As a multiplication operator, $\Delta^{1/2}$ is self-adjoint on its natural domain $\CD$ and positive non-singular because $\delta_G$ maps into $(0,\infty)$. The modular relation \eqref{eq:JDeltaCommutation} follows by direct computation:
 \begin{align*}
        (J\Delta^{1/2} Jf)(g)
        &=
        \sqrt{\delta_G(g^{-1})}\overline{(\Delta^{1/2} Jf)(g^{-1})}
        =
        \delta_G(g^{-1})\overline{(Jf)(g^{-1})}
        \\
        &=
        \delta_G(g^{-1})\sqrt{\delta_G(g)}f(g)
        =
        \delta_G(g)^{-1/2}f(g)
        =
        (\Delta^{-1/2} f)(g).
    \end{align*}
 Thus $S:=J\Delta^{1/2}$ is a closed antilinear involution, and its fixed-point space $\Fix(S)$, which we see to coincide with $H$ \eqref{eq:HaarStandardSubspace}, is a standard subspace with the claimed modular data by Corollary~\ref{corollary:StandardSubspacesAndOneParameterGroups}. The statement about unimodularity is immediate from Proposition~\ref{proposition:CenterAndMaximalAbelianness}.
 \end{proof}

This example explains the origin of the name ``modular theory'' and the notation~$\Delta$ for the modular operator: Both derive from the modular function, which is typically denoted~$\Delta$.

\subsection{Positive contractions and the graph model}\label{section:ContractionsAndGraphs}

In this section we begin with another construction method for standard subspaces, which will turn into a general description of $\Std(\Hil)$ in terms of positive contractions and graphs. By positive contraction we mean a positive non-singular contraction on a Hilbert space $\Hil$, i.e. an element of $\E_+(\Hil)$\footnote{In physics, an operator $B\in B(\Hil)$ satisfying $0\leq B\leq1$ is called an \emph{effect}, hence the symbol $\E_+(\Hil)$ for the injective (strictly positive) effects on $\Hil$.}\glsadd{EplusH}
\begin{align}\label{eq:EplusH}
    \E(\Hil)
    &:=
    \{B\in B(\Hil) \col 0\leq B\leq1\},\\
    \E_+(\Hil)
    &:=
    \{B\in\E(\Hil) \col \ker B=\{0\}\}.
\end{align}

\begin{example}[Standard subspaces from positive non-singular contractions]\label{example:PositiveContractions}
 Let $B\in\E_+(\Hil)$ be a positive contraction, and split $\Hil$ according to $\Hil=\Kil_1\oplus\Kil_<$, where $\Kil_1:=\ker(B-1)$ is the fixed-point space of $B$ and $\Kil_<:=\Kil_1^\perp$ its orthogonal complement. Furthermore, let $\bar\Kil_<$ denote the complex conjugate Hilbert space of $\Kil_<$, and pick a conjugation $J_1$ on $\Kil_1$. On
 \begin{align}
        \Hil_B:=\Kil_<\oplus\bar\Kil_<\oplus\Kil_1,
    \end{align}
    define
    \begin{align}\label{eq:DeltaB}
        \Delta_B
        &:=
        B_<\oplus\bar B_<^{-1}\oplus1,
        \\
        J_{B} : \xi_1\oplus \bar\xi_2\oplus \eta
        &\mapsto \xi_2\oplus \bar\xi_1\oplus J_1\eta,
    \end{align}
    where $B_<=B|_{\Kil_<}$ is the restriction, and $\bar B_<$ the conjugate operator, given by $\bar B_<\bar \xi=\overline{B_<\xi}$.

    Then $\Delta_B$ is positive non-singular and $J_{B}$ is a conjugation on $\Hil_B$. As $J_{B} \Delta_B J_{B} = \Delta_B^{-1}$,
    \begin{align}\label{eq:HB}
        H_{B}:=\Fix(J_{B}\Delta_B^{1/2})
    \end{align}
     is a standard subspace of $\Hil_B$ by Corollary~\ref{corollary:StandardSubspacesAndOneParameterGroups}~\ref{item:JDelta}.
\end{example}

The standard subspace $H_B$ \eqref{eq:HB} is factorial if and only if $B$ does not have eigenvalue~$1$. In this case, it does also not depend on a choice of conjugation on $\Kil_1=\ker(B-1)$, but only on~$B$. In the general case, choosing two different conjugations $J_1,\tilde J_1$ on $\ker(B-1)$ results in two unitarily equivalent standard subspaces because all conjugations (all maximal abelian standard subspaces) are unitarily equivalent (Prop.~\ref{proposition:AbelianSpaces}). This justifies our notation $H_B$, $J_B$ instead of $H_{B,J_1}$, $J_{B,J_1}$, up to unitary equivalence.

 The next theorem takes this observation one step further and shows that unitary equivalence classes of standard subspaces are labelled by unitary equivalence classes of non-singular positive contractions. As the space on which a positive contraction lives has to be doubled up in the sense of Example~\ref{example:PositiveContractions} to give the Hilbert space in which the corresponding standard subspace lives, we will consider the family (category) of {\em all} complex Hilbert spaces. We will write $\bStd/\U$ for the family of equivalence classes of standard subspaces in arbitrary Hilbert spaces, where unitarily equivalent standard subspaces are identified, and $\E_+/\U$ for the family of all injective positive contractions on arbitrary Hilbert spaces, where unitarily similar contractions are identified. The category of standard subspaces in all complex Hilbert spaces will be studied in more detail in Section~\ref{section:Operations}.

\begin{theorem}[Standard subspaces and positive contractions up to unitary equivalence]\label{theorem:StdAndContractions}\booktheoremindex{theorem:StdAndContractions}{Standard subspaces and positive contractions up to unitary equivalence}\glsadd{restrictedModular}\index{positive contraction!classification of standard subspaces}\index{classification of standard subspaces!by positive contractions}
 Let $H\in\Std(\Hil)$. Using the spectral subspaces from \eqref{eq:SpectralProjectionsDelta}, put
 \begin{align}\label{eq:StdToE}
        \Delta_{H,\leq}:=\Delta_H|_{\Hil^H_\leq}\in\E_+(\Hil^H_\leq).
    \end{align}
    \begin{enumerate}
        \item\label{item:PC-equivariance} {
        For every $V\in\AU(\Hil)$,
        \begin{align}\label{eq:RestrictedModularCovariance}
            \Hil^{VH}_\leq=V\Hil^H_\leq,
            \qquad
            \Delta_{VH,\leq}=V\Delta_{H,\leq}V^{-1},
        \end{align}
 where the latter operator acts on $V\Hil^H_\leq$. Thus the assignment $H\mapsto(\Hil^H_\leq,\Delta_{H,\leq})$ is equivariant. }

        \item\label{item:PC-quotient} Dividing by the action of the unitary groups, the map
        \begin{align}
            \bStd/\U&\to \E_+/\U,
            \qquad
            [H]\mapsto [\Delta_{H,\leq}],
        \end{align}
        is well-defined and bijective.

        \item\label{item:PC-HB} Every standard subspace is unitarily equivalent to one of the form $H_B$ \eqref{eq:HB}.
    \end{enumerate}
\end{theorem}

 \begin{proof}
    {
    The restriction $\Delta_{H,\leq}$ is bounded by $1$, positive and injective, so it belongs to $\E_+(\Hil^H_\leq)$.

    \ref{item:PC-equivariance} Proposition~\ref{proposition:ModularDataProperties}~\ref{item:JDEquivariance} gives $\Delta_{VH}=V\Delta_HV^{-1}$. Taking its spectral projection for $(0,1]$ gives $E^{\Delta_{VH}}_\leq=VE^{\Delta_H}_\leq V^{-1}$, and hence \eqref{eq:RestrictedModularCovariance}. This holds also for antiunitary $V$, since the characteristic function of $(0,1]$ is real-valued.

    \ref{item:PC-quotient} The same covariance argument for a unitary between two possibly different ambient Hilbert spaces proves that the map on equivalence classes is well-defined. For $H_B$ from Example~\ref{example:PositiveContractions}, we have
    \begin{align*}
        \Hil^{H_B}_\leq=\Kil_<\oplus\{0\}\oplus\Kil_1\subset\Hil_B,
    \end{align*}
    and $\Delta_{H_B,\leq}$ is unitarily equivalent to $B_<\oplus1$ on $\Kil_<\oplus\Kil_1$, hence to $B$. This proves surjectivity.

    We also record explicitly that unitarily equivalent contractions give unitarily equivalent model subspaces. Indeed, if $W$ is unitary and $WBW^{-1}=\widetilde B$, then $W$ maps the $<$ and $1$ spectral subspaces of $B$ onto those of $\widetilde B$. Write $W_<$ and $W_1$ for these restrictions, and choose the conjugation for the second model as $\widetilde J_1=W_1J_1W_1^{-1}$. Then $W_<\oplus\overline{W_<}\oplus W_1$ intertwines the modular data in~\eqref{eq:DeltaB}, and hence maps $H_B$ onto $H_{\widetilde B}$. Other choices of conjugation give equivalent models, as observed before the theorem.

    To prove injectivity and~\ref{item:PC-HB}, it therefore remains to show that $H$ is unitarily equivalent to $H_B$ for $B=\Delta_{H,\leq}$. In this case $\Kil_<=\Hil^H_<$ and $\Kil_1=\Hil^H_1$, and we choose $J_1=J_H|_{\Hil^H_1}$. The spectral decomposition and the definition of $\Hil_B$ give
    \begin{align*}
        \Hil
        &=
        \Hil^H_<\oplus\Hil^H_>\oplus\Hil^H_1,
        \\
        \Hil_{B}
        &=
        \Hil^H_<\oplus\overline{\Hil^H_<}\oplus\Hil^H_1.
    \end{align*}
    Since $J_H\Hil^H_>=\Hil^H_<$, the map $Y:\Hil^H_>\to\overline{\Hil^H_<}$, $Y\xi:=\overline{J_H\xi}$, is unitary. Thus $U=\id\oplus Y\oplus\id:\Hil\to\Hil_B$ is unitary. For $\xi=\xi_<\oplus\overline{\eta_<}\oplus\xi_1\in\Hil_B$, we have
    \begin{align*}
        U^*\xi=\xi_<\oplus J_H\eta_<\oplus\xi_1.
    \end{align*}
    By $J_H\Delta_HJ_H=\Delta_H^{-1}$, this vector belongs to $\dom\Delta_H$ precisely when $\eta_<\in\dom B_<^{-1}$, which is equivalent to $\xi\in\dom\Delta_B$. On this domain,
    \begin{align*}
        U\Delta_HU^*\xi
        &=U(B_<\xi_<\oplus\Delta_HJ_H\eta_<\oplus\xi_1)\\
        &=B_<\xi_<\oplus\overline{B_<^{-1}\eta_<}\oplus\xi_1
        =\Delta_B\xi .
    \end{align*}
    For every $\xi\in\Hil_B$,
    \begin{align*}
        UJ_HU^*\xi
        &=U(\eta_<\oplus J_H\xi_<\oplus J_H\xi_1)\\
        &=\eta_<\oplus\overline{\xi_<}\oplus J_H\xi_1=J_B\xi .
    \end{align*}
    Consequently $US_HU^*=J_B\Delta_B^{1/2}$, including equality of domains, and $U$ maps $H=\Fix S_H$ onto $H_B$. This proves injectivity and part~\ref{item:PC-HB}.
}
\end{proof}

The most important corollary of this theorem is that standard subspaces with unitarily equivalent modular operators are unitarily equivalent, the modular conjugation does not have any effect on the unitary equivalence classes.

 \medskip

If one prefers a classification of standard subspaces in a fixed Hilbert space by a family of operators on that fixed Hilbert space $\Hil$, one may consider $H\mapsto\Delta_H$ as a map onto positive injective operators $A$ satisfying $A\cong A^{-1}$, or $H\mapsto\log\Delta_H$ as a map onto self-adjoint operators that are unitarily similar to their own negative.

\medskip

 As an interesting corollary to this construction, we can now prove the fact that the unitary and (anti-)unitary orbit of any closed real subspace coincide, as already mentioned in Section~\ref{section:RealSubspaces}.

\begin{corollary}[Unitary and antiunitary equivalences of closed real subspaces]\label{corollary:AntiunitaryAndUnitaryImages}
    \leavevmode
    \begin{enumerate}
        \item\label{item:conju} Every closed real subspace $H\in\Real(\Hil)$ admits a conjugation $C\in\Con(\Hil)$ such that $CH=H$.

 \item\label{item:AUvsU} If $K,H\in\Real(\Hil)$ are antiunitarily equivalent, i.e. $K=VH$ for some antiunitary operator $V$, then they are unitarily equivalent, i.e. $K=UH$ for some unitary operator~$U$.\looseness=-1 \end{enumerate}
\end{corollary}
\begin{proof}
 \ref{item:conju} We use the complex orthogonal decomposition $H=\Hil_1\oplus\{0\}\oplus H_0$, with $H_0$ standard, see Lemma~\ref{lemma:DecomposeClosedRealSubspace}. As the first two summands are clearly invariant under conjugations, we may restrict to the case that $H$ is a standard subspace. By Theorem~\ref{theorem:StdAndContractions}, we may assume $H=H_B$ for some $B\in\E_+(\Kil)$ on a complex Hilbert space $\Kil$. As a bounded self-adjoint operator, there exists a measure space $(X,\mu)$ such that $B$ is unitarily equivalent to $(\tilde B f)(x)=\varphi(x)f(x)$, $f\in L^2(X,\mu)$, for some real function $\varphi\in L^\infty(X,\mu)$. We emphasize that this is also true for non-separable $\Kil$, in which case $\mu$ may fail to be $\sigma$-finite \cite[Thm.~3.11]{Kubrusly:2020}. Pointwise complex conjugation $\tilde C$ on $L^2(X,\mu)$ is a conjugation commuting with $\tilde B$, and this results in a conjugation fixing $H$.

    \ref{item:AUvsU} is a direct consequence of \ref{item:conju}.
\end{proof}

The classification of standard subspaces in terms of positive contractions has taught us that i) conjugate spaces naturally show up, and ii) the modular conjugation $J_H$ of $H$ plays a less significant role than the modular operator $\Delta_H$: Changing $J_H$ to another conjugation $\tilde J_H$ satisfying the modular relation $\tilde J_H\Delta_H\tilde J_H=\Delta_H^{-1}$ leads to a standard subspace $\tilde H$ with modular data $\tilde J_H,\Delta_H$ that is unitarily equivalent to $H$.

With these two lessons learned, we now describe an alternative perspective that exchanges the conjugate spaces with a modified complex structure. We will consider the space $\widehat\Hil=\Hil\oplus\Hil$ as a real vector space, and equip it with the {\em non-canonical} complex structure and scalar product (see also Section~\ref{section:Polarizers} for more information on complex structures and complexifications)
\begin{align}\label{eq:IComplexStructure}
    I(\xi\oplus\eta)
    &:=
    (i\xi)\oplus(-i\eta),\\
    \langle\xi_1\oplus\xi_2,\eta_1\oplus\eta_2\rangle
    &:=
    \langle\xi_1,\eta_1\rangle+\overline{\langle\xi_2,\eta_2\rangle}.
    \label{eq:DoubledScalarProduct}
\end{align}

\begin{lemma}[The doubled Hilbert space $\boldsymbol{\widehat\Hil}$]
    \label{lemma:DoubledHilbertSpace}
    \leavevmode
    \begin{enumerate}
        \item\label{item:DoubledHilbertSpace} $\widehat\Hil$ is a complex Hilbert space with the complex structure $I$ \eqref{eq:IComplexStructure} and scalar product \eqref{eq:DoubledScalarProduct}.

        \item\label{item:EmbeddingIntoDoubledHilbertSpace}
        For every conjugation $C\in\Con(\Hil)$, the map
        \begin{align}\label{eq:UC}
            U_C:\Hil\longrightarrow\widehat\Hil,\qquad
            U_C\xi:=\frac{1}{\sqrt{2}}(\xi\oplus C\xi)
        \end{align}
 is complex-linear and isometric.

        \item\label{item:DoubledConjugation}
        The flip
        \begin{align}\label{eq:Jhat}
            \widehat J:\widehat\Hil\longrightarrow\widehat\Hil,\qquad
            \widehat J(\xi\oplus\eta):=\eta\oplus\xi
        \end{align}
        is a conjugation.
    \end{enumerate}
\end{lemma}
\begin{proof}
    All claims follow by direct and straightforward calculation.
\end{proof}

In $\widehat\Hil$, graphs of positive operators induce standard subspaces.

\begin{example}[Standard subspaces from graphs]
    \label{example:StandardSubspacesFromGraphs}
 Let $A$ be a positive self-adjoint complex-linear injective operator on $\Hil$. Then the graph $\Gamma(A)\subset\widehat\Hil$ is a standard subspace, with complex span $\dom S_{\Gamma(A)}=\dom A\oplus\dom A^{-1}$ and modular data
 \begin{align}\label{eq:GraphData}
        S_{\Gamma(A)}(\xi\oplus\eta)
        =
        A^{-1}\eta\oplus A\xi,
        \qquad
        \Delta_{\Gamma(A)}
        =A^2\oplus A^{-2},
        \qquad
        J_{\Gamma(A)}=\widehat J.
    \end{align}
\end{example}
\begin{proof}
 We read \eqref{eq:GraphData} as the definition of an operator $\Delta:=A^2\oplus A^{-2}$. We note that $\Delta$ has trivial kernel and is self-adjoint positive, commutes with $I$ (i.e. it is complex-linear), and satisfies $\widehat J\Delta\widehat J=\Delta^{-1}$. Hence the assumptions of Corollary~\ref{corollary:StandardSubspacesAndOneParameterGroups}~\ref{item:JDelta} are satisfied, and we conclude that $\widehat H:=\Fix(\widehat J\Delta^{1/2})$ is standard with the claimed modular data.

    It remains to show $\widehat H=\Gamma(A)$. Noting that $\dom\Delta^{1/2}=\dom A\oplus\dom A^{-1}$, we pick $\xi\in\dom A$, $\eta\in\dom A^{-1}$, and compute
    \begin{align*}
        \widehat J\Delta^{1/2}(\xi\oplus\eta)-\xi\oplus\eta
        =
        (A^{-1}\eta-\xi) \oplus (A\xi-\eta),
    \end{align*}
    which vanishes if and only if $\eta=A\xi$, i.e. if $\xi\oplus\eta\in\Gamma(A)$. Hence $\widehat H=\Gamma(A)$.
\end{proof}

We will refer to standard subspaces of this form as \emph{graph standard subspaces}\index{graph standard subspace (Example~\ref{example:StandardSubspacesFromGraphs})}.

 Apart from standard subspaces with centre of finite odd dimension, all standard subspaces are graph standard subspaces, with a unique positive contraction up to unitary equivalence.

\begin{theorem}[The graph model of a standard subspace]
    \label{theorem:GraphStandardSubspaces}\booktheoremindex{theorem:GraphStandardSubspaces}{Graph model of a standard subspace}\index{classification of standard subspaces!by graphs of positive operators}
 Let $H\in\Std(\Hil)$. Use the spectral subspaces $\Hil^H_<$ and $\Hil^H_1$ from \eqref{eq:SpectralProjectionsDelta}, put $\Delta_{H,<}:=\Delta_H|_{\Hil^H_<}$, and denote the double of $\Hil^H_<$ by $\widehat{\Hil}^H_<$.
 \begin{enumerate}
        \item\label{item:UH} The map
 \begin{align}
            U_H:\widehat{\Hil}^H_<\oplus\Hil^H_1
            &\longrightarrow\Hil,
            \qquad
            U_H\bigl((\xi_1\oplus\xi_2)\oplus\eta\bigr)
            :=\xi_1+J_H\xi_2+\eta
            \nonumber
        \end{align}
 is unitary, and
        \begin{align}\label{eq:UHmatching}
            U_H\bigl(\Gamma(\Delta_{H,<}^{1/2})\oplus Z(H)\bigr)=H.
        \end{align}

        \item\label{item:FactorialUH} Every factorial standard subspace is unitarily equivalent to a graph standard subspace, namely the graph standard subspace of the positive injective contraction $\Delta_{H,<}^{1/2}$.

 \item\label{item:AbelianPartUH} A standard subspace is unitarily equivalent to a graph standard subspace (without additional abelian summand) if and only if its centre $Z(H)$ is not of finite odd dimension.
 \end{enumerate}
\end{theorem}
\begin{proof}
 \ref{item:UH} The complex structure of $\widehat{\Hil}^H_<\oplus\Hil^H_1$ is $I\oplus i$, where $I=i\oplus(-i)$ is the non-canonical complex structure on $\widehat{\Hil}^H_<$ \eqref{eq:IComplexStructure} and $i$ the usual complex structure on $\Hil^H_1$. Then
 \begin{align}
        U_H(I\oplus i)=(\id\oplus J_H\oplus\id)(i\oplus(-i)\oplus i)=iU_H,
    \end{align}
 so $U_H$ is complex-linear. The conjugation $J_H$ in $U_H$ implies that $U_H$ is isometric (note the definition of the scalar product on $\widehat{\Hil}^H_<$). Furthermore, the spectral decomposition $\Hil=\Hil^H_<\oplus\Hil^H_>\oplus\Hil^H_1$ and $J_H\Hil^H_<=\Hil^H_>$ imply that $U_H$ is surjective, hence unitary.

    The space $\Gamma(\Delta_{H,<}^{1/2})\oplus Z(H)$ consists of all vectors of the form $\xi\oplus\Delta_H^{1/2}\xi\oplus\zeta$, with $\xi\in\Hil^H_<$ and $\zeta\in Z(H)$. Under $U_H$, these vectors are mapped to $(1+S_H)\xi+\zeta\in H$. Conversely, given $h\in H$, we may split it as $h=f+z$ with $z\in Z(H)$ and $f\in F(H)=Z(H)^\perp\cap H$ (Prop.~\ref{proposition:DecomposeHIntoMaximallyAbelianAndFactorPart}). Choosing $\zeta:=z$ and $\xi:=E^{\Delta_H}_<f$, we have
 \begin{align}
        U_H(\xi\oplus\Delta_H^{1/2}\xi\oplus z)
        =
        E^{\Delta_H}_<f+S_HE^{\Delta_H}_<f+z
        =
        E^{\Delta_H}_<f+E^{\Delta_H}_>f+z
        =
        f+z=h.
    \end{align}
    Hence $U_H(\Gamma(\Delta_{H,<}^{1/2})\oplus Z(H))=H$.

 \ref{item:FactorialUH} Factorial spaces have trivial centre $Z(H)=0$, hence the claim is clear from \eqref{eq:UHmatching}.

    \ref{item:AbelianPartUH} {
Suppose that $H$ is unitarily equivalent to $\Gamma(A)$. By \eqref{eq:GraphData} its graph model has modular operator $A^2\oplus A^{-2}$. For positive injective $A$ we have $\ker(A^2-1)=\ker(A^{-2}-1)=\ker(A-1)$. Intersecting the corresponding eigenspace with the graph, Proposition~\ref{proposition:CenterAndMaximalAbelianness}~\ref{item:centerH} gives
    \begin{align*}
        Z(\Gamma(A))
        &=\Gamma(A)\cap\ker(\Delta_{\Gamma(A)}-1)
        =\{\xi\oplus \xi\col \xi\in\ker(A-1)\}.
    \end{align*}
    Hence $\dim_\Rl Z(H)=2\dim_\Cl\ker(A-1)$, which rules out finite odd dimension.
}

    Conversely, suppose that the dimension of $Z(H)$ is infinite or finite and even. Choose a complex Hilbert space $\CK$ with
    $2\dim_{\Cl}\CK=\dim_{\Rl}Z(H)$.
    The graph $\Gamma(\id_{\CK})$ is a maximal abelian
    standard subspace with that real dimension, so it is unitarily
    equivalent to $Z(H)$ in $Z(H)+iZ(H)$. Representing the factorial part $F(H)$ of $H$ as a graph standard subspace, $F(H)\cong\Gamma(A)$, we then have $H=F(H)\oplus Z(H)\cong\Gamma(A)\oplus\Gamma(\id_{\CK})=\Gamma(A\oplus\id_{\CK})$. (See Section~\ref{section:DirectSumsAndIntegrals} for more on direct sums of standard subspaces.)
\end{proof}

{
Whenever the graph model exists, its defining operator can be chosen as the positive injective contraction
\begin{align*}
    A=\Delta_{H,<}^{1/2}\oplus 1_{\CK},
    \qquad 2\dim_\Cl\CK=\dim_\Rl Z(H).
\end{align*}
This contraction is unique up to unitary equivalence. Indeed, for any positive injective contraction $A$ giving such a model, the restriction of $A^2\oplus A^{-2}$ to its spectral subspace for $(0,1)$ recovers $A^2|_{\ker(A-1)^\perp}$ up to unitary equivalence. Part~\ref{item:AbelianPartUH} and its proof determine the dimension of $\ker(A-1)$ from $Z(H)$. In particular, $\ker(A-1)=\{0\}$ exactly when $H$ is factorial. Zero may belong to $\spec(A)$, in which case $A^{-1}$ and the Tomita operator of $\Gamma(A)$ are unbounded.
}

Also the graph of the unrestricted modular operator $\Delta_H^{1/2}$ is unitarily equivalent to a standard subspace: The unitary
\begin{align}\label{eq:GraphDoublingUnitary}
    W_H:\Hil\oplus\Hil&\longrightarrow\widehat\Hil,
    \\
    W_H(\xi\oplus\eta)
    &:=
    \frac1{\sqrt2}
    \bigl((\xi+i\eta)\oplus(J_H\xi+iJ_H\eta)\bigr)
    \nonumber
\end{align}
from the ordinary complex direct sum $\Hil\oplus\Hil$ into $\widehat\Hil$ maps $H\oplus H$ onto $\Gamma(\Delta_H^{1/2})$, i.e. $\Gamma(\Delta_H^{1/2})$ corresponds to a doubled standard subspace. See Section~\ref{section:DirectSumsAndIntegrals} for more on direct sums of standard subspaces.

\bigskip
For later reference, we record here also a construction that uses the graph Hilbert space of the modular operator, namely the complex Hilbert space with scalar product
\begin{align}
    \Hil_H&:=\dom\Delta_H^{1/2},
    \\
    \langle\xi,\eta\rangle_H
    &:=
    \langle\xi,\eta\rangle+
    \langle\Delta_H^{1/2}\xi,\Delta_H^{1/2}\eta\rangle
    =
    \langle(1+\Delta_H)^{1/2}\xi,
           (1+\Delta_H)^{1/2}\eta\rangle
\end{align}
and usual complex structure.

\begin{proposition}[The graph Hilbert space of the modular operator]
    \label{proposition:GraphHilbertSpaceOfDelta}\index{graph Hilbert space of the modular operator}
    Let $H\in\Std(\Hil)$.
    \begin{enumerate}
        \item\label{item:SConjugation}
        The Tomita operator $S_H$, considered as an everywhere defined
        operator on $\Hil_H$, is a conjugation.

        \item\label{item:HTurnsAbelian}
        The subspace $H\subset\Hil_H$ is a maximal abelian standard
        subspace. Its Tomita operator and modular conjugation are both
        $S_H$, and its modular operator is $1$.
    \end{enumerate}
\end{proposition}
\begin{proof}
    \ref{item:SConjugation}
    Since $\Delta_H^{1/2}$ is closed, its domain is complete for
    the graph norm. Moreover, $\Hil_H=\dom S_H=H+iH$,
    $S_H(\Hil_H)=\Hil_H$, and $S_H^2=1$ on this space.
    Using $S_H=J_H\Delta_H^{1/2}$, we obtain
    \begin{align*}
        \|S_H\xi\|_H^2
        &=
        \|S_H\xi\|^2+\|S_H(S_H\xi)\|^2
        =\|S_H\xi\|^2+\|\xi\|^2
          =\|\xi\|_H^2.
    \end{align*}
    Thus $S_H$ is an antilinear isometric involution on $\Hil_H$,
    and hence a conjugation.

    \ref{item:HTurnsAbelian} Also as a subspace of $\Hil_H$, we have $H=\Fix(S_H)$. Hence $H$ is maximal abelian by Proposition~\ref{proposition:CenterAndMaximalAbelianness}.
\end{proof}

\section{Characterization of modular data and standard subspaces}\label{section:Characterization}

By the Tomita-Takesaki Theorem, every standard subspace $H$ has a canonical unitary one-parameter group of automorphisms and a canonical conjugation mapping $H$ onto $H'$. As the definition of these objects involves the non-explicit polar decomposition of the Tomita operator, it is often not straightforward to explicitly compute them for a given standard subspace. It is therefore helpful to know properties that characterize the modular group within all unitary one-parameter groups leaving $H$ invariant, and properties that characterize the modular conjugations within all conjugations mapping $H$ onto $H'$. Such characterizations will be derived in this section.

\subsection{The KMS condition}\label{section:KMS}

It will turn out that the characterization of the modular group is closely connected to the so-called KMS condition, defined below. The abbreviation KMS stands for Kubo, Martin, and Schwinger, who studied the von Neumann algebraic version of this condition in the context of thermal equilibrium states. The definition that we give is a standard subspace version of the original KMS condition. The parameter $\beta$ that it involves is nonetheless referred to as ``inverse temperature'', and we refer to Section~\ref{section:KMSApplication} and the Notes and Remarks in Section~\ref{section:NotesAndRemarksChapter1} for explanations of this terminology.

 Whereas our tools so far were entirely based in Hilbert space theory, the KMS condition will introduce complex analysis. In particular, we will frequently be concerned with bounded analytic (holomorphic) functions on strip regions in the complex plane.

Given $a,b\in\Rl$, $a<b$, and $\beta\in\Rl$, we write \glsadd{Stripab}\glsadd{Stripb}
\begin{align}\label{eq:Strips}
    \Strip_{a,b}
    :=
    \Rl+i(a,b),
    \qquad
    \Strip_\beta
    :=
    \begin{cases}
        \Strip_{0,\beta} & \beta>0,\\
        \Rl & \beta=0\\
        \Strip_{\beta,0} & \beta<0
    \end{cases}.
\end{align}

 and consider the {\em Hardy space}\index{Hardy space} of all holomorphic bounded functions on $\Strip_{\beta}$,\glsadd{Hinfty}
\begin{align}
    \Hinfty(\Stripb)
    &:=
    \{f:\Stripb\to\Cl \col f\,\text{analytic and bounded}\}.
\end{align}
 Any $f\in\Hinfty(\Stripb)$ has pointwise almost everywhere $L^\infty$-boundary values at $\Rl$ and $\Rl+i\beta$. In Appendix~\ref{section:HinftyStrip}, some relevant background material on $\Hinfty(\Stripb)$ is collected, see in particular Theorem~\ref{theorem:HinftyBoundaryValues}. We will be interested in the subspace\glsadd{HbC}
 \begin{align}
    \HinftyC(\Strip_{a,b})
    &:=
    \{f:\overline{\Strip_{a,b}}\to\Cl \col f\,\text{continuous},\,f|_{\Strip_{a,b}}\in\Hinfty(\Strip_{a,b})\}
\end{align}
of functions in $\Hinfty(\Stripb)$ that extend continuously to the closed strip $\overline{\Stripb}$. A further restriction comes from a symmetry under an antilinear involution.

 \begin{definition}[Symmetric functions]\label{def:SymmetricFunction}
    \leavevmode
    \begin{enumerate}
 \item\label{item:StripReflection}\index{strip reflection}\glsadd{stripStar} For a function $f:\Stripb\to\Cl$, we define
 \begin{align}
            f^*:\Stripb\to\Cl,\qquad \zeta\mapsto \overline{f(i\beta+\overline\zeta)},
        \end{align}
        and call $f$ \emph{symmetric} if $f=f^*$.

 \item\label{item:StripSymmetricSpace}\index{symmetric function on a strip}\glsadd{stripSymmetric} The subspace of symmetric functions in $\HinftyC(\Strip_\beta)$ is denoted $\HinftyCs(\Strip_\beta)$.
 \end{enumerate}
\end{definition}

 We note that $f\mapsto f^*$ is an antilinear isometric involution on the complex Banach space~$\HinftyC(\Stripb)$. As $\HinftyC(\Stripb)$ is not a Hilbert space, $f\mapsto f^*$ is however not a Tomita operator. Geometrically speaking, this involution amounts to a reflection about the midline $\Im\zeta=\beta/2$ of $\Stripb$. In view of the complex conjugation of the function values, symmetric functions take real values on the mid line $\Rl+\frac{i\beta}{2}$. This also explains our convention for defining the KMS condition at $\beta=0$.

We have now introduced all prerequisites to define the KMS condition.

\begin{definition}[KMS condition for real subspaces]\label{definition:KMSCondition}\index{KMS!condition (Def.~\ref{definition:KMSCondition})}
    A unitary one-parameter group $U(t)$, $t\in\Rl$, is said to satisfy the {\em KMS condition w.r.t. a real subspace $H\in\SubR(\Hil)$ at inverse temperature $\beta\in\Rl\backslash\{0\}$} if for every $h_1,h_2\in H$, there exists a symmetric function  $f_{h_1,h_2}\in\HinftyCs(\Stripb)$ such that
 \begin{align}\label{eq:KMSBoundaryCondition}
        f_{h_1,h_2}(t)
        =
        \langle h_1,U(t)h_2\rangle,
        \qquad t\in\Rl.
    \end{align}
 For $\beta=0$, we define the KMS condition by demanding $\langle h_1,U(t)h_2\rangle\in\Rl$ for every $t\in\Rl$.
\end{definition}

 \begin{remark}[Remarks on the KMS condition]\label{remark:KMS}
    \leavevmode
    \begin{enumerate}
 \item\label{item:KMSOppositeBoundary} As any $f\in\HinftyC(\Stripb)$ is uniquely determined by its boundary value $f|_\Rl$ by the Schwarz Reflection Principle, the KMS condition can be rephrased as follows: Given any $h_1,h_2\in H$, the continuous function $f_{h_1,h_2}:\Rl\to\Cl$, $f_{h_1,h_2}(t):=\langle h_1,U(t)h_2\rangle$, is required to have an analytic continuation to $\Stripb$, with continuous extension to the closed strip, and the boundary value at $\Rl+i\beta$ is required to be
		\begin{align}\label{eq:KMSOppositeBoundary}
            f_{h_1,h_2}(t+i\beta)=\langle U(t)h_2,h_1\rangle.
		\end{align}

		\item\label{item:KMSZeroTemperatureParameter} For $\beta=0$, the KMS condition is very special: $\langle h_1,U(t)h_2\rangle\in\Rl$ for all $h_1,h_2\in H$ is equivalent to $U(t)H\subset H'$ for all $t\in\Rl$. Setting $t=0$ implies $H\subset H'$. If $H$ is cyclic, it is then standard in case $H$ is closed, and maximal abelian $H=H'$ by Proposition~\ref{proposition:AbelianSpaces}.  We will not consider the KMS condition for $\beta=0$ from now on.

		\item\label{item:DeltaAnalytic} If $U$ satisfies the KMS condition w.r.t. $H$ at inverse temperature $\beta$, then $V(t):=U(ct)$ (for some $c\neq0$), satisfies the KMS condition w.r.t. $H$ at inverse temperature $\beta/c$. Hence the inverse temperature can be eliminated by rescaling the one-parameter group. In particular, its sign can be flipped by switching from $U(t)$ to $U(-t)$.
	\end{enumerate}
\end{remark}

When checking the KMS condition for a given one-parameter group and standard subspace, it is often convenient to work on a dense subspace with particularly nice properties. The following lemma shows that this is enough.

 \begin{lemma}[The KMS condition can be checked on a dense subset]\label{lemma:KMSDenseIsEnough}
    Let $H\in\Real(\Hil)$ and $U(t)$, $t\in\Rl$, a unitary one-parameter group that satisfies the KMS condition on a dense subset $D\subset H$, i.e. the functions $f$ in Definition~\ref{definition:KMSCondition} exist and have the required boundary values \eqref{eq:KMSBoundaryCondition} and \eqref{eq:KMSOppositeBoundary} (for some $\beta$), but only for $h_1,h_2\in D$. Then $U$ satisfies the KMS condition on all of $H$.
\end{lemma}
 \begin{proof}
 The three-lines theorem (Theorem~\ref{theorem:HinftyBoundaryValues}~\ref{item:3lines}), implies that any $f\in\HinftyC(\Stripb)$ satisfies
 \begin{align}\label{eq:ThreeLinesEstimate}
			\sup_{\zeta\in\Strip_\beta}|f(\zeta)|
			=
			\sup_{\zeta\in\partial\Strip_\beta}|f(\zeta)|.
		\end{align}
		For the functions appearing in the KMS condition, we have $|f(t)|=|\langle h_1,U(t)h_2\rangle|\leq\|h_1\|\|h_2\|$ and $|f(t+i\beta)|=|\langle U(t)h_2,h_1\rangle|\leq\|h_1\|\|h_2\|$ for $t\in\Rl$, so \eqref{eq:ThreeLinesEstimate} implies the uniform bound $|f(\zeta)|\leq\|h_1\|\|h_2\|$ for all $\zeta\in\Stripb$.

 Now suppose that the KMS condition holds for all $h_1,h_2\in D$. Given $k_1,k_2\in H$ and approximating sequences $h_j^{(n)}\to k_j$, $j=1,2$, with $h_j^{(n)}\in D$ for all $n\in\Nl$, we then have $f_{h_1^{(n)},h_2^{(n)}}\to f_{k_1,k_2}$ uniformly in $\overline{\Stripb}$ as $n\to\infty$. As uniform limits of holomorphic functions are holomorphic, this implies $f_{k_1,k_2}\in\HinftyC(\Stripb)$ and the boundary values \eqref{eq:KMSBoundaryCondition} and \eqref{eq:KMSOppositeBoundary}.
 \end{proof}

 \begin{theorem}[KMS characterization of the modular group]\label{theorem:KMS}\booktheoremindex{theorem:KMS}{KMS characterization of the modular group}\index{KMS characterization of the modular group}\index{KMS!condition (Def.~\ref{definition:KMSCondition})|seealso{KMS characterization of the modular group}}
	\leavevmode
	\begin{enumerate}
		\item\label{item:ModularGroupsSatisfiesKMS} Let $H\in\Std(\Hil)$. Then the modular group $\Delta_H^{it}$, $t\in\Rl$, satisfies the KMS condition w.r.t. $H$ at inverse temperature $\beta=-1$.
		\item\label{item:KMSCharacterizesModularGroup} Let $H\in\Real(\Hil)$ be a closed real subspace and $U$ a one-parameter group satisfying
		\begin{enumerate}[label=\emph{(\roman*)},ref=(\roman*)]
            \item\label{item:KMSInvariance} $U(t)H=H$, $t\in\Rl$, and
            \item\label{item:KMSHypothesis} the KMS condition w.r.t. $H$, at inverse temperature $\beta\neq0$.
		\end{enumerate}
		 Then $H$ is separating. If $H$ is cyclic, it is standard, and $U(t)=\Delta_H^{-it/\beta}$, $t\in\Rl$.
	\end{enumerate}
\end{theorem}
 \begin{proof}
	\ref{item:ModularGroupsSatisfiesKMS} Let $h_1\in H$ and $h_2\in D:=H\cap\dom\Delta_H$. By Lemma~\ref{lemma:DeltaDomains}, $D$ is dense in $H$, so by Lemma~\ref{lemma:KMSDenseIsEnough}, it is enough to verify the KMS condition for these vectors.

	For $h_1\in H$, $h_2\in D$, we claim that the function $f:\overline{\Strip_{-1}}\to\Cl$, $f(t):=\langle h_1,\Delta_H^{it}h_2\rangle$, lies in $\HinftyC(\Strip_{-1})$. Indeed, $f$ is well-defined, bounded, and continuous because $h_2\in\dom\Delta_H\subset\dom\Delta_H^{-\Im(z)}$ for $-1\leq\Im z\leq 0$ (Lemma~\ref{lemma:DeltaDomains}~\ref{item:DomDeltapOrder}).

	For analyticity, we use Morera's Theorem from complex analysis. Let $\gamma$ be any triangular curve in $\Strip_{-1}$. Then, by Fubini's Theorem,
	\begin{align*}
		\oint_\gamma f(z)\,dz
		=
		\oint_\gamma \int_0^\infty \la^{iz}\,dE^{\Delta_H}_{h_1,h_2}(\la)\,dz
		=
		\int_0^\infty \left(\oint_\gamma \la^{iz}\,dz\right)\,dE^{\Delta_H}_{h_1,h_2}(\la)
		=0,
	\end{align*}
	because $z\mapsto \la^{iz}$ is holomorphic.

 The crucial part is now to verify the opposite-boundary condition \eqref{eq:KMSOppositeBoundary}. We have
 \begin{equation*}
	\begin{aligned}[t]
		f(t-i)
		&=
		\langle h_1,\Delta_H^{it}\Delta_H h_2\rangle &
		\\
		&=
		\langle \Delta_H^{1/2}h_1,\Delta_H^{1/2}\Delta_H^{it}h_2\rangle
		\qquad &&\text{because } h_1\in\dom\Delta_H^{1/2}
		\\
		&=
		\langle J_H h_1,J_H\Delta_H^{it}h_2\rangle
		\qquad &&\text{by Tomita's Theorem and definition of $H$}
		\\
		&=
		\langle \Delta_H^{it}h_2,h_1\rangle
		\qquad &&\text{by antiunitarity of } J_H.
	\end{aligned}
	\end{equation*}

 \ref{item:KMSCharacterizesModularGroup} We first show that $H$ is separating under the given circumstances. So let $h_2,\tilde h_2\in H$ such that $h_2=i\tilde h_2$. Then $f(t):=\langle h_1,U(t)h_2\rangle=i\langle h_1,U(t)\tilde h_2\rangle=:ig(t)$ satisfy for $t\in\Rl$
	\begin{align*}
		f(t+i\beta)
		&=
		\langle U(t)h_2,h_1\rangle
		=
		-i\langle U(t)\tilde h_2,h_1\rangle
		,\\
		ig(t+i\beta)
		&=
		i\langle U(t)\tilde h_2,h_1\rangle.
	\end{align*}
	As $f$ and $ig$ are uniquely fixed by their restrictions to $\Rl$, and coincide there, we conclude $\langle U(t)\tilde h_2,h_1\rangle=0$. Setting $t=0$ and $h_1=\tilde h_2$ then shows $\tilde h_2=0$, i.e. $H$ is separating.

	We now assume that $H$ is cyclic, hence standard, and next want to show that its modular unitaries are given by rescaling $U$, namely $U(t)=\Delta_H^{-it/\beta}$. By hypothesis~\ref{item:KMSHypothesis}, $U(t)$ satisfies the KMS condition w.r.t. $H$ at inverse temperature $\beta\neq0$, hence $V(t):=U(-\beta t)$ satisfies the KMS condition w.r.t. $H$ at inverse temperature $-1$. By \ref{item:ModularGroupsSatisfiesKMS}, $\Delta_H^{it}$ satisfies the same condition. Furthermore, we have $V(t)H=H$ for all $t\in\Rl$ (hypothesis~\ref{item:KMSInvariance}), and since the modular group of $V(t)H$ is $V(t)\Delta^{is}_HV(t)^*=\Delta_H^{is}$, $s\in\Rl$, these two one-parameter groups commute.

	As in the proof of Proposition~\ref{prop:ClassificationOfPassiveOneParameterGroups}, there exists a common core $\CD=(\CD\cap H)+i(\CD\cap H)$ for the self-adjoint generators $L_H=\log\Delta_H$ and $P$ (the generator of $V$), which is also a common core for $\Delta_H$ and $e^P$ by construction. For $h_1,h_2\in\CD\cap H$, we have
 \begin{align*}
		\langle h_1,\Delta_H h_2\rangle = \langle h_2,h_1\rangle
		=
		\langle h_1,e^P h_2\rangle,
	\end{align*}
	this follows by evaluating the KMS function at $t=0$. Hence $\Delta_H=e^P$, so $\Delta_H^{it}=e^{itP}=V(t)=U(-\beta t)$, as claimed.
\end{proof}

 \subsection[Reflection positivity]{Reflection positivity}\label{section:ReflectionPositivity}

We now turn to the characterization of the modular conjugation $J_H$.

 \begin{proposition}[Characterization of $\boldsymbol{J_H}$ by reflection positivity]\label{proposition:ReflectionPositivity}\index{reflection positivity}
	Let $H\in\Std(\Hil)$. Then the modular conjugation $J_H$ is the unique conjugation $J$ on $\Hil$ such that
	\begin{enumerate}[label=\emph{(\roman*)},ref=(\roman*)]
		\item\label{item:ReflectionComplement} $JH\supset H'$ and
		\item\label{item:ReflectionPositive} $J$ is reflection positive w.r.t. $H$, namely $\langle h,Jh\rangle\geq0$ for all $h\in H$.
	\end{enumerate}
\end{proposition}
 \begin{proof}
 For $J=J_H$, \ref{item:ReflectionComplement} holds by Tomita's Theorem, and \ref{item:ReflectionPositive} follows from the positivity of the modular operator~$\Delta_H$,
 \begin{align*}
		\langle h,J_Hh\rangle=\langle h,\Delta_H^{1/2}h\rangle\geq0.
	\end{align*}
 Conversely, if $J$ satisfies~\ref{item:ReflectionPositive}, then for all $h_1,h_2\in H$,
 \begin{align*}
		0
		&\leq
		\langle h_1+h_2,J(h_1+h_2)\rangle
		=
		\langle h_1,Jh_1\rangle+
		\langle h_2,Jh_2\rangle
		+
		2\langle h_1,Jh_2\rangle,
	\end{align*}
 which implies $\langle h_1,Jh_2\rangle\in\Rl$. This shows $JH\subset H'$, and therefore $JH=H'$ by~\ref{item:ReflectionComplement}.

	Now for all $h_1,h_2\in H$, we have
	\begin{align*}
		\langle h_1+ih_2,JS_H( h_1+ih_2)\rangle
		&=
		\langle  h_1+ih_2, Jh_1+iJh_2\rangle
		=
		\langle h_1,Jh_1\rangle
		+
		\langle h_2,Jh_2\rangle
		\geq0.
	\end{align*}
 This implies that the densely defined operator $JS_H$ is symmetric and positive. Furthermore, $JH=H'$ implies $S_H^*=JS_HJ$, hence $(JS_H)^*=S_H^*J=JS_H$. Thus $JS_H$ is positive selfadjoint. Uniqueness of the polar decomposition of $JS_H=JJ_H\Delta_H^{1/2}$ then implies $JJ_H=1$, i.e. $J=J_H$.
 \end{proof}

 The term ``reflection positivity'' seems adequate for condition~\ref{item:ReflectionPositive}, but is actually meant to refer to the reflection positivity of Euclidean quantum field theory, the link of which to modular theory is not yet understood. In the Notes and Remarks, we link to literature on reflection positivity and modular theory.

\begin{example}[Modular data of Hilbert-Schmidt standard subspaces]\label{example:ModularDataOfHilbertSchmidt}
    In Example~\ref{example:HilbertSchmidt}, we considered a Hilbert-Schmidt operator $\Om\in\CI_2(\Hil)$ with $\ker\Om=\ker\Om^*=\{0\}$ and the corresponding standard subspace $H=\overline{\ell(B(\Hil)_\sa)\Om}$, with symplectic complement $H'=\overline{r(B(\Hil)_\sa)\Om}$. We claim that the modular data of $H$ take the same form as in the finite-dimensional case \eqref{eq:ModularDataMatrixExample}, namely, $X\in\CI_2(\Hil)$,
    \begin{align}\label{eq:ModularDataHilbertSchmidtExample}
        J_H(X)=U^*X^*U^*,
        \qquad
        \Delta_H^{it}(X)
        =
        \rho^{it}XU\rho^{-it}U^*,
    \end{align}
    where $\Om^*=U\rho^{1/2}$ is the polar decomposition of $\Om^*$.
\end{example}

 \begin{proof}
    The map $J(X):=U^*X^*U^*$ is easily checked to be a conjugation on $\CI_2(\Hil)$. Recalling that $U^*\Om^*=\rho^{1/2}=\Om U$, we have for any $A=A^*\in B(\Hil)$
    \begin{align*}
        J(A\Om)
        =
        U^*\Om^*AU^*
        =
        \Om\,UAU^*,
    \end{align*}
 and analogously $J(\Om A)=U^*AU\,\Om$. Taking into account the continuity of $J$, this implies $JH=\overline{r(B(\Hil)_\sa)\Om}=H'$, thus verifying Proposition~\ref{proposition:ReflectionPositivity}~\ref{item:ReflectionComplement}.

    Given $A\in B(\Hil)_\sa$, we furthermore have
    \begin{align*}
        \langle A\Om,J(A\Om)\rangle_2
        &=
        \Tr(\Om^*AU^*\Om^*AU^*)
        \\
        &=
        \Tr(\rho^{1/2}A\rho^{1/2}A)
        \\
        &=
        \Tr\left(\left|\rho^{1/4}A\rho^{1/4}\right|^2\right)
        \geq0,
    \end{align*}
 verifying Proposition~\ref{proposition:ReflectionPositivity}~\ref{item:ReflectionPositive}. Thus $J=J_H$.

    We next prove the formula for the modular group, using Theorem~\ref{theorem:KMS}. Let $V(t)(X):=\rho^{it}XU\rho^{-it}U^*$, $X\in\CI_2(\Hil)$, $t\in\Rl$ \eqref{eq:ModularDataHilbertSchmidtExample}. Then $V(t)$ is a strongly continuous unitary one-parameter group on $\CI_2(\Hil)$. For $B\in B(\Hil)_\sa$ and $X\in\CI_2(\Hil)$, we have
    \begin{align*}
        V(t)(B\Om)
        &=
        \rho^{it}B\Om U\rho^{-it}U^*
        \\
        &=
        \rho^{it}B\rho^{1/2}\rho^{-it}U^*
        \\
        &=
        \rho^{it}B\rho^{-it}\rho^{1/2}U^*
        \\
        &=
        \rho^{it}B\rho^{-it}\Om,
    \end{align*}
 which is again of the form of a self-adjoint bounded operator left multiplied onto $\Om$. By continuity of $V(t)$, this implies $V(t)H\subset H$. As this inclusion holds for $t$ and $-t$, and $V(-t)=V(t)^{-1}$, we arrive at $V(t)H=H$. This verifies the invariance hypothesis of Theorem~\ref{theorem:KMS}~\ref{item:KMSCharacterizesModularGroup}\,\ref{item:KMSInvariance}.

    For the KMS hypothesis~\ref{item:KMSHypothesis}, we consider
    \begin{align}
		D:=\{A\Om\col A^*=A=E^{\rho}([n^{-1},\infty))AE^{\rho}([n^{-1},\infty))\;\text{for some } n\in\Nl\}
		.
    \end{align}
    This real space is contained in $H$ and dense in Hilbert-Schmidt norm, so it is enough to verify the KMS condition for vectors in $D$ (Lemma~\ref{lemma:KMSDenseIsEnough}).

    For $A,B\in D$, the function
     \begin{align*}
		f:\Cl\to\Cl, \qquad f(z):=\langle A\Om,V(z)(B\Om)\rangle_2
		=
		\Tr(\Om^*A\rho^{iz}B\rho^{-iz}\Om)
		=
		\Tr(\rho^{1-iz}A\rho^{iz}B)
     \end{align*}
	 is entire. Because of the spectral cutoff in $A=E^{\rho}([n^{-1},\infty))AE^{\rho}([n^{-1},\infty))$ and $B=E^{\rho}([n^{-1},\infty))BE^{\rho}([n^{-1},\infty))$, these operators are of finite rank. For $z\in\Strip_{-1}$, this yields the bound
	 \begin{align*}
		|f(z)|
		&\leq
		\|\rho^{1-iz}A\|_2\|\rho^{iz}B\|_2
		\leq
		\|\rho^{1-iz}\|\|\rho^{iz}\|\|A\|_2\|B\|_2
		\leq
		\|A\|_2\|B\|_2
	 \end{align*}
	 and bounded on $\Strip_{-1}$. Its boundary value at $\Rl-i$ is
 \begin{align*}
        f(t-i)
        &=
        \Tr(\rho^{-it}A\rho^{it}\rho B)
        =
        \Tr(\rho \rho^{it} B\rho^{-it}A)
        =
        \langle V(t)(B\Om),A\Om\rangle_2,
    \end{align*}
    and the proof is finished.
\end{proof}

\section{The standard subspace net of the \boldmath{$U(1)$}-current}\label{section:U1current}

 This section has a different character than the previous ones: It does not develop the theory of standard subspaces further, but is instead dedicated to a detailed discussion of an example family of standard subspaces. These spaces have their origin in quantum field theory, where they are collectively known as the \emph{$U(1)$-current}. This name will not be obvious from the description given here, and we refer to the Notes and Remarks for further explanations. Independently of the connection to quantum field theory, analysing these standard subspaces will show how to work with concepts developed so far -- for instance, the KMS and reflection positivity conditions, and the projections associated with a standard subspace -- in non-trivial situations.

We will consider a family of standard subspaces $H(I)\subset\Hil$ indexed by subsets $I\subset\Rl$. The map $I\mapsto H(I)$ will display all main features of a quantum field theory in vacuum representation, namely (with the terminology defined below) isotony, covariance (under the affine group), locality, positive energy and Haag duality. The setting will therefore be much richer than considered up to now: Instead of a single standard subspace, we will deal with several such spaces that are related by inclusions, intersections, and group actions. In many ways, this section is therefore a precursor of a more general analysis of such nets in Chapter 3.

The spaces we are interested in can be described from various different but equivalent perspectives. In Section~\ref{section:U1HalfPlane}, we start with a description that is close to the field-theoretic picture and naturally associates standard subspaces with subsets of the real line via functions with suitable support properties. This will lead us to standard subspaces $H(I)$ in $\Hil=L^2(\Rl_+,p\,dp)$. The most important spaces corresponding to half-lines like $I=\Rl_+$ will receive particular attention, and we will determine their modular data in Theorem~\ref{theorem:U1C+}.

For a full understanding of the $U(1)$-current, it is very helpful to consider the spaces $H(I)$ also in different, unitarily equivalent, Hilbert spaces. We will present these in Section~\ref{section:U1DifferentSpaces}. This includes in particular Hilbert spaces of analytic functions (Hardy spaces) on which the analytically continued modular group and conjugation act in a more transparent manner than on the $L^2$-spaces of their real boundary values. In this description, we will partially draw on material in Appendix~\ref{section:Hardy2Strip} on Hardy spaces.

 \subsection{The \boldmath{$U(1)$}-current in \boldmath{$L^2(\Rl_+,p\,dp)$}}\label{section:U1HalfPlane}

 Let $\Sch(\Rl)$\glsadd{Schwartz} denote the space of complex-valued Schwartz functions on $\Rl$. We consider its real subspace $\Sch_\Rl(\Rl)$ of real-valued functions, and the ``local'' subspaces
 \begin{align}
    \Sch_{(\Rl)}(I)
    :=
    \{f\in\Sch_{(\Rl)}(\Rl) \col \supp(f)\subset I\}
\end{align}
of real or complex functions with support inside a given subset\footnote{This implies $\Sch_{(\Rl)}(I)=\{0\}$ if $I$ has no interior points.} $I\subset\Rl$.
For an open set $O\subset\Rl^d$, we write $C_c^\infty(O)$\glsadd{compactTestFunctions} for the smooth complex-valued functions with compact support in $O$, and $C_{c,\Rl}^\infty(O)$ for the real-valued ones.

 We also consider the \emph{affine group $\Aff(\Rl)$ of $\Rl$}\glsadd{affineGroup}, defined as the group generated by the transformations\footnote{Abstractly speaking, $\Aff(\Rl)=\Rl\rtimes_\alpha\Rl^\times$ is the semidirect product of $\Rl^\times$ acting on $\Rl$ by multiplication (dilation) $\alpha_a(b)=ab$. It is a two-dimensional Lie group with two connected components. The component containing the identity is $\Aff_+(\Rl)=\Rl\rtimes_\alpha\Rl_+$. This group is often called the ``$ax+b$ group'', but we avoid this name because we typically use ``$a$'' to denote the translations instead of the dilations.}
 \begin{align}
    (\pm e^t,a):\Rl\to\Rl,\quad x\mapsto \pm e^t x+a,
\end{align}
 i.e. by translations $x\mapsto x+a$, dilations $x\mapsto e^{t}x$, and the reflection $x\mapsto-x$. We will use the notation $(\pm e^t,a)$ when the parameters need to be emphasized, and $\gamma\in\Aff(\Rl)$ when talking about a general group element. We also introduce the notation
 \begin{align}
    j&:=(-1,0):x\mapsto -x,\\
    \Aff_\pm(\Rl)&:=\{(\pm e^t,a)\col t,a\in\Rl\}
\end{align}
 for the reflection $j$\glsadd{affineReflection}, the normal subgroup $\Aff_+(\Rl)$\glsadd{positiveAffine} of orientation-preserving transformations, and the coset $\Aff_-(\Rl)=j\Aff_+(\Rl)$ of orientation-reversing transformations.

We also adopt the usual quantum field theory terminology to refer to maps between two partially ordered sets that are inclusion-preserving (monotone w.r.t. inclusions) as \emph{isotonous} maps or \emph{nets}.

\begin{lemma}[The net of local spaces in $\boldsymbol{\Sch_\Rl(\Rl)}$]\label{lemma:SchAffCovariance}
    \leavevmode
    \begin{enumerate}
        \item\label{item:SNetIsotony} The map $I\mapsto\Sch_\Rl(I)$ from subsets of $\Rl$ to subspaces of $\Sch_\Rl(\Rl)$ is isotonous,
        \begin{align}
            I_1\subset I_2 \implies \Sch_\Rl(I_1) \subset \Sch_\Rl(I_2).
        \end{align}

        \item\label{item:SNetAffineSymmetry} The affine group $\Aff(\Rl)$ acts on this net covariantly via the action
        \begin{align}
            \alpha:\Aff(\Rl)\times\Sch_\Rl(\Rl)&\to\Sch_\Rl(\Rl),
            \\
            \alpha_{\gamma}(f)&:=\pm f\circ\gamma^{-1},\qquad \gamma\in\Aff_\pm(\Rl),\label{eq:AffActsOnSchwartz}
            \\
            \alpha_\gamma(\Sch_\Rl(I))
            &=
            \Sch_\Rl(\gamma(I)).
            \label{eq:SNetCovariance}
        \end{align}

        \item\label{item:SNetNotStandard} Considered as real subspaces of $L^2(\Rl)$ via the inclusion embedding $\Sch_\Rl(\Rl)\hookrightarrow L^2(\Rl)$, the space $\overline{\Sch_{\Rl}(I)}\in\Gr_\Rl(L^2(\Rl))$ is separating for any $I\subset\Rl$, but not cyclic if $\Rl\backslash I$ has non-zero measure.
    \end{enumerate}
\end{lemma}
\begin{proof}
    \ref{item:SNetIsotony} is obvious from the definition. In \ref{item:SNetAffineSymmetry}, $\alpha$ is clearly an action, and the covariance property \eqref{eq:SNetCovariance} follows from $\supp\alpha_\gamma(f)=\gamma(\supp f)$, $f\in\Sch_\Rl(\Rl)$, $\gamma\in\Aff(\Rl)$.

    \ref{item:SNetNotStandard} As $\overline{\Sch_\Rl(I)}$ consists of real-valued functions, it is separating w.r.t. the complex structure (multiplication by $i$) of $L^2(\Rl)$. In case $\Rl\backslash I$ has non-zero measure, there exists a non-zero $L^2$-function $\xi$ supported on it. Then $\xi\perp(\Sch_\Rl(I)+i\Sch_\Rl(I))$, hence $\overline{\Sch_\Rl(I)}$ is not cyclic.
\end{proof}

 \begin{remark}[The sign of the reflection]\label{remark:U1AffineReflectionSign}
    The $\pm$ sign in the definition \eqref{eq:AffActsOnSchwartz} of the action of $\Aff(\Rl)$ on $\Sch_\Rl(\Rl)$ might be unexpected. The reason for this is that later we will mostly be interested in derivatives $f'$ of test functions $f\in\Sch_\Rl(\Rl)$, and $(\alpha_jf)'(x)=f'(-x)$ yields the natural action on derivatives.
\end{remark}

Item~\ref{item:SNetNotStandard} tells us that this construction does not yet fit into our standard subspace setting. To arrive at standard subspaces, we will embed this net into a different Hilbert space, using the Fourier transform\glsadd{fullFourier}
 \begin{align}\label{eq:FourierTransform}
    \tilde f(p):=(\EuF f)(p):=(2\pi)^{-1/2}\int_\Rl f(x)\,e^{ipx}\,dx.
\end{align}
 We consider the real-linear map\glsadd{positiveFourier}
\begin{align}\label{eq:U1Embedding}
    \EuF_+:\Sch_\Rl(\Rl)\to L^2(\Rl_+,p\,dp), \qquad
    f
    \mapsto
    \psi_f
    :=\tilde f|_{\Rl_+}
    .
\end{align}
This map is well-defined and injective because a real function $f$ satisfies $\tilde f(-p)=\overline{\tilde f(p)}$. By the same symmetry, the range of $\EuF_+$ contains $C_{c}^\infty(\Rl_+)$, so $\EuF_+$ is an embedding with dense range.

Our goal is to understand the properties of the images of the real local spaces $\Sch_\Rl(I)$.

 \begin{definition}[The $\boldsymbol{U(1)}$-current net in $\boldsymbol{L^2(\Rl_+,p\,dp)}$]\label{def:U1Net}\index{U(1) current net@$U(1)$ current net}\glsadd{localCurrent}
    The \emph{net of standard subspaces of the $U(1)$-current in $L^2(\Rl_+,p\,dp)$} is the net
    \begin{align}\label{eq:U1Net}
        I\mapsto H(I)
        :=
        \overline{\EuF_+(\Sch_\Rl(I))}
        \in
        \Real(L^2(\Rl_+,p\,dp))
        ,\qquad I\subset\Rl.
    \end{align}
\end{definition}

Note that \eqref{eq:U1Net} is really a net, i.e.
\medskip
 \begin{description}
 \item[\bfseries(Isotony)]\index{U(1) current net@$U(1)$ current net!isotony} $I_1\subset I_2\implies H(I_1)\subset H(I_2)$,
\end{description}
 \medskip
holds, because $I\mapsto\Sch_\Rl(I)$ has this property. The standard subspace property anticipated in the name of this net will be established below.

We start our investigation of \eqref{eq:U1Net} with the covariance properties.

 \begin{lemma}[Affine covariance of the $\boldsymbol{U(1)}$-net]\label{lemma:U1AffCovariance}\index{U(1) current net@$U(1)$ current net!affine covariance}
    \leavevmode
    \begin{enumerate}
        \item\label{item:UAffRep} The map
        \begin{align}
            U:\Aff(\Rl)&\to\AU(L^2(\Rl_+,p\,dp)),
            \\
            (U(e^{t},a)\psi)(p)
            &:=
            e^{ipa}e^{t}\psi(e^{t}p)
            \\
            (U(j)\psi)(p)
            &:=
            -\overline{\psi(p)}
        \end{align}
        defines a strongly continuous (anti-)unitary representation. $U|_{\Aff_+(\Rl)}$ is irreducible.

        \item\label{item:U1AffCovariance}\emph{\bfseries(Covariance)} The net \eqref{eq:U1Net} transforms covariantly under $U$: For any $\gamma\in\Aff(\Rl)$
        \begin{align}\label{eq:AffIntertwining}
            U(\gamma)\circ\EuF_+
            &=
            \EuF_+\circ\alpha_\gamma,\\
            U(\gamma)H(I)
            &=
            H(\gamma(I)),\qquad I\subset\Rl.
            \label{eq:U1AffCovariance}
        \end{align}
    \end{enumerate}
\end{lemma}
 \begin{proof}
    By direct calculation, one verifies that $U(e^t,a)$ is unitary for any $t,a\in\Rl$, and $U(j)$ is antiunitary. Also the relations
    \begin{align}
        U(e^t,a)U(e^{t'},a')
        &=
        U(e^{t+t'},a+e^{t}a'),
        \\
        U(j)U(e^t,a)U(j)
        &=
        U(e^t,-a),\qquad U(j)^2=1,
    \end{align}
    are checked by straightforward calculation. Hence $U$ defines an (anti-)unitary representation of $\Aff(\Rl)$ in which $\Aff_+(\Rl)$ is represented unitarily and $\Aff_-(\Rl)$ antiunitarily.

    The translations $(U(1,a))_{a\in\Rl}$ clearly form a strongly continuous unitary one-parameter group. The same is true for the dilations $(U(e^t,0))_{t\in\Rl}$: This can be checked for example by verifying continuity of $t\mapsto\langle\psi,U(e^t,0)\phi\rangle$ for $\psi,\phi\in C_c^\infty(\Rl_+)$. This implies that $U$ is a strongly continuous representation.

 As the generator $(P\psi)(p)=p\psi(p)$ of the translation group $(U(1,a))_{a\in\Rl}$ is multiplicity-free, the commutant of this one-parameter group is the maximal abelian von Neumann algebra $\{M_\varphi\col\varphi\in L^\infty(\Rl_+)\}$, where $M_\varphi\psi=\varphi\psi$ denotes the operator multiplying with $\varphi$. Such a multiplication operator commutes with dilations $(U(e^{2\pi t},0))_{t\in\Rl}$ if and only if $\varphi$ is dilation-invariant and therefore constant. This shows $U(\Aff_+(\Rl))'=\Cl1$, which is equivalent to the irreducibility of $U|_{\Aff_+(\Rl)}$.

    For the covariance claims, let $f\in\Sch_\Rl(\Rl)$ and $p\geq0$. We have
    \begin{align*}
        (\EuF_+\alpha_{(e^t,a)}(f))(p)
        &=
        \frac{1}{\sqrt{2\pi}}\int_\Rl f(e^{-t}(x-a))\,e^{ipx}\,dx
        \\
        &=
        \frac{1}{\sqrt{2\pi}}\int_\Rl f(e^{-t}x)\,e^{ip(x+a)}\,dx
        =
        (U(e^t,a)\EuF_+ f)(p)
    \end{align*}
    and
    \begin{align*}
        (\EuF_+\alpha_{j}(f))(p)
        &=
        -\frac{1}{\sqrt{2\pi}}\int_\Rl f(-x)\,e^{ipx}\,dx
        \\
        &=
        -\tilde f(-p)
        \\
        &=
        (U(j)\EuF_+f)(p).
    \end{align*}
    This proves \eqref{eq:AffIntertwining}. In view of Lemma~\ref{lemma:SchAffCovariance}~\ref{item:SNetAffineSymmetry} and (anti-)unitarity of $U$, this also implies \eqref{eq:U1AffCovariance}.
\end{proof}

 \begin{remark}[Failure of Reeh-Schlieder property without positive energy]\label{remark:U1TranslationEnergy}
 When embedding $\Sch_\Rl(\Rl)$ into $L^2(\Rl)$ via the inclusion map, we similarly get a representation $V$ of $\Aff(\Rl)$ that acts covariantly on the net $I\mapsto\Sch_\Rl(I)$. In this representation, dilations are not implemented unitarily. The essential difference to $U$ is however that the representation $V$ represents translations as the strongly continuous one-parameter group
 \begin{align}\label{eq:TranslationsV}
        (V(1,a)f)(x)=f(x-a),\qquad f\in L^2(\Rl),
    \end{align}
 the generator of which is not positive (it is the quantum-mechanical momentum operator $i\partial_x$). In contrast, the self-adjoint generator of $(U(1,a))_{a\in\Rl}$ is $(P\psi)(p)=p\psi(p)$, $\psi\in L^2(\Rl_+,p\,dp)$, a \emph{positive} operator. For this reason, $U$ is called a \emph{positive-energy representation}.
 \end{remark}

Having settled the isotony and covariance features of the net \eqref{eq:U1Net}, our aim is to establish the following theorem.

 \begin{theorem}[The net of the $\boldsymbol{U(1)}$-current on $\boldsymbol{L^2(\Rl_+,p\,dp)}$]\label{theorem:U1C+}\booktheoremindex{theorem:U1C+}{Net of standard subspaces of the U(1) current}
    \leavevmode
	\begin{enumerate}
        \item\label{item:U1Locality}\index{U(1) current net@$U(1)$ current net!locality} \emph{\bfseries(Locality)} The net \eqref{eq:U1Net} is \emph{local} in the sense
        \begin{align}
            I_1\cap I_2=\emptyset
            \implies
            H(I_1)\subset H(I_2)'.
        \end{align}

        \item\label{item:CycSep} \emph{\bfseries(Standardness)} The space $H(I)$ is cyclic if $I$ contains an interior point, and separating if $\Rl\backslash I$ contains an interior point.

        \item\label{item:Duality}\index{U(1) current net@$U(1)$ current net!duality}\glsadd{currentHalfLines}\emph{\bfseries(Duality)} $H_\pm:=H(\Rl_\pm)$ are factorial standard subspaces and symplectic complements of each other,
        \begin{align}
            H_+'=H_-.
        \end{align}

        \item\label{item:ModDat}\emph{\bfseries(Geometric modular action)}

        The modular data $\Delta_+,J_+$ of $H_+$ are given by
        \begin{align}\label{eq:U1CGMAgroup}
            \Delta_+^{it}
            &=
            U(e^{-2\pi t},0),
            \\
            J_+
            &=
            U(j),
            \label{eq:U1CGMAconjugation}
        \end{align}
        where $U$ denotes the $\Aff(\Rl)$-representation of Lemma~\ref{lemma:U1AffCovariance}.
	\end{enumerate}
\end{theorem}

Since the full proof is quite long and uses several different ideas, we split it up into several parts.

We first compute what the scalar product of $L^2(\Rl_+,p\,dp)$ looks like when pulled back to the real Schwartz space $\Sch_\Rl(\Rl)$.

\begin{lemma}[Symplectic structure and locality of the $\boldsymbol{U(1)}$-net]\label{lemma:SymplecticStructureAndLocality}
    \leavevmode
    \begin{enumerate}
        \item\label{item:SympForm} The real and imaginary parts of the scalar product of $L^2(\Rl_+,p\,dp)$ pull back to the following positive bilinear and symplectic forms on the real space $\Sch_\Rl(\Rl)$, respectively: For any $f,g\in\Sch_\Rl(\Rl)$,
        \begin{align}\label{eq:RealPartonSchR}
            \Re\langle\psi_f,\psi_g\rangle
            &=
            \frac{1}{2}\int_\Rl\overline{\tilde f(p)}\tilde g(p)\,|p|\,dp
            .
            \\
            \Im\langle\psi_f,\psi_g\rangle
            &=
            \frac12\int_\Rl f(x)g'(x)\,dx
            =
            -\frac12\int_\Rl f'(x)g(x)\,dx.
        \end{align}

        \item\label{item:U1NetLocal} The net is local, i.e. the claim in Theorem~\ref{theorem:U1C+}~\ref{item:U1Locality} holds.
        \end{enumerate}
\end{lemma}
\begin{proof}
    \ref{item:SympForm} Let $f,g\in\Sch_\Rl(\Rl)$. Then $\overline{\tilde f(p)}=\tilde f(-p)$, $\overline{\tilde g(p)}=\tilde g(-p)$, and we obtain
    \begin{align}
        \Im\langle\psi_f,\psi_g\rangle
        &=
        \frac{1}{2i}\left[\int_0^\infty\overline{\tilde f(p)}\tilde g(p)\,p\,dp
        -
        \int_0^\infty\tilde f(p)\overline{\tilde g(p)}\,p\,dp
        \right]
        \\
        &=
        \frac{1}{2i}\left[\int_0^\infty\overline{\tilde f(p)}\tilde g(p)\,p\,dp
        -
        \int_0^\infty\overline{\tilde f(-p)}\tilde g(-p)\,p\,dp
        \right]
        \\
        &=
        \frac{1}{2i}\left[\int_0^\infty\overline{\tilde f(p)}\tilde g(p)\,p\,dp
        +
        \int_{-\infty}^0\overline{\tilde f(p)}\tilde g(p)\,p\,dp
        \right]
        \\
        &=
        \frac{1}{2}\int_\Rl\overline{\tilde f(p)}[-ip\tilde g(p)]\,dp
        \\
        &=
        \frac{1}{2}\int_\Rl f(x)g'(x)\,dx.
    \end{align}
    For the real part, the difference in sign results in
    \begin{align*}
        \Re\langle\psi_f,\psi_g\rangle
        &=
        \frac{1}{2}\left[\int_0^\infty\overline{\tilde f(p)}\tilde g(p)\,p\,dp
        +
        \int_0^\infty\tilde f(p)\overline{\tilde g(p)}\,p\,dp
        \right]
        \\
        &=
        \frac{1}{2}\left[\int_0^\infty\overline{\tilde f(p)}\tilde g(p)\,p\,dp
        -
        \int_{-\infty}^0\overline{\tilde f(p)}\tilde g(p)\,p\,dp
        \right]
        \\
        &=
        \frac{1}{2}\int_\Rl\overline{\tilde f(p)}\tilde g(p)\,|p|\,dp
        .
    \end{align*}

 \ref{item:U1NetLocal} If two real Schwartz functions $f,g$ have supports in disjoint sets $I_1,I_2$, part~\ref{item:SympForm} shows $\Im\langle\psi_f,\psi_g\rangle=0$. Hence $\EuF_+(\Sch_\Rl(I_1))\subset\EuF_+(\Sch_\Rl(I_2))'$, which implies $H(I_1)\subset H(I_2)'$ by taking the symplectic complement twice.
 \end{proof}

We now prove part~\ref{item:CycSep} of Theorem~\ref{theorem:U1C+} by a so-called Reeh-Schlieder argument (see Notes and Remarks).

\bigskip

\noindent
{\bfseries Proof of standardness (Theorem~\ref{theorem:U1C+}~\ref{item:CycSep})}\\
\noindent
    Let $I\subset\Rl$ have an interior point, and $\xi\perp H(I)$. We have to show $\xi=0$. There exists a bounded interval $I_0\subset I$ and $\eps>0$ such that $x+a\in I$ for all $x\in I_0$ and all $a\in(-\eps,\eps)$. This means that $U(1,a)H(I_0)\subset H(I)$ for $|a|<\eps$, and $\xi$ satisfies for any $f\in C_{c,\Rl}^\infty(I_0)$
	\begin{align*}
		0
		&=
		\langle\xi,U(1,a)\EuF_+(f)\rangle
		=
		\int_0^\infty\overline{\xi(p)}\widetilde{f}(p)\,e^{iap}\,p\,dp,\qquad |a|<\eps.
	\end{align*}
 As a function of $a\in\Rl$, this integral admits a bounded continuous extension to the closed upper half-plane $\overline{\Cl_+}$ which is analytic in the open half-plane $\Cl_+$. As this function vanishes on the interval $(-\eps,\eps)\subset\Rl$ at the boundary of $\Cl_+$,  it must vanish everywhere by the Schwarz Reflection Principle. The injectivity of the Fourier transform then implies $\overline{\xi(p)}\widetilde{f}(p)p=0$ for almost all $p\in\Rl_+$. As the Fourier transform of a compactly supported non-zero function, $\widetilde f$ vanishes only on a set of measure zero, this implies $\xi=0$. We conclude $\{0\}=H(I)^\perp=(H(I)+iH(I))^\rperp$, hence $H(I)$ is cyclic.

	If $\Rl\backslash I$ contains an interior point, $H(\Rl\backslash I)$ is cyclic and therefore $H(\Rl\backslash I)'$ is separating. By locality, $H(I)\subset H(\Rl\backslash I)'$, so $H(I)$ is separating as well.\hfill$\square$

 \bigskip

 It should be noted that the positivity of the generator of translations is the reason why this argument works, and fails for the non-positive-energy representation obtained by embedding the net into $L^2(\Rl)$.

\bigskip

The proof of the geometric modular action claims in Theorem~\ref{theorem:U1C+}~\ref{item:ModDat} is the most difficult part of this theorem (it will imply the duality claim in Theorem~\ref{theorem:U1C+}~\ref{item:Duality} directly).

We split the proof of part~\ref{item:ModDat} and begin with the modular group by verifying the KMS condition. As a preparation, we note the following standard estimate.

\begin{lemma}[A Fourier transform bound]\label{lemma:FourierBound}
    Let $f\in C_{c,\Rl}^\infty(\Rl_+)$ and $N\in\Nl$. Then there exists $C_{f,N}>0$ such that
    \begin{align}\label{eq:ftiBound}
        |\tilde f(w)|\leq \frac{C_{f,N}}{(1+|w|)^N},\qquad \Im(w) \geq0.
    \end{align}
\end{lemma}
\begin{proof}
    Note that $\tilde f$ is an entire analytic function because $f$ is compactly supported. By integration by parts, we have for any $w\in\Cl$, $\Im(w)>0$,
    \begin{align*}
        \sqrt{2\pi}|\tilde f(w)|
        =
        \left|\int_\Rl f(x)\,(iw)^{-N}\partial_x^N\,e^{iwx}dx\right|
        \leq
        \frac{\|f^{(N)}\|_1}{|w|^N},
    \end{align*}
    because the support of $f$ in $\Rl_+$ and $\Im(w)>0$ imply that the exponential $|e^{iwx}|=e^{-\Im(w)x}$ is bounded by $1$. We use this estimate for $|w|\geq1$, and for small $|w|$ with $\Im(w)\geq0$, we estimate directly $\sqrt{2\pi}|\tilde f(w)|\leq\|f\|_1$. Combining these two bounds yields \eqref{eq:ftiBound}.
\end{proof}

\bigskip

\noindent
{\bfseries Proof of formula \eqref{eq:U1CGMAgroup} for the modular group of $\boldsymbol{H_+}$.}\\
    \noindent
 To prove the claimed form of the modular group, we will use Theorem~\ref{theorem:KMS} and verify the KMS condition. We know already that $V_t:=U(e^{-2\pi t},0)$ is a strongly continuous unitary one-parameter group (Lemma~\ref{lemma:U1AffCovariance}~\ref{item:UAffRep}) that leaves $H_+$ invariant (Lemma~\ref{lemma:U1AffCovariance}~\ref{item:U1AffCovariance}) because the half-line $\Rl_+$ is dilation-invariant. This verifies the invariance hypothesis of Theorem~\ref{theorem:KMS}~\ref{item:KMSCharacterizesModularGroup}\,\ref{item:KMSInvariance}.

    The other hypothesis is the KMS condition~\ref{item:KMSHypothesis}. It is enough to check it on a dense subspace (Lemma~\ref{lemma:KMSDenseIsEnough}), which we choose as $D:=\EuF_+(C_{c,\Rl}^\infty(\Rl_+))\subset H_+$. That is, given any $f,g\in C_{c,\Rl}^\infty(\Rl_+)$, we have to show that the ``KMS function''
 \begin{align}\label{eq:KMSfunction}
        \begin{aligned}
            k(t)
            &:=
            \langle\psi_f,V_t\psi_g\rangle
            \\
            &=
            \langle V_{-t/2}\psi_f,V_{t/2}\psi_g\rangle
            \\
            &=
            \int_0^\infty \tilde f(-e^{\pi t}p)\tilde g(e^{-\pi t}p)\,p\,dp.
        \end{aligned}
    \end{align}
    has a bounded analytic continuation to the strip $\Strip_{-1}$, and satisfies the KMS boundary condition
    \begin{align}\label{eq:KMSBoundaryConditionInProof}
        k(t-i)=\overline{k(t)},\qquad t\in\Rl.
    \end{align}
    Thanks to the compact support of $f$ and $g$, the integrand of \eqref{eq:KMSfunction} is entire in $t\in\Cl$. For $0\leq s\leq 1$, $t\in\Rl$, $p>0$, we furthermore note
    \begin{align*}
        \Im(e^{-\pi(t-is)}p)=\sin(\pi s)e^{-\pi t}p\geq0,
        \qquad
        \Im(-e^{\pi(t-is)}p)=\sin(\pi s)e^{\pi t}p\geq0.
    \end{align*}
    Hence we may apply Lemma~\ref{lemma:FourierBound} and obtain
    \begin{align*}
        |\tilde f(-e^{\pi (t-is)}p)\tilde g(e^{-\pi (t-is)}p)\,p|
        &\leq
        \frac{C_{f,N}C_{g,N}\,p}{(1+e^{-\pi t}p)^N(1+e^{\pi t}p)^N}
        \leq
        \frac{C_{f,N}C_{g,N}\,p}{(1+p^2)^N},
    \end{align*}
    which is integrable over $p\in\Rl_+$ for $N>1$. Hence
    \begin{align*}
        k(t-is)
        &:=
        \int_0^\infty \tilde f(-e^{\pi (t-is)}p)\tilde g(e^{-\pi (t-is)}p)\,p\,dp,\qquad t\in\Rl,\;0\leq s\leq1,
    \end{align*}
    is well-defined and bounded. Standard dominated convergence arguments show that $k$ is analytic in $\Strip_{-1}$ and bounded on its closure, i.e. $k\in\Hardy^\infty_{C}(\Strip_{-1})$, and of course it restricts to the KMS function \eqref{eq:KMSfunction} for $s=0$.

    At $s=1$, we find
 \begin{align*}
        k(t-i)
        =
        \int_0^\infty \tilde f(e^{\pi t}p)\tilde g(-e^{-\pi t}p)\,p\,dp.
    \end{align*}
 Comparison with \eqref{eq:KMSfunction} shows $k(t-i)=\langle \psi_g,V_{-t}\psi_f\rangle=\overline{k(t)}$, which shows \eqref{eq:KMSBoundaryConditionInProof} and finishes the proof.

 We know that the centre of $H_+$ is given by the vectors in $H_+$ that are invariant under $\Delta_+^{it}$ (Proposition~\ref{proposition:CenterAndMaximalAbelianness}). But the dilation one-parameter group $(V_t)_{t\in\Rl}$ has only $0$ as a fixed point in $L^2(\Rl_+,p\,dp)$. Thus $H_\pm$ are factorial standard subspaces.\qed 

\bigskip

We now present the first strategy for proving the claimed formula \eqref{eq:U1CGMAconjugation} for the modular conjugation of $H_+$.

\medskip

\noindent
{\bfseries First proof of duality (Theorem~\ref{theorem:U1C+}~\ref{item:Duality}) and of formula \eqref{eq:U1CGMAconjugation} for modular conjugation of~$\boldsymbol{H_+}$.}\\
\noindent Let $\xi\in L^2(\Rl_+,p\,dp)$ and $f\in C_{c,\Rl}^\infty(\Rl_+)$, i.e. $\psi_f\in H_+$. As $H_+\subset\dom\Delta_+^{1/2}$, we know that
\begin{align}\label{eq:dinproof}
    d_1(t)
    &:=
    \langle\xi,\Delta_+^{it}\psi_f\rangle
    =
    \langle\xi,U(e^{-2\pi t},0)\psi_f\rangle
    \\
    =
    d_2(t)
    &:=
    \int_0^\infty \overline{\xi(p)}e^{-2\pi t}\tilde f(e^{-2\pi t}p)\,p\,dp,\qquad t\in\Rl.
\end{align}
The function $d_1$ has a continuous continuation to the strip $\overline{\Strip_{-1/2}}$ which is analytic in its interior, and satisfies $d_1(-\frac{i}{2})=\langle\xi,\Delta_+^{1/2}\psi_f\rangle=\langle\xi,J_+\psi_f\rangle$, where we have used $\Delta_+^{1/2}\psi_f=J_+\psi_f$ because $\psi_f\in H_+$. On the other hand, the integrand of $d_2$ is entire analytic. By the same estimates as in the proof of the KMS condition, it follows that inserting complex values of $t$ in this integral gives a continuous extension of $d_2$ to $\overline{\Strip_{-1/2}}$ which is analytic in the interior. This function satisfies
\begin{align*}
    d_2\left(-\tfrac{i}{2}\right)
    &=
    -\int_0^\infty \overline{\xi(p)}\tilde f(-p)\,p\,dp
    =
    \int_0^\infty \overline{\xi(p)}(U(j)\tilde f)(p)\,p\,dp
\end{align*}
because $\tilde f(-p)=\overline{\tilde f(p)}$. As $d_1$ and $d_2$ coincide on $\Rl$, their analytic continuations coincide as well, and we conclude
\begin{align}
    J_+\psi_f=U(j)\psi_f,\qquad f\in C_{c,\Rl}^\infty(\Rl_+).
\end{align}
 Taking into account that $\psi_f$ runs over a cyclic space, this shows $J_+=U(j)$.

The duality relation $H_+=H_-'$ is now a direct consequence because the Tomita-Takesaki theorem tells us that $J_+$ maps $H_+$ onto $H_+'$, and reflection covariance tells us that $U(-1,0)$ maps $H_+$ onto $H_-$, so
\begin{align}
    H_+'=J_+H_+=U(j)H_+=H_-.
\end{align}
This finishes the proof. \hfill$\square$

\bigskip

With minor changes, this proof can be extended to show that for $f\in C_{c,\Rl}^\infty(\Rl_+)\subset H_+$, we have
\begin{align}
    (\Delta_+^{i(t-is)}\psi_f)(p)
    =
    e^{-2\pi(t-is)}\tilde f(e^{-2\pi(t-is)}p),\qquad 0\leq s\leq\tfrac12,\;t\in\Rl,\;p>0,
\end{align}
making use of the analytic continuation of $\tilde f$. We will later see that a similar picture emerges for general vectors in $\dom\Delta_+^{1/2}$.

 \medskip

 To also show how to work with reflection positivity, we add an alternative proof based on Proposition~\ref{proposition:ReflectionPositivity}. We first verify the reflection positivity condition~\ref{item:ReflectionPositive}.

\begin{lemma}[Reflection Positivity]\label{lemma:ReflectionPositivity}\index{reflection positivity}
    The conjugation $U(j)$ is reflection positive on $H_+$, i.e.
    \begin{align}
        \langle h,U(j)h\rangle
        \geq 0,\qquad h\in H_+.
    \end{align}
\end{lemma}
 \begin{proof}
    By continuity, it is enough to verify the claim for $h=\psi_f$, $f\in\Sch_\Rl(\Rl_+)$.  Let $f\in\Sch_\Rl(\Rl_+)$. By regularization and Fubini's Theorem, the scalar product in question is
 \begin{align*}
    \langle\psi_f,U(j)\psi_f\rangle
    &=
    -\int_0^\infty \overline{\tilde f(p)^2}p\,dp
    \\
    &=
    -\lim_{\eps\downarrow0}\int_0^\infty \overline{\tilde f(p)^2}e^{-\eps p}p\,dp
    \\
    &=
    -\frac{1}{2\pi}\lim_{\eps\downarrow0}\int_0^\infty f(x)\int_0^\infty f(y)\int_0^\infty e^{-ip(x+y)}e^{-\eps p}p\,dp\,dy\,dx
    .
\end{align*}
 The inner integral is
\begin{align}
    \int_0^\infty p\,e^{-p(\eps+i(x+y))}\,dp
    &=
    \frac{1}{(\eps+i(x+y))^2}
    \longrightarrow
    -\frac{1}{(x+y)^2}\quad \text{as }\;\eps\to0
    \\
    &=
    -\int_0^\infty k\,e^{-k(x+y)}\,dk,\qquad x,y>0.
\end{align}
We therefore obtain
\begin{align*}
    \langle\psi_f,U(j)\psi_f\rangle
    &=
    \frac{1}{2\pi}\int_0^\infty k\left(\int_0^\infty f(x)e^{-kx}\,dx\right)^2dk
    \geq0.
\end{align*}
\end{proof}

We now give the second proof of duality and the formula $J_+=U(j)$ for the modular conjugation.

\bigskip

\noindent
{\bfseries Second proof of duality (Theorem~\ref{theorem:U1C+}~\ref{item:Duality}) and of formula \eqref{eq:U1CGMAconjugation} for modular conjugation of~$\boldsymbol{H_+}$.}\\
\noindent
Let $a\geq0$. For vectors $h\in H_+\subset\dom\Delta_+^{1/2}$ and $h'\in H_+'\subset\dom\Delta_+^{-1/2}$, the $L^2(\Rl_+,p\,dp)$-valued functions $z\mapsto\Delta_+^{-iz}h$ and $z\mapsto\Delta_+^{iz}h'$ are analytic and bounded on the strip $\Strip_{1/2}$. The positivity of the generator $P$ of the translation group implies that
\begin{align}
    \Strip_{1/2}
    \ni
    z\mapsto U(1,e^{2\pi z}a)=e^{ie^{2\pi\Re(z)}\cos(2\pi \Im(z))aP}e^{-e^{2\pi\Re(z)}\sin(2\pi\Im(z))aP}
    \in
    B(L^2(\Rl_+,p\,dp))
    \notag \end{align}
is an operator-valued bounded analytic function. Hence
\begin{align}
    c:\overline{\Strip_{1/2}}\to\Cl,\qquad
    c(z)
    &:=
    \langle\Delta_+^{-i\overline z}h',U(1,e^{2\pi z}a)\Delta_+^{-iz}h\rangle
\end{align}
is bounded and analytic in $\Strip_{1/2}$. For real $z=t$, we may use the fact that we know that $\Delta_+^{-it}$ coincides with dilations \eqref{eq:U1CGMAgroup}. This results in
\begin{align}
    c(t)
    &=
    \langle h',U(e^{-2\pi t},0)U(1,e^{2\pi t}a)U(e^{2\pi t},0)h\rangle
    =
    \langle h',U(1,a) h\rangle,
\end{align}
 and we conclude that $c|_\Rl$ is constant. As $c$ is analytic and bounded, it follows that $c$ is constant on the whole strip $\overline{\Strip_{1/2}}$. Hence
 \begin{align*}
    \langle h',U(1,a)h\rangle
    =
    c(0)
    =
    c\left(\tfrac{i}{2}\right)
    =
    \langle \Delta_+^{-1/2}h',U(1,-a)\Delta_+^{1/2}h\rangle.
\end{align*}
 The right-hand entry $U(1,-a)\Delta_+^{1/2}h=U(1,-a)J_+h$ of the scalar product lies in $H_+'$ because (note $a\geq0$ and Lemma~\ref{lemma:U1AffCovariance}~\ref{item:U1AffCovariance})
 \begin{align*}
    U(1,a)H_+\subset H_+\implies U(1,-a)H_+\supset H_+
    \implies
    U(1,-a)H_+'=(U(1,-a)H_+)'\subset H_+'.
\end{align*}
Thus $U(1,-a)J_+h$ lies in $\dom\Delta_+^{-1/2}$, and $\Delta_+^{-1/2}$ acts on it as $J_+$. This means
\begin{align}
    \langle h',U(1,a)h\rangle
    =
    c\left(\tfrac{i}{2}\right)
    =
    \langle h',J_+U(1,-a)J_+h\rangle.
\end{align}
As $H_+,H_+'$ are cyclic, this implies the crucial exchange relation
\begin{align}
    U(1,a)=J_+U(1,-a)J_+,
\end{align}
which by taking adjoints we easily see to hold for $a\in\Rl$, not just $a\geq0$.

Now consider unitary $W:=J_+U(j)$. Then $W$ commutes with dilations, and because $J_+$ and $U(-1,0)$ act in the same manner on translations, we find that $W$ commutes with translations as well. But the dilations and translations generate the irreducible $\Aff_+(\Rl)$-representation, which implies that $W$ is a multiple of the identity, i.e. $U(j)=wJ_+$ for some $w\in\Cl$, $|w|=1$.

To show $w=1$, we use reflection positivity. From Lemma~\ref{lemma:ReflectionPositivity} and Proposition~\ref{proposition:ReflectionPositivity} we know that $U(j)$ and $J_+$ are both reflection positive on $H_+$. For non-zero $h\in H_+$, this gives
\begin{align*}
    w
    =
    \frac{\langle h,U(j)h\rangle}{\langle h,J_+h\rangle}\geq0,
\end{align*}
and $w=1$ follows. The duality relation $H_+=H_-'$ follows from this exactly as in the first proof. \hfill$\square$

 \bigskip

As an illustration of cutting projections (Def.~\ref{definition:cuttingprojection}), we now work out how this projection looks like for the factorial standard subspace $H_+$.

\begin{lemma}[Test functions cut off at zero]\label{lemma:CutFunctions}
    Let $f\in C_{c,\Rl}^\infty(\Rl)$, and $f_\pm:=\chi_{\pm}f$, where $\chi_\pm=\chi_{\Rl_\pm}$ is the characteristic function of $\Rl_\pm\subset\Rl$, and $\psi_{f_\pm}:=\widetilde{f_\pm}|_{\Rl_+}$.
    \begin{enumerate}
        \item\label{item:Cutf0} $\psi_{f_\pm}\in L^2(\Rl_+,p\,dp) \iff f(0)=0$.
        \item\label{item:Cutf0HH'} If $f(0)=0$, then $\psi_{f_\pm}\in H_\pm$ and $\psi_f=\psi_{f_+}+\psi_{f_-}$.
    \end{enumerate}
\end{lemma}
\begin{proof}
 \ref{item:Cutf0} As $f_++f_-=f$, it is enough to consider $f_+$. Since $f_+\in L^1(\Rl)$, we have $\psi_{f_+}\in C_b(\Rl)$, so $\psi_{f_+}\in L^2(\Rl_+,p\,dp)$ is equivalent to $p\mapsto p|\widetilde{f_+}(p)|^2$ being integrable at $p\to\infty$. The derivative of $f_+$ in the sense of distributions is ${f_+}'=\chi_+f'+f(0)\delta_0$. If $f(0)=0$, this yields $\widetilde{f_+}(p)=\frac{i}{p}\widetilde{(f')_+}(p)$. As $\widetilde{f'}\in L^2(\Rl)$, we see that $p\mapsto p|\widetilde{f_+}(p)|^2$ is integrable at $p\to\infty$ in this case. If $f(0)\neq0$, we have $\widetilde{f_+}(p)=\frac{i}{\sqrt{2\pi}\,p}f(0)+\frac{i}{p}\widetilde{(f')_+}(p)$ With $f'\in L^1(\Rl)$, the second term goes to zero faster than $p^{-1}$ as $p\to\infty$, hence $\widetilde{f_+}(p)\sim\frac{i}{\sqrt{2\pi}\,p}f(0)$ as $p\to\infty$, and $\int_0^\infty p|\widetilde{f_+}(p)|^2\,dp$ is logarithmically divergent.

    \ref{item:Cutf0HH'} For $f(0)=0$, we find sequences $f_n^\pm\in C_{c,\Rl}^\infty(\Rl)$ with $\supp f_n^+\subset(0,\infty)$ and $\supp f_n^-\subset(-\infty,0)$ such that $f_n^\pm\to f_\pm$ in the norm of the Sobolev space of order $1$ by smoothly cutting $f$ down to $0$ in a neighbourhood of $x=0$. This implies $\psi_{f_n^\pm}\to\psi_{f_\pm}$ in the topology of $L^2(\Rl_+,p\,dp)$. As $\psi_{f_n^+}\in H_+$ and $\psi_{f_n^-}\in H_-$, the claim follows.
\end{proof}

The following corollary explains the name ``cutting projection''.

 \begin{corollary}[The cutting projection of the half-line space]\label{proposition:U1Cutting}
    The cutting projection $P_{H_+}$ of~$H_+$ is the closure of the map
    \begin{align}\label{eq:CutH+-}
        \psi_f\mapsto\psi_{f_+},\qquad f\in C_{c,\Rl}^\infty(\Rl),\quad f(0)=0.
    \end{align}
\end{corollary}
 \begin{proof}
    Let us consider the real spaces
    \begin{align*}
        R&:=\{\psi_{f}\col f\in C_{c,\Rl}^\infty(\Rl),\quad f(0)=0\},\\
        K_\pm&:=\{\psi_f\col f\in C_{c,\Rl}^\infty(\Rl),\quad \supp f\subset\Rl_\pm\}.
    \end{align*}
    We note $\overline{K_\pm}=H_\pm$ and $K_++K_-\subset R$. The map \eqref{eq:CutH+-} is an extension of the projection $P_{K_+,K_-}$ (cf. Lemma~\ref{lemma:ProjectionsAndPairsOfIndependentSpaces}) to the larger domain $R$. As $H_+\cap H_-=\{0\}$ by factoriality of $H_+$ (we have $H_-=H_+'$), it follows that \eqref{eq:CutH+-} closes to the cutting projection $P_{H_+}=P_{H_+,H_-}$.
\end{proof}

\subsection{The $\boldsymbol{U(1)}$-current in different $\boldsymbol{L^2}$- and Hardy spaces}\label{section:U1DifferentSpaces}

As mentioned before, we now describe the $U(1)$-current on several other Hilbert spaces. In order not to overburden our notation, we will often refer to unitarily equivalent objects by the same symbols and indicate by writing $H_+\subset L^2(\Rl_+,p\,dp)$ or $H_+\subset L^2(\Rl_+,\tfrac{dp}{p})$ in which Hilbert space a standard subspace (or representation) is considered.

 \begin{lemma}[Unitarily equivalent versions of the $\boldsymbol{\Aff}$-representation]\label{lemma:UnitaryEquivalencesAff}
    \leavevmode
    \begin{enumerate}
        \item\label{item:U1ModelUnitaries} The maps
        \begin{alignat}{2}
             L^2(\Rl_+,p\,dp)&\to L^2(\Rl_+,\tfrac{dp}{p}),\qquad \psi &&\mapsto \varphi,\qquad \varphi(p):=ip\psi(p)
             ,\\
             L^2(\Rl_+,\tfrac{dp}{p})&\to L^2(\Rl,d\te),\qquad \psi &&\mapsto f,\qquad f(\te):=\psi(e^\te)\iff\psi(p)=f(\log p),\notag\\L^2(\Rl,d\te)&\to L^2(\Rl,d\xi),\qquad f&&\mapsto \tilde f,
        \end{alignat}
        are unitaries (with the sign convention of the Fourier transform $\EuF f=\tilde f$ as in \eqref{eq:FourierTransform}).

        \item\label{item:U1AffineModelActions} When transported with the unitaries above, the $\Aff(\Rl)$-representation $U$ of Lemma~\ref{lemma:U1AffCovariance} acts according to\footnote{The translations on $L^2(\Rl,d\xi)$ take a less explicit form, so we do not write them down here.}
        \begin{alignat}{4}
            \text{{On }} &\boldsymbol{L^2(\Rl_+,\tfrac{dp}{p})}\,\text{{:\quad}} (U(e^t,a)\varphi)(p)&&=e^{ipa}\varphi(e^t p), \qquad &&(U(j)\varphi)(p)&&=\overline{\varphi(p)},\\
            \text{{On }} &\boldsymbol{L^2(\Rl,d\te)}\,\text{{:\quad}} (U(e^t,a)f)(\te)&&=e^{iae^\te}f(\te+t), \qquad &&(U(j)f)(\te)&&=\overline{f(\te)},\\
            \text{{On }}&\boldsymbol{L^2(\Rl,d\xi)}{:\quad} (U(e^t,0)\tilde f)(\xi)&&=e^{-it\xi}\tilde f(\xi), \qquad &&(U(j)\tilde f)(\xi)&&=\overline{\tilde f(-\xi)}.
        \end{alignat}
    \end{enumerate}
\end{lemma}
\begin{proof}
    The proof is a straightforward calculation that is omitted here.
\end{proof}

In each of these Hilbert spaces, we may consider the net $H(I)$ constructed in the previous section, and in particular the half-line standard subspace $H_+=H(\Rl_+)$ that is uniquely fixed in terms of its modular group and conjugation,
 \begin{align}
    \Delta_+^{it}=U(e^{-2\pi t},0),\qquad J_+=U(j).
\end{align}

\subsubsection*{The $\boldsymbol{U(1)}$-current in $\boldsymbol{L^2(\Rl,d\xi)}$}

A direct description of the vectors in $H_+$ is most easy in $L^2(\Rl,d\xi)$ because the modular operator is a multiplication operator, $(\Delta_+^{1/2} g)(\xi)=e^{\pi\xi}g(\xi)$. As in Example~\ref{example:StandardSubspacesFromMeasures}, we have
\begin{align}
    H_+
    &=
    \left\{g\in L^2(\Rl,d\xi)
    \col
          \int_\Rl e^{2\pi\xi}|g(\xi)|^2\,d\xi<\infty,\qquad
          \overline{g(-\xi)}=e^{\pi\xi}g(\xi)
    \right\},\\
    \dom\Delta_+^{1/2}
    &=
    H_++iH_+,
    \\
    (\Delta_+^{iz}g)(\xi)
    &=
    e^{2\pi iz\xi}g(\xi),\qquad z\in\overline{\Strip_{-1/2}},\quad g\in H_++iH_+.
\end{align}

In order to distinguish $\dom\Delta_+^{iz}$ from unitarily equivalent versions of this domain in other Hilbert spaces, we introduce the notation
 \begin{align}
    \widetilde\CD_\beta
    \label{eq:U1WeightedFourierDomain}
    &:=
    \left\{g\in L^2(\Rl,d\xi)
    \col
          \int_\Rl e^{2\beta\xi}|g(\xi)|^2\,d\xi<\infty
    \right\}
    \\
    &=
    \left\{g:\Rl\to\Cl\,\text{ measurable } \col \sup_{\la}\int_\Rl e^{2\la\xi}|g(\xi)|^2d\xi<\infty \right\}
    ,
\end{align}
such that $\dom\Delta_+^{iz}=\widetilde\CD_{-2\pi\Im(z)}$. Here the supremum in \eqref{eq:U1WeightedFourierDomain} runs over $0\leq\la\leq\beta$ for $\beta>0$ and over $\beta\leq\la\leq0$ for $\beta<0$.

\subsubsection*{The $\boldsymbol{U(1)}$-current in $\boldsymbol{L^2(\Rl,d\te)}$ and Hardy spaces on strips}

To arrive at a similarly explicit description of $H_+$ and the action of $\Delta_+^{iz}$, $z\in\Cl$, on $L^2(\Rl,d\te)$, we have to characterize
\begin{align}
    \CD_\beta:=\EuF^{-1}\widetilde\CD_\beta \subset L^2(\Rl,d\te),
\end{align}
the inverse Fourier transform of the dense subspace $\widetilde\CD_\beta$. This is best done in terms of Hardy spaces on the strip.

\begin{definition}[Hardy space on the strip]\label{def:U1StripHardySpace}
 Let $\beta\in\Rl\backslash\{0\}$. The {\em Hardy space on $\Strip_\beta$} is the space $\Hardy^2(\Strip_\beta)$\glsadd{HardyStrip}\index{Hardy space!on strip} of all analytic functions $F:\Strip_\beta\to\Cl$ such that
    \begin{align}\label{eq:U1BanachNormHardy}
        \|F\|^2_{\Hardy^2(\Strip_\beta)}
        :=
        \sup_{\la}\int_\Rl |F(\te+i\la)|^2\,d\te<\infty,
    \end{align}
 where the supremum runs over $0<\la<\beta$ for $\beta>0$ and over $\beta<\la<0$ for $\beta<0$.
\end{definition}

 This space of analytic functions is studied in the appendix in Section~\ref{section:Hardy2Strip}. We will need the following facts (see Theorem~\ref{theorem:HardyFunctionsAppendix} for the proof).

\begin{theorem}[Analytic properties of Hardy space functions]\label{theorem:HardyFunctionsInText}\booktheoremindex{theorem:HardyFunctionsInText}{Analytic properties of Hardy space functions (current net)}
    Let $\beta\neq0$.
    \begin{enumerate}

        \item\label{item:U1HardyBoundaryValues1} For any $F\in\Hardy^2(\Strip_\beta)$, the limits
        \begin{align}
            F_0(\te):=\lim_{\la\to0}F(\te+i\la),\qquad
            F_\beta(\te):=\lim_{\la\to\beta}F(\te+i\la)
        \end{align}
        exist as norm limits in $L^2(\Rl)$. In particular, $F_0,F_\beta\in L^2(\Rl)$.

        \item\label{item:U1HardyBoundaryValues2} The maps $\Hardy^2(\Strip_\beta)\ni F\mapsto F_0\in L^2(\Rl)$ and $\Hardy^2(\Strip_\beta)\ni F\mapsto F_\beta\in L^2(\Rl)$ are injective.

        \item\label{item:U1HardyFourierTransform} Given any $g\in\widetilde\CD_\beta$ \eqref{eq:U1WeightedFourierDomain}, the function
        \begin{align}\label{eq:U1gtoF}
            F:\Strip_\beta\to\Cl,\quad
            F(\te+i\la):=
            F_\la(\te):=
            \frac{1}{\sqrt{2\pi}}\int_\Rl e^{-i\te\xi}e^{\la\xi}g(\xi)\,d\xi
        \end{align}
        lies in $\Hardy^2(\Strip_\beta)$. Conversely, any $F\in\Hardy^2(\Strip_\beta)$ is of the form \eqref{eq:U1gtoF} for some $g\in\widetilde\CD_\beta$. One has
        \begin{align}
            F(\te+z)
            =
            F_{\Im(z)}(\te+\Re(z))
            =
            \EuF^{-1}(\Delta_+^{-iz/2\pi} g)(\te),\qquad z\in\overline{\Strip_{\beta}},
        \end{align}
        to be understood in the sense of boundary values for $\Im(z)\in\{0,\beta\}$.
    \end{enumerate}
\end{theorem}

With this theorem, we may identify the inverse Fourier transform of $\widetilde\CD_\beta$ as the space of all boundary values $F_0$ of Hardy functions $F\in\Hardy^2(\Stripb)$ at the boundary $\Rl$ of $\Stripb$, and in view of the injectivity of the boundary value maps, we may identify this space with $\Hardy^2(\Stripb)$. The identification of the holomorphic functions $F\in\Hardy^2(\Stripb)$ with their boundary values $F_0\in L^2(\Rl,d\te)$ is usually suppressed in the notation.

 \begin{corollary}[The half-line space in $\boldsymbol{L^2(\Rl,d\te)}$ and Hardy spaces on the strip]\label{corollary:U1currentStandardSubspace}\index{U(1) current net@$U(1)$ current net!half-line localization}
    Consider the half-line standard subspace $H_+\in\Std(L^2(\Rl,d\te))$.
    \begin{enumerate}
        \item\label{item:Delta+Rapidity} The domain of $\Delta_+^{c}$, $c\neq0$, is
        \begin{align}
            \dom\Delta_+^c
            &=
            \CD_{2\pi c}
            \cong\Hardy^2(\Strip_{2\pi c})
            ,
        \end{align}
        and on $F_0$ (boundary value of $F\in\Hardy^2(\Strip_{2\pi c})$), the modular operator acts according to
        \begin{align}
            (\Delta_{+}^{iz}F_0)(\te)
            &=
            F(\te-2\pi z),\qquad \te\in\Rl,\qquad z\in\overline{\Strip_{-c}}.
            \label{eq:U1DeltaHbetaAction}
        \end{align}

        \item\label{item:H+Rapidity} The standard subspace is
        \begin{align}
            H_+
            &=
            \{F_0\col F\in\Hardy^2(\Strip_{\pi}),\quad  \overline{F(\te+i\pi)}=F(\te)\,\text{ a.e. in } \te\in\Rl \},
        \end{align}
        and its symplectic complement is
        \begin{align}
            H_+'=H_-
            &=
            \{F_0\col F\in\Hardy^2(\Strip_{-\pi}),\quad  \overline{F(\te-i\pi)}=F(\te)\,\text{ a.e. in } \te\in\Rl \}.
        \end{align}
    \end{enumerate}
\end{corollary}

All these statements follow directly from Theorem~\ref{theorem:HardyFunctionsInText}.

 We know from Proposition~\ref{proposition:GraphHilbertSpaceOfDelta} that the domain of $\Delta_H^{1/2}$ with its graph norm is a Hilbert space on which the Tomita operator acts as a conjugation. In the case at hand, this establishes a Hilbert space structure on the Hardy space. Working with general strip width $\beta$ again, we obtain the following corollary.

\begin{corollary}[The Hardy space $\boldsymbol{\Hardy^2(\Strip_\beta)}$ as a Hilbert and Banach space]\label{corollay:HardyHilbertSpace}
    Let $\beta\neq0$.
 \begin{enumerate}
        \item\label{item:U1HardyHilbert} $\Hardy^2(\Strip_\beta)$ is a Hilbert space with the scalar product
        \begin{align}\label{eq:HardyStripScalarProduct}
            \langle F,G\rangle_{\Hardy^2(\Strip_\beta)}
            :=
            \langle F_0,G_0\rangle_{L^2(\Rl)}
            +
            \langle F_\beta,G_\beta\rangle_{L^2(\Rl)}.
        \end{align}
        This identifies $\Hardy^2(\Strip_\beta)$ as the graph-norm Hilbert space $\dom\Delta_{+}^{\beta/2\pi}$.

        \item\label{item:U1HardyTomita} On this Hilbert space, the Tomita operator $S:=J_+\Delta_+^{\beta/2\pi}$ is a conjugation, and acts as the ${}^*$-operation from Definition~\ref{def:SymmetricFunction}, namely
        \begin{align}\label{eq:U1HardyTomita}
            S:\Hardy^2(\Strip_\beta)\to\Hardy^2(\Strip_\beta),\qquad
            (SF)(\zeta)
            :=
            \overline{F(i\beta+\overline\zeta)}.
        \end{align}
        In particular, the standard subspace $H_\beta:=\Fix(S_\beta)$ is
        \begin{align}
            H_\beta
            \cong
            \Hardy^2_{\sa}(\Stripb)
            =
            \{F\in \Hardy^2(\Stripb) \col F=F^*\},
        \end{align}
        where the identification proceeds via the boundary value $F_0=F|_\Rl$ of $F\in \Hardy^2(\Stripb)$.

    \end{enumerate}
 \end{corollary}
\begin{proof}
 \ref{item:U1HardyHilbert} In Corollary~\ref{corollary:U1currentStandardSubspace}, we have shown $\Hardy^2(\Strip_{\beta})=\dom\Delta_+^{\beta/2\pi}$. Thus $\Hardy^2(\Strip_\beta)$ is a Hilbert space in the graph norm of $\Delta_{+}^{\beta/2\pi}$ (Proposition~\ref{proposition:GraphHilbertSpaceOfDelta}). Identifying $\Hardy^2(\Strip_\beta)$ with $\CD_\beta=\dom\Delta_{+}^{\beta/2\pi}$, the scalar product inducing this graph norm is
 \begin{align*}
        \langle F,G\rangle_{\Hardy^2(\Strip_\beta)}
        &=
        \langle F,G\rangle_{L^2(\Rl)}
        +
        \langle \Delta_{+}^{\beta/2\pi}F,\Delta_{+}^{\beta/2\pi}G\rangle_{L^2(\Rl)}
        \\
        &=
        \langle F_0,G_0\rangle_{L^2(\Rl)}
        +
        \langle F_\beta,G_\beta\rangle_{L^2(\Rl)}.
    \end{align*}

 \ref{item:U1HardyTomita} We know from Proposition~\ref{proposition:GraphHilbertSpaceOfDelta} that the Tomita operator $S=J_+\Delta_+^{\beta/2\pi}$ is a conjugation on the graph Hilbert space $\Hardy^2(\Strip_\beta)$. It acts according to (with $F\in\Hardy^2(\Strip_\beta)$)
 \begin{align}
        (SF_0)(\te)=\overline{F_0(\te+i\beta)}=\overline{F_0(\te+i\beta)},\qquad \te\in\Rl.
    \end{align}
    In terms of the analytic continuation $F$, this is the map $F\mapsto F^*$, $F^*(\zeta)=\overline{F(\overline\zeta+i\beta)}$, $\zeta\in\Strip_\beta$.

 \end{proof}

 Note that the Hilbert space norm on $\Hardy^2(\Stripb)$ introduced here is different, but equivalent to, the Banach space norm~\eqref{eq:U1BanachNormHardy} (Theorem~\ref{theorem:HardySpace}~\ref{item:HardyBanach}).

\subsubsection*{The $\boldsymbol{U(1)}$-current in $\boldsymbol{L^2(\Rl_+,\frac{dp}{p})}$ and Hardy spaces on half-planes}

We move on to describe the $U(1)$-current in the Hilbert space $L^2(\Rl_+,dp/p)$, which is related to $L^2(\Rl,d\te)$ by the change of variables $p=e^\te$. To that end, we introduce the radial Hardy space on the upper half-plane.

\begin{definition}[The radial Hardy space on the upper half-plane]\label{def:U1RadialHardySpace}\glsadd{Cplus}
    The {\em radial Hardy space on the upper half-plane}\index{Hardy space!radial on $\Cl_+$} is the space $\Hardy^2_\rad(\Cl_+)$\glsadd{H2radC+} of all analytic functions $\Phi:\Cl_+\to\Cl$ satisfying
    \begin{align}\label{eq:BanachNormHardyRadial}
        \|\Phi\|^2_{\Hardy^2_\rad(\Cl_+)}
        :=
        \sup_{0<\alpha<\pi}\int_0^\infty|\Phi(e^{i\alpha}r)|^2\,\frac{dr}{r}<\infty.
    \end{align}
\end{definition}

We call this space the {\em radial} Hardy space on $\Cl_+$ because the $L^2$-integration is taken over the rays $e^{i\alpha }r$, $r>0$, pointing radially from the origin (and the dilation-invariant measure $\frac{dr}{r}$ is used). See Appendix~\ref{section:Hardy2Halfplane} for a comparison with a different Hardy space on the upper half-plane.

\begin{proposition}[Properties of the radial Hardy space on $\boldsymbol{\Cl_+}$]\label{prop:Hardy2UpperHalfPlane}
    \leavevmode
    \begin{enumerate}
        \item\label{item:StripC+Identification} The map
        \begin{align}
            \Hardy^2(\Strip_\pi)\to\Hardy^2_\rad(\Cl_+),\qquad F\mapsto\Phi:=F\circ\log
            \iff
            F=\Phi\circ\exp
        \end{align}
        is a complex-linear bijection. Here $\log$ denotes the principal branch of the logarithm.

        \item\label{item:U1RadialPointwiseBound} Any $\Phi\in\Hardy^2_\rad(\Cl_+)$ satisfies the pointwise bounds
        \begin{align}\label{eq:U1RadialPointwiseBounds}
            |\Phi(p)|
             \leq
            \sqrt{\frac2\pi}\|\Phi\|_{\Hardy^2_\rad(\Cl_+)} \sqrt{\frac{|p|}{\Im p}}
            ,\qquad
            p\in\Cl_+.
        \end{align}

        \item\label{item:U1RadialL2BoundaryValues} For any $\Phi\in\Hardy^2_\rad(\Cl_+)$, the limits
        \begin{align}
            \Phi_+(r)=\lim_{\alpha\downarrow0}\Phi(e^{i\alpha}r),\qquad
            \Phi_-(r)=\lim_{\alpha\uparrow\pi}\Phi(e^{i\alpha}r)
        \end{align}
        exist as norm limits in $L^2(\Rl_+,\frac{dr}{r})$. In particular, $\Phi_\pm\in L^2(\Rl_+,\frac{dr}{r})$, and the boundary value map $\Hardy^2_\rad(\Cl_+)\ni\Phi\mapsto\Phi_+\in L^2(\Rl_+,\frac{dp}{p})$ is injective.

        \item\label{item:U1CartesianDistributionalBoundaryValues} For any $\Phi\in\Hardy^2_\rad(\Cl_+)$, the limit
        \begin{align}
            \Phi_0(p)=\lim_{q\downarrow0}\Phi(p+iq), \qquad p\in\Rl,
        \end{align}
        exists in the sense of\footnote{That is, for any $g\in\Sch(\Rl)$, the limit $\Phi_0(g):=\lim_{q\downarrow0}\int_\Rl g(p)\Phi(p+iq)dp$ exists and defines a tempered distribution.} $\Sch'(\Rl)$.

        \item\label{item:HardyRadialHilbert} $\Hardy^2_\rad(\Cl_+)$ is a Hilbert space with the scalar product
        \begin{align}
            \langle \Phi_1,\Phi_2\rangle_{\Hardy^2_\rad(\Cl_+)}
            :=
            \langle \Phi_{1,+},\Phi_{2,+}\rangle_{L^2(\Rl_+,dp/p)}
            +
            \langle \Phi_{1,-},\Phi_{2,-}\rangle_{L^2(\Rl_+,dp/p)}
            .
        \end{align}
        With this scalar product, the map in \ref{item:StripC+Identification} is unitary.
    \end{enumerate}
\end{proposition}
 \begin{proof}
 \ref{item:StripC+Identification} The exponential function is a biholomorphic map $\exp:\Strip_\pi\to\Cl_+$, with inverse $\log$. As this map sends the lines $\Rl+i\alpha$ to the rays $e^{i\alpha}\Rl_+$, it is clear that $\Hardy^2(\Strip_\pi)$ \eqref{eq:U1BanachNormHardy} and $\Hardy^2_\rad(\Cl_+)$ \eqref{eq:BanachNormHardyRadial} are in bijection.

    Write $p=re^{i\alpha}$ with $r>0$ and $0<\alpha<\pi$. Applying
    Theorem~\ref{theorem:HardyFunctionsAppendix}~\ref{item:HardyPointWiseBound}
    to $F(\log r+i\alpha)=\Phi(p)$ gives \ref{item:U1RadialPointwiseBound},
    since $\sin\alpha\leq\min\{\alpha,\pi-\alpha\}$ and $\Im p=r\sin\alpha$.
    The norm limits in \ref{item:U1RadialL2BoundaryValues} follow from
    Theorem~\ref{theorem:HardyFunctionsAppendix}~\ref{item:HardyBoundaryValues1}
    by the substitution $r=e^\te$, for which $dr/r=d\te$.
    Injectivity follows from part~\ref{item:HardyBoundaryValues2} of the same theorem.

    For \ref{item:U1CartesianDistributionalBoundaryValues}, choose a holomorphic
    primitive $G$ of $\Phi$ on $\Cl_+$. By \eqref{eq:U1RadialPointwiseBounds},
    $|\Phi(p+iq)|\leq C(1+|p|)^{1/2}q^{-1/2}$ for $p\in\Rl$ and $0<q\leq1$.
    In particular,
    \begin{align*}
        G(p+iq)=G(p+i)-i\int_q^1\Phi(p+it)\,dt
    \end{align*}
    has a locally uniform limit $G_0(p)$ as $q\downarrow0$.
    Integrating the same estimate along the line $\Im z=1$ gives
    $|G(p+i)|\leq C'(1+|p|)^{3/2}$. The displayed identity gives this bound,
    up to a constant, uniformly for $0<q\leq1$. Dominated convergence against
    each Schwartz function now gives $G(\cdot+iq)\to G_0$ in $\Sch'(\Rl)$.
    Differentiating in the sense of distributions gives
    $\Phi(\cdot+iq)\to\partial_p G_0$, as claimed.

    \ref{item:HardyRadialHilbert} The scalar product given here amounts to transporting the scalar product $\scpr_{\Hardy^2(\Strip_\pi)}$ \eqref{eq:HardyStripScalarProduct} to $\Hardy^2_\rad(\Cl_+)$, which implies both claims.
\end{proof}

As in the case of the strip, we may view $\Hardy^2_\rad(\Cl_+)$ as a (dense) subspace of $L^2(\Rl_+,\frac{dp}{p})$ by identifying Hardy functions with their boundary values on $\Rl_+$.

With these results, it is easy to characterize $H_+$ and its modular data in $L^2(\Rl_+,\frac{dp}{p})$.

 \begin{corollary}[$\boldsymbol{H_+}$ in $\boldsymbol{L^2(\Rl_+,\frac{dp}{p})}$]\label{cor:U1RadialHalfLine}\index{U(1) current net@$U(1)$ current net!half-line localization}
    Consider the half-line standard subspace $H_+\in\Std(L^2(\Rl_+,\frac{dp}{p}))$.
    \begin{enumerate}
        \item\label{item:Delta+HalfPlane} The domain of $\Delta_+^{1/2}$ is
        \begin{align}
            \dom\Delta_+^{1/2}
            &=
            \Hardy^2_\rad(\Cl_+)
            ,
        \end{align}
        and on $\Phi_+$ (boundary value of $\Phi\in\Hardy^2_\rad(\Cl_+)$), the modular operator acts according to
        \begin{align}
            (\Delta_{+}^{iz}\Phi_+)(r)
            &=
            \Phi(e^{-2\pi z}r),\qquad r\in\Rl_+,\qquad z\in\overline{\Strip_{-1/2}}.
            \label{eq:DeltaH+ActionHalfPlane}
        \end{align}

        \item\label{item:H+HalfPlane} The standard subspace is
        \begin{align}
            H_+
            &=
            \{\Phi_+\col \Phi\in\Hardy^2_\rad(\Cl_+),\quad  \overline{\Phi_-(r)}=\Phi_+(r)\,\text{ a.e. in } r\in\Rl_+ \},
        \end{align}
        and its symplectic complement is
        \begin{align}
            H_+'=H_-
            &=
            \{\Phi_+\col \overline{\Phi_+}\in H_+\}.
        \end{align}
    \end{enumerate}
\end{corollary}
 \begin{proof}
    \ref{item:Delta+HalfPlane} is clear by the unitary equivalence of $L^2(\Rl,d\te)$ and $L^2(\Rl_+,\frac{dp}{p})$ (Lemma~\ref{lemma:UnitaryEquivalencesAff}) which maps the Hardy subspace $\Hardy^2(\Strip_{\pi})$ onto the Hardy subspace $\Hardy^2_\rad(\Cl_+)$ (Proposition~\ref{prop:Hardy2UpperHalfPlane}).

    For \ref{item:H+HalfPlane}, we only need to note that in the sense of boundary values,
    \begin{align*}
        (\Delta_+^{1/2}\Phi_+)(r)=\Phi(-r)=\Phi_-(r),\qquad r>0,
    \end{align*}
 and that $J_+$ acts by pointwise complex conjugation. This explains the formulas for $H_+$ and $H_-$, and we note that $H_-+iH_-$ can be identified with the radial Hardy space $\Hardy^2_\rad(\Cl_-)$ on the {\em lower} half-plane $\Cl_-$.
 \end{proof}

\subsubsection*{Description of local spaces by distributional support}

We have now seen various perspectives on the $U(1)$-current. However, its standard subspaces are always described as closures of other spaces. Here we comment on a closure-free description, which is sometimes helpful.

We continue to work in $\Hil=L^2(\Rl_+,\tfrac{dp}{p})$, where test functions $f\in\Sch_{\Rl}(\Rl)$ define Hilbert space vectors $\varphi_f(p):=ip\tilde f(p)$, $p>0$. For $\psi\in\Hil$, let $\psi^{\mathrm{h}}$ denote its Hermitian extension to the line, given by
\begin{align}
    \psi^{\mathrm h}:\Rl\to\Cl,\qquad
    \psi^{\mathrm h}(p)
    :=
    \begin{cases}
        \psi(p) & p>0\\
        \overline{\psi(-p)} & p<0
    \end{cases}
    .
\end{align}
We consider $\psi^{\mathrm h}$ as a distribution and its distributional inverse Fourier transform $T_\psi:=\EuF^{-1}\psi^{\mathrm{h}}$. Note that $T_\psi$ is a real distribution, and $T_{\varphi_f}=-f'$ holds.

 \begin{proposition}[Localization by distributional support]\label{lemma:HDistributionalCharacterization}\index{localization by distributional support}\index{U(1) current net@$U(1)$ current net!half-line localization}
    Let $\psi\in L^2(\Rl_+,\tfrac{dp}{p})$.
    \begin{enumerate}
        \item For every open set $I\subset\Rl$,
        \begin{align}
            \psi\in H(I)' \quad&\Longleftrightarrow\quad T_\psi|_I=0.
            \label{eq:U1ComplementBySupport}
        \end{align}

        \item For every open half-line or bounded open interval $I\subset\Rl$,
        \begin{align}
            \psi\in H(I) \quad&\Longleftrightarrow\quad \supp T_\psi\subset\overline I.
            \label{eq:U1LocalizationBySupport}
        \end{align}
        In particular, $H((a,b))=H((a,\infty))\cap H((-\infty,b))$ for $a<b$.
    \end{enumerate}
\end{proposition}
 \begin{proof}
    a) For $f\in\Sch_\Rl(\Rl)$, our Fourier convention \eqref{eq:FourierTransform} gives
    \begin{align*}
        T_\varphi(f)
        &=2\Re\int_0^\infty \overline{\varphi(p)}\,\tilde f(p)\,dp
          =2\Im\langle\varphi,\varphi_f\rangle_{\Hil}.
    \end{align*}
    The space of vectors $\varphi_f$ with $f\in C_{c,\Rl}^\infty(I)$ is dense in $H(I)$ \eqref{eq:U1Net}. This proves
    \eqref{eq:U1ComplementBySupport}.

    b) For half-lines,
    \eqref{eq:U1LocalizationBySupport} now follows from half-line duality
    (Theorem~\ref{theorem:U1C+}~\ref{item:Duality}) and translation covariance. For an interval $I=(a,b)=(a,\infty)\cap(-\infty,b)$, the implication ``$\Rightarrow$'' in \eqref{eq:U1LocalizationBySupport} follows by isotony.

    For the other implication, suppose $\supp T_\varphi\subset[a,b]$, and pick $\delta>0$. We first show that \(\varphi\in H(I_\delta)\), where \(I_\delta=(a-\delta,b+\delta)\). Choose nonnegative $\rho\in C_c^\infty((-1,1))$ with integral one, and set $\rho_\eps(x):=\eps^{-1}\rho(\frac{x}{\eps})$, for $0<\varepsilon<\delta$. The convolution \(g_\varepsilon=\rho_\varepsilon*T_\varphi\) is real and smooth, with support in \([a-\varepsilon,b+\varepsilon]\). Its Fourier transform is
\[
\widetilde g_\varepsilon(p)=m(\varepsilon p)\varphi(p),
\qquad m(t)=\int_{\mathbb R}\rho(x)e^{itx}\,dx,\qquad p>0.
\]
Since \(|m|\leq1\) and \(m(\varepsilon p)\to1\), dominated convergence gives \(\widetilde g_\varepsilon|_{\mathbb R_+}\to\varphi\) in \(L^2(\mathbb R_+,dp/p)\).

As \(\widetilde g_\varepsilon\) is continuous at zero, a nonzero value there would make its norm in \(L^2(\mathbb R_+,dp/p)\) infinite. Thus \(\widetilde g_\varepsilon(0)=0\), namely \(\int g_\varepsilon=0\). Its primitive
\[
f_\varepsilon(x)=-\int_{-\infty}^x g_\varepsilon(y)\,dy
\]
therefore has its support contained in \([a-\varepsilon,b+\varepsilon]\). Hence \(f_\varepsilon\in C_c^\infty(I_\delta,\mathbb R)\), and \(g_\varepsilon=-f_\varepsilon'\) gives \(\widetilde g_\varepsilon|_{\mathbb R_+}=\varphi_{f_\varepsilon}\in H(I_\delta)\). Taking the limit proves \(\varphi\in H(I_\delta)\).

Finally, let \(\gamma_\delta\) be the increasing affine map taking \(I_\delta\) onto \(I=(a,b)\). By covariance, \(U(\gamma_\delta)\varphi\in H(I)\). As \(\delta\downarrow0\), these maps tend to the identity, so \(U(\gamma_\delta)\varphi\to\varphi\) by strong continuity. Closedness of \(H(I)\) gives the claim.
\end{proof}

\begin{remark}[Hermitian extensions and embeddings into $\boldsymbol{\widehat\Hil}$]\label{remark:U1SupportDoubledHilbertSpace}
    The Hermitian extension $\psi^{\mathrm h}$ also recovers the embedding of
    Lemma~\ref{lemma:DoubledHilbertSpace}~\ref{item:EmbeddingIntoDoubledHilbertSpace}.
    Indeed, in the present realization $C:=U(j)$ is pointwise complex
    conjugation. Identifying the second summand of $\widehat\Hil$ with the
    negative momentum axis gives $\widehat\Hil\cong L^2(\Rl,\tfrac{dp}{|p|})$
    with complex structure $(I\Phi)(p)=i\,\mathrm{sign}(p)\Phi(p)$.
    Formula~\eqref{eq:UC} then reads $U_C\varphi=\varphi^{\mathrm{h}}/\sqrt{2}$.
\end{remark}

\section{Standard subspaces and complex structures}\label{section:ComplexStructures}

The interesting properties of a standard subspace $H\in\Std(\Hil)$ derive from its interplay with the complex structure of the surrounding Hilbert space $\Hil$. Indeed, the very definition of the crucial cyclic and separating properties, i.e. $\overline{H+iH}=\Hil$ and $H\cap iH=\{0\}$, explicitly refers to $\Hil$ and its complex structure $i$. Neither the ambient space $\Hil$ nor its complex structure $i$ is intrinsic to $H$, but so far were assumed to be given from the outset.

In this section, we reverse our perspective and consider standard subspaces as real Hilbert spaces in their own right. This is possible because~$H$ is a closed real subspace of~$\Hil$, equipped with the real scalar product $\Rescpr$, and complete w.r.t. this scalar product because it induces the same norm as the original complex scalar product. Furthermore, $H$ carries a pre-symplectic form coming from the imaginary part of the complex scalar product, and the relation between these two structures is connected to the modular data of $H$.

In Section~\ref{section:Polarizers}, we review some basic material regarding complex structures, symplectic forms, and polarizers, and then study the polarizer of a standard subspace. It turns out that a standard subspace may equivalently be viewed as a polarized real Hilbert space, or as a real Hilbert space with an orthogonal one-parameter group (Theorem~\ref{thm:PolarizedRealHilbertSpacesAreStandardSubspaces}). Section~\ref{section:complexifications} discusses constructions of standard subspaces by complexification. Important examples of such a construction are the local standard subspaces of a Klein-Gordon field, which will be presented later, in Section~\ref{section:KleinGordonSpaces}.

 \subsection{Polarizers and complex structures}\label{section:Polarizers}

We begin by recalling some well-known facts about complex structures, (pre-)sym\-plec\-tic forms, and polarizers.

\index{complex structure!general (Def.~\ref{def:ComplexStructures})}
 \index{complex structure!orthogonal (Def.~\ref{def:ComplexStructures})}\index{symplectic form (Def.~\ref{def:ComplexStructures})}\index{pre-symplectic form (Def.~\ref{def:ComplexStructures})}\index{polarizer!general@general (Def.~\ref{def:ComplexStructures})}
 \begin{definition}[Complex structures, (pre-)symplectic forms, polarizers]\label{def:ComplexStructures}
    Let $V$ be a real vector space.
 \begin{enumerate}
        \item\label{item:ComplexStructureOnRealSpace} A \emph{complex structure} on $V$ is a real-linear map $I:V\to V$ such that $I^2=-1$.
        \item\label{item:OrthogonalComplexStructure} A complex structure on a real pre-Hilbert space $V$ (i.e. $V$ carries a real scalar product $\scpr_\Rl$) is called \emph{orthogonal} if
        \begin{align}
            \langle Iv,Iw\rangle_\Rl=\langle v,w\rangle_\Rl,\qquad
            v,w\in V.
        \end{align}

        \item\label{item:PresymplecticForm} A \emph{pre-symplectic form} on $V$ is a real bilinear antisymmetric map $s:V\times V\to\Rl$. The kernel of a pre-symplectic form $s$ is
        \begin{align}
            \ker s
            :=
            \{v\in V\col s(v,w)=0\,\forall w\in V\}
            =
            \{v\in V\col s(w,v)=0\,\forall w\in V\}
            ,
        \end{align}
        and a \emph{symplectic form} is a pre-symplectic form that is non-degenerate, i.e. has trivial kernel $\ker s=\{0\}$.
        \item\label{item:PolarizerOfForm} If $V$ is a real Hilbert space with scalar product $\scpr_\Rl$ and $s$ a continuous pre-symplectic form on $V$, the \emph{polarizer} (of $V,\scpr_\Rl,s$) is the unique real-linear operator $D:V\to V$ such that
        \begin{align}
            \langle v,Dw\rangle_\Rl
            =
            s(v,w)
            ,\qquad v,w\in V.
        \end{align}
    \end{enumerate}
 \end{definition}

\begin{remark}[Orthogonal complex structures and symplectic forms]\label{remark:OrthogonalComplexStructures}
    \leavevmode
 \begin{enumerate}
        \item\label{item:ComplexScalarMultiplication} A complex structure on a real vector space $V$ gives it the structure of a complex vector space $(V,I)$ by $(x+iy)v:=xv+yIv$, $x,y\in\Rl$, $v\in V$.

        \item\label{item:ComplexifiedScalarProduct} Given an orthogonal complex structure on a real pre-Hilbert space $(V,\scpr_\Rl)$, there exists a unique complex scalar product $\scpr$ on the complex space $(V,I)$ such that its real part is $\scpr_\Rl$, namely
        \begin{align}\label{eq:ComplexifiedScalarProduct}
            \langle v,w\rangle
            =
            \langle v,w\rangle_\Rl-i\langle v,Iw\rangle_\Rl,\qquad v,w\in V.
        \end{align}
        The norms induced by $\scpr_\Rl$ and $\scpr$ are identical. The complex structure $I$ extends uniquely to a complex structure $\bar I$ on the Hilbert space $\overline{V}$, and turns $\overline V$ into a complex Hilbert space.

        \item\label{item:SymplecticFormOfComplexStructure} An orthogonal complex structure on a real pre-Hilbert space defines a continuous symplectic form
        \begin{align}
            s(v,w)
            =
            \Im\langle v,w\rangle
            =
            -\langle v,Iw\rangle_\Rl,\qquad v,w\in V.
        \end{align}

        \item\label{item:AdjointOfComplexStructure} An orthogonal complex structure $I$ on a real pre-Hilbert space $(V,\scpr_\Rl)$ has an adjoint $I^*:V\to V$, given by $I^*=-I=I^{-1}$.

        \item\label{item:EquivalenceOfOrthogonalComplexStructures} Any two orthogonal complex structures $I_1,I_2$ on a real Hilbert space $V$ are equivalent, i.e. there exists an orthogonal operator $O$ such that $OI_1O^{-1}=I_2$. To see that, choose complex orthonormal bases $(e_j^{(k)})_j$ of $(V,I_k)$, $k=1,2$, so that $\{e_j^{(k)},I_ke_j^{(k)}\}_j$ is an orthonormal basis of the real Hilbert space $V$. Then the map $O$ defined by $ Oe_j^{(1)}:=e_j^{(2)}$, $O(I_1e_j^{(1)}):=I_2e_j^{(2)}$ is orthogonal and satisfies $OI_1O^{-1}=I_2$.
    \end{enumerate}
 \end{remark}

 \begin{lemma}[Continuous (pre-)symplectic forms and polarizers]\label{lemma:Polarizer}
    Let $(V,\scpr_\Rl)$ be a real Hilbert space.
    \begin{enumerate}
        \item\label{item:PolarizerFromS} Let $s$ be a continuous pre-symplectic form on $V$. Then its polarizer $D\in B(V)$ satisfies $D^*=-D$ and $\ker D=\ker s$.

        \item\label{item:SFromPolarizer} Let $D\in B(V)$ with $D^*=-D$.
 \begin{enumerate}[(i)]
            \item\label{item:PolarizerNormal} $D$ is normal, its polar decomposition $D=I|D|$ commutes, i.e. $I|D|=|D|I$, and the phase $I:(\ker D)^\perp\to(\ker D)^\perp$ is an orthogonal complex structure on $(\ker D)^\perp$.\glsadd{IH}
            \item\label{item:FormOfSkewAdjointOperator} The map
            \begin{align}
                s:V\times V\to\Rl,\qquad s(v,w):=\langle v,Dw\rangle_\Rl
            \end{align}
            is a continuous pre-symplectic form, with polarizer $D$.
        \end{enumerate}
 \end{enumerate}
\end{lemma}
\begin{proof}
    \ref{item:PolarizerFromS} $D^*=-D$ follows from the antisymmetry of $s$. The inclusion $\ker D\subset\ker s$ is obvious, and the reverse inclusion follows from the non-degeneracy of $\scpr_\Rl$.

    \ref{item:SFromPolarizer} The antisymmetry $D^*=-D$ implies that $D$ is normal, and therefore the phase $I$ of its polar decomposition is an isometry from $(\ker D)^\perp$ to $(\ker D^*)^\perp=(\ker D)^\perp$, which commutes with $|D|$ on this space because $D$ is normal. Thus, on $(\ker D)^\perp\oplus\ker D$,
    \begin{align*}
        D^*
        &=
        (I|D||_{{\ker D}^\perp}\oplus0)^*
        =
        |D|I^*\oplus0
        =
        I^*|D||_{{\ker D}^\perp}\oplus0,\\
        -D
        &=
        -I|D||_{{\ker D}^\perp}\oplus0,
    \end{align*}
    and $D^*=-D$ implies $I^*=-I=I^{-1}$, i.e. $I$ is an orthogonal complex structure on $(\ker D)^\perp$. This proves (i), and (ii) is obvious.
\end{proof}

This lemma may be summarized by saying that there is a canonical bijection between continuous pre-symplectic forms on $V$ and bounded operators $D:V\to V$ satisfying $D^*=-D$. In this bijection, symplectic forms correspond to injective polarizers.

\bigskip

We also recall that any real vector space $V$ has a canonical complexification\index{complexification!canonical},
\begin{align}\label{eq:RealForm}
    V^{\Cl}
    =
    V\otimes_{\Rl}\Cl
    \cong
    V\oplus V.
\end{align}
Here the isomorphism $v\otimes \la\mapsto (\Re(\la)v,\Im(\la)v)$ to the direct sum is in the sense of real vector spaces. We will denote vectors in the direct sum as $\xi=(\xi_+,\xi_-)=\xi_+\oplus\xi_-$. This amounts to considering the real space $V\oplus V$ with the complex structure
\begin{align}
    I
    &:=
    \begin{pmatrix}
        0&-1\\
        1&0
    \end{pmatrix},\qquad
    (I\xi)_\pm=\mp\xi_\mp.
\end{align}
If $V$ carries a real scalar product $\scpr_\Rl$, so does its real double $V\oplus V$. Denoting the scalar product on $V\oplus V$ by the same symbol $\scpr_\Rl$, the complex scalar product on~$V^\Cl$ corresponding to $I$ is \eqref{eq:ComplexifiedScalarProduct}
\begin{align}
    \langle\xi,\eta\rangle
    &=
    \langle\xi,\eta\rangle_{\Rl}-i\langle\xi,I\eta\rangle_{\Rl}
    \\
    &=
    \langle\xi_+,\eta_+\rangle_{\Rl}+\langle\xi_-,\eta_-\rangle_{\Rl}+i\langle\xi_+,\eta_-\rangle_{\Rl}-i\langle\xi_-,\eta_+\rangle_{\Rl}
    .
\end{align}
Let us now assume that $V$ is a real Hilbert space. Then we may view $V$ as a closed real subspace of the complex Hilbert space $V^\Cl$ according to
\begin{align}
    V
    \cong
    V\otimes_\Rl\Rl
    \cong
    V\oplus 0
    \subset
    V\oplus V
    \cong V^\Cl.
\end{align}
From this perspective, $IV\cong0\oplus V$ and $V+IV\cong V^\Cl$, so $V\in\Std(V^\Cl)$ becomes a standard subspace of its complexification. The canonical conjugation
\begin{align}
    \Gamma:V^\Cl\to V^\Cl,\qquad \Gamma(v\otimes\la)=v\otimes\overline\lambda,
    \qquad
    \Gamma(\xi_+,\xi_-)=(\xi_+,-\xi_-),
\end{align}
 has fixed-point space $\Fix\Gamma=V$. Hence the Tomita operator of $V$ is $S_V=\Gamma$ (Proposition~\ref{proposition:StandardSubspacesVsClosedInvolutions}), and it follows that $V\in\Std(V^\Cl)$ is maximal abelian (Proposition~\ref{proposition:AbelianSpaces}).

\bigskip

This simple observation shows that any real Hilbert space may be embedded into a complex Hilbert space as a maximal abelian standard subspace. In order to understand embeddings as a standard subspace with non-trivial modular group, we need to provide additional data. An important structure in this context is the polarizer of a standard subspace, which we consider next.

As pointed out at the beginning of this section, any closed real subspace $H\in\Real(\Hil)$ of a complex Hilbert space is a real Hilbert space with scalar product $\Rescpr$. Furthermore,
\begin{align}\label{eq:sH}\glsadd{sH}
    s_H:H\times H\to\Rl,\qquad s_H(h_1,h_2)
    :=
    \Im\langle h_1,h_2\rangle
\end{align}
is a continuous pre-symplectic form. This suggests the following definition.

\begin{definition}[The polarizer of $\boldsymbol{H\in\Real(\Hil)}$]\label{def:PolarizerOfStandardSubspace}\index{polarizer!of $H$ (Def.~\ref{def:PolarizerOfStandardSubspace})}\glsadd{polarizer}
 Let $H\in\Real(\Hil)$. The \emph{polarizer} $D_H$ of $H$ is the polarizer of $(H,\Rescpr,s_H)$, i.e. the unique real-linear map $D_H:H\to H$ satisfying
 \begin{align}
        \Re\langle h_1,D_Hh_2\rangle
        =
        \Im\langle h_1,h_2\rangle,\qquad h_1,h_2\in H.
    \end{align}
\end{definition}

\begin{proposition}[Properties of the polarizer]\label{proposition:proppolirizer}
Let $H\in \Real(\Hil)$ have polarizer $D_H$ and real orthogonal projection $E_H$. Then
\begin{enumerate}
    \item\label{item:DHBounded} $D_H$ is bounded, with $\|D_H\|\leq1$.
	\item\label{item:DHEH} $D_H=-\left.E_Hi\right|_H$.
	\item\label{item:DHKernel} $\ker(D_H)=H\cap H'$. In particular, the pre-symplectic form $s_H$ \eqref{eq:sH} is symplectic if and only if $H$ is factorial.
	\item\label{item:DHSeparating} $\ker(D_H^2+1)=H\cap iH$. In particular, $\ker(D_H^2+1)=\{0\}$ if and only if $H$ is separating.
\end{enumerate}
In case $H\in \Std(\Hil)$ is standard with modular operator $\Delta_H$, we have furthermore
\begin{enumerate}\setcounter{enumi}{4}
	\item\label{item:DHDeltaH} $D_H=-i\left.\left(\frac{1-\Delta_H}{1+\Delta_H}\right)\right|_H$.
	\item\label{item:DHPolar} The polar decomposition of the polarizer is $D_H=I_H|D_H|$, with
	\begin{align}\label{eq:DHPolar}
        I_H
        &=
        -i\left(E^{\Delta_H}([0,1))-E^{\Delta_H}((1,\infty))\right)|_H,
        \qquad
        |D_H|
        =
        \left.\frac{|1-\Delta_H|}{1+\Delta_H}\right|_H.
	\end{align}
 \item\label{item:DHInverse} If \, $\ker(D_H)=0$, the polarizer has a densely defined closed inverse
 \begin{align}
        D_H^{-1}:D_H(H)\to H,
        \qquad
        D_H^{-1}=i\left.\left(\frac{1+\Delta_H}{1-\Delta_H}\right)\right|_{D_H(H)}.
	\end{align}
    In particular, $D_H$ has a bounded inverse if and only if $\Delta_H$ has a spectral gap ($1\not\in\spec(\Delta_H)$, see Def.~\ref{def:SpectralGap}).
\end{enumerate}
\end{proposition}

 \begin{proof}
    \ref{item:DHBounded} follows immediately from the Cauchy-Schwarz Inequality.

    \ref{item:DHEH} A simple calculation yields for any $h_1,h_2\in H$
    \begin{align*}
        \Re\langle h_1,D_H h_2\rangle=\Im\ip{h_1}{h_2}=\Re\ip{h_1}{-ih_2}=\Re\ip{h_1}{-E_Hih_2}.
    \end{align*}

    \ref{item:DHKernel} Let $h\in H$. Then $h\in\ker D_H$ is equivalent to $0=\Re\ip{\tilde{h}}{D_H h}=\Im\ip{\tilde{h}}{h}$ for all $\tilde{h}\in H$, i.e. to $h\in H'$.

    \ref{item:DHSeparating} From \ref{item:DHEH} we get $D_H^2=\left.E_HiE_{H}i\right|_H=-\left.E_HE_{iH}\right|_H$. Hence
    \begin{align*}
        \ker(D_H^2+1)=\{h\in H\col E_HE_{iH}h=h\},
    \end{align*}
    and $H\cap iH\subset\ker(D_H^2+1)$ follows. Conversely, $E_HE_{iH}h=h$ implies $\|h\|=\|E_HE_{iH}h\|\leq\|E_{iH}h\|\leq\|h\|$, hence $h=E_{iH}h$ and $E_{iH}h=E_HE_{iH}h$, i.e. $\ker(D_H^2+1)\subset H\cap iH$.

 \ref{item:DHDeltaH} According to Proposition~\ref{proposition:orthogonalprojectionmodulardata}, we  may express the projection $E_H$ in terms of the modular data of $H$ as $E_H=(1+S_H)(1+\Delta_H)^{-1}$. This implies for $h\in H$
    \begin{align*}
        D_H h
        &=
        -E_H ih
        \\
        &=
        -(1+S_H)(1+\Delta_H)^{-1}ih
        \\
        &=
        -i(1-S_H)(1+\Delta_H)^{-1}h
        \\
        &=
        -i\left((1+\Delta_H)^{-1}-(1+\Delta_H^{-1})^{-1}S_H\right)h
        \\
        &=
        -i\left((1+\Delta_H)^{-1}-(1+\Delta_H^{-1})^{-1}\right)h
        \\
        &=
        -i\left(\frac{1-\Delta_H}{1+\Delta_H}\right)h
        .
    \end{align*}

    \ref{item:DHPolar} By \ref{item:DHDeltaH}, we have $|D_H|^2=\left(\frac{1-\Delta_H}{1+\Delta_H}\right)^2$, hence $|D_H|=\frac{|1-\Delta_H|}{1+\Delta_H}$. Note that this operator preserves $H$ by application of Lemma~\ref{lemma:DeltaDomains}~\ref{item:HInvariance}. The proposed operator $I_H$ commutes with $\Delta_H$, and, in view of the modular relation and the antilinearity of $J_H$, also with the modular conjugation. Hence also $I_H$ preserves $H$. It is then clear that $I_H|D_H|=D_H$ holds.

    By definition, $I_H$ is a partial isometry with kernel $E^{\Delta_H}(\{1\})H=H\cap H'=\ker(D_H)$ (here we have used Prop.~\ref{proposition:CenterAndMaximalAbelianness} and part~\ref{item:DHKernel}) and range $(\ker D_H)^\perp=\overline{\Ran D_H}$. This proves the claim by uniqueness of the polar decomposition.

    \ref{item:DHInverse} follows directly from \ref{item:DHDeltaH}.
\end{proof}

\begin{corollary}[Factorial standard subspaces are complex Hilbert spaces]\label{corollary:FactorialSpacesAreComplexHilbert}
    Let $H\in\Std(\Hil)$ be a factorial standard subspace.
    \begin{enumerate}
        \item\label{item:InternalI} The phase $I_H$ of the polarizer $D_H$ is an orthogonal complex structure on $H$, called the \emph{internal complex structure of $H$}.\index{complex structure!internal to $H$ (Cor.~\ref{corollary:FactorialSpacesAreComplexHilbert})}
        \item\label{item:InternalSP} $H$ is a complex Hilbert space with complex structure $I_H$ and \emph{internal complex scalar product}\index{complex scalar product!internal to $H$ (Cor.~\ref{corollary:FactorialSpacesAreComplexHilbert})}
        \begin{align}
            \langle h_1,h_2\rangle_H
            &:=
            \Re\langle h_1,h_2\rangle-i\Im\left\langle h_1,\left(E^{\Delta_H}([0,1))-E^{\Delta_H}((1,\infty))\right)\,h_2\right\rangle
            ,
            \quad
            h_1,h_2\in H.
            \notag \end{align}
    \end{enumerate}
\end{corollary}
\Needspace{6\baselineskip}
\begin{proof}
    \ref{item:InternalI} For a general standard subspace, Lemma~\ref{lemma:Polarizer} says that $I_H$ is an orthogonal complex structure on $(\ker D_H)^\perp=Z(H)^{\perp_\Rl}\cap H=F(H)$, the factorial part of $H$ (see Prop.~\ref{proposition:DecomposeHIntoMaximallyAbelianAndFactorPart}). Hence it is an orthogonal complex structure on $H$ in case $H$ is factorial, $H=F(H)$.

    \ref{item:InternalSP} As $H$ is a real Hilbert space with orthogonal complex structure, it is a complex Hilbert space w.r.t. the complex scalar product (cf. Remark~\ref{remark:OrthogonalComplexStructures}~\ref{item:ComplexifiedScalarProduct}), $h_1,h_2\in H$,
    \begin{align*}
        \langle h_1,h_2\rangle_H
        &:=
        \Re\langle h_1,h_2\rangle-i\Re\langle h_1,I_Hh_2\rangle
        \\
        &=
        \Re\langle h_1,h_2\rangle-i\Im\left\langle h_1,\left(E^{\Delta_H}([0,1))-E^{\Delta_H}((1,\infty))\right)\,h_2\right\rangle
        .
    \end{align*}
\end{proof}

As an example, we revisit a special case of the standard subspaces considered in Example~\ref{example:StandardSubspacesFromMeasures}.

\begin{example}[The internal Hilbert structure of Example~\ref{example:StandardSubspacesFromMeasures}]
    In $\Hil=L^2(\Rl,dx)$, we consider the standard subspace $H$ given by modular group $(\Delta_H^{it}f)(x)=e^{itx}f(x)$ and $(J_Hf)(x)=\overline{f(-x)}$, i.e.
    \begin{align*}
        H
        =
        \left\{
            f\in L^2(\Rl,dx) \col \int_\Rl e^{x}|f(x)|^2dx<\infty,\quad f(-x)=e^{x/2}\overline{f(x)},\; \text{for } \text{almost all }\,x
        \right\}.
    \end{align*}
    Note that $H$ is factorial because $\Delta_H$ has empty point spectrum. As a complex Hilbert space, $H$ can be identified with $\Hil_H:=L^2(\Rl_+,(1+e^x)dx)$, via
    \begin{align*}
        U:\Hil_H\to H,\qquad
        (U\xi)(x)
        :=
        \begin{cases}
            \xi(x) & x\geq0,\\
            e^{-x/2}\overline{\xi(-x)}    & x<0
        \end{cases}.
    \end{align*}
 It is easy to check that $U$ is well-defined, i.e. really maps into $H$, bijective, and isometric w.r.t. the norm of $L^2(\Rl,dx)$ on $H$. While $U$ is not complex-linear w.r.t. the usual complex structure of $L^2(\Rl,dx)$, it is complex-linear w.r.t. the internal complex structure $I_H$ of $H$ \eqref{eq:DHPolar}, which is easily seen to be
 \begin{align}
        (I_Hf)(x)=i\sign(x)f(x).
    \end{align}
    It follows that $U$ is unitary when $H$ is given the internal complex scalar product
    \begin{align*}
        \langle h_1,h_2\rangle_H
        &=
        \Re\langle h_1,h_2\rangle_{L^2(\Rl,dx)}-i\Re\langle h_1,I_Hh_2\rangle_{L^2(\Rl,dx)}
        \\
        &=
        \Re\langle h_1,h_2\rangle_{L^2(\Rl,dx)}+i\Im\langle h_1,\sign\cdot h_2\rangle_{L^2(\Rl,dx)}.
    \end{align*}
\end{example}

\subsection{The internal structure of standard subspaces}\label{section:InternalStructureOfH}

We have already noticed that any real Hilbert space $V$ can be seen as a maximal abelian standard subspace in its canonical complexification $V^\Cl=V\otimes_\Rl\Cl$. We now want to equip $V$ with additional structure that allows us to view it as a standard subspace in a complex Hilbert space, with non-trivial modular operator. In the previous section, we have seen that the polarizer $D_H$ of a standard subspace $H\in\Std(\Hil)$ on the one hand allows us to reconstruct the imaginary part of the scalar product of $\Hil$ from its real part, i.e. the scalar product of $H$ as a real Hilbert space, and on the other hand encodes the modular operator $\Delta_H$. Hence we are led to consider a candidate for a polarizer of $V$ as such additional structure.

 \index{polarization of a real Hilbert space (Def.~\ref{def:PolarizedHilbertSpace})}\index{polarized real Hilbert space (Def.~\ref{def:PolarizedHilbertSpace})}\index{standard subspace realization of a polarized real Hilbert space (Def.~\ref{def:PolarizedHilbertSpace})}
\begin{definition}[Polarized real Hilbert spaces]\label{def:PolarizedHilbertSpace}
    \leavevmode
    \begin{enumerate}
        \item\label{item:PolarizedHilbertSpace} A \emph{polarization} of a real Hilbert space $V$ is an operator $D\in B_\Rl(V)$ such that $D^*=-D$, $\|D\|\leq1$, and $\ker(D^2+1)=\{0\}$. A \emph{polarized real Hilbert space} is a pair $(V,D)$ consisting of a real Hilbert space $V$ and a polarization.
 \item\label{item:StandardSubspaceRealizationOfPolarizedHilbertSpace} A \emph{standard subspace realization} of a polarized Hilbert space $(V,D)$ is a pair $(\Hil,\Phi)$ consisting of a complex Hilbert space $\Hil$ and a real-linear map $\Phi:V\to\Hil$ such that\footnote{Note that $\Phi(V)$ is automatically closed because $V$ is complete and $\Phi$ is isometric by~\ref{item:RealizationPreservesRealProduct}.}
        \begin{enumerate}[(i)]
            \item\label{item:RealizationPreservesRealProduct} $\langle v,w\rangle_V=\Re\langle\Phi(v),\Phi(w)\rangle_{\Hil}$, $v,w\in V$,
            \item\label{item:RealizationIsStandard} $\Phi(V)\in\Std(\Hil)$,
            \item\label{item:RealizationTransportsPolarizer} $D_{\Phi(V)}=\Phi\circ D\circ\Phi^{-1}:\Phi(V)\to\Phi(V)$. Here $\Phi^{-1}$ exists on $\Phi(V)$ because $\Phi$ is injective by~\ref{item:RealizationPreservesRealProduct}.
        \end{enumerate}
 Two standard subspace realizations $(\Hil_1,\Phi_1)$, $(\Hil_2,\Phi_2)$ of a polarized real Hilbert space $(V,D)$ are called unitarily equivalent if there exists a unitary $U:\Hil_1\to\Hil_2$ such that $U|_{\Phi_1(V)}=\Phi_2\circ\Phi_1^{-1}|_{\Phi_1(V)}$.
    \end{enumerate}
\end{definition}

At this point it is clear that any standard subspace $H$ defines a polarized real Hilbert space $(H,D_H)$, and the inclusion map $\Phi:H\to\Hil$ is a standard subspace realization. Below we will see that any polarized Hilbert space has a unique standard subspace realization, up to unitary equivalence (Theorem~\ref{thm:PolarizedRealHilbertSpacesAreStandardSubspaces}).

Before we prove that, we point out that a polarized real Hilbert space can be equivalently viewed as a real orthogonal dynamical system.

\index{real orthogonal dynamical system (Def.~\ref{def:RealOrthogonalDynamicalSystem})}\index{orthogonal dynamics (Def.~\ref{def:RealOrthogonalDynamicalSystem})}

\begin{definition}[Real orthogonal dynamical systems]\label{def:RealOrthogonalDynamicalSystem}
    A \emph{real orthogonal dynamical system} is a real Hilbert space $V$ together with an \emph{orthogonal dynamics}, namely a strongly continuous orthogonal one-parameter group $(U(t))_{t\in\Rl}$ of $V$.
\end{definition}

Every standard subspace $H\in\Std(\Hil)$ defines a real orthogonal dynamical system with the dynamics $U(t)=\Delta_H^{it}$ given by the modular group of $H$, thanks to the Tomita-Takesaki Theorem (Thm.~\ref{theorem:Tomita}). Its generator is
\begin{align}
    A=\left.i\log\Delta_H\right|_H,\qquad \dom A=H\cap\dom\log\Delta_H.
\end{align}

 The generator~$A$ of a general orthogonal dynamics $(U(t))_{t\in\Rl}$, i.e. $U(t)=e^{tA}$, is a skew-adjoint ($A^*=-A$) not necessarily bounded operator on $V$. Conversely, every skew-adjoint operator $A$ defines an orthogonal dynamics by $U(t):=e^{tA}$.

\begin{lemma}[Polarizations and orthogonal dynamics]\label{lemma:PolarizationsAndOrthogonalDynamics}
    Let $V$ be a real Hilbert space. Polarizations $D$ and orthogonal dynamics $U(t)=e^{tA}$, $t\in\Rl$, on $V$ are in bijection via
    \begin{align}
        D=I_A\tanh\frac{|A|}{2}
        \iff
        A=2I_D\arctanh|D|,
        \qquad
        \dom A=\dom\arctanh|D|
    \end{align}
    where $A=I_A|A|$ and $D=I_D|D|$ are the polar decompositions. In case $H$ is a standard subspace, this bijection maps its polarizer $D_H$ to the generator $A=i\log\Delta_H|_{H\cap\dom\log\Delta_H}$ of the modular group.
\end{lemma}
\begin{proof}
    Let $A$ be a skew-adjoint operator on $V$, with polar decomposition $A=I_A|A|$. As $A$ is normal, its phase $I_A$ commutes with the spectral projections of $|A|$, and furthermore $I_A^*=-I_A$, $I_A^2=-P_{(\ker A)^\perp}$. Define $D:=I_A\tanh\frac{|A|}{2}$. Then \(D^*=-D\) and \(\|D\|\le1\). Since \(\tanh(0)=0\),
    \begin{align*}
        1+D^2
        &=
        1-\tanh^2\frac{|A|}{2}=\sech^2\frac{|A|}{2}.
    \end{align*}
    As $\sech^2$ is strictly positive, spectral calculus gives $\ker(1+D^2)=\{0\}$. Thus $D$ is a polarization. Moreover,
    \begin{align*}
        |D|=\tanh\frac{|A|}{2},
        \qquad
        \ker D=\ker A,\qquad
        I_D=I_A.
    \end{align*}

 Conversely, let $D$ be a polarization with polar decomposition $D=I_D|D|$. We have $0\leq|D|\leq1$. Since $\ker(1-|D|^2)=\ker(1+D^2)=\{0\}$, the absolute value $|D|$ does not have eigenvalue~$1$. Therefore $\arctanh|D|$ is a densely defined positive self-adjoint operator, with kernel equal to $\ker D$. We define $A:=2I_D\arctanh|D|$.

    The phase $I_D$ vanishes on $\ker D$ and is an orthogonal complex structure on the orthogonal complement $(\ker D)^\perp$ (Lemma~\ref{lemma:Polarizer}~\ref{item:SFromPolarizer}). As $I_D$ commutes with the functional calculus of $|D|$ by normality, we have that $I_D$ preserves the domain of $\arctanh|D|$, and, using the decomposition $V=\ker D\oplus(\ker D)^\perp$,
    \begin{align*}
        A^*=(2I_D\arctanh|D|)^*=2\arctanh|D|\cdot I_D^*
        =
        -2I_D\arctanh|D|=-A,
    \end{align*}
 i.e. $A$ is skew-adjoint, with polar decomposition $I_A=I_D$, $|A|=2\arctanh|D|$.

    The two constructions are inverses of each other by functional calculus.

    In case $H\in\Std(\Hil)$ is a standard subspace, Proposition~\ref{proposition:proppolirizer} gives
    \begin{align*}
        D_H=-i\,\frac{1-\Delta_H}{1+\Delta_H}\bigg|_H
        =i\tanh\left(\frac12 \log\Delta_H\right)\bigg|_H.
    \end{align*}
    The absolute value and phase of the real skew-adjoint generator $A_H:=i\log\Delta_H|_H$ are the restrictions of $|\log\Delta_H|$ and $i\sign(\log\Delta_H)$, respectively. Thus
        \begin{align*}
        I_{A_H}\tanh\frac{|A_H|}{2}
        &=i\,\sign(\log\Delta_H)\tanh\left(\frac12|\log\Delta_H|\right)\bigg|_H
        =
        i\tanh\left(\frac12\log\Delta_H\right)\bigg|_H
        =
        D_H.
    \end{align*}
    This proves the last assertion.
\end{proof}

We may therefore view a polarized real Hilbert space as a real orthogonal dynamical system and vice versa. We next show that these structures can also equivalently be viewed as standard subspaces.

 \begin{theorem}[Polarized real Hilbert spaces are standard subspaces]\label{thm:PolarizedRealHilbertSpacesAreStandardSubspaces}\booktheoremindex{thm:PolarizedRealHilbertSpacesAreStandardSubspaces}{Standard subspace realization of polarized real Hilbert spaces}\index{classification of standard subspaces!by polarized real Hilbert spaces}
    Any polarized real Hilbert space $(V,D)$ has a standard subspace realization, unique up to unitary equivalence.
\end{theorem}
 \begin{proof}
 We first show the existence of the realization. To that end, consider the canonical complexification $V^\Cl=V\otimes_\Rl\Cl$ of $V$, with canonical complex structure $I$ and scalar product $\scpr_\Cl$. The map $D$ has a unique complex-linear extension $D_\Cl\cong D\oplus D$ to $V^\Cl\cong V\oplus V$, and we consider the sesquilinear form
 \begin{align}
        \langle\xi,\eta\rangle_D:=\langle\xi,(1+ID_\Cl)\eta\rangle_\Cl,\qquad
        \xi,\eta\in V^\Cl.
    \end{align}
    As $D_\Cl$ has norm $\|D_\Cl\|=\|D\|\leq1$ and satisfies $D_\Cl^*=-D_\Cl$, it follows that $1+ID_\Cl$ is positive. To show that it has trivial kernel, suppose $0=(1+ID_\Cl)(v\oplus w)=(v-Dw)\oplus(w+Dv)$. This implies $D^2w=-w$ and $D^2v=-v$, hence $v=w=0$ by our assumption $\ker(D^2+1)=\{0\}$. Thus $\scpr_D$ is a scalar product on $V^\Cl$, and we define $\Hil_D$ as the completion of $V^\Cl$ in the norm given by $\scpr_D$. Although the new norm may be weaker than the original norm on $V^\Cl$, its restriction to $V$ is unchanged, so $V$ remains closed in $\Hil_D$.

 For $v_1,v_2\in V$, we have
    \begin{align*}
        \langle v_1,v_2\rangle_D
        =
        \langle v_1,v_2\rangle_V+i\langle v_1,Dv_2\rangle_V,
    \end{align*}
 so the real part of $\scpr_D$ restricts to $\scpr_\Rl$ on $V\times V$. Thus $V$ is a closed real subspace of $\Hil_D$, and we define $\Phi:V\to\Hil_D$ to be the inclusion map, which preserves real scalar products (item~\ref{item:RealizationPreservesRealProduct} in Definition~\ref{def:PolarizedHilbertSpace}~\ref{item:StandardSubspaceRealizationOfPolarizedHilbertSpace}).

    The space $V$ is separating because $IV\cong0\oplus V$ intersects $V$ trivially, and cyclic because its complex span is $V^\Cl$, which is dense in $\Hil_D$ by construction. Thus $V\in\Std(\Hil_D)$, verifying item~\ref{item:RealizationIsStandard} in Definition~\ref{def:PolarizedHilbertSpace}~\ref{item:StandardSubspaceRealizationOfPolarizedHilbertSpace}.

    For the last property~\ref{item:RealizationTransportsPolarizer} of a standard subspace realization, we note that for $v,w\in V$, we have
 \begin{align*}
        \Re\langle\Phi(v),[\Phi\circ D\circ\Phi^{-1}]\Phi(w)\rangle_D
        &=
        \Re\langle\Phi(v),\Phi(Dw)\rangle_D
        =
        \langle v,Dw\rangle_V
        =
        \Im\langle\Phi(v),\Phi(w)\rangle_D.
    \end{align*}
    Hence, by definition of the polarizer of a standard subspace, $\Phi\circ D\circ\Phi^{-1}=D_{\Phi(V)}$. This proves that $(\Hil_D,\Phi)$ is a standard subspace realization of $(V,D)$.

    We next show the uniqueness claim. Let $(\Hil_1,\Phi_1)$ and $(\Hil_2,\Phi_2)$ be two standard subspace realizations of $(V,D)$. We consider the map $U_0:=\Phi_2\circ\Phi_1^{-1}:\Phi_1(V)\to\Phi_2(V)$, and have to show that it extends to a unitary $U:\Hil_1\to\Hil_2$.

    For $v,w\in V$, we have
    \begin{align*}
        \Re\langle U_0\Phi_1(v),U_0\Phi_1(w)\rangle_{2}
        &=
        \Re\langle\Phi_2(v),\Phi_2(w)\rangle_2
        \\
        &=
        \langle v,w\rangle_V
        \\
        &=
        \Re\langle\Phi_1(v),\Phi_1(w)\rangle_1,
    \end{align*}
    and
    \begin{align*}
        \Im\langle U_0\Phi_1(v),U_0\Phi_1(w)\rangle_{2}
        &=
        \Im\langle \Phi_2(v),\Phi_2(w)\rangle_{2}
        \\
        &=
        \Re\langle \Phi_2(v),D_{\Phi_2(V)}\Phi_2(w)\rangle_{2}
        \\
        &=
        \Re\langle \Phi_2(v),\Phi_2(Dw)\rangle_{2}
        \\
        &=
        \langle v,Dw\rangle_V
        \\
        &=
        \Re\langle\Phi_1(v),\Phi_1(Dw)\rangle_1
        \\
        &=
        \Re\langle\Phi_1(v),D_{\Phi_1(V)}\Phi_1(w)\rangle_1
        \\
        &=
        \Im\langle\Phi_1(v),\Phi_1(w)\rangle_1.
    \end{align*}
    Thus $U_0$ preserves real and imaginary parts of scalar products.

 As the spaces $\Phi_1(V)$, $\Phi_2(V)$ are standard, they are in particular separating. Thus the complex-linear extension of $U_0$, namely $U:\Cl\Phi_1(V)\to\Cl\Phi_2(V)$, $\Phi_1(v)+i\Phi_1(w)\mapsto \Phi_2(v)+i\Phi_2(w)$, is well-defined, densely defined, isometric and has dense range. It therefore extends to a unitary $\Hil_1\to\Hil_2$ which gives the claimed equivalence between the realizations $(\Hil_1,\Phi_1)$ and $(\Hil_2,\Phi_2)$.
 \end{proof}

This theorem provides us with a new, internal, perspective on standard subspaces: Instead of cyclic separating closed real subspaces embedded into a complex Hilbert space, we may equivalently view them as polarized real Hilbert spaces, dropping the embedding. In view of Lemma~\ref{lemma:PolarizationsAndOrthogonalDynamics}, we may also view real orthogonal dynamical systems as standard subspaces.

 \subsection{Standard subspaces and complexifications}\label{section:complexifications}

 We now describe constructions of standard subspaces that are based on a pair of closed real spaces and lead to interesting construction methods.

In the first step, we consider subspaces of a real Hilbert space $V$ and then pass to its canonical complexification $V^\Cl\cong V\oplus V$, with $I(v_1\oplus v_2)=(-v_2)\oplus v_1$. The projections in part~\ref{item:HdoubleModularData} below refer to the notation introduced in Lemma~\ref{lemma:ProjectionsAndPairsOfIndependentSpaces}.

\begin{proposition}[\bfseries{Standard subspaces in complexifications}]\label{proposition:StandardSubspacesInComplexifications}
    Let $V$ be a real Hilbert space, $V_\pm\subset V$ two closed real subspaces, and
    \begin{align}
        H:=V_+\oplus V_-
        \subset V^{\Cl}.
    \end{align}
    \begin{enumerate}
        \item\label{item:HdoubleSymplecticComplement} $H\in\Real(V^{\Cl})$ is a closed real subspace with symplectic complement
        \begin{align}
            H'
            =
            V_-^\rperp\oplus V_+^\rperp.
        \end{align}
        \item\label{item:HdoubleStandard} $H$ is separating if and only if $V_+\cap V_-=\{0\}$, and cyclic if and only if $V_-+V_+$ is dense in $V$ (equivalently, $V_+^\rperp\cap V_-^\rperp=\{0\}$).

        \item\label{item:HdoubleModularData} If $H$ is standard (i.e. $V_+\cap V_-=V_+^\rperp\cap V_-^\rperp=\{0\}$), then its Tomita operator is
        \begin{align}\label{eq:HdoubleS}
            S_H
            &=
            2P_{V_+\oplus V_-,V_-\oplus V_+}-1
            \\
            &=
            \begin{pmatrix}
                2P_{V_+,V_-}-1 & 0\\
                0 & 2P_{V_-,V_+}-1
            \end{pmatrix}
            =
            \begin{pmatrix}
                2P_{V_+,V_-}-1 & 0\\
                0 & -(2P_{V_+,V_-}-1)
            \end{pmatrix}
            .
        \end{align}

        \item\label{item:UniversalityOf2SpaceConstruction} Every standard subspace $H\in\Std(\Hil)$ is of this form: Choosing a conjugation $C\in\Con(\Hil)$ with $CH=H$, define $V:=\Fix C$ and $V_+:=H\cap V$, $V_-:=(iH)\cap V$. Then
        \begin{align}
            \Std(V^\Cl) \ni V_+\oplus V_- \cong V_++iV_-=H \in\Std(\Hil).
        \end{align}
    \end{enumerate}
\end{proposition}
\begin{proof}
    \ref{item:HdoubleSymplecticComplement} $H\in\Real(V^{\Cl})$ is clear by construction. Its symplectic complement is
    \begin{align}
        H'
        =IH^\rperp
        =
        I(V_+^\rperp\oplus V_-^\rperp)
        =
        V_-^\rperp\oplus V_+^\rperp.
    \end{align}
    \ref{item:HdoubleStandard} $H$ is separating if and only if
    \begin{align*}
        H\cap IH
        &=
        (V_+\oplus V_-)\cap (V_-\oplus V_+)
        =
        (V_+\cap V_-)\oplus(V_-\cap V_+)
    \end{align*}
    coincides with $\{0\}$, which is equivalent to $V_+\cap V_-=\{0\}$. For cyclicity, we know that $H$ is cyclic if and only if $H'$ is separating, which in view of \ref{item:HdoubleSymplecticComplement} is equivalent to $V_-^\rperp\cap V_+^\rperp=\{0\}$. By taking real orthogonal complements, this condition is equivalent to $V_++V_-$ being a dense subspace of $V$.

    \ref{item:HdoubleModularData} For $H=V_+\oplus V_-$ standard, we know $S_H=2P_{H,iH}-1=2P_{V_+\oplus V_-,V_-\oplus V_+}-1$ from \eqref{eq:SHvsPHiH} in general. As $V_+$ and $V_-$ are independent, the projections $P_{V_\pm,V_\mp}$ are well-defined, and with $v_\pm,w_\pm\in V_\pm$, we have
    \begin{align*}
        P_{V_+\oplus V_-,V_-\oplus V_+}\left[\begin{pmatrix}
                                              v_+\\ v_-
                                             \end{pmatrix}
                                             +
                                             \begin{pmatrix}
                                              w_-\\ w_+
                                             \end{pmatrix}
                                             \right]
                                             =
                                             \begin{pmatrix}
                                              v_+\\ v_-
                                             \end{pmatrix}
                                             =
                                             \begin{pmatrix}
                                              P_{V_+,V_-}(v_++w_-)\\ P_{V_-,V_+}(v_-+w_+)
                                             \end{pmatrix}
    \end{align*}
    This proves the claimed formula.

 \ref{item:UniversalityOf2SpaceConstruction} For the existence of $C$, see Corollary~\ref{corollary:AntiunitaryAndUnitaryImages}~\ref{item:conju}. The space $V$ is a real Hilbert space with two closed subspaces $V_{\pm}$, and in $V^{\Cl}$, we clearly have $V_++iV_-\subset H$. To show the opposite inclusion, we use invariance of $H$ under $C$. This yields for any $h\in H$
    \begin{align*}
        h=\tfrac{1}{2}(h+Ch)
          +i\tfrac{1}{2i}(h-Ch)
        \in V_++iV_-.
    \end{align*}
    Thus $V_+\oplus V_-\cong V_++iV_-=H$.
\end{proof}

In applications, one also meets the situation where $V_\pm$ are not subspaces of a Hilbert space to begin with, but rather a pair of vector spaces in duality in such a way that a corresponding Hilbert space structure and complexification emerge naturally.

\index{dual pair!of real spaces (Def.~\ref{def:DualPairGeneral})}
\begin{definition}[Dual pair of real spaces]\label{def:DualPairGeneral}
    A \emph{dual pair of real vector spaces with a compatible positive isomorphism} $(V_\pm,b,A)$ (a \emph{dual pair}, for short) consists of two real vector spaces $V_\pm$, a non-degenerate bilinear form $b:V_+\times V_-\to\Rl$, and a linear bijection $A:V_+\to V_-$, such that
    \begin{align}
        b(v_+,Aw_+)
        &=
        b(w_+,Av_+),\qquad v_+,w_+\in V_+,
        \\
        b(v_+,Av_+)&>0,\qquad v_+\neq0.
    \end{align}
\end{definition}

\begin{lemma}[Completions and complexifications of dual pairs]\label{lemma:DualPairCompletion}
    Let $(V_\pm,b,A)$ be a dual pair.
    \begin{enumerate}
        \item\label{item:DualPairCompletion}
        The maps
        \begin{align}
            \langle v_+,w_+\rangle_+
            &:=
            b(v_+,Aw_+),\qquad
            \langle v_-,w_-\rangle_-
            :=
            b(A^{-1}v_-,w_-),
            \qquad v_\pm,w_\pm\in V_\pm,
            \notag \end{align}
        are scalar products on $V_\pm$, and $A:V_+\to V_-$ is an isometric isomorphism of pre-Hilbert spaces. It extends to a unitary $\overline A:\overline{V_+}\to\overline{V_-}$ between the Hilbert space completions $\overline{V_\pm}$ of~$V_\pm$, and $b$ extends to the continuous bilinear pairing
        \begin{align}
            \overline b&:\overline{V_+}\times\overline{V_-}\to\Rl,\\
            \overline b(\xi_+,\xi_-)
            &=
            \langle \xi_+,\overline A^{-1}\xi_-\rangle_+
            =
            \langle\overline A\xi_+,\xi_-\rangle_-.
        \end{align}

        \item\label{item:DualPairComplexification}
        Consider the direct sum $V_+\oplus V_-$, equipped with the real scalar product $\scpr_{A,b,\Rl}:=\scpr_+\oplus\scpr_-$. Then
        \begin{align}\label{eq:DefIA}
            I_A
            &:=
            \begin{pmatrix}
                0&A^{-1}\\
                -A&0
            \end{pmatrix}
        \end{align}
        is an orthogonal complex structure, with associated complex scalar product
        \begin{align}\label{eq:AbScalarProduct}
            \langle v,w\rangle_{A,b}
            &:=
            \langle v,w\rangle_{A,b,\Rl}
            -i\langle v,I_Aw\rangle_{A,b,\Rl}
            \\
            &=
            b(v_+,Aw_+)+b(A^{-1}v_-,w_-)
            -ib(v_+,w_-)+ib(w_+,v_-).
        \end{align}
        Denoting the resulting complex pre-Hilbert space by $V_{A,b}$, its completion $\overline{V_{A,b}}$ has underlying real Hilbert space $\overline V_+\oplus\overline V_-$, with complex structure $\overline I_A=I_{\overline A}$, and
        \begin{align}
            U:\overline{V_{A,b}}
            &\longrightarrow(\overline V_+)^{\mathbb C},\\
            U(\xi_+\oplus\xi_-)
            &:=\xi_+\oplus(-\overline A^{-1}\xi_-)
        \end{align}
        is a unitary onto the canonical complexification of the real Hilbert space $\overline V_+$.
    \end{enumerate}
\end{lemma}
\begin{proof}    \ref{item:DualPairCompletion} The assumptions on $b$ and $A$ imply that $\langle\cdot,\cdot\rangle_{\pm}$ are symmetric and positive definite.
    Since $\langle Av_+,Aw_+\rangle_-
        =
        b(v_+,Aw_+)
        =
        \langle v_+,w_+\rangle_+$ for $v_+,w_+\in V_+$, and $A$ is bijective, $A$ is an isometric isomorphism. It therefore extends uniquely
    to an orthogonal isomorphism $\overline A$ between the completions. Likewise, $b$ extends uniquely to a continuous bilinear pairing on
    the completions, and its asserted formulas follow by continuity.

    \ref{item:DualPairComplexification} Direct calculation gives $I_A^2=-1$ and orthogonality of $I_A$. By
    Remark~\ref{remark:OrthogonalComplexStructures}, it determines the complex scalar product in the
    statement. Its explicit expression follows from
    \begin{align}
        \langle v,I_Aw\rangle_{A,b,\Rl}
        &=
        \langle v_+,A^{-1}w_-\rangle_+
        -\langle v_-,Aw_+\rangle_-
        =
        b(v_+,w_-)-b(w_+,v_-).
    \end{align}
    As the induced norm satisfies $\|v_+\oplus v_-\|_{A,b}^2=        \|v_+\|_+^2+\|v_-\|_-^2$, the underlying real Hilbert space of the completion is
    $\overline V_+\oplus\overline V_-$, and the continuous extension
    of $I_A$ is the stated operator~$I_{\overline A}$.
    Let $I(x\oplus y)=(-y)\oplus x$ be the canonical complex
    structure on $(\overline V_+)^{\mathbb C}$. Then
    \begin{align}
        U\overline I_A(\xi_+\oplus\xi_-)
        &=
        \overline A^{-1}\xi_-\oplus\xi_+
        =
        IU(\xi_+\oplus\xi_-),
    \end{align}
 so $U$ is complex-linear. Direct calculation shows that $U$ is isometric and surjective, hence unitary.
 \end{proof}

Dual pairs can be defined on the basis of suitable positive operators. The following definition introduces the necessary terminology.

\index{dual pair!defined by positive operator (Def.~\ref{def:PositiveOperatorDualPair})}
\begin{definition}[The dual pair defined by a positive operator]
    \label{def:PositiveOperatorDualPair}
    Let $\mathcal H_{\Rl}$ be a real Hilbert space and let
    $A:\operatorname{dom}A\to\mathcal H_{\Rl}$ be a positive
 self-adjoint injective operator. The \emph{dual pair defined by $A$} is $(V_\pm,b,A)$, given by
 \begin{align}
        V_+&:=\dom A,
        &
        V_-&:=\Ran A=\dom A^{-1},
        \\
        b&:V_+\times V_-\to\Rl,
        &
        b(v_+,v_-)&:=\langle v_+,v_-\rangle_{\Hil_{\Rl}}
        .
    \end{align}
\end{definition}

The implicit claim in this definition is that $(V_\pm,b,A)$ is a dual pair. Indeed, both $\dom A$ and $\Ran A$ are dense in $\Hil_{\Rl}$, the latter
because $(\Ran A)^\perp=\ker A=\{0\}$. Hence $b$ is non-degenerate. The compatibility and positivity follow from $b(v_+,Aw_+)=\langle A^{1/2}v_+,A^{1/2}w_+\rangle_{\Hil_{\Rl}}$. In particular, the norms induced on $V_\pm$ are
\begin{align}
    \|v_\pm\|_\pm
    &:=
    \|A^{\pm1/2}v_\pm\|_{\mathcal H_{\Rl}}.
\end{align}

In the following, we denote the completions of $V_\pm$ in these norms (which need not be subspaces of $\Hil_{\Rl}$) by $\Hil_A^\pm$, and write $\Hil_{A,b}$ for the complex Hilbert space $\overline{V_{A,b}}$ (with underlying real Hilbert space $\Hil_A^+\oplus\Hil_A^-$ and complex structure $\overline I_A$).

 \begin{theorem}[Standard subspaces and dual pairs given by a positive operator]
 \label{theorem:PositiveOperatorStandardSubspaces}\booktheoremindex{theorem:PositiveOperatorStandardSubspaces}{Standard subspaces and dual pairs given by a positive operator}
    Consider the dual pair $(V_\pm,b,A)$ defined by a positive self-adjoint injective operator $A$ on a real Hilbert space. Let $K\subset\dom A\cap\dom A^{-1}$ be a subspace, and define
 \begin{align}
        K_A&:=\overline K^{\|\cdot\|_+}\oplus \overline K^{\|\cdot\|_-}\subset\Hil_{A,b}
        .
    \end{align}
    \begin{enumerate}
        \item\label{item:KAStandard} $K_A$ is standard if and only if $\overline K^{\|\cdot\|_-}\cap\overline{AK}^{\|\cdot\|_-}=\{0\}$ and $\overline{K+AK}^{\|\cdot\|_-}=\Hil_A^-$, or, equivalently, $\overline K^{\|\cdot\|_+}\cap\overline{A^{-1}K}^{\|\cdot\|_+}=\{0\}$ and $\overline{K+A^{-1}K}^{\|\cdot\|_+}=\Hil_A^+$ hold.

        \item\label{item:PositiveOperatorModularData} Let $K_A$ be standard. Then the two closed spaces $\overline K^{\|\cdot\|_+},\overline{A^{-1}K}^{\|\cdot\|_+}\in\Gr_{\Rl}(\Hil_A^+)$ are independent, and we may consider the closed projection $P:=P_{\overline K^{\|\cdot\|_+},\overline{A^{-1}K}^{\|\cdot\|_+}}$ and closed involution $T_A:=2P-1$ (see Prop.~\ref{proposition:twosubspacesprojection} and Lemma~\ref{lemma:InvolutionsAndProjections}).

        Writing $T_A=V_A|T_A|$ for the polar decomposition, we have (on $\Hil_A^+\oplus\Hil_A^-$)
        \begin{alignat}{2}
            S_{K_A}
            &=\begin{pmatrix}
                T_A&0\\
                0&-\overline A T_A\overline A^{-1}
            \end{pmatrix},
            &\quad \dom S_{K_A}
            &=\dom T_A\oplus\overline A\dom T_A,
            \label{eq:PositiveOperatorTomita}\\
            \Delta_{K_A}
            &=\begin{pmatrix}
                T_A^*T_A&0\\
                0&\overline A T_A^*T_A\overline A^{-1}
            \end{pmatrix},
            &\quad\dom\Delta_{K_A}
            &=\dom(T_A^*T_A)
            \oplus\overline A\dom(T_A^*T_A),
            \label{eq:PositiveOperatorDelta}\\
            J_{K_A}
            &=\begin{pmatrix}
                V_A&0\\
                0&-\overline A V_A\overline A^{-1}
            \end{pmatrix}.&&
            \label{eq:PositiveOperatorJ}
        \end{alignat}
        In terms of the real orthogonal projections $E_{\pm}\in B(\Hil_A^\pm)$ with ranges $\overline{K}^{\|\cdot\|_\pm}$, we have
        \begin{align*}
            T_A^*T_A
            &=2\bigl(E_++\overline A^{-1}E_-\overline A\bigr)^{-1}-1,
            \qquad
            \dom(T_A^*T_A)
            =\Ran\bigl(E_++\overline A^{-1}E_-\overline A\bigr)
            \\
            V_A
            &=\sign
            \bigl(E_+-\overline A^{-1}E_-\overline A\bigr).
        \end{align*}

 \item\label{item:UniversalityOf2DualPairConstruction}\index{classification of standard subspaces!by dual pairs (factorial case)} Every factorial standard subspace $H\in\Std(\Hil)$ is unitarily equivalent to one of this form. With a conjugation $C$ such that $CH=H$, the real Hilbert space $\Hil_{\Rl}:=\Fix C$, and the non-negative Borel function
 \begin{align}\label{eq:SpectralDualPairFunction}
            a:\Rl_+\to\Rl,\qquad
            a(\la)
            :=
            \frac{1+\sqrt \la}{1-\sqrt \la}\cdot\chi_{(0,1)}(\la)+\frac{\sqrt \la-1}{\sqrt \la+1}\cdot\chi_{(1,\infty)}(\la)
            ,
        \end{align}
        the operator
        \begin{align}
            A:=a(\Delta_H)|_{\dom a(\Delta_H)\cap\Hil_{\Rl}}
        \end{align}
 is positive self-adjoint injective on $\Hil_{\Rl}$, and the space
 \begin{align}
            K:=\{\xi\in\Hil_{\Rl}\cap\dom a(\Delta_H):J_H\xi=\xi\}
        \end{align}
        lies in $\dom A\cap\dom A^{-1}$. The space $K_A$ constructed from these data is standard, and unitarily equivalent to $H$.
    \end{enumerate}
\end{theorem}
\begin{proof}
    \ref{item:KAStandard} Noting $\overline{A}^{\pm1}\,\overline{K}^{\|\cdot\|_{\pm}}=\overline{A^{\pm1}K}^{\|\cdot\|_{\mp}}$, we have
    \begin{align*}
        K_A\cap I_{\overline A}K_A
        &=
        K_A\cap\left(
            (\overline A)^{-1}\overline{K}^{\|\cdot\|_-}\oplus \overline A\,\overline{K}^{\|\cdot\|_+}
        \right)
        \\
        &=
        \overline A^{-1}\left(\overline{AK}^{\|\cdot\|_-}\cap\overline{K}^{\|\cdot\|_-}\right)\oplus \left(\overline{K}^{\|\cdot\|_-}\cap\overline{A K}^{\|\cdot\|_-}\right),
    \end{align*}
    which equals $\{0\}$ if and only if $\overline K^{\|\cdot\|_-}
    \cap\overline{AK}^{\|\cdot\|_-}=\{0\}$. This shows the separating claim for $K_A$. For cyclicity, we have similarly
    \begin{align*}
        \overline{K_A+I_{\overline A}K_A}
        &=
        \overline{K+A^{-1}K}^{\|\cdot\|_+}
        \oplus
        \overline{K+AK}^{\|\cdot\|_-}
        ,
    \end{align*}
    which coincides with the full Hilbert space if and only if $\overline{K+AK}^{\|\cdot\|_-}=\Hil_A^-$.

    \ref{item:PositiveOperatorModularData} We use the unitary $U:\Hil_A^+\oplus\Hil_A^-\to\Hil_A^+\oplus\Hil_A^+$, $U=\id\oplus-\overline{A}^{-1}$ from Lemma~\ref{lemma:DualPairCompletion}~\ref{item:DualPairComplexification} to map $K_A$ to $UK_A=\overline{K}^{\|\cdot\|_+}\oplus\overline{A^{-1}K}^{\|\cdot\|_+}$. According to Proposition~\ref{proposition:StandardSubspacesInComplexifications}~\ref{item:HdoubleModularData},
    the Tomita operator of $UK_A$ is
    \begin{align}
        S_{UK_A}
        =U S_{K_A}U^{-1}
        &=
        \begin{pmatrix}
            T_A&0\\
            0&-T_A
        \end{pmatrix}
        =
        \begin{pmatrix}
            V_A&0\\
            0&-V_A
        \end{pmatrix}
        \begin{pmatrix}
            |T_A|&0\\
            0&|T_A|
        \end{pmatrix},
    \end{align}
    where the last expression is the polar decomposition. Conjugation with $U^*$ gives the claimed formulas for $S_{K_A}$, $\Delta_{K_A}$, and $J_{K_A}$.

    To also prove the expressions in terms of the projections $E_\pm$, we note
    \begin{align}
        E_{K_A}
        &=
        \begin{pmatrix}
            E_+&0\\
            0&E_-
        \end{pmatrix},
        \\
        E_{iK_A}
        &=
        -I_{\overline A}E_{K_A}I_{\overline A}
        =
        \begin{pmatrix}
            \overline A^{-1}E_-\overline A&0\\
            0&\overline A E_+\overline A^{-1}
        \end{pmatrix}.
    \end{align}
    Recalling $J_{K_A}=\sign(E_{K_A}-E_{iK_A})$ and $(1+\Delta_{K_A})^{-1}=\frac12(E_{K_A}+E_{iK_A})$ from Prop.~\ref{proposition:orthogonalprojectionmodulardata}~\ref{item:JviaE}, the claimed formulae now follow by comparing the upper-left blocks of these matrices.

    \ref{item:UniversalityOf2DualPairConstruction} We first show that $A$ and $K$ satisfy the requirements of the dual pair construction.

 By definition of the function $a$, we have $a(\la)>0$ for all $\la\neq1$, and $a(1)=0$. As $H$ is factorial, $a(\Delta_H)$ is positive self-adjoint and injective. The conjugation $C$ fixing $H$ means that it commutes with $\Delta_H$ and $J_H$ (for the existence of $C$, see Corollary~\ref{corollary:AntiunitaryAndUnitaryImages}~\ref{item:conju}). Since $C$ commutes with $\Delta_H$, each spectral projection $E^{\Delta_H}(B)$ preserves $\Hil_{\Rl}=\Fix C$. The restrictions $E_{\Rl}(B):=E^{\Delta_H}(B)|_{\Hil_{\Rl}}$ therefore form a spectral measure on $\Hil_{\Rl}$. Its spectral integral $\int a(\lambda)\,dE_{\Rl}(\lambda)$ is precisely $A$, with domain $\Hil_{\Rl}\cap\dom a(\Delta_H)$, so $A$ is positive and self-adjoint and inherits injectivity from $a(\Delta_H)$. The space $K$ lies by definition in $\dom A$.

    Since $C$ also commutes with $J_H$, we may restrict $J_H$ to $\Hil_{\Rl}$ as well. The function $a$ satisfies $a(\la^{-1})=a(\la)^{-1}$, $\la\neq1$, by definition, and therefore the restriction $(J_H)|_{\Hil_{\Rl}}$ intertwines $A$ and $A^{-1}$. As $K$ consists of $J_H$-fixed vectors, this implies $K\subset\dom A^{-1}$. Hence the dual pair defined by $A$ exists, and $K$ satisfies the assumptions of the above standard subspace construction.

    We claim that $Y:\Hil_{A,b}=\Hil_A^+\oplus\Hil_A^-\to\Hil$,
    \begin{align}
        Y(v_+\oplus v_-)
        :=
        R(A^{1/2}v_+-iA^{-1/2}v_-),\qquad
        R
        :=
        \tfrac{1}{\sqrt{2}}(2E^{\Delta_H}_<-1+J_HC)
    \end{align}
    is a well-defined unitary mapping $K_A$ onto $H$. Here $A^{\pm1/2}$ denote the isometric extensions of these operators to $\Hil_A^\pm$.

    Writing
    $\eps:=2E^{\Delta_H}_<-1=E^{\Delta_H}_<-E^{\Delta_H}_\geq$ as a shorthand, factoriality and the modular relation imply $J_H\eps J_H=-\eps$. Since $C$ commutes with $J_H$ and $\Delta_H$, the operators
 $\eps$ and $J_HC$ are anticommuting complex-linear unitary involutions. This implies that  $R=(\eps+J_HC)/\sqrt{2}$ is a unitary involution.

    The map
    \begin{align*}
        W_0:\dom A\oplus\dom A^{-1}\to\Hil,
        \qquad
        W_0(v_+\oplus v_-)
        :=
        A^{1/2}v_+-iA^{-1/2}v_-
    \end{align*}
 is well defined because $\dom A^{\pm1}\subset\dom A^{\pm1/2}$. One checks directly that $W_0I_A=iW_0$ holds, i.e. $W_0$ is complex-linear. Using the fact that $A^{1/2}v_+$ and $A^{-1/2}v_-$ have real scalar product because they lie in $\Fix C$, one can also check directly that $W_0$ is isometric, namely
 \begin{align}
        \|W_0(v_+\oplus v_-)\|^2=\|A^{1/2}v_+\|^2+\|A^{-1/2}v_-\|^2=\|v_+\|_+^2+\|v_-\|_-^2.
    \end{align}
 Reading $A^{\pm1/2}$ as their continuous isometric extensions $A^{\pm1/2}:\Hil_A^\pm\to\Fix C$, it thus becomes apparent that $W_0$ extends to a complex-linear isometry $W:\Hil_{A,b}\to\Hil$.

    To show that $W$ is surjective, we consider the sequence of projections $E_n:=E^A([n^{-1},n])$. For $\xi\in E_n\Fix C$, we have $A^{\mp1/2}\xi\in\dom A^{\pm1}$ and $\xi=A^{\pm1/2}(A^{\mp1/2}\xi)$. Thus complex linear combinations of such vectors lie in the range of $W$. As $E_n\to1$ strongly as $n\to\infty$ because $A$ is injective, this shows that $W$ is surjective. Hence $W$ and $Y=RW$ are unitary.

    It remains to show $YK_A=H$. To this end, we note another important property of the function $a$, namely
    \begin{align}
        \frac{2a(\la)}{1+a(\la)^2}
        =
        \frac{|1-\la|}{1+\la},\qquad \la>0.
    \end{align}
    Thus the bounded injective operator
    \begin{align}
        \delta:=\left(\frac{|1-\Delta_H|}{1+\Delta_H}\right)^{1/2}
    \end{align}
    is related to $A$ on $\Hil_{\Rl}$ by
    \begin{align*}
        \delta^2=\frac{2A}{1+A^2}.
    \end{align*}
    Furthermore, $\delta$ commutes with $C$ and $J_H$. For $k\in K$, the relations $J_Hk=k$ and $J_HAJ_H=A^{-1}$ therefore
    give
    \begin{align*}
        \|\delta k\|_+^2
        =
        \|\delta k\|_-^2
        =
        \frac12
        \langle k,\delta^2(A+A^{-1})k\rangle
        =
        \|k\|^2.
    \end{align*}
    We now consider
    \begin{align}
        K_0&:=\Fix C\cap\Fix J_H \supset K.
    \end{align}
    Approximating with the spectral projections $E^A([n^{-1},n])$, one sees that $K$ is dense in $K_0$ and that $\delta K$ is dense in $K$ for both weighted norms ($\delta^{-1}$ is bounded on each of these spectral subspaces). Hence $\delta$ extends to real orthogonal isomorphisms
    \begin{align*}
        \delta:K_0
        \longrightarrow
        \overline K^{\|\cdot\|_\pm}.
    \end{align*}

    We now make use of the real orthogonal isomorphism
    (Proposition~\ref{proposition:ModularDataProperties}~\ref{item:FixJvsFixS})
    \begin{align}
        B_H:\Fix J_H\to H,\qquad B_H=\sqrt{2}(1+\Delta_H)^{-1/2}.
    \end{align}
    Since $C$ commutes with $J_H$ and $\eps$, whereas
    $J_H\eps=-\eps J_H$, we have the real orthogonal decomposition
    \begin{align*}
    \Fix J_H=K_0\oplus i\eps K_0.
    \end{align*}
    Finally, the defining scalar formula for $a$ gives
    \begin{align*}
        RA^{1/2}\delta k&=B_Hk,
        &
        RA^{-1/2}\delta k&=B_H\eps k,
        \qquad k\in K_0,
    \end{align*}

    where $B_H$ stands for $\sqrt{2}(1+\Delta_H)^{-1/2}$ for general vectors. The operators $A^{\pm1/2}\delta$ are defined everywhere and bounded, since
    \begin{align}
        A^{1/2}\delta=\left(\frac{2A^2}{1+A^2}\right)^{1/2},\qquad A^{-1/2}\delta=\left(\frac{2}{1+A^2}\right)^{1/2}.
    \end{align}

 It follows that
    \begin{align*}
        Y(\delta k_1\oplus\delta k_2)
        =
        B_H(k_1-i\eps k_2),
        \qquad k_1,k_2\in K_0.
    \end{align*}
    The left-hand side parametrizes $Y(K_A)$, while the
    right-hand side parametrizes $B_H(\Fix J_H)=H$.
    Hence $Y(K_A)=H$, and $K_A$ is standard.
\end{proof}

 In the construction in part~\ref{item:UniversalityOf2DualPairConstruction}, the complex-linear operator $a(\Delta_H)$ is determined directly by the modular operator $\Delta_H$ of the given standard subspace $H$. The choice of $C$ is not canonical, but choosing different $C$ only results in a unitarily equivalent construction.

We also note that non-zero maximal abelian standard subspaces cannot be realized by the dual pair from a positive operator construction: If $k\in K\backslash\{0\}$, both $k\oplus0$ and $0\oplus k$ lie in $K_A$, but $\Im\langle k\oplus0,0\oplus k\rangle_{A,b}=-\|k\|_{\Hil_{\Rl}}^2\neq0$ (cf.~\eqref{eq:AbScalarProduct}). Our assumption of factoriality in part~\ref{item:UniversalityOf2DualPairConstruction} rules out this case. In general, $K_A$ being factorial is equivalent to the completed pairing remaining non-degenerate on the two weighted closures of $K$.

For connections of the construction in part~\ref{item:UniversalityOf2DualPairConstruction} to related constructions in the literature, see the Notes and Remarks.

\bigskip

Theorem~\ref{theorem:PositiveOperatorStandardSubspaces} will be applied in Section~\ref{section:KleinGordonSpaces} to describe the localized subspaces of a free Klein-Gordon field.

\section{Operations with standard subspaces}\label{section:Operations}

So far we have investigated standard subspaces $H\in\Std(\Hil)$ of a fixed Hilbert space $\Hil$. In this section, we investigate operations with standard subspaces, possibly in different Hilbert spaces, which allow us to construct new standard subspaces within new Hilbert spaces.  These operations include direct sums, direct integrals, and tensor products of standard subspaces. 

We begin with direct sums in Section~\ref{section:DirectSumsAndIntegrals}, where also the more involved direct integrals are presented. The more subtle tensor products of standard subspaces are the subject of Section~\ref{section:TensorProducts}. Tensor products allow us in particular to give a characterization of the modular Hamiltonian of a fixed standard subspace $H$ in terms of complete passivity, which involves the tensor powers $H^{\otimes n}$, $n\in\Nl$. 

As an application of direct integrals, we consider standard subspaces that are multiplicity-free, in the sense that their modular operator is multiplicity-free, in Section~\ref{section:MultiplicityFreeSpaces}. These spaces can be decomposed into direct integrals of two-dimensional standard subspaces.

\subsection{Direct sums and direct integrals}\label{section:DirectSumsAndIntegrals}

We consider a family of Hilbert spaces $\{\H_a\}_{a\in A}$ over the same field $\Kl$, which will always be either $\Rl$ or $\Cl$. Then, the Cartesian product $\prod_{a\in A} \Hil_a$, i.e., the space of all functions $\Phi:A\to \bigcup_{a\in A} \Hil_a$ such that $\Phi(a)\in \Hil_a$ for all $a\in A$, is a vector space over $\Kl$ and the \index{direct!sum}\emph{direct sum} Hilbert space is defined as follows:

\begin{definition}[\bfseries{External direct sum of Hilbert spaces}]\label{definition:DirectSum}
	The \emph{direct sum} Hilbert space of a family $\{\H_a\}_{a\in A}$ of Hilbert spaces over a field $\Kl$ is the Hilbert space
	\begin{align}
		\bigoplus_{a\in A} \H_a
		=
		\left\{\Phi\in\prod_{a\in A}\Hil_a \col \sum_{a\in A}\|\Phi(a)\|_{\Hil_a}^2 < \infty\right\},
	\end{align}
	provided with the inner product $\ip{\Phi}{\Psi}=\sum_{a\in A} \ip{\Phi(a)}{\Psi(a)}_{\Hil_a}$.
\end{definition}

From Definition \ref{definition:DirectSum}, given a family of Hilbert spaces $\{\H_a\}_{a\in A}$ and a family of subsets  $\{H_a\}_{a\in A}$ with $H_a\subset \Hil_a$ it is easy to see that we can define
$$\bigoplus_{a\in A} H_a:=\left\{\Phi\in \bigoplus_{a\in A} \Hil_a \col \Phi(a)\in H_a\right\},$$ and with this definition the following holds.
\begin{proposition}[Direct sums of subspaces]\label{prop:basicdirectsum} Let $\{\H_a\}_{a\in A}$ be a family of Hilbert spaces over a field $\Kl$, and define $H:=\bigoplus_{a\in A} H_a$ and $\Hil:=\bigoplus_{a\in A} \H_a$ . Then
\begin{enumerate}
	\item\label{item:DirectSumSubspace} $H_a\in \Sub(\H_a)\, \forall a\in A\Leftrightarrow H\in \Sub\left(\Hil\right)$;
	\item\label{item:DirectSumRealSubspace} $H_a\in \Sub_\Rl(\H_a)\,\forall a\in A \Leftrightarrow H\in \Sub_\Rl\left(\Hil\right)$;
	\item\label{item:DirectSumClosedSubspace} $H_a\in \Gr(\H_a)\,\forall a\in A \Leftrightarrow H\in \Gr\left(\Hil\right)$;
	\item\label{item:DirectSumClosedRealSubspace} $H_a\in \Gr_\Rl(\H_a)\,\forall a\in A \Leftrightarrow H\in \Gr_\Rl\left(\Hil\right)$;
	\item \label{item:sumstandard} $H_a\in \Std(\H_a)\,\forall a\in A\Leftrightarrow H\in \Std\left(\Hil\right)$, in case, $\Kl=\Cl$.
\end{enumerate}
\end{proposition}
\begin{proof}
	Items~\ref{item:DirectSumSubspace} and~\ref{item:DirectSumRealSubspace} follow straightforwardly from the definition. Items~\ref{item:DirectSumClosedSubspace} and~\ref{item:DirectSumClosedRealSubspace} follow from noticing the direct sum is itself a Hilbert space and that the evaluation map $\bigoplus_{a\in A} \H_a=\Hil \ni \Phi\mapsto \Phi(a)\in\H_a$ is bounded. Finally, for item~\ref{item:sumstandard}, we just observe that $H+iH\supset\left\{\Phi\in \bigoplus_{a\in A} (H_a+iH_a) \col \Phi(a)= 0 \text{ for all but finitely many } a\in A\right\}$, therefore $\bigoplus_{a\in A} \overline{H_a+iH_a}= \overline{H+iH}$, and $H\cap iH=\left(\bigoplus_{a\in A} H_a\right)\cap i\left(\bigoplus_{a\in A} H_a\right)=\bigoplus_{a\in A} (H_a\cap iH_a)$.
\end{proof}

We are particularly interested in item~\ref{item:sumstandard}, the direct sum of standard subspaces, and isolate it in a definition.

\begin{definition}[\bfseries{External standard subspace direct sum}]\label{definition:StdDirectSum}
	The \emph{direct sum} of a family of standard subspaces $\{H_a\subset \H_a\}_{a\in A}$, where $\{\H_a\}_{a\in A}$ is a family of Hilbert spaces over $\Cl$, is the standard subspace	$\bigoplus_{a\in A} H_a\in \Std\left(\bigoplus_{a\in A} \H_a\right)$
\end{definition}

Note that the standard subspace direct sum is not merely the direct sum of the standard subspaces as a (real) Hilbert space, but also involves a compatible direct sum of their ambient complex Hilbert spaces. Statements of the form ``$H=\bigoplus_a H_a$ as a direct sum of standard subspaces'' are nonetheless unambiguous: By cyclicity, the complex Hilbert spaces must be $\Hil=\overline{H+iH}$, $\Hil_a=\overline{H_a+iH_a}$, and the condition is that $\Hil=\bigoplus_a\Hil_a$ is a complex orthogonal decomposition of $\Hil$, and $H_a\in\Std(\Hil_a)$, $H\in\Std(\Hil)$.

Given a family of operators $\{T_a\}_{a\in A}$, where $T_a:\Hil_a\supset \dom(T_a)\to \Hil_a$, we can define on the domain 
\begin{align*}
	\dom\left(\bigoplus_{a\in A}T_a\right):=\left\{\Phi\in \bigoplus_{a\in A}\dom(T_a) \col \bigoplus_{a\in A}T_a\Phi(a)\in \bigoplus_{a\in A} \Hil_a\right\}\subset \bigoplus_{a\in A}\Hil_a
\end{align*}
 the direct sum operator \index{operator!direct sum} 
\begin{align*}
	\bigoplus_{a\in A}T_a:\dom\left(\bigoplus_{a\in A}T_a\right)\to \bigoplus_{a\in A}\Hil_a, \quad \left(\bigoplus_{a\in A}T_a\right)(\Phi)(b)=T_b\Phi(b).
\end{align*}
Operators of the form above are also called \emph{decomposable operators} \index{operator!decomposable}.
A direct sum $\bigoplus_{a\in A}T_a$ is bounded if and only if $\{T_a\}_{a\in A}$ is a uniformly bounded family in which case it holds $\|\bigoplus_{a\in A}T_a\|=\sup_{a\in A}\|T_a\|$. Furthermore, if $T_a$ is densely defined for all $a\in A$, $\left(\bigoplus_{a\in A}T_a\right)^\ast=\bigoplus_{a\in A}T_a^\ast$ and, in particular, $\bigoplus_{a\in A}T_a$ is closable if and only if $T_a$ is closable for all $a\in A$, in which case $\overline{\bigoplus_{a\in A}T_a}=\bigoplus_{a\in A}\overline{T_a}$ holds.

Focusing now on the direct sum of a family of standard subspaces, it is natural to ask about the modular data of the direct sum, which can be determined by a straightforward calculation.

\begin{lemma}[Modular data of direct sums]\label{lemma:ModDataDirectSum}
    Let $\{\H_a\}_{a\in A}$ be a family of complex Hilbert spaces and $H_a\in \Std(\H_a)$ for all $a\in A$. Then, the Tomita operator and the modular data of $H:=\bigoplus_{a\in A} H_a\in\Std(\bigoplus_{a\in A}\Hil_a)$ are
    \begin{align}
        S_H=\bigoplus_{a\in A} S_{H_a} \quad, \quad \Delta^{it}_H  = \bigoplus_{a\in A} \Delta^{it}_{H_a}\quad and \qquad  J_H  = \bigoplus_{a\in A}  J_{H_a}\, .
    \end{align}
    Moreover, its symplectic complement is
    \begin{align}
        H' = \bigoplus_{a\in A} H'_a .
    \end{align}
\end{lemma}
\begin{proof}
	On one hand, the straightforward calculation
	\begin{align*}
		\big(S_H(h_1+ih_2)\big)(a)=(h_1-ih_2)(a)=h_1(a)-ih_2(a)=S_{H_a}\big((h_1+ih_2)(a)\big)
	\end{align*}
	immediately shows $S_H\subset\bigoplus_{a\in A} S_{H_a}$. On the other hand, $\dom(\bigoplus_{a\in A} S_{H_a})=\bigoplus_{a\in A} \dom(S_{H_a})=\bigoplus_{a\in A} H_a+iH_a=H+iH$. Hence $S_H=\bigoplus_{a\in A} S_{H_a}$, and therefore
	$S_H=\bigoplus_{a\in A} J_{H_a}\bigoplus_{a\in A} \Delta_{H_a}^{\nicefrac{1}{2}}$ is the polar decomposition of $S_H$ and the claimed formula for the modular data follows.
	
	For the last claim, we have $S_{H'}=S_H^\ast=\bigoplus_{a\in A} S_{H_a}^\ast=\bigoplus_{a\in A} S_{H_a'}=S_{\bigoplus_{a\in A} H_a'}$, and thus $H' = \bigoplus_{a\in A} H'_a$.
\end{proof}

Notice that, so far, we have adopted the viewpoint of the external direct sum. That raises the question of what is the \emph{internal} direct sum of standard subspaces, i.e. the decomposition of a standard subspace into a direct sum.

\begin{definition}[(internal) Standard subspace direct sum]\label{def:internalStandardSubspaceDirecSum}
	Given $H\in \Std(\H)$, we say that $H$ is the \emph{standard subspace direct sum} of the subspaces $\{H_a\}_{a\in A}$, where $H_a\subset \Hil$, if there exists a family of pairwise orthogonal complex Hilbert spaces $\H_a\subset \H$ with $H_a\in \Std(\H_a)$, $\bigoplus_{\a \in A}\H_a=\Hil$ and $\bigoplus_{\a \in A}H_a=H$.
	Each component $H_a$ in the direct sum decomposition of $H$ is called a \emph{direct summand}\index{direct summand} standard subspace.
\end{definition}

Aiming at a standard subspace decomposition, it is natural to ask when a closed real subspace $H\in \Std(\Hil)$ is a direct sum standard subspace having $K\subset H$ as a summand. The answer to this question is the content of the next lemma.

\begin{proposition}[Direct summands of a standard subspace]\label{prop:DirectSummandCriteria}
	Let $H\in\Std(\Hil)$ and let $K\subset H$ be a closed real subspace.
	Set $\CK:=K^\Cl=\overline{K+iK}$, and let $E$ be the orthogonal
	projection onto $\CK$. Then, the following are equivalent.
	\begin{enumerate}
		\item\label{item:DirectSummandInternal} $K$ is an internal direct summand of $H$.
		\item\label{item:DirectSummandModularGroup} $\Delta_H^{it}K=K$ for every $t\in\Rl$.
		\item\label{item:DirectSummandConjugation} $J_H K=K^\prime\cap \CK$.
		\item\label{item:DirectSummandProjection} $EH=K$.
	\end{enumerate}
	In this case the complementary summand of $K$ is $H\cap \CK^\perp$, i.e.
	\begin{align}
		H=K\oplus(H\cap \CK^\perp)
	\end{align}
	as a direct sum of standard subspaces.
\end{proposition}
\begin{proof}
	If $K$ is an internal direct summand, Lemma~\ref{lemma:ModDataDirectSum}
	gives $\Delta_H^{it}|_\CK=\Delta_{K}^{it}$, showing that
	\ref{item:DirectSummandInternal} implies \ref{item:DirectSummandModularGroup}.
	
	Assuming \ref{item:DirectSummandModularGroup}, we have $\Delta_H^{it}\CK=\CK$
	for all $t\in\Rl$. Moreover, $K$ is cyclic in $\CK$ by definition
	of $\CK$, and separating since $K\subset H$, hence $K$ is standard in $\CK$. Therefore, the restricted
	group $\Delta_H^{it}\restr{\CK}$ preserves $K$ and inherits the KMS condition at $\beta=-1$
	from $H$. Theorem~\ref{theorem:KMS} thus gives
	$\Delta_H^{it}\restr{\CK}=\Delta_{K}^{it}$. In particular, for $h\in K$,
	\begin{align*}
		J_Hh=\Delta_H^{1/2}h=\Delta_{K}^{1/2}h=J_{K}h.
	\end{align*}
	Hence $J_H K=J_{K}K=K^{\prime}\cap \CK$, proving \ref{item:DirectSummandConjugation}. Note that $K'\cap \CK=:K^{\prime_\CK}$ is the symplectic complement of $K$ inside $\CK$.
	
	Starting now from \ref{item:DirectSummandConjugation}, we have $K^{\prime}\cap \CK=J_H K\subset J_H H=H'$. By taking $h\in H$ and $k\in K^{\prime}\cap \CK$, and observing that $Ek=k$ we obtain that $\Im\langle Eh,k\rangle=\Im\langle h,k\rangle=0$. Hence, 
	$$Eh\in (K^{\prime}\cap \CK)'\cap \CK=K^{\prime_\CK \prime_\CK}=K,$$
	where we have used Lemma~\ref{lemma:CyclicSeparating}~\ref{item:DoubleComplement}.We have then established $EH\subset K$, proving~\ref{item:DirectSummandProjection}.
	
	Finally, assume \ref{item:DirectSummandProjection}. Then $EH=K$ and
	$(1-E)H=H\cap \CK^\perp$, from which follows that $H=K\oplus(H\cap \CK^\perp)$.
	The first summand is clearly standard in $\CK$ and second summand standard in $\CK^\perp$ because being contained in $H$ implies it is separating, and
	having $(1-E)(H+iH)$ as its complex span implies it is cyclic. This proves \ref{item:DirectSummandInternal}.
\end{proof}

\begin{corollary}[Decomposition of standard subspaces with discrete $\boldsymbol{\sigma(\Delta_H)}$]\label{Cor:directsumdecomposition}
	Let $H\in\Std(\Hil)$ be such that its modular operator is discrete in the sense that its eigenvectors span a dense subspace of $\Hil$. Then there exists a family $\{H_j\}_{j\in J}$ of two-dimensional factorial standard subspaces such that
	\begin{align}
		H=Z(H)\oplus\bigoplus_{j\in J}H_j
	\end{align}
	as a direct sum of standard subspaces. If $\Hil$ is separable, $J$ is countable.
\end{corollary}
\begin{proof}
	That $H=Z(H)\oplus F(H)$ is the standard subspace direct sum, where $F(H)$ is the factorial part, is precisely the content of Proposition \ref{proposition:DecomposeHIntoMaximallyAbelianAndFactorPart}. Now, let $\Lambda:=\spec(\Delta_{H})\cap (0,1)$. By the modular relation and the hypothesis on the eigenspaces $\Hil_{\la}:=\ker(\Delta_H-\la)$, we have
	\begin{align}
		\Hil=\Hil_1\oplus\bigoplus_{\la\in\Lambda}(\Hil_{\la}\oplus J_H\Hil_{\la}).
	\end{align}
	For each $\la\in\Lambda$, let $\{b_{\la,j}\}_{j\in J_{\la}}$ be an orthonormal basis of $\Hil_{\la}$. Then, setting
	$$H_{\la,j}=\rlspn\{b_{\la,j}+\lambda^{\nicefrac{1}{2}}J_Hb_{\la,j},ib_{\la,j}-i\lambda^{\nicefrac{1}{2}}J_Hb_{\la,j}\}$$
	we easily see that, for all $\la\in\Lambda$ and $j\in J_{\la}$, $H_{\la,j}\subset H$, and
	$H_{\la,j}^\Cl=\spn\{b_{\la,j},J_H b_{\la,j}\}$. Moreover, one can easily check that $\Delta_H^{it}H_{\la,j}=H_{\la,j}$ for all $t\in \Rl$. Therefore, it follows from Proposition \ref{prop:DirectSummandCriteria}~\ref{item:DirectSummandModularGroup}, that
	$F(H)=\bigoplus_{\la\in \Lambda,j \in J_{\la}} H_{\la,j}$ as a standard subspace direct sum, since $(\bigoplus_{\la\in \Lambda,j \in J_{\la}} H_{\la,j})^\perp\cap F(H)=\{0\}$.
\end{proof}

Any maximal abelian standard subspace is a standard subspace direct sum of one-dimensional standard subspaces. Hence this corollary in particular informs us that every finite-dimensional standard subspace is a direct sum of one-dimensional and two-dimensional standard subspaces.

It is natural to wonder if there is a decomposition similar to the one in Corollary \ref{Cor:directsumdecomposition} for standard subspaces which do not have only point spectrum. For answering that we need to go beyond direct sums and define \index{direct!integral} \emph{direct integral} as anticipated, but for defining the direct integral of a family of Hilbert spaces we need first to introduce the concept of measurable fields. For more details, see \cite[Part II, Chapter 1]{Dixmier:1981} and \cite[Chapter IV.8]{Takesaki:2001}.

\begin{definition}[Measurable field of Hilbert spaces]
	Let $(\Lambda,\Sigma,\mu)$ be a $\sigma$-finite measure space,  $\{\H(\l)\}_{\l\in \Lambda}$ be a family of Hilbert spaces over the same field $\Kl$, and $\FM\subset \prod_{\l\in \Lambda}\H(\lambda)$ a subspace satisfying:
	\begin{enumerate}
		\item\label{item:MeasurableFieldNorm} For every $\xi\in \mathfrak M$, the function $\Lambda\ni \l\mapsto \|\xi(\l)\|^2$ is a $\mu$-measurable function;
		\item\label{item:MeasurableFieldMaximality} For every $\eta\in \prod \H(\l)$, if the function $\l \in \Lambda \mapsto \ip{\xi(\l)}{\eta(\l)}_{\Hil(\l)}$ is measurable for every $\xi \in \mathfrak M$, then $\eta\in \mathfrak M$;
		\item\label{item:MeasurableFieldCountableTotalSet} There exists a countable subset $\{\xi_j\}_{j\in \Nl}$ of $\mathfrak M$ such that $\{\xi_j(\l)\}_{j\in \Nl}$ is total in $\H(\l)$ for every $\l\in \Lambda$.
	\end{enumerate}
	Then, $(\{\H(\l)\}_{\l\in \Lambda},\FM)$ is called a measurable field of Hilbert spaces \index{measurable!field} and the elements of $\FM$ are called measurable fields.
\end{definition}

Having the definition of a measurable field of Hilbert spaces, the basic definitions of direct integrals of Hilbert spaces and standard subspaces follow the same structure of the direct sum of Hilbert spaces and standard subspaces, respectively.
\begin{definition}[Direct integral]
	Let $(\Lambda,\Sigma,\mu)$ be a $\sigma$-finite measure space and  $(\{\H(\l)\}_{\l\in \Lambda},\FM)$ be a measurable field of Hilbert spaces. The direct integral of such a measurable field is the vector space	
	\begin{align}
		\int^\oplus_\Lambda \H(\l)d\mu :=\left\{\xi\in \FM \col \int_\Lambda \|\xi(\lambda)\|_{\Hil(\l)}^2 d\mu<\infty \right\}\Big/\sim \,,
	\end{align} 
	where $\xi\sim \eta$ if and only if $\xi(\l)=\eta(\l)$ $\mu$-almost everywhere, provided with the inner product $\ip{\xi}{\eta}:=\int_\Lambda \ip{\xi(\l)}{\eta(\l)}_{\Hil(\l)}d\mu$, which is well defined thanks to polarization identity and Cauchy-Schwarz inequality.
\end{definition}

We also define, for a family of subsets  $\{H_a\}_{a\in A}$ with $H_a\subset \Hil_a$, the direct integral
\begin{equation}\label{eq:directintegralsubsets}\int_\Lambda^\oplus H(\l)d\mu:=\left\{[\xi]_\sim\in \int_\Lambda^\oplus \Hil(\l)d\mu \col \xi(\lambda)\in H(\l) \, \mu\text{-almost everywhere}\right\}.
\end{equation}
We notice that, as a special case, a family $\{H(\l)\subset \Hil(\lambda)\}_{\l \in \Lambda}$ closed subspaces is family of Hilbert spaces itself, and $(\{H(\lambda)\}_{\lambda\in \Lambda},\FM\cap\prod_{\lambda\in \Lambda}H(\lambda))$ is a measurable field of Hilbert spaces if we can choose the countable family $\{\xi_j\}_{j\in\Nl} \in \FM\cap\prod_{\lambda\in \Lambda}H(\lambda)$ of measurable fields such that $\Span\{\xi_j(\l)\}_{j\in \Nl}$ is dense in
$H(\l)$ for all $\lambda\in \Lambda$. That entails the direct integral Hilbert space of the family $\{H(\lambda)\}_{\lambda \in \Lambda}$ coincides with the subset direct integral defined in equation \eqref{eq:directintegralsubsets}, i.e.,
\begin{equation}\label{dirint}
	\int^\oplus_\Lambda H(\l)d\mu \subset \int^\oplus_\Lambda \H(\l)d\mu\,
\end{equation}
is in fact a closed subspace.
Moreover, if $H(\l)\in\Std(\H(\l))$ for all $\l\in \Lambda$, then we have a \emph{measurable field of standard subspaces} and $\int^\oplus_\Lambda H(\l)d\mu\in\Std\left(\int^\oplus_\Lambda \Hil(\l)d\mu\right)$ is called the \emph{direct integral of the standard subspaces} $H(\l)$.

Similar to the decomposable operators with respect to a direct sum, given a family of operators $\{T(\l):\H(\l)\ni \dom(T(\lambda))\to \H(\l)\}_{\lambda\in \Lambda}$ we can define on the domain
\begin{align*}
	\dom\left(\int_\Lambda^\oplus T(\l)d\mu\right)&:=\left\{\Phi\in \int_\Lambda^\oplus\dom(T(\l))d\mu \col \int_\Lambda^\oplus T(\l)\Phi(\l)d\mu\in\int^\oplus_\Lambda \H(\l)d\mu\right\}\\
	&\subset \int_\Lambda^\oplus\Hil(\l)d\mu
\end{align*} 
and the \emph{direct integral operator}\index{operator!direct integral} by
\begin{align*}
	\int_\Lambda^\oplus T(\l)d\mu:\dom\left(\int_\Lambda^\oplus T(\l)d\mu\right)\to\int_\Lambda^\oplus\Hil(\l)d\mu\,, \quad \int_\Lambda^\oplus T(\l)d\mu\,\Phi:=\int_\Lambda^\oplus T(\l)\Phi(\l)d\mu.
\end{align*}

As in the case of the direct sum, we have that $\int_\Lambda^\oplus T(\l)d\mu\in B\left(	\int^\oplus_\Lambda \H(\l)d\mu\right)$ if and only if $\lambda\mapsto\|T(\l)\| \in L^\infty(\Lambda, \mu)$. Furthermore, $\left(\int_\Lambda^\oplus T(\l)d\mu\right)^\ast=\int_\Lambda^\oplus T(\l)^\ast d\mu$.

\begin{proposition}[Direct integrals of standard subspaces]
Let $(\Lambda,\mu)$ be a $\sigma$-finite measure space, $\{\H(\l)\}_{\l\in \Lambda}$ be a measurable field of Hilbert spaces and $(\{H(\l)\}_{\l\in \Lambda},\FM\cap \prod_{\lambda\in \Lambda}H(\lambda))$ a measurable field of standard subspaces $H(\l)\in \Std(\Hil(\l))$. Then, $H = \int^\oplus_\Lambda H(\l)d\mu$ is a standard subspace, and we have
\begin{align}
S_H=\int_\Lambda^\oplus S_{H(\l)}d\mu \,, \quad \Delta^{is}_H = \int^\oplus_\Lambda \Delta^{is}_{H(\l)} d\mu\, ,\, s\in\RR\, ,\qquad J_H = \int^\oplus_\Lambda J_{H(\l)} d\mu\, .
\end{align}
Moreover, $H'=\int_\Lambda^\oplus H(\l)'d\mu$.
\end{proposition}
\begin{proof}
Since $\overline{H+iH}= \int^\oplus_\Lambda \overline{H(\l)+i H(\l)}d\mu$ it is clear that $H\in \Std\left(\int^\oplus_\Lambda \H(\l)d\mu\right)$ and the remaining part of the proof is an adaptation of Lemma \ref{lemma:ModDataDirectSum}: we can directly see that
\begin{align*}
	S_H\left(\int_\Lambda^\oplus h_1(\l)d\mu+i\int_\Lambda^\oplus h_2(\l)d\mu\right)&=\int_\Lambda^\oplus h_1(\l)d\mu-i\int_\Lambda^\oplus h_2(\l)d\mu\\
	&=\int_\Lambda^\oplus (h_1-ih_2)(\l)d\mu\\
	&=\int_\Lambda^\oplus S_{H(\l)}(h_1+ih_2)(\l)d\mu,
\end{align*}
which immediately shows that $S_H=\int_\Lambda^\oplus S_{H(\l)d\mu}$. Therefore
$S_H=\int_\Lambda^\oplus J_{H(\l)}d\mu \int_\Lambda^\oplus \Delta_{H(\l)}^{\nicefrac{1}{2}}d\mu$ is the polar decomposition of $S_H$ and the claimed formula for the modular data follows from the uniqueness of the polar decomposition.

For the last claim, we have $S_{H'}=S_H^\ast=\int_\Lambda^\oplus S_{H(\l)}^\ast d\mu=\int_\Lambda^\oplus S_{H(\l)'}=S_{\int_\Lambda^\oplus H(\l)'d\mu}$, and thus $H' = \int_\Lambda^\oplus H(\l)'$.
\end{proof}

\subsection{Tensor products}\label{section:TensorProducts}

We now consider tensor products. On the level of complex Hilbert spaces $\Hil,\CK$, we work with the canonical complex Hilbert space tensor product $\Hil\otimes\CK$, with scalar product fixed by $\langle\xi_1\otimes\eta_1,\xi_2\otimes\eta_2\rangle_{\Hil\otimes\CK}=\langle\xi_1,\xi_2\rangle_{\Hil}\langle\eta_1,\eta_2\rangle_{\CK}$. For complex subspaces $\CU\subset\Hil$, $\CV\subset\CK$, we write $\CU\odot\CV:=\Span\{u\otimes v\col u\in\CU,\,v\in\CV\}$ for the algebraic tensor product (so $\Hil\otimes\CK=\overline{\Hil\odot\CK}$), and similarly
\begin{align}
	H\odot K=\rlspn\{h\otimes k\col h\in H,\,k\in K\}\subset \Hil\otimes \K \quad \text{and} \quad H\otimes K :=\overline{H\odot K}
\end{align}
for real-linear subspaces $H\subset\Hil$, $K\subset\CK$.\index{tensor product!standard subspaces}\glsadd{standardTensor}

Viewing $H\odot K$ as a real-linear subspace of the complex Hilbert space $\Hil\otimes \K$ circumvents the somehow cumbersome distinction between the real tensor product $H\otimes_\Rl K$\index{tensor product!real Hilbert spaces}\glsadd{realTensor}, which is defined as the real tensor product of the real Hilbert spaces $H$ and $K$, and the subspace $H\odot K$ of the complex tensor product, thanks to the universal properties in the complex tensor product, namely,
$$i(h\otimes k)=(ih)\otimes k = h \otimes (ik)\in\Hil \otimes \K.$$
Apart from the next result where the relation between these two spaces will be studied, we will give preference to the complex inner product $H\otimes K\subset \H\otimes \K$ due to its technical advantages.

\begin{theorem}[Comparison of the tensor products $\otimes_{\Rl}$ and $\otimes$]\label{theorem:TensorProductComparison}\booktheoremindex{theorem:TensorProductComparison}{Comparison of real and ambient complex tensor products}
	Let $H\in \Gr_\Rl(\H)$ and $K\in \Gr_\Rl(\K)$. Then
	\begin{enumerate}
		\item\label{item:TensorComparisonMap} $h\otimes_\Rl k\mapsto h\otimes k$ defines a bounded linear map \, $\iota:H\odot_\Rl K\to H\odot K$ with dense range.
		\item\label{item:TensorComparisonInjectivity} $\iota^\ast\iota=1-D_H\otimes_\Rl D_K$ is injective if and only if either $H$ or $K$ is separating.
		\item\label{item:TensorComparisonUnitary}  In case either $H$ or $K$ is separating, $U (H\otimes_\Rl K)=H\otimes K$, where $U$ is the unitary operator showing up in the polar decomposition of $\iota=U\sqrt{1-D_H\otimes_\Rl D_K}$ .
	\end{enumerate}
\end{theorem}
\begin{proof}
\ref{item:TensorComparisonMap} The map $h\otimes_\Rl k\mapsto h\otimes k $ is well defined thanks to the universal property of the tensor product and it is clear that we can extend this map to the algebraic tensor product \, $\iota:H\odot_\Rl K\to H\odot K$.
Now,
\begin{align*}
	\ip{\iota(h_1\otimes_\Rl k_1)}{\iota(h_2\otimes_\Rl k_2)}_{H\otimes K}&=\Re\ip{h_1\otimes k_1}{h_2\otimes k_2}\\
	&=\Re\ip{h_1}{h_2}\Re\ip{k_1}{k_2}-\Im\ip{h_1}{h_2}\Im\ip{k_1}{k_2}\\
	&=\Re\ip{h_1}{h_2}\Re\ip{k_1}{k_2}-\Re\ip{h_1}{D_H h_2}\Re\ip{k_1}{D_K k_2}\\
	&=\ip{h_1 \otimes_\Rl k_1}{(1-D_H\otimes_\Rl D_K) h_2\otimes_\Rl k_2}_{H\otimes_\Rl K},
\end{align*}
where $\ip{\cdot}{\cdot}=\ip{\cdot}{\cdot}_{\H\otimes \K}$. Hence, by using bilinearity of the inner product, we see that $\iota$ is bounded ($\|\iota\|^2=\|1-D_H\otimes_\Rl D_K\|$) and, therefore, it has an extension $\iota: H\otimes_\Rl K\to H\otimes K$. The exact same calculation yields to 
$\iota^\ast\iota=(1-D_H\otimes_\Rl D_K)$. Finally, the range of $\iota$ is dense because it contains the algebraic tensor product.

\ref{item:TensorComparisonInjectivity} Let $\xi\in \ker\left(1-D_H\otimes_\Rl D_K\right)$. Then, $\|\xi\|=\|(D_H\otimes_\Rl D_K)\xi\|\leq \|(D_H\otimes_\Rl 1)\xi\|\leq \|\xi\|$ and,
hence, $0=\|\xi\|^2-\|(D_H\otimes_\Rl 1)\xi\|^2=\ip{\xi}{\left((1-D_H^\ast D_H)\otimes_\Rl 1\right)\xi}_{H\otimes_\Rl K}$. Now, since $1-D_H^\ast D_H\geq 0$, we conclude from Proposition \ref{proposition:proppolirizer}~\ref{item:DHSeparating} that
$$\xi\in \ker\left((1-D_H^\ast D_H)\otimes_\Rl1\right)=\ker(1-D_H^\ast D_H)\otimes_\Rl K=(H\cap iH)\otimes_\Rl K.$$

An analogous calculation also implies that  $\xi\in H\otimes_\Rl (K\cap iK)$. Therefore, $\xi\in (H\cap iH)\otimes_\Rl (K\cap iK)$ and, hence, if either $H$ or $K$ is separating, it follows that $\iota^\ast\iota: H\otimes_\Rl K\to H\otimes_\Rl K$ is injective.

For the converse, if neither $H$ or $K$ is separating, take $h\in H\cap iH\setminus\{0\}$ and $k\in K\cap iK\setminus\{0\}$, which satisfy $D_H^2h=-h$ and $D_K^2k=-k$ due to Proposition \ref{proposition:proppolirizer}~\ref{item:DHEH}. Now, set $\xi=h\otimes_\Rl k+D_H h\otimes_\Rl D_K k$. We first claim $\xi\neq 0$: in fact, $h$ and $D_H h$ are linearly independent in $H$ because $h=\alpha D_H h$ for some $\alpha\in \Rl$ would imply $D_H h=\alpha D_H^2  h=-\alpha h$, which requires $h=-\alpha^2 h$. Now,we can easily calculate
$(D_H\otimes_\Rl D_K)\xi=D_H h\otimes_\Rl D_K k+D_H^2 h\otimes_\Rl D_K^2 k=\xi$, which shows {$\iota^*\iota$} is not injective.

\ref{item:TensorComparisonUnitary} The bounded operator $\iota$ has a polar decomposition $\iota=U\sqrt{1-D_H\otimes_\Rl D_K}$ with unitary operator $U:H\otimes_\Rl K\to H\otimes K$ if and only if either $H$ or $K$ is separating. From the properties of the polar decomposition it follows that $U (H\otimes_\Rl K)=H\otimes K$.
\end{proof}

We recall that, given two densely defined closable (anti)linear operators $A:\dom(A)\subset \Hil\to \Hil$ and $B:\dom(B)\subset \K\to \K$, the (anti)linear operator $A\odot B:\dom(A)\odot \dom(B)\to \Hil\otimes \K$ defined by $A\odot B(a\otimes b)=Aa\otimes Bb$ for all $a\in \dom(A)$ and $b\in \dom(B)$ is again densely defined and closable due to the fact that $A^\ast\odot B^\ast\subset(A\odot B)^\ast$ is densely defined. Therefore, we define $A\otimes B:=\overline{A\odot B}$, for which $(A\otimes B)^\ast=A^\ast\otimes B^\ast$ holds. For details of the tensor product of two closable operators, we refer to \cite[Section VIII]{ReedSimon:1972}.

\begin{proposition}[Projections and tensor products]\label{prop:projectiontensorproduct}
	Let $H\in \Std(\Hil)$ and $K\in \Std(\K)$. Then, $P_{H\odot K,i(H\odot K)}$ is a closable densely defined projection and
	\begin{equation}
		\overline{P_{H\odot K,i(H\odot K)}}=P_{H\otimes K,i(H\otimes K)}=2^{-1}(S_H\otimes S_K+1).
	\end{equation}
\end{proposition}
\begin{proof}
	A straightforward calculation yields
	\begin{align*}
		2^{-1}(S_H\odot S_K+1)(h_1\otimes k_1+i(h_2\otimes k_2))&=2^{-1}(S_H\otimes S_K+1)(h_1\otimes k_1+(ih_2)\otimes k_2)\\
		&=h_1\otimes k_1\\
		&=P_{H\odot K,i(H\odot K)}(h_1\otimes k_1+i(h_2\otimes k_2)).
	\end{align*}
	Hence, $P_{H\odot K,i(H\odot K)}= 2^{-1}(S_H\odot S_K+1)$ and, since the operator on the right-hand side is closable, $P_{H\odot K,i(H\odot K)}$ is closable. Moreover, on one hand, Proposition \ref{proposition:twosubspacesprojection}~\ref{item:2spacesprojectionclosable} tells us that $\overline{P_{H\odot K,i(H\odot K)}}=P_{H\otimes K,i(H\otimes K)}$ and, on the other hand, $\overline{2^{-1}(S_H\odot S_K+1)}=2^{-1}(S_H\otimes S_K+1)$.
\end{proof}

\begin{corollary}[Modular data of tensor products]\label{cor:com}
	Let $H\in \Std(\Hil)$ and $K\in \Std(\K)$. Then
	\begin{enumerate}
		\item\label{item:stdtensor} $H\otimes K\in \Std(\Hil\otimes \K)$ and $S_{H\otimes K}=S_H\otimes S_K$.
		\item\label{item:symptensor} $(H\otimes K)'=H'\otimes K'$.
		\item\label{item:modulartensor} $\Delta_{H\otimes K}=\Delta_H\otimes \Delta_K$ and $J_{H\otimes K}=J_H\otimes J_K$.
	\end{enumerate}
\end{corollary}
\begin{proof}
	\ref{item:stdtensor} It is obvious from Proposition \ref{prop:projectiontensorproduct} and Lemma \ref{lemma:InvolutionsAndProjections}~\ref{item:PvsS} that $S_{H\otimes K}=S_H\otimes S_K$ holds.
	
	\ref{item:symptensor} From item~\ref{item:stdtensor} $S_{(H\otimes K)'}=S_{H\otimes K}^\ast=(S_H\otimes S_K)^\ast=S_H^\ast\otimes S_K^\ast=S_{H'}\otimes S_{K'}=S_{H'\otimes K'}.$
	
	\ref{item:modulartensor} It trivially follows that $\Delta_{H\otimes K}=S_{H\otimes K}^\ast S_{H\otimes K}=(S_H^\ast\otimes S_K^\ast) (S_H\otimes S_K)=\Delta_H\otimes \Delta_K$ and $S_{H\otimes K}=S_H\otimes S_K=(J_H\otimes J_K)(\Delta_H^{\nicefrac{1}{2}}\otimes \Delta_K^{\nicefrac{1}{2}})$.
\end{proof}

Note that the analogue of Corollary \ref{cor:com} for real orthogonal complements is not true in general, namely
$(H\otimes K)^{\perp_\RR} \neq H^{\perp_\RR}\otimes K^{\perp_\RR}$; in fact, notice that $(H\otimes K)^{\perp_\RR}=i(H\otimes K)'=i(H'\otimes K')$ while $(H^\rperp\otimes K^\rperp)=(iH')\otimes (iK')=H'\otimes K'$.

\subsubsection{Complete passivity}\label{sec:completepassivity}

Having properly defined tensor products of standard subspaces, we are able to define and characterize the modular Hamiltonian in terms of complete passivity. Recall that a strongly continuous unitary one-parameter group $U(t)=e^{itA}$ is passive w.r.t. a standard subspace $H$ if $U(t)H=H$ for all $t\in\Rl$, and $\langle h,Ah\rangle\leq0$ for all $h\in\dom(A)\cap H$ (Definition~\ref{def:passivity}). Also recall that this is equivalent to $U(t)$ commuting with $J_H$ and $\Delta_H^{is}$ for all $t,s$, and $AL_H\geq0$ (Proposition~\ref{prop:ClassificationOfPassiveOneParameterGroups}).

\begin{definition}[Complete passivity]\label{def:compassivity}\index{complete passivity}
	Let $H\in \Std(\H)$. A strongly continuous unitary one-parameter group $U(t)=e^{itA}$ is {\em completely passive with respect to $H$} if $U(t)^{\otimes n}$ is passive with respect to $H^{\otimes n}$ for all $n\in\Nl$.
\end{definition}

Note that the selfadjoint generator of the tensor power $U(t)^{\otimes n}$ is the closure of the sum $A\otimes1\otimes\ldots\otimes 1+1\otimes A\otimes\ldots\otimes 1+\ldots+1\otimes 1\otimes \ldots\otimes A$.

\begin{theorem}[Characterization of $\boldsymbol{L_H}$ by complete passivity]\label{theorem:competepassivity<=>modulargroup}\booktheoremindex{theorem:competepassivity<=>modulargroup}{Complete passivity and the modular group}
	Let $H\in \Std(\Hil)$ and let $U(t)=e^{itA}$ be a strongly continuous unitary one-parameter group. If $H$ is maximal abelian, $U$ is completely passive w.r.t. $H$ if and only if $U(t)H=H$ for all $t\in\Rl$. If $H$ is not maximal abelian, $U$ is completely passive with respect to $H$ if, and only if, $A=\lambda L_H$ for some $\lambda \geq 0$.
\end{theorem}
\begin{proof}
	The statement in the maximal abelian case is clear from Proposition~\ref{prop:ClassificationOfPassiveOneParameterGroups}. The groups generated by $A=\la L_H$, $\la\geq0$, are $(e^{itA})^{\otimes n}=(\Delta_H^{i\la t})^{\otimes n}$, the positively rescaled modular group of $H^{\otimes n}$. Hence $U$ is completely passive in this case by Proposition~\ref{prop:PassivityOfDelta}.

	Conversely, assume that $U$ is completely passive w.r.t $H$. Complete passivity implies passivity. By Proposition~\ref{prop:ClassificationOfPassiveOneParameterGroups}~\ref{item:passivejointsign}, $A$ and $\log\Delta_H$ commute strongly and their joint spectrum satisfies $\spec(A,L_H)\subset \{(a,b)\in \Rl^2 \col ab\geq 0 \}$. Now, using that $U(t)$ is completely passive with respect to $H$, we obtain that
	$$\spec(A,L_H)+\cdots+\spec(A,L_H)\subset\{(a,b)\in \Rl^2 \col ab\geq 0 \},$$where we used the above form of generators of tensor powers of one-parameter groups. Therefore, if $(a_1,b_1),(a_2,b_2)\in \spec(A,L_H)$, it must hold that for all $n,m\in \Nl$
	\begin{equation}\label{eq:combinationspectrum}
		n(a_1,b_1)+m(a_2,b_2)=(na_1+ma_2,nb_1+mb_2)\in\{(a,b)\in \Rl^2 \col ab\geq 0 \}.
	\end{equation}
	Thanks to Proposition~\ref{prop:ClassificationOfPassiveOneParameterGroups}~\ref{item:passivejointsign}, we have that $J_HAJ_H=-A$ and also $J_HL_HJ_H=-L_H$. Hence, similarly to the discussion in Section~\ref{section:SpectrumDelta}, we obtain that $\spec(A,L_H)=-\spec(A,L_H)$. Therefore \eqref{eq:combinationspectrum} also holds for $n,m \in \Zl$. Hence, if $(a_1,b_1)$ and $(a_2,b_2)$ are not collinear, we would have linear combinations $n(a_1,b_1)+m(a_2,b_2)$, $n,m\in \Zl$ in all four quadrants of $\Rl^2$, contradicting \eqref{eq:combinationspectrum}. Therefore, $\spec(A,L_H)$ must be contained in a straight line. If $H$ is not maximal abelian, $L_H\neq0$, and $A=\lambda L_H$ for some $\lambda\geq 0$ follows.
\end{proof}

\subsubsection{The real $C^*$-tensor category of standard subspaces}

We note here without detailed proof that the family of all standard subspaces in all ambient complex Hilbert spaces, form a real $C^*$-tensor category $\bStd$. Its morphisms from $H\in\Std(\Hil)$ to $K\in\Std(\K)$ are the bounded complex-linear maps $T:\Hil\to\K$ satisfying $TJ_H=J_KT$ and $T\Delta_H^{it}=\Delta_K^{it}T$ for all $t\in\Rl$. These norm-closed real spaces of operators are stable under composition and Hilbert-space adjoints, and inherit the $C^*$-identity $\|T^*T\|=\|T\|^2$. By Corollary~\ref{cor:com}~\ref{item:modulartensor}, the usual Hilbert-space tensor product defines tensor products of objects and morphisms, with unit $\Rl\in\Std(\Cl)$ and the canonical associativity and unit identifications.\glsadd{StdCategory}\index{tensor category of standard subspaces}

\subsection{Multiplicity-free standard subspaces and their classification}\label{section:MultiplicityFreeSpaces}

	We now apply the spectral theory and direct integrals developed above to the classification of standard subspaces. The basic idea is simple: the modular operator $\Delta_H$ is a positive self-adjoint operator whose spectrum is symmetric under the inversion $q\mapsto q^{-1}$, while the modular conjugation $J_H$ implements this symmetry. Thus, the spectral representation of $\Delta_H$, together with the action of $J_H$, provides a natural model for $H$.
	
	We begin by making this correspondence precise. Consider the $C^*$-algebra $C_0(\Rl_+)$ of continuous functions on $\Rl_+$ vanishing at $0$ and at $+\infty$. The inversion $q\mapsto q^{-1}$ induces an involutive anti-automorphism
	\begin{align*}
		j:C_0(\Rl_+)\to C_0(\Rl_+),
		\qquad
		(jf)(q):=\overline{f(q^{-1})}.
	\end{align*}
	A \emph{symmetric representation} of $C_0(\Rl_+)$ on a complex Hilbert space $\Hil$ is a pair $(\pi,J)$, where $\pi$ is a representation of $C_0(\Rl_+)$ on $\Hil$ and $J$ is a conjugation on $\Hil$, satisfying
	\begin{equation}\label{eq:SymmetricRepresentation}
		J\pi(f)J=\pi(jf),
		\qquad f\in C_0(\Rl_+).
	\end{equation}
	The terminology reflects the fact that $J$ exchanges the spectral values $q$ and $q^{-1}$.

\begin{lemma}[Standard subspaces and symmetric representations]\label{corrHP}\index{classification of standard subspaces!by symmetric representations}
	Let $\Hil$ be a complex Hilbert space. Then
	\begin{align}
		H\mapsto(\pi_H,J_H),
		\qquad
		\pi_H(f):=f(\Delta_H),
	\end{align}
	is a bijection between $\Std(\Hil)$ and the nondegenerate symmetric representations of $C_0(\Rl_+)$ on $\Hil$.
\end{lemma}
\begin{proof}
	If $H\in\Std(\Hil)$, the pair $(\pi_H,J_H)$ is a nondegenerate symmetric representation because of the modular relation.

	Conversely, given a symmetric nondegenerate representation $(\pi, J)$, by the Borel functional calculus, there is a unique positive self-adjoint operator $\Delta_\pi$ on $\Hil$ such that $\Delta_\pi\pi(f)=\pi(qf)$, with $(qf)(q) = qf(q)$, for every compactly supported $f\in C_0(\Rl_+)$. Then, as $J\Delta_\pi J = \Delta_\pi^{-1}$, we have $S_\pi := J \Delta^{1/2}_\pi\in \Inv(\Hil)$, and $H_\pi=\Fix(S_\pi)\in\Std(\Hil)$ is the unique standard subspace associated to $S_\pi$ (see Proposition \ref{proposition:StandardSubspacesVsClosedInvolutions}).
\end{proof}
The inverse of the bijection in Lemma \ref{corrHP} will be denoted by
\[
(\pi, J) \mapsto H_\pi\, ,
\]
although $H_\pi$ depends on $J$ as well.

We now come to the definition of multiplicity-free spaces.

\begin{definition}[Multiplicity-free standard subspaces]\label{mfdef}
A standard subspace $H\in \Std(\Hil)$ is called {\em multiplicity-free} if $\pi_H$ has a cyclic vector, i.e. if there exists $\xi\in \Hil$ such that $\overline{\pi_H\left(C_0(\Rl_+)\right)\xi}=\overline{\{f(\Delta_H)\xi \col f\in C_0(\Rl_+)\}}=\Hil$.
\end{definition}
Note that $H$ being multiplicity-free requires $\Hil$ to be separable because $C_0(\Rl_+)$ is separable and  $\pi_H\big(C_0(\Rl_+)\big)\xi$ is dense in $\Hil$.

\begin{definition}[Symmetric Radon measures on $\boldsymbol{\Rl_+}$]
	A Radon measure $\mu$ on $\Rl_+$ is called \emph{symmetric} if $\mu(E^{-1}) = \mu(E)$ for every Borel set $E\subset \Rl_+$, where $E^{-1}=\{x^{-1}\in \Rl_+ \col x\in E\}$. Given two symmetric Radon measures $\mu,\nu$ on $\Rl_+$, we write $\nu \prec \mu$ if $\nu$ is absolutely continuous w.r.t. $\mu$, and $\nu \cong \mu$ is $\nu$ is equivalent to $\mu$, that is $\nu \prec \mu$ and $\mu \prec \nu$.
    \glsadd{precMeasures}
\end{definition}

The definition of symmetric measure and the example below are closely related to Example~\ref{example:StandardSubspacesFromMeasures}.

\begin{example}[The standard subspace $H_\mu$]\label{example:Hmu}
Let $\mu$ be a symmetric Radon measure on $\Rl_+$ and $\Delta_\mu$ the operator of multiplication by $q$ on $L^2(\Rl_+, d\mu)$, namely,
\begin{align}
	\dom\Delta_\mu
	:=
	\big\{f (q)\in L^2(\Rl_+, d\mu) : qf (q)\in L^2(\Rl_+, d\mu)\big\},
	\qquad
	(\Delta_\mu f)(q)=qf(q),
\end{align}
and let $J_\mu$ be the conjugation $(J_\mu f)(q):=\overline{f(q^{-1})}.$ Then
\begin{align}
	H_{\mu}
	:=
	\Fix(J_\mu\Delta_\mu^{1/2})\in\Std(L^2(\Rl_+,d\mu)
	=
	\big\{f \in L^2(\Rl_+, d\mu) : \bar f(q^{-1}) = q^{\nicefrac{1}{2}} f(q)\big\}
\end{align}
is a multiplicity-free standard subspace of $L^2(\Rl_+,d\mu)$. Its associated symmetric representation is
\begin{equation}\label{pmu}
\pi_\mu(f)g = fg\, ,\quad f\in C_0(\Rl_+)\, , \ g\in L^2(\Rl_+, d\mu).
\end{equation}
\end{example}
\begin{proof}
	The modular relation $J_\mu\Delta_\mu J_\mu = \Delta_\mu^{-1}$ holds because
	\[
(J_\mu\Delta_\mu J _\mu f)(q) = \overline{(\Delta_\mu J_\mu f)(q^{-1})} = q^{-1} f(q) =  (\Delta_\mu^{-1} f)(q).
\]
Hence $H_\mu$ is standard. The statement on $\pi_\mu$ is clear by definition of this representation. Any non-vanishing function in $L^2(\Rl_+,d\mu)$ is cyclic for $\pi_\mu$, so $H_\mu$ is multiplicity-free.
\end{proof}

Recall that the notation $K\cong H$ means that the standard subspaces $K$ and $H$ are unitarily equivalent. Below we also use the notation $K\prec H$ to express that $K$ is unitarily equivalent to a direct summand of $H$.
\glsadd{precStandardSubspaces}
\begin{proposition}[Multiplicity-free standard subspaces and measure classes]\label{prop:mf}\index{classification of standard subspaces!by symmetric measure classes}
Let $\nu,\mu$ be symmetric Radon measures on $\Rl_+$. Then
\begin{align}
	H_\nu\prec H_\mu \iff \nu \prec \mu,\qquad H_\nu\cong H_\mu \iff \nu \cong \mu.
\end{align}
\end{proposition}

This proposition can be proven directly by considering the Radon-Nikodým derivative of the two measures. We will however obtain it as a corollary of Theorem \ref{thm:mf}. Note that this proposition states that the unitary equivalence class of $H_\mu$ depends only on the measure class $[\mu]$ of $\mu$.  

\begin{proposition}[Multiplicity free standard subspaces and symmetric measures]\label{propmf}\index{multiplicity-free standard subspace}
Let $\Hil$ be a separable Hilbert space and $H\in\Std(\Hil)$. Then, the following are equivalent:
\begin{enumerate}
\item\label{item:MFCyclic} $H$ is multiplicity-free;
\item\label{item:MFMaximalAbelian}  The von Neumann algebra $\M_H := \pi_H(C_0(\Rl_+))''$ is maximal abelian;
\item\label{item:MFFixedCyclic}  There exists a $J_H$-invariant vector $\xi\in \Hil$ cyclic for $\M_H$;
\item\label{item:MFMeasure}  There exists a symmetric Radon measure $\mu$ on $\Rl_+$ such that $H$ is unitarily equivalent to $H_\mu$.
\end{enumerate}
\end{proposition}
\begin{proof}
\ref{item:MFCyclic} $\Rightarrow$ \ref{item:MFMaximalAbelian}. By the cyclic spectral theorem, a cyclic representation of $C_0(\Rl_+)$ is unitarily equivalent to multiplication on $L^2(\Rl_+,\nu)$ for a finite measure $\nu$. In this representation,~$\M_H$ is the algebra of multiplication operators by functions in $L^\infty(\Rl_+,\nu)$, which is maximal abelian.

\ref{item:MFMaximalAbelian} $\Rightarrow$ \ref{item:MFCyclic}. By separability, decompose $\Hil$ into a finite or countable orthogonal sum of cyclic reducing subspaces for $\M_H$:
\begin{align*}
 \Hil=\bigoplus_k\Hil_k,\qquad
 \Hil_k=\overline{\M_H\xi_k},\qquad \|\xi_k\|=1.
\end{align*}
The orthogonal projections $P_k$ onto these subspaces belong to $\M_H'=\M_H$. Consequently, $\xi:=\sum_k2^{-k}\xi_k$ is cyclic for $\M_H$: since $P_k\xi=2^{-k}\xi_k$, its cyclic subspace contains every $\Hil_k$.

\ref{item:MFCyclic} $\Rightarrow$ \ref{item:MFFixedCyclic}. Let $\xi$ be cyclic. The spectral projections of $\Delta_H$ satisfy $J_HE^H_<J_H=E^H_>$ and $J_HE^H_1J_H=E^H_1$, and $J_H\M_H J_H=\M_H$. The vector $\xi_<:=E^H_<\xi$ is cyclic for the restriction of $\M_H$ to $E^H_<\Hil$. Therefore $J_H\xi_<$ is cyclic for its restriction to $E^H_>\Hil$.

Moreover, $\dim E^H_1\Hil\leq1$, because $\M_H$ acts by scalars there and $\xi$ is cyclic. Choose a $J_H$-fixed unit vector $\eta_1$ in this subspace if it is nonzero, and put $\eta_1=0$ otherwise. Then
\begin{align*}
 \eta:=\xi_<+J_H\xi_<+\eta_1
\end{align*}
satisfies $J_H\eta=\eta$. Its three spectral components are cyclic in their respective subspaces. Since $E^H_<,E^H_1,E^H_>\in\M_H$, this proves that $\eta$ is cyclic for $\M_H$.

\ref{item:MFFixedCyclic} $\Rightarrow$ \ref{item:MFMeasure}. Let $\eta$ be $J_H$-fixed and cyclic, and define the finite Radon measure
\begin{align*}
 \mu(B):=\langle\eta,E^{\Delta_H}(B)\eta\rangle.
\end{align*}
The relation $J_HE^{\Delta_H}(B)J_H=E^{\Delta_H}(B^{-1})$ implies $\mu(B^{-1})=\mu(B)$, so $\mu$ is symmetric. The map $V:C_0(\Rl_+)\longrightarrow\Hil$, $Vf:=f(\Delta)\eta$ extends to a unitary $V:L^2(\Rl_+,\mu)\to\Hil$. It intertwines the multiplication operator $\Delta_\mu$ with $\Delta$. Furthermore, writing $(jf)(q):=\overline{f(q^{-1})}$, we obtain $J_HVf=J_Hf(\Delta_H)\eta=(jf)(\Delta_H)\eta=VJ_\mu f$. Thus $V$ intertwines both modular operators and both modular conjugations. Taking the fixed-point spaces of their Tomita operators gives $VH_\mu=H$.

\ref{item:MFMeasure} $\Rightarrow$ \ref{item:MFCyclic}. The multiplication representation on $L^2(\Rl_+,\mu)$ is cyclic: any square-integrable function that is nonzero almost everywhere is a cyclic vector. Such a function exists because a Radon measure on $\Rl_+$ is $\sigma$-finite. Hence $H_\mu$, and therefore $H$, is multiplicity-free.
\end{proof}
We may equivalently describe $H_\mu$ as follows. Let $\mathbb C^2$ be the two-dimensional Hilbert space and $\big\{\Hil(q), q\in (0,1]\big\}$ the constant field of Hilbert spaces over the measure space $\big((0,1], \mu\big)$ such that $\Hil(q) = \mathbb C^2$ if $q <1$ and $\Hil(1) = \mathbb C$. Let $H_q \in\Std(\Hil(q))$ be given as in the \eqref{eq:Hlambda}, $q <$, and  $\H(1) = \mathbb C$, $H(1) = \RR$. With $\mathfrak M$ the real linear span of the measurable fields $q \mapsto h_k(q)$, with $h_k(q)$ given by \eqref{eq:2dBasis},  $k=1,2$, $q\in (0,1)$, and $h_k(1)\in \Rl_+$, the family $\big\{H(q)\in \Std(\H(q)), q\in (0,1]\big\}$ is a measurable field of standard subspaces and
\begin{equation}\label{Hdec}
H = \int^\oplus_{(0,1]} H(q) d\mu \, .
\end{equation}
Then $H$ is unitarily equivalent to $H_\mu$.

\begin{theorem}[Direct sum decomposition into multiplicity-free spaces]\label{thm:mf}\booktheoremindex{thm:mf}{Spectral multiplicity classification of standard subspaces}\index{multiplicity-free standard subspace}\index{classification of standard subspaces!by spectral multiplicity (separable case)}
Let $H\in\Std(\Hil)$, with $\Hil$ separable. There exists a sequence  $\mu_\infty, \mu_1, \mu_2, \dots$ such that $\mu_n$ is a symmetric Radon measure on $\Rl_+$  (possibly $\mu_n = 0$), the measures $\mu_n$ are mutually singular and
\begin{equation}\label{dec}
H \cong \bigoplus_n n H_{\mu_n}\, .
\end{equation}
If $H \cong \bigoplus_n nH_{\mu'_n}$ is another decomposition as above, then $[\mu_n] = [\mu'_n]$.
\end{theorem}
\begin{proof}
Let $\Delta_{H,\leq} = \Delta_H|_{\Hil_{(0,1]}}$\glsadd{restrictedModular}, with $\Hil_I$ the spectral subspace of $\Delta_H$ relative to $I\subset \Rl_+$.
By the spectral multiplicity theorem for representations of commutative C$^*$-algebras, the representation $\pi_{\leq}: f\in C_0((0,1])\mapsto \pi_H(f)|_{\Hil_{(0,1]}}$ decomposes as
\begin{equation}\label{dec2}
\pi_{\leq} = \bigoplus_{n\in \Nl\cup\{\infty\}} n \rho_{\l_n}\, ,
\end{equation}
where $\{\l_n\}_{n\in \Nl\cup\{\infty\}}$ are mutually singular Radon measures on $(0,1]$, and the representation $\pi_{\l_n}$ of $C_0((0,1])$ on $L^2((0,1],d\l_n)$ is given by
\[
\rho_{\l_n}(f)g = fg\, ,\quad f\in C_0((0,1]), \ g\in L^2((0,1],\l_n).
\]

Define now $\mu_n$ to be the measure on $\Rl_+$ such that $\mu_n(E) = \l_n(E)$ for all Borel sets $E\subset (0,1]$, and $\mu_n(E) = \l_n(E^{-1})$ for all Borel sets $E\subset(1,\infty)$, where $E^{-1}=\{x^{-1}\in \Rl_+ \col x\in E\}$. Then the standard subspace
$H_{\mu_n}$ is the one associated with $\Delta_{\mu_n}$ by Theorem \ref{theorem:StdAndContractions}.
\end{proof}

As a corollary, we can represent any standard subspace on a separable space as a direct integral of one- and two-dimensional standard subspaces (Example~\ref{example:2dStandardSubspaces}).

\begin{corollary}[Direct integral decomposition of standard subspaces]\label{cor:multifree}
Let $H\in\Std(\Hil)$, with $\H$ separable. There exists a Radon measure $\mu$ on $\Rl_+$ such that
\begin{equation}\label{dirdec}
H \cong \int_{(0,1]}^\oplus n(q) H_q d\mu\, ,
\end{equation}
where $H_q$, $q\neq1$, is the two-dimensional standard subspace of
$\mathbb C^2$ with $q\in \text{sp}(\Delta_{H_q})$, and $H_1=\Rl\in\Std(\Cl)$. Here $n(q) = 1,2, \dots$ or $n = \infty$, and the decomposition \eqref{dirdec} depends only on the measure class of $\mu$.
\end{corollary}
\begin{proof}
We start with the canonical decomposition \eqref{dec}, where we may assume that the measures $\mu_n$ are finite and $\mu_n(\Rl_+) \leq 2^{-n}$, $n=1,2\dots$. Let $\{X_n\}$ be a measurable partition of $\Rl_+$ such that $\mu_n(X_n) = \mu_n(\Rl_+)$, and set $\mu = \sum_n \mu_n$. Then $\big\{nH(q)\in \Std(n\H(q), q \in (0,1]\big\}$, is a measurable field of standard subspaces and by \eqref{Hdec} $H \cong \int^\oplus_{(0,1]} n(qH(q) d\mu(q)$.
\end{proof}

\section{Applications}\label{section:Chapter1Applications}

In this section we collect some applications of the theory developed so far, mostly to quantum field theory. The different sections are largely independent. In contrast to the rest of Chapter~\ref{chapter:Fundamentals}, some background in operator algebras and quantum physics is required here.

In Section~\ref{section:KMSApplication}, we use the KMS condition to compute the KMS states of a type I factor. In Section~\ref{section:KleinGordonSpaces}, we describe the net of standard subspaces underlying the free Klein-Gordon field on Minkowski space of dimension $d\geq1+1$. In Section~\ref{section:Entropy}, we introduce an abstract notion of entropy in standard subspace language. We also explain how this entropy looks like for the Klein-Gordon spaces of Section~\ref{section:KleinGordonSpaces}. In Section~\ref{section:Hadamard}, we give an application of the tensor product of standard subspaces to the Hadamard product of square matrices.

\subsection{The KMS states of a type I factor}\label{section:KMSApplication}

The purpose of this section is to explain the relation of the KMS condition for standard subspaces to the KMS condition for $C^*$-$W^*$-dynamical systems and to thermal equilibrium states \cite{BratteliRobinson:1987}.

\begin{definition}[$\boldsymbol{C^*}$- and $\boldsymbol{W^*}$-dynamical systems]\label{def:DynamicalSystem}
    \leavevmode
 \begin{enumerate}
        \item\label{item:CstarSystem}\index{C-star dynamical system@$C^*$-dynamical system} A {\em $C^*$-dynamical system} $(\A,\alpha)$ consists of a $C^*$-algebra $\A$ and a one-parameter group $(\alpha_t)_{t\in\Rl}\in\Aut\A$ of automorphisms of $\A$ such that $t\mapsto\alpha_t(A)$ is continuous for every $A\in\A$.
        \item\label{item:WstarSystem}\index{W-star dynamical system@$W^*$-dynamical system} A {\em $W^*$-dynamical system} $(\M,\alpha)$ consists of a von Neumann algebra $\M$ and a one-parameter group $(\alpha_t)_{t\in\Rl}\in\Aut\M$ of automorphisms of $\M$, such that $t\mapsto\varphi(\alpha_t(A))$ is continuous for all $A\in\M$ and all normal functionals $\varphi\in\M_*$.
    \end{enumerate}
 \end{definition}

The interpretation of the algebra $\A$ is that of the observable algebra of a physical system, and the interpretation of the one-parameter group of automorphisms is that of {\em dynamics} ({\em time evolution}).

 \begin{definition}[KMS states]\index{KMS!state}\glsadd{omega}
    Let $(\A,\alpha)$ be a $C^*$- or $W^*$-dynamical system and let $\beta\in\Rl\backslash\{0\}$. A state $\omega$ of $\A$, assumed normal in the $W^*$-case, is called a {\em $\beta$-KMS state} if for every $A,B\in\A$, there exists a bounded continuous function $F_{A,B}$ on the closed strip $\overline\Strip_\beta$ \eqref{eq:Strips}, analytic in its interior, such that for all $t\in\Rl$
    \begin{align}
        F_{A,B}(t)=\omega(A\alpha_t(B)),\qquad
        F_{A,B}(t+i\beta)=\omega(\alpha_t(B)A).
    \end{align}
\end{definition}

In the terminology of symmetric functions (Definition~\ref{def:SymmetricFunction}), the KMS condition asks $F_{A,B}\in\HinftyC(\Stripb)$ and $F_{A,B}^*=F_{A^*,B^*}$. The similarity to Definition~\ref{definition:KMSCondition} is apparent.

\medskip

As an example, we here consider a complex Hilbert space $\Hil$ and the von Neumann algebra $\M:= B(\Hil)$ (a type I factor) as our observable algebra. To complete it to a $W^*$-dynamical system, note that for a type I factor, all automorphisms are inner, and the continuity assumption of Definition~\ref{def:DynamicalSystem}~\ref{item:WstarSystem} means precisely that
\begin{align}\label{eq:InnerDynamics}
    \alpha_t(A)=U(t)AU(-t),\qquad A\in B(\Hil),\quad t\in\Rl,
\end{align}
 for a strongly continuous unitary one-parameter group $(U(t))_{t\in\Rl}$ \cite[Sect.~1.1, pp.~2--3]{Arveson:2003}. Writing $U(t)=e^{itP}$, the selfadjoint generator $P$ (the {\em Hamiltonian}) is unique up to an additive real constant.

Also recall that the normal states $\om\in B(\Hil)_*$\glsadd{Mpredual} of $B(\Hil)$ are in bijection with the density matrices (positive trace class operators with unit trace) $\rho\in\CD(\Hil)$\glsadd{rho}\glsadd{Dens} via $\om(A)=\Tr(\rho A)$\glsadd{Tr}.

\begin{proposition}[The normal KMS states of $\boldsymbol{B(\Hil)}$]\index{KMS!state, normal}\index{Gibbs density matrix}
    Let $U(t)=e^{itP}$, $t\in\Rl$, be a strongly continuous unitary one-parameter group on $\Hil$, and let $\beta\in\Rl\backslash\{0\}$. The $W^*$-dynamical system with observable algebra $\M=B(\Hil)$ and dynamics \eqref{eq:InnerDynamics} given by $U$ has a KMS state at inverse temperature $\beta\neq0$ if and only if $e^{-\beta P}$ is trace class. If this is the case, the KMS state is unique and faithful, and explicitly given by
    \begin{align}\label{eq:GibbsDensityMatrix}
        \rho
        =
        \frac{e^{-\beta P}}{\Tr(e^{-\beta P})}
        .
    \end{align}
\end{proposition}

 \begin{proof}
 Let $\om=\Tr(\rho\,\cdot\,)$ be a KMS state of $B(\Hil)$ at inverse temperature $\beta\neq0$. Setting $A=1$ in the KMS condition, we observe that the function $f(t)=\om(\alpha_t(B))$ lies in $\HinftyC(\Stripb)$ and is $i\beta$-periodic. As $\beta\neq0$, this implies that $f$ has an entire analytic bounded continuation which is constant by Liouville's Theorem. So we conclude that the KMS condition implies $\alpha$-invariance of the state, namely $\Tr(\rho U(t)AU(t)^*)=\Tr(U(t)^*\rho U(t)A)$ is independent of $t\in\Rl$ for any $A\in B(\Hil)$. This implies that $\rho$ and $U(t)$ commute.

    It also follows that $\rho$ is injective. To see that, let $\xi\in\Hil$ with $\langle\xi,\rho\xi\rangle>0$, and let $\eta\in\ker\rho$. Then $A:=|\xi\rangle\langle\eta|$ satisfies $\rho A^*=0$, and therefore
 \begin{align*}
        \om(A^*A)
        &=
        \Tr(\rho A^*A)=0.
    \end{align*}
 By the KMS condition, this implies $0=\om(AA^*)=\Tr(\rho AA^*)=\|\eta\|^2\langle\xi,\rho\xi\rangle$, i.e. $\eta=0$. Hence $\rho$ is injective, and $\om$ is faithful.

 The operator $\Om_\om:=\rho^{1/2}$ is a Hilbert-Schmidt operator with $\ker\Om_\om=\ker\Om_\om^*=\{0\}$. As we have learned in Example~\ref{example:HilbertSchmidt}, $H:=\overline{\ell(B(\Hil)_\sa)\Om_\om}$ is a standard subspace of $\Hil_\om$, with modular group $\Delta_H^{it}(X)=\rho^{it}X\rho^{-it}$ (Example~\ref{example:ModularDataOfHilbertSchmidt}, note that here $\Om$ is positive so the unitary $U$ appearing there is here the identity $U=1$).

    The unitary one-parameter group $U_\om(t)(X):=U(-\beta t)XU(-\beta t)^*$ leaves $H$ invariant because $U(t)$ commutes with $\rho$. Furthermore, by the KMS assumption, for $A,B\in B(\Hil)_\sa$, the function
    \begin{align*}
        f(t)
        :=
        \om(A\alpha_{-\beta t}(B))
        =
        \Tr(\rho A U(-\beta t)BU(\beta t))
        =
        \langle A\Om_\om,U_\om(t)(B\Om_\om)\rangle_2
    \end{align*}
    lies in $\HinftyC(\Strip_{-1})$ and satisfies the KMS boundary condition
    \begin{align*}
        f(t-i)
        =
        \om(\alpha_{-\beta t}(B)A)
        =
        \langle U_\om(t)(B\Om_\om),A\Om_\om\rangle_2.
    \end{align*}

 As $\ell(B(\Hil)_\sa)\Om_\om$ is a dense subspace of $H$, this shows that $U_\om(t)=\Delta_H^{it}$ is the modular group (Lemma~\ref{lemma:KMSDenseIsEnough}, Theorem~\ref{theorem:KMS}). By Example~\ref{example:ModularDataOfHilbertSchmidt}, this implies $U_\om(t)(X)=U(-\beta t)XU(\beta t)=\rho^{it}X\rho^{-it}$ for all $X\in\CI_2(\Hil)$, namely
    \begin{align}
        \rho^{it} = e^{ict}\,U(-\beta t) = e^{ict}\,e^{-i\beta t P}
    \end{align}
 for some constant $c\in\Rl$. Thus $\rho=e^ce^{-\beta P}$. If $e^{-\beta P}$ is not trace class, this is a contradiction, i.e. no KMS states exist. If $e^{-\beta P}$ is trace class, taking the trace gives the claimed formula for $\rho$.

    Conversely, let $\rho=(\Tr(e^{-\beta P}))^{-1}e^{-\beta P}$ (assuming $e^{-\beta P}$ is trace class). We have to show that $\om$ is indeed a KMS state. This argument works precisely as in the proof of Example~\ref{example:ModularDataOfHilbertSchmidt}.
 \end{proof}

 The density matrix \eqref{eq:GibbsDensityMatrix} has the Gibbs form well-known from equilibrium states at temperature $1/\beta$ in statistical mechanics, which explains a link between the KMS condition and thermal equilibrium. While Gibbs states appear mostly in finite-volume situations (reflected in the trace-class condition on $e^{-\beta P}$), the KMS condition remains valid in the thermodynamic limit as a characterization of equilibrium \cite{BratteliRobinson:1997}.

\subsection{The local standard subspaces of the Klein--Gordon field}\label{section:KleinGordonSpaces}

In this section we describe the net of local standard subspaces of the free Klein-Gordon field of mass $m>0$ on Minkowski spacetime $\Rl^{d+1}$ of spacetime dimension $d+1$, $d\in\Nl$. Our treatment here is preliminary, much more can and will be said about this model in future editions.

The following example is an application of the construction of standard subspaces from the dual pair given by a positive operator, see Definition~\ref{def:PositiveOperatorDualPair} and Theorem~\ref{theorem:PositiveOperatorStandardSubspaces}.

We consider the real Hilbert space $\Hil_{\Rl}=L^2_{\Rl}(\Rl^n)$, and the Laplace operator $\Delta$ (as an expectation, $\Delta$ does {\em not} denote a modular operator here!) on its domain in $\Hil_{\Rl}$. The operator
\begin{align}
    A&:=(-\Delta+m^2)^{1/2}\ge m
\end{align}
acts in the Fourier (momentum space) picture as the multiplication operator multiplying with the function $\om:\Rl^d\to\Rl$, $\omega(\bp):=(|\bp|^2+m^2)^{1/2}$. This shows directly that $A$ is selfadjoint, positive, injective, and unbounded with domain of selfadjointness
\begin{align}
    \dom A
    &=
    \left\{
        f\in L^2_{\Rl}(\Rl^d):
        \int_{\Rl^d}\omega(\bp)^2|\tilde f(\bp)|^2\,d\bp<\infty
        \right\}
    \end{align}
    We are therefore in the situation of Definition~\ref{def:PositiveOperatorDualPair}. Note that the inverse $A^{-1}$ is bounded, with $\|A^{-1}\|\le m^{-1}$ and $\dom A^{-1}=L^2_{\Rl}(\Rl^d)$.

    The norms from Definition~\ref{def:PositiveOperatorDualPair}    are
    \begin{align}
        \|f\|_\pm^2
        &=
        \int_{\Rl^d}\omega(\bp)^{\pm1}|\tilde f(\bp)|^2\,d\bp,
        \qquad f\in\dom A^{\pm1}.
    \end{align}
    Their completions identify with the real Sobolev spaces\glsadd{Sobolev}
    \begin{align}
        \mathcal H_A^\pm&=\Sob^{\pm1/2}_{\Rl}(\Rl^d).
    \end{align}
    We equip these spaces with the displayed $m$-dependent norms,
    which are equivalent to their usual Sobolev norms. The $L^2$-scalar product extends to the continuous duality
    pairing
    \begin{align}
        \overline b&:
        \Sob^{1/2}_{\Rl}(\Rl^d)\times \Sob^{-1/2}_{\Rl}(\Rl^d)
        \to\Rl.
    \end{align}
    The extension $\overline A:\Sob^{1/2}_{\Rl}(\Rl^d)\to \Sob^{-1/2}_{\Rl}(\Rl^d)$ is the Fourier multiplier with symbol $\omega$.

    We will write $\Hil_m:=\Hil_{A,b}$
    for the complex Hilbert space constructed in     Definition~\ref{def:PositiveOperatorDualPair}. Its underlying real Hilbert space is $\Sob^{1/2}_{\Rl}(\Rl^d)\oplus \Sob^{-1/2}_{\Rl}(\Rl^d)$, and its complex structure is $\overline I_A$ \eqref{eq:DefIA}.

    \bigskip

    The physical interpretation of this setup is that the functions $f\oplus g\in\Hil_m$ describe the time zero Cauchy data of a solution of the Klein--Gordon Equation $(\square+m^2)F=0$, with $\square=\partial_0^2-\Delta$ the D'Alambert operator and $F:\Rl^{d+1}\to\Rl$ a smooth function. The physical meaning of $A$, and hence its Fourier multiplier $\om$, is the energy of a Klein--Gordon particle, reflecting the relativistic dispersion relation of a massive particle.

    Hence $\Rl^d$ corresponds to the time-zero surface $\{0\}\times\Rl^d\subset\Rl^{d+1}$ of Minkowski spacetime.

    To describe localization properties, we consider an open subset $B\subset\Rl^d$ and the corresponding test function space
    \begin{align}
        K(B)
        :=
        C_{c,\Rl}^\infty(B).
    \end{align}
    Fourier transform maps $K(B)$ into the Schwartz space, which shows $K(B)\subset\dom A$. As $\dom A^{-1}=\Hil_{\Rl}$, this shows that $K(B)$ satisfies the initial requirement of Theorem~\ref{theorem:PositiveOperatorStandardSubspaces}.

    The corresponding real subspace of $\Hil_m$ that is considered in Theorem~\ref{theorem:PositiveOperatorStandardSubspaces} (denoted $K_A$ there) is here
    \begin{align}
        H_m(B)
        :=
        \overline{K(B)}^{\|\cdot\|_+}
        \oplus
        \overline{K(B)}^{\|\cdot\|_-}
        \subset\Hil_m.
    \end{align}

    \begin{proposition}[Local standard subspaces of the Klein--Gordon field]\label{prop:KleinGordonDualPair}\label{example:KleinGordonDualPair}\index{Klein-Gordon field@Klein--Gordon field!local standard subspaces}

    Let $B\subset\Rl^d$ be open. When $B$ is nonempty, $H_m(B)$ is cyclic. If $\Rl^d\backslash B$ has interior points, $H_m(B)$ is separating. In particular, if $B$ is nonempty and bounded, $H_m(B)$ is standard.
\end{proposition}
\begin{proof}
    We have to verify the conditions in Theorem~\ref{theorem:PositiveOperatorStandardSubspaces}.     We will use the following anti-locality property\index{anti-locality}: for every nonempty open set $B\subset\Rl^d$,
    \begin{align}
        u\in \Sob^{1/2}_{\Rl}(\Rl^d),\quad
        u|_B=0,\quad
        (\overline Au)|_B=0
        &\quad\Longrightarrow\quad u=0.
        \label{eq:KleinGordonAntilocality}
    \end{align}
    The restrictions are understood in the sense of distributions.
    This is a special case of the anti-locality theorem in
    \cite{Murata:1973}, which applies to tempered distributions.

    Let $\Rl^d\backslash B$ have interior points. To prove that $H_m(B)$ is separating, we verify the trivial-intersection condition in Theorem~\ref{theorem:PositiveOperatorStandardSubspaces}~\ref{item:KAStandard}. Let
    \begin{align}
        u&\in
        \overline{K(B)}^{\|\cdot\|_+}
        \cap
        \overline{A^{-1}K(B)}^{\|\cdot\|_+}.
    \end{align}
    Then $u\in\overline{K(B)}^{\|\cdot\|_+}$ and
    $\overline Au\in\overline{K(B)}^{\|\cdot\|_-}$.
    Convergence in either of these Sobolev spaces implies
    convergence in the sense of distributions. Since all
    approximating test functions are supported in $B$, it follows
    that
    \begin{align}
        \operatorname{supp}u&\subset\overline B,
        &
        \operatorname{supp}\overline Au&\subset\overline B.
    \end{align}
    Both distributions therefore vanish on the nonempty open set
    $\Rl^n\setminus\overline B
    =\operatorname{int}(\Rl^n\setminus B)$.
    By \eqref{eq:KleinGordonAntilocality}, $u=0$.

    To prove cyclicity, we assume that $B$ is not empty and verify    the density condition in Theorem~\ref{theorem:PositiveOperatorStandardSubspaces}~\ref{item:KAStandard}.
    Let $u\in\mathcal H_A^+=\Sob^{1/2}_{\Rl}(\Rl^d)$ be orthogonal,  with respect to $\langle\cdot,\cdot\rangle_+$, to
    $K(B)+A^{-1}K(B)$. For every $f\in K(B)$, we then have
    \begin{align}
        0&=\langle u,f\rangle_+
        =\overline b(f,\overline Au),
        \\
        0&=\langle u,A^{-1}f\rangle_+
        =\langle u,f\rangle_{L^2}.
    \end{align}
    The first identity says that $(\overline Au)|_B=0$ as a
    distribution, and the second says that $u|_B=0$.
    Since $B$ is nonempty and open,
    \eqref{eq:KleinGordonAntilocality} again gives $u=0$.
    Hence $K(B)+A^{-1}K(B)$ is dense in $\mathcal H_A^+$,
    proving cyclicity of $H_m(B)$.
\end{proof}

The family $(H_m(B))_B$ of standard subspaces of $\Hil_m$ describes the time-zero localization of a free massive Klein--Gordon quantum particle. The standardness of the spaces $H_m(B)$ is equivalent to the Reeh--Schlieder property of the corresponding free quantum field theory. As the net of the $U(1)$-current in Section~\ref{section:U1current}, the net $B\mapsto H_m(B)$ has various interesting inclusion, separation, covariance, and duality properties. These will be discussed in detail in Chapter~4.

\subsection{The entropy operator of a standard subspace}\label{section:Entropy}

In this section we introduce the concept of entropy of a vector $\psi\in\Hil$ relative to a standard subspace $H\in\Std(\Hil)$. Its name comes from the relative entropy of a coherent state with respect to the vacuum when both states are restricted to a local Weyl algebra acting on the Bose Fock space over $\Hil$.

The standard subspace entropy will be defined in terms of functional calculus of the modular operator $\Delta_H$ and also uses the cutting projection (Definition~\ref{definition:cuttingprojection}). We begin with the following preparatory lemma.

\begin{lemma}[Functional calculus and the cutting projection]\label{lemma:fmapingin H+H'}
Let $H\in\Std(\Hil)$ be factorial, and let $h:(0,\infty)\to\Cl$ be a Borel function. Define the two functions $g_1,g_2:(0,\infty)\backslash\{1\}\to\Cl$,
\begin{align*}
    g_1(\lambda)=\frac{h(\lambda)}{1-\lambda},\qquad
    g_2(\lambda)=\sqrt\lambda\,g_1(\lambda).
\end{align*}
If $\psi$ belongs to the domains of $h(\Delta_H)$, $g_1(\Delta_H)$ and $g_2(\Delta_H)$, then $h(\Delta_H)\psi\in H+H'$ and
\begin{align}\label{eq:EntropyCuttingMultipliers}
    P_Hh(\Delta_H)\psi=g_1(\Delta_H)\psi+J_Hg_2(\Delta_H)\psi .
\end{align}
If $g_1$ and $g_2$ are bounded, the right hand side gives a bounded real-linear
extension of $P_Hh(\Delta_H)$ to $\Hil$.
\end{lemma}
\begin{proof}
By factoriality, $1$ is not an eigenvalue of $\Delta_H$, so the values of
$g_1$ and $g_2$ at $1$ are immaterial. Let $E_n$ be the spectral projections
of $\Delta_H$ for $[1/n,n]\setminus(1-1/n,1+1/n)$, $n\geq2$.
For $E_n\psi$, Proposition~\ref{proposition:propertiescutting}~\ref{item:PHformula} gives \eqref{eq:EntropyCuttingMultipliers} by expanding
$(1+S_H)(1-\Delta_H)^{-1}$ and using $S_H=J_H\Delta_H^{1/2}$.
The projections increase strongly to $1$. As $n\to\infty$, each of the three multipliers applied to $E_n\psi$
converges in norm to its value on $\psi$, by the domain assumptions. Closedness of $P_H$
now proves the claim.
\end{proof}

For the entropy operator we use the positive functions
\begin{align*}
    g_1(\lambda)=\frac{\log\lambda}{\lambda-1},\qquad
    g_2(\lambda)=\sqrt\lambda\,g_1(\lambda)\quad(\lambda\ne1),\qquad
    g_1(1)=g_2(1)=1,
\end{align*}
where the values at $1$ are given by continuity. They satisfy
\begin{align}
    g_1(\lambda^{-1})=\lambda g_1(\lambda),\qquad
    g_2(\lambda^{-1})=g_2(\lambda),\qquad \lambda>0.
\end{align}
The function $g_2$ is bounded, whereas $g_1(\la)\to\infty$ for $\la\to0$.

For a factorial standard subspace $H$ with modular Hamiltonian $L_H=\log\Delta_H$, we apply Lemma~\ref{lemma:fmapingin H+H'} with $h(\lambda)=i\log\lambda$. Its two multipliers are $-ig_1$ and $-ig_2$. Since $g_1(\lambda)=O(1+|\log\lambda|)$, the lemma applies on $\dom L_H$, and antilinearity of $J_H$ gives
\begin{align}\label{eq:EntropyCuttingLog}
 P_H iL_H\psi=-i\bigl(g_1(\Delta_H)-J_Hg_2(\Delta_H)\bigr)\psi,
 \qquad \psi\in\dom L_H .
\end{align}

In the following definition, we drop the assumption of factoriality on $H$.

\begin{definition}[Entropy operator and entropy of a vector]\label{def:entropyoperator}
Let $H\in\Std(\Hil)$.
\begin{enumerate}
  \item\label{item:EntropyOperator} The \emph{entropy operator}\index{entropy operator of $H$}\glsadd{entropyOperator} is the closed real-linear operator
    \begin{align*}
       \E_H:=g_1(\Delta_H)-J_Hg_2(\Delta_H),
       \qquad \dom\E_H=\dom g_1(\Delta_H).
    \end{align*}
  \item\label{item:EntropyVector} The \emph{entropy of a vector $\psi\in\Hil$ relative to $H$}\index{entropy of a vector}\glsadd{entropyvector} is the extended quadratic form
    \begin{align}\label{eq:defentropyvector}
       \E_H(\psi):=
       \begin{cases}
       \|g_1(\Delta_H)^{1/2}\psi\|^2
           -\Re\ip{\psi}{J_Hg_2(\Delta_H)\psi},
             &\psi\in\dom g_1(\Delta_H)^{1/2},\\
       +\infty,&\text{otherwise}.
       \end{cases}
    \end{align}
\end{enumerate}
\end{definition}

Various properties of the entropy operator and the entropy quadratic form are collected in the following proposition.

\begin{proposition}[The entropy operator and its quadratic form]\label{prop:EntropyOperatorProperties}
Let $H\in\Std(\Hil)$.
\begin{enumerate}
    \item\label{item:EntropyOperatorAdjoint} The operator $\E_H$ is real selfadjoint. If $H$ is factorial, then $\E_H=\overline{iP_H iL_H}$. If $\Delta_H\psi=\psi$, then $\E_H\psi=(1-J_H)\psi$.

    \item\label{item:EntropySquareForm} The entropy admits the following expressions.
    \begin{align}\label{eq:EntropyExpressions}
        \E_H(\psi)=
        \begin{cases}
            \Re\ip{\psi}{\E_H\psi},
                &\psi\in\dom\E_H,\\
            \tfrac12\|(1-J_H)g_1(\Delta_H)^{1/2}\psi\|^2,
                &\psi\in\dom g_1(\Delta_H)^{1/2}.
        \end{cases}
    \end{align}
    In particular, $\E_H(\psi)\geq0$ for all $\psi\in\Hil$.

    If $H$ is factorial and $\psi=h+h'\in\dom L_H$, with $h\in H$, $h'\in H'$, then
    \begin{align}
        \E_H(\psi)
        =
        -\ip{h}{L_Hh}.
    \end{align}

    \item\label{item:EntropyFinitenessAndKernel} For $\psi\in\Hil$,
    \begin{alignat}{2}
        \E_H(\psi)&<\infty &\qquad &\Longleftrightarrow\qquad
            \int_{(0,1)}|\log\lambda|\,dE^{\Delta_H}_\psi(\lambda)<\infty,\\
        \E_H(\psi)&=0      &       &\Longleftrightarrow\qquad \psi\in H'.
    \end{alignat}
\end{enumerate}
\end{proposition}
\begin{proof}
    We use the shorthand notations $A:=g_1(\Delta_H)$, $B:=A^{1/2}$ and $C=g_2(\Delta_H)$. We also use the spectral projections $F_n:=E^{\Delta_H}_{[1/n,n]}$, which increase strongly to $1$ as $n\to\infty$. They commute with $J_H$ by the modular relation.

\ref{item:EntropyOperatorAdjoint}
Since $g_2(\lambda^{-1})=g_2(\lambda)$, the operator $C$ commutes with $J_H$. Both are bounded and selfadjoint on the real Hilbert space, so their product $J_HC$ is selfadjoint there.
Thus $\E_H=A-J_HC$ is selfadjoint on $\dom A$.
If $H$ is factorial, \eqref{eq:EntropyCuttingLog} identifies $iP_H iL_H$ with the restriction of $\E_H$ to $\dom L_H$.
For $\psi\in\dom A$, $F_n\psi$ belongs to $\dom L_H$ and converges to $\psi$ in the graph norm of $A$, hence also in that of $\E_H$. Thus $\dom L_H$ is a core for $\E_H$, and $\E_H=\overline{iP_H iL_H}$ follows.

On the fixed space of $\Delta_H$, both $A$ and $C$ act as the identity.

\ref{item:EntropySquareForm}
For $\psi\in\dom A$, we have $\|B\psi\|^2=\ip{\psi}{A\psi}$, so the first formula follows from \eqref{eq:defentropyvector}. On $F_n\Hil$, the identity $g_1(\lambda^{-1})=\lambda g_1(\lambda)$ gives $BJ_HB=J_HC$. Expanding the square therefore gives
\begin{align*}
    \tfrac12\|(1-J_H)B\eta\|^2
       =\|B\eta\|^2-\Re\ip{\eta}{J_HC\eta},\qquad \eta\in F_n\Hil.
\end{align*}
For $\psi\in\dom B$, take $\eta=F_n\psi$ and pass to the limit, using $BF_n\psi=F_nB\psi\to B\psi$.
This proves the second formula and non-negativity on the form domain. Outside this domain the entropy is $+\infty$ by definition.

For the last claimed formula, suppose $H$ is factorial and write $\psi=h+h'$ with $h=P_H\psi\in\dom L_H$ and $h'\in H'$.
By \ref{item:EntropyFinitenessAndKernel}, $\E_H(h')=0$, so $h'\in\dom B$ and $(1-J_H)Bh'=0$ by the second formula in square formula.
Since $h\in\dom L_H\subset\dom\E_H$, that formula gives $\E_H(\psi)=\E_H(h)$.
Differentiating $\Delta^{it}h\in H$ at $t=0$ gives $iL_Hh\in H$.
Thus \eqref{eq:EntropyCuttingLog} gives $\E_Hh=iP_H iL_Hh=-L_Hh$, and the first line of \ref{item:EntropySquareForm} yields $\E_H(h)=-\ip{h}{L_Hh}$.

\ref{item:EntropyFinitenessAndKernel}
Finiteness means $\psi\in\dom B$, or equivalently that $g_1$ is integrable with respect to the spectral measure $E^{\Delta}_\psi$.
Now $g_1(\lambda)\sim|\log\lambda|$ as $\lambda\downarrow0$, and $g_1$ is bounded on $[\varepsilon,\infty)$ for every $\varepsilon>0$. This proves the first equivalence.

For the second, recall that $H'$ is the fixed space of $S_{H'}=J_H\Delta_H^{-1/2}$
(Proposition~\ref{proposition:ModularDataProperties}~\ref{item:JDComplement}).
On $F_n\Hil$, $B$ is boundedly invertible and $J_HB=\Delta_H^{1/2}BJ$, again by the inversion identity for $g_1$. Hence for $\eta\in F_n\Hil$,
\begin{align*}
    J_HB\eta=B\eta
       \quad\Longleftrightarrow\quad J_H\eta=\Delta_H^{-1/2}\eta
       \quad\Longleftrightarrow\quad \eta\in H'.
\end{align*}
If $\E_H(\psi)=0$, then $J_HB\psi=B\psi$ by \ref{item:EntropySquareForm}.
As $F_n$ commutes with $J_H$ and $B$, each $F_n\psi$ lies in $H'$. This space is closed, so $\psi\in H'$.
Conversely, if $\psi\in H'$, then $\psi\in\dom\Delta_H^{-1/2}\subset\dom B$ since $g_1(\lambda)\leq\max\{1,\lambda^{-1}\}$.
Each $F_n\psi$ still lies in $H'$ because $F_n$ commutes with $S_{H'}$ on its domain.
Thus $J_HBF_n\psi=BF_n\psi$. Passing to the limit and applying \ref{item:EntropySquareForm} gives $\E_H(\psi)=0$.
\end{proof}

Let us mention two further interesting properties of the entropy operator: For any standard subspace $H$, the entropy depends only on the class of $\psi$ modulo $H'$. More precisely,
\begin{align}\label{eq:EntropyLocality}
    \E_H(\psi+h')=\E_H(\psi),\qquad \psi\in\Hil,\quad h'\in H'.
\end{align}
Indeed, put $B=g_1(\Delta_H)^{1/2}$. By Proposition~\ref{prop:EntropyOperatorProperties}~\ref{item:EntropyFinitenessAndKernel} and \eqref{eq:EntropyExpressions}, every $h'\in H'$ belongs to $\dom B$ and satisfies $(1-J_H)Bh'=0$. Thus $\psi+h'\in\dom B$ exactly when $\psi\in\dom B$. On this domain the square in \eqref{eq:EntropyExpressions} is unchanged by adding $h'$. Outside it both entropies are infinite.

The second observation is that the entropy operators of $H$ and its symplectic complement $H'$ also determine the expectation of the modular Hamiltonian, namely
\begin{align}\label{eq:EntropyComplementDifference}
    \E_{H'}(\psi)-\E_H(\psi)=\ip{\psi}{L_H\psi},
    \qquad \psi\in\dom L_H.
\end{align}
Indeed, $\Delta_{H'}=\Delta_H^{-1}$ and $J_{H'}=J_H$ by Proposition~\ref{proposition:ModularDataProperties}~\ref{item:JDComplement}. On $\dom L_H=\dom L_{H'}$ both entropies are finite. In \eqref{eq:defentropyvector} the terms involving $J_H$ cancel because $g_2(\lambda^{-1})=g_2(\lambda)$. The difference of the remaining terms gives $\ip{\psi}{L_H\psi}$ since $g_1(\lambda^{-1})-g_1(\lambda)=\log\lambda$.

\subsection{Field and momentum contributions to entropy}\label{section:FieldMomentumEntropy}

We now consider the entropy operator for factorial standard subspaces represented in the dual pair formalism of Theorem~\ref{theorem:PositiveOperatorStandardSubspaces}. In this direct sum representation, the cutting projection and the entropy operator are diagonal. This applies in particular to the local standard subspaces $H_m(B)$ of the Klein--Gordon field (Section~\ref{section:KleinGordonSpaces}), where the components describe the field and its time derivative on the time-zero surface.

This section will be added in the next edition.

\subsection{The Hadamard product and a matrix inequality}\label{section:Hadamard}

Here we apply the comparison of real and complex tensor products from Section~\ref{section:TensorProducts} to the Hadamard (entrywise) product of matrices.

We write $A_{nm}$ for the entries of $A\in M_N(\Cl)$, where $1\leq n,m\leq N$. We call $A$ positive, and write $A\geq 0$, if $A$ is positive semidefinite.

\begin{lemma}[Gram matrices]\label{lemmaS}
    A matrix $A\in M_N(\Cl)$ is positive if and only if there are vectors $\xi_1,\ldots,\xi_N$ in a complex Hilbert space $\Hil$ such that $A_{nm}=\ip{\xi_n}{\xi_m}$. For any such representation, $\rank(A)=\dim\spn\{\xi_1,\ldots,\xi_N\}$.
\end{lemma}
\begin{proof}
    Given such vectors, define $T:\Cl^N\to\Hil$ by $Tz:=\sum_n z_n\xi_n$. Then $A=T^*T$, since both matrices have entries $\ip{\xi_n}{\xi_m}$ in the canonical basis of $\Cl^N$. In particular,
    \begin{equation}\label{qf}
        \ip{z}{Az}=\sum_{n,m}\bar z_n z_m A_{nm}=\|Tz\|^2.
    \end{equation}
    Thus $A\geq0$ and $\ker A=\ker T$. By rank-nullity, $\rank(A)=\dim\Ran T=\dim\spn\{\xi_1,\ldots,\xi_N\}$.

    Conversely, if $A\geq0$, take $\xi_n:=A^{1/2}e_n$, where $e_1,\ldots,e_N$ are the canonical basis vectors of $\Cl^N$. Then $\ip{\xi_n}{\xi_m}=\ip{e_n}{Ae_m}=A_{nm}$.
\end{proof}

Recall that the \emph{Hadamard product} of two matrices $A,B\in M_N(\Cl)$ is the entrywise product,
\begin{align}
    (A\circ B)_{nm}:=A_{nm}B_{nm}.
\end{align}
The Hadamard product is associative.

\begin{theorem}[Schur product theorem]\label{thmSchur}\booktheoremindex{thmSchur}{Schur product theorem}\index{Schur product theorem}\index{Hadamard product|see{Schur product theorem}}
Let $A,B\in M_N(\Cl)$ be positive. Then also $A\circ B$ is positive.
\end{theorem}
\begin{proof}
By Lemma~\ref{lemmaS}, write $A_{nm}=\ip{\xi_n}{\xi_m}$ and $B_{nm}=\ip{\eta_n}{\eta_m}$ for vectors in complex Hilbert spaces $\Hil$ and $\Kil$, respectively. In the complex tensor product $\Hil\otimes\Kil$ we have
\[
    (A\circ B)_{nm}=\ip{\xi_n\otimes\eta_n}{\xi_m\otimes\eta_m}.
\]
Thus $A\circ B$ is itself a Gram matrix and is positive by Lemma~\ref{lemmaS}.
\end{proof}

We now apply Theorem~\ref{theorem:TensorProductComparison} to the Gram representations of two positive matrices.

\begin{proposition}[A bound for the sum of Hadamard product entries]\label{prop:HadamardSumBound}
    Let $A,B\in M_N(\Cl)$ be positive. Then
    \begin{equation}\label{inmat}
        0\leq \sum_{n,m=1}^N A_{nm}B_{nm}
        \leq 2\sum_{n,m=1}^N (\Re A_{nm})(\Re B_{nm}).
    \end{equation}
\end{proposition}
\begin{proof}
    By Lemma~\ref{lemmaS}, write $A_{nm}=\ip{\xi_n}{\xi_m}$ and $B_{nm}=\ip{\eta_n}{\eta_m}$ for vectors in complex Hilbert spaces $\Hil$ and $\Kil$, respectively. Let $H\subset\Hil$ and $K\subset\Kil$ be their respective real linear spans. Both are finite-dimensional and hence closed.

    In the real Hilbert tensor product $H\otimes_\Rl K$, set $u:=\sum_{n=1}^N\xi_n\otimes_\Rl\eta_n$. The comparison map $\iota:H\otimes_\Rl K\to H\otimes K\subset\Hil\otimes\Kil$ of Theorem~\ref{theorem:TensorProductComparison} sends $u$ to $\sum_{n=1}^N\xi_n\otimes\eta_n$. The two tensor-product scalar products therefore give
    \begin{align*}
        \|\iota u\|^2 &= \sum_{n,m=1}^N A_{nm}B_{nm},\\
        \|u\|^2 &= \sum_{n,m=1}^N (\Re A_{nm})(\Re B_{nm}).
    \end{align*}
    By the same theorem, $\iota^*\iota=1-D_H\otimes_\Rl D_K$. Since the polarizers $D_H$ and $D_K$ are contractions, $\|\iota\|^2\leq 2$, which proves \eqref{inmat}.
\end{proof}

One can also see the upper bound directly by applying Theorem~\ref{thmSchur} to $A\circ\overline B$, where $\overline B$ is the (positive) entrywise conjugate of $B$.

\section{Notes and Remarks}\label{section:NotesAndRemarksChapter1}

\subsubsection*{Section~\ref{section:RealSubspaces} (\nameref{section:RealSubspaces})}

The presentation of the Grassmannian $\Gr(\Hil)$ and the isomorphic projection lattice $\CP(\Hil)$ is standard material that can be found in many books on functional analysis or quantum logic \cite[Ch.~4]{Redei:1998}. Closely related to our discussion of the most general projections in Def.~\ref{def:Projection} and Prop.~\ref{proposition:twosubspacesprojection} is the work of Rieffel and van Daele \cite{RieffelvanDaele:1977}, and that of Ando \cite{Ando:2013}. For so-called standard pairs (linearly independent subspaces $H,K\subset\Hil$ with $\overline{H+K}=\Hil$) and strongly standard pairs (satisfying even $H+K=\Hil$), Figliolini and Guido \cite[Sect.~3]{FiglioliniGuido:2000} have carried out a similar analysis, focussing on the involutions $S_{H,K}$ \eqref{eq:SHK} instead of the projections $P_{H,K}$ \eqref{eq:def2subspaceprojection}. We warn the reader that the term ``standard pair'' nowadays carries a different meaning \cite{LongoWitten:2010}.

The terminology of cyclic and separating spaces (Def.~\ref{definition:CyclicSeparating}) is inspired by its connection to cyclic and separating vectors for a von Neumann algebra $\M\subset B(\Hil)$. We will discuss this in Chapter~2, but already here note that a vector $\Om\in\Hil$ is cyclic for $\M$ if and only if $H:=\overline{\M_{\sa}\Om}\in\Real(\Hil)$ is cyclic, and if it is cyclic, then it is separating if and only if $H$ is separating \cite{Longo:2008_2}.

The word ``complement'' in the context of symplectic complement is unfortunate because it is not a complement, but too widely used to change it. The term ``symplectic orthogonal'' is also used by some authors.

\subsubsection*{Section~\ref{section:StandardSubspacesBasics} (\nameref{section:StandardSubspacesBasics})}

Decompositions of closed real subspaces as in Lemma~\ref{lemma:DecomposeClosedRealSubspace} have appeared in various forms in the literature, for example in \cites{Araki:1963,FiglioliniGuido:1994,Longo:2008_2}.

The introductory Example~\ref{example:2dStandardSubspaces} of two-dimensional standard subspaces is also covered in the work of Figliolini and Guido \cite[Sect.~3]{FiglioliniGuido:1994}, who discuss it with the help of modular theory. The purpose of our presentation is rather to uncover the modular structure directly from the definition.

Examples~\ref{example:MatrixStandardSubspace} and \ref{example:HilbertSchmidt}, together with their later modular calculations, are standard examples arising from the GNS representation of a matrix algebra or type I factor from a density matrix state \cite{BratteliRobinson:1987,Takesaki:2003}.

\subsubsection*{Section~\ref{section:ModularData} (\nameref{section:ModularData})}

The historical context of modular theory is Tomita's original work and Takesaki's systematic presentation of it  \cite{Tomita:1967,Takesaki:1970}. The standard subspace version of this can be found in \cite{Longo:2008_2}, \cite[Prop.~2.4]{Guido:2011}, \cite[Prop.~1.6 and Lemma~1.7]{Neeb:2026b}.

The description of the real orthogonal projection $E_H$ in terms of modular data (Proposition~\ref{proposition:orthogonalprojectionmodulardata}) first appeared in the work of Figliolini and Guido \cite{FiglioliniGuido:1994}. The cutting projection (Proposition~\ref{proposition:propertiescutting}) has been introduced by Ciolli, Longo and Ruzzi in \cite{CiolliLongoRuzzi:2019}.

The origin of passivity lies in (quantum) statistical mechanics. Pusz and Woronowicz \cite{PuszWoronowicz:1978} have characterized complete passivity by ground states or equilibrium KMS states (including the zero-inverse-temperature case in their convention). The standard-subspace condition in Definition~\ref{def:passivity} is a one-particle analogue expressed through a quadratic form on $H$ by Longo and Morsella \cite{LongoMorsella:2023}.

The standard subspaces emerging from locally compact groups are the origin of modular theory and explain the name ``modular'' (from the modular function).

The realization of standard subspaces in terms of graphs is due to Neeb \cite[Sect.~1.3, Prop.~1.12]{Neeb:2026b}, who gives the graph construction and its modular data. The angle decomposition into two-dimensional fibres in \cite[Sect.~3, proof of Props.~3.1--3.2]{FiglioliniGuido:1994} is a related spectral model. The graph Hilbert space of $\Delta_H^{1/2}$ is taken from Lechner et. al. \cite{LechnerLiQueffelecPiazza:2018}.

\subsubsection*{Section~\ref{section:Characterization} (\nameref{section:Characterization})}

In physics, the abbreviation KMS stands for the three names Kubo--Martin--Schwinger. Kubo worked on linear response theory \cite{Kubo:1957} and Martin and Schwinger on many-particle theory \cite{MartinSchwinger:1959}. The conncection of these physics works to operator algebras was carried out by Haag, Hugenholtz and Winnink \cite{HaagHugenholtzWinnink:1967}, and is nowadays the usual operator-algebraic formulation of equilibrium states \cite{BratteliRobinson:1997}. The KMS condition generalizes Gibbs states (see Section~\ref{section:KMSApplication}) to infinite systems, in which the Hamiltonian $P$ is such that $e^{-\beta P}$ is not trace class. Both for Gibbs and KMS states, the significance of the parameter $\beta$ is that of inverse temperature (in natural units). We refer to \cite{BratteliRobinson:1997} for detailed information about KMS states in physics.

In mathematics, Takesaki realized that the ``modular condition'' (the KMS condition at inverse temperature $\beta=-1$) characterizes the modular group of a von Neumann algebra w.r.t. a standard vector \cite{Takesaki:1970}. We refer to \cite{Takesaki:2003} and \cite{Thomsen:2023} for detailed information about KMS weights in mathematics.

The terminology {\em symmetric function} (Definition~\ref{def:SymmetricFunction}) refers to the symmetry underlying the Schwarz Reflection Principle \cite{LongoWitten:2010}. For a characterization of the KMS condition by positive definite kernels, see Neeb and Ólafsson \cite[Thm.~2.6]{NeebOlafsson:2016}. Symmetric functions on the strip that have absolute value $1$ at the boundary, i.e. symmetric inner functions, play an important role as endomorphisms of standard pairs \cite{LongoWitten:2010} and scattering functions of integrable quantum field theories \cite{Lechner:2008,AbdallaAbdallaRothe:2001}.

The name {\em reflection positivity} or {\em Osterwalder--Schrader positivity} has its origin in Euclidean quantum field theory. Here reflection positivity is one of the axioms that permit the analytic continuation of the Schwinger functions to Wightman functions (``Wick rotation''), i.e. the reconstruction of the physical Hilbert space and positive-energy time evolution \cite{OsterwalderSchrader:1973,OsterwalderSchrader:1975}. A systematic mathematical account is \cite{NeebOlafsson:2018}. The relation between modular theory in the Euclidean setting is best understood at finite temperature, see \cite{GerardJkel:2003,JaffeJaekelMartinez:2014}.

\subsubsection*{Section~\ref{section:U1current} (\nameref{section:U1current})}

The real Schwartz space $\Sch_\Rl(\Rl)$ is a symplectic space when equipped with the symplectic form
\begin{align}
    s(f,g)
    =
    \frac12\int_\Rl f(x)g'(x)\,dx,
\end{align}
which is invariant under the translation action $f\mapsto f_a$, $f_a(x)=f(x-a)$. This is a classical abelian current algebra for the Lie algebra $\Rl$ of $\Rl$, which is also the Lie algebra of $U(1)$ (the latter point of view becomes clearer when considering this on the circle).

The structure behind the constructions in this section is the quantization of this symplectic system, called $U(1)$-current. One proceeds to the Weyl algebra over this symplectic space and looks for quasifree states on it in which the translations are implemented with positive energy. This results in a Fock space with one-particle space one of the unitarily equivalent choices described in this section, for instance $L^2(\Rl_+,dp/p)$. The corresponding quantum field $\phi(f)$, $f\in\Sch_\Rl(\Rl)$, is meant to create the single particle state $\tilde f|_{\Rl_+}$ from the vacuum. This has however only finite norm if $\tilde f$ vanishes at $p=0$, so the field $\phi(f)$ is ill-defined in general. What is well defined is its derivative (the current), i.e. $-\phi'(f)=\phi(f')$, which creates the function $p\mapsto -ip\tilde f(p)$ from the vacuum.

From the perspective of a massless free field theory on two-dimensional Minkowski space, this current is one chiral component of the derivative of the field. Thus the variable $x$ of $\Sch_\Rl(\Rl)$ should be regarded as a lightlike coordinate, which also explains the locality properties encountered in Theorem~\ref{theorem:U1C+} (vanishing symplectic form for disjoint supports).

\medskip

The $U(1)$-current is one of the most basic models of (conformal) quantum field theory, and we describe in this section its one-particle structure. Our treatment shows some general features of quantum field theory: The proof of standardness of this net is modelled after the Reeh-Schlieder Theorem \cite{ReehSchlieder:1961}, originally discovered in Wightman quantum field theory \cite{StreaterWightman:1964}. The main point for this is the combination of isotony and positivity of the energy, together with an additivity property that is valid in our case.

The fact that the modular data of $H_+$ act ``geometrically'' is a special case of the so-called Bisognano-Wichmann Theorem \cite{BisognanoWichmann:1975,BisognanoWichmann:1976}, which explains that the modular data of von Neumann algebras (or standard subspaces) localized in wedges have this feature (in a Wightman QFT). In our one-dimensional picture, wedges degenerate to half-lines. A more detailed analysis of the Bisognano-Wichmann property for standard subspaces will appear in a future version of this book. For now we refer to the literature \cite{Mund:2001,Morinelli:2018,MorinelliNeeb:2023}.

For an explicit form of the translations on $L^2(\Rl,\xi)$, see Koot \cite{Koot:2025}. Hardy Hilbert spaces as graph Hilbert spaces of modular operators are treated in \cite{LechnerLiQueffelecPiazza:2018}. The strategy of describing the local subspaces via support properties of distributions is taken from Lechner and Longo \cite{LechnerLongo:2014}.

\subsubsection*{Section~\ref{section:ComplexStructures} (\nameref{section:ComplexStructures})}

While polarizers are a standard concept in the context of complex structures and complexifications, polarizers of standard subspaces were first studied systematically by Longo in \cite{Longo:2022_2}, where most of the properties listed in Proposition~\ref{proposition:proppolirizer} were shown. The observation that factorial standard subspaces are complex Hilbert spaces (Corollary~\ref{corollary:FactorialSpacesAreComplexHilbert}) is original to this work. The central Theorem~\ref{thm:PolarizedRealHilbertSpacesAreStandardSubspaces} of this section is connected to several previous investigations: Shlyakhtenko showed that any real orthogonal dynamical system defines a standard subspace \cite{Shlyakhtenko:1997}. The uniqueness of the standard subspace realization is closely connected to Kay's work on one-particle structures \cite{Kay:1985-2}. See also \cite{NeebOlafsson:2017,Longo:2022_2} for related existence and uniqueness statements.

The dual pair construction of Proposition~\ref{proposition:StandardSubspacesInComplexifications} is known from free quantum field theories, see in particular \cite{FiglioliniGuido:1989}. It is extended here to a classification theorem.

\subsubsection*{Section~\ref{section:Operations} (\nameref{section:Operations})}

The definition of tensor products of standard subspaces via tensor products of Tomita operators was given by Longo, Morinelli, and Rehren in \cite{LongoMorinelliRehren:2015}. The proof that complete passivity was equivalent to the KMS condition first appeared in \cite{PuszWoronowicz:1978}. In the context of standard subspaces, Longo and Morsella proved Theorem \ref{theorem:competepassivity<=>modulargroup} in \cite{LongoMorsella:2023}.

Standard references for direct integrals are \cite{Dixmier:1981,Takesaki:2001}.  A relevant application to standard subspaces appears in \cite[Sect.~3, proof of Props.~3.1--3.2]{FiglioliniGuido:1994}. For an application of direct integrals of standard subspaces, see Morinelli and Tanimoto \cite{MorinelliTanimoto:2019}.

\subsubsection*{Section~\ref{section:Chapter1Applications} (\nameref{section:Chapter1Applications})}

For a general discussion of equilibrium (KMS) states in an operator-algebraic framework, see \cite{BratteliRobinson:1997}. The original account is the work of Haag, Hugenholtz, and Winnink \cite{HaagHugenholtzWinnink:1967}.

\bigskip

Some early sources discussing the standard subspaces of a Klein--Gordon field are \cite{Araki:1963,EckmannOsterwalder:1973,LeylandsRobertsTestard:1978}, see also \cite{Guido:2011}, and \cite{BrunettiGuidoLongo:2002} for an approach emphasizing the underlying representation of the Poincaré group. This approach has been vastly generalized to other Lie groups by Morinelli and Neeb \cite{MorinelliNeeb:2023}.

The antilocality result of fractional elliptic operators, see \cite{Murata:1973}. A modern precise massless result is contained in \cite[Thm.~1.2]{GhoshSaloUhlmann:2020}. The relation of the Klein--Gordon standard subspaces to localization observables is discussed in \cite{LechnerdeOliveira:2026}.

The entropy operator of a standard subspace and its quadratic form are due to Longo and Ciolli--Ruzzi--Longo \cite{Longo:2019,CiolliLongoRuzzi:2019}. In subsequent work by Longo and Morinelli \cite{LongoMorinelli:2024}, the locality formula \eqref{eq:EntropyLocality} and the identity \eqref{eq:EntropyComplementDifference} have been proven. These quantities are motivated by Araki's relative entropy for normal states of von Neumann algebras \cite{Araki:1976,Araki:1977}.

Bostelmann, Cadamuro and Del Vecchio studied general CCR algebras and separate factorial, abelian and nonseparating contributions to entropy \cite{BostelmannCadamuroVecchio:2021}, related to our standard subspace discussion in Proposition~\ref{prop:EntropyOperatorProperties}. For further discussion of the mass dependence of local modular operators and entropies, we refer to \cite{Longo:2022_2, Longo:2023Correction,BostelmannCadamuroMinz:2023, LongoMorsella:2023}.

\vfill

\pagebreak
\appendix
\pagebreak

\chapter*{Appendix}

 Section~\ref{section:HardySpaces} collects some facts about Hardy spaces on strips and half-planes used in Chapter~\ref{chapter:Fundamentals}. 

\subsection*{Contents of Appendix}
\addcontentsline{toc}{chapter}{Appendix}
\setcounter{chapter}{1}
\setcounter{section}{0}

\etocsettocstyle{}{}
\etocsetnexttocdepth{subsection} \localtableofcontents

\section{Hardy spaces}\label{section:HardySpaces}

Hardy spaces $\Hardy^p(G)$ are spaces of functions that are analytic on a non-empty open connected simply connected domain $G\subsetneq\Cl$ and satisfy certain $L^p$-integrability conditions along a family of curves tending to the boundary of $G$, see \cite{Duren:1970,SteinWeiss:1971,Krantz:1951} for some general references.

In the context of modular theory, one is mostly interested in the powers $p=\infty$ and $p=2$, and as regions $G$, strips, half-planes, and sometimes the unit disc, denoted as follows:
\begin{align}\label{eq:regions}
    \Strip_{a,b}
    &:=
    \Rl+i(a,b),\quad a<b,
    \qquad
    \Strip_\beta
    :=
    \begin{cases}
        \Strip_{0,\beta} & \beta>0,\\
        \Strip_{\beta,0} & \beta<0
    \end{cases},
    \\
    \Cl_\pm
    &:=
    \{z\in\Cl\col\pm\Im(z)>0\},
    \\
    \Disc
    &:=
    \{z\in\Cl\col|z|<1\}.
\end{align}

\subsection{Bounded analytic functions on a strip}\label{section:HinftyStrip}

 This section currently only contains one result that we use. It will be extended in later editions.

\medskip

For $\beta\neq0$, let $\Hinfty(\Stripb)$ denote the Banach space of bounded holomorphic functions on $\Stripb$, with the supremum norm. For $\la$ strictly between $0$ and $\beta$, write $f_\la(\te):=f(\te+i\la)$.

\pagebreak

\begin{theorem}[Boundary values and the three-lines theorem]\label{theorem:HinftyBoundaryValues}\booktheoremindex{theorem:HinftyBoundaryValues}{Boundary values and the three-lines theorem}\index{boundary values on a strip}\index{three-lines theorem}
    Let $f\in\Hinfty(\Stripb)$.
    \begin{enumerate}
        \item\label{item:LinftyBoundaryValues} $f$ has a non-tangential boundary value $f_0,f_\beta\in L^\infty(\Rl)$. As $\la\to 0$ and $\la\to\beta$ from inside the strip, $f_\la\to f_0$ and $f_\la\to f_\beta$, respectively, almost everywhere and in the weak-star topology of $L^\infty(\Rl)$.
        \item\label{item:HinftyInjectivity} Each boundary value function $f_0,f_\beta$ determines $f$ uniquely.
        \item\label{item:3lines} For $0<t<1$,
        \begin{align*}
            \|f_{t\beta}\|_\infty
            \leq \|f_0\|_\infty^{1-t}\|f_\beta\|_\infty^t.
        \end{align*}
        In particular, $\|f\|_\infty=\max\{\|f_0\|_\infty,\|f_\beta\|_\infty\}$.
    \end{enumerate}
\end{theorem}

For the disc forms of the boundary-limit and uniqueness theorems, see \cite[Thms.~11.23, 11.32 and 17.18]{RudinRCA:1987}. The corresponding strip statements use a conformal map from the strip onto the disc. For the three-lines estimate with $L^\infty$ boundary traces, see \cite[eq.~(1.4)]{CarvalhoCorso:2026}. The version for functions continuous on the closed strip, used in Lemma~\ref{lemma:KMSDenseIsEnough}, is \cite[Thm.~12.8]{RudinRCA:1987}.

\subsection{Hardy Hilbert spaces on strips}\label{section:Hardy2Strip}

 We begin with the Hardy space on the strip $\Strip_\beta=\Rl+i(0,\beta)$ (with $\beta>0$). This space is not as prominently treated in the Hardy space literature as the versions for the unit disc and the upper half-plane (see, however, the Notes and Remarks), so we provide detailed proofs.

\begin{definition}[Hardy space on the strip]
    Let $\beta>0$. The {\em Hardy space on $\Strip_\beta$} is the space $\Hardy^2(\Strip_\beta)$\glsadd{HardyStrip}\index{Hardy space!on strip} of all analytic functions $F:\Strip_\beta\to\Cl$ such that
    \begin{align}\label{eq:AppBanachNormHardy}
        \|F\|^2_{\Hardy^2(\Strip_\beta)}
        :=
        \sup_{0<\la<\beta}\int_\Rl |F(\te+i\la)|^2\,d\te<\infty.
    \end{align}
\end{definition}

For $F\in\Hardy^2(\Stripb)$, the restrictions
 \begin{align}
    F_\la : \Rl\to\Cl,\qquad F_\la(\te):=F(\te+i\la),\qquad 0<\la<\beta,
\end{align}
 are in particular smooth functions in $L^2(\Rl)$. In the following theorem about Hardy functions, we use the unitary $L^2$-Fourier transform with the sign convention
\begin{align}
        \EuF:L^2(\Rl)\to L^2(\Rl),\qquad (\EuF f)(\te):=\tilde f(\te):=\frac{1}{\sqrt{2\pi}}\int_\Rl e^{-i\te\xi}f(\xi)\,d\xi.
\end{align}
 Here the integral is understood in the usual $L^2$-sense, namely as the $L^2$-limit of the functions $\te\mapsto\frac{1}{\sqrt{2\pi}}\int_{-R}^R e^{-i\te\xi}f(\xi)\,d\xi$ as $R\to\infty$. We will also use the dense subspace of $L^2(\Rl)$ defined by\footnote{The inclusion ``$\supset$'' in \eqref{eq:AppWeightedFourierDomain} follows by considering $\la=0$ and $\la=\beta$, and the inclusion ``$\subset$'' follows by splitting the integral $\int_\Rl e^{2\la\xi}|g(\xi)|^2d\xi$ at $\xi=0$ and elementary estimates.}
\glsadd{fourierBoundaryDomain}
 \begin{align}\label{eq:FourierCbeta}
    \widetilde\CD_\beta
    &:=
    \left\{g\in L^2(\Rl) \col \int_\Rl e^{2\beta\xi}|g(\xi)|^2d\xi<\infty \right\}
    \\
    &=
    \left\{g:\Rl\to\Cl\,\text{ measurable } \col \sup_{0\leq\la\leq\beta}\int_\Rl e^{2\la\xi}|g(\xi)|^2d\xi<\infty \right\}.
    \label{eq:AppWeightedFourierDomain}
\end{align}

 \pagebreak

\begin{theorem}[Analytic properties of Hardy space functions]\label{theorem:HardyFunctionsAppendix}\booktheoremindex{theorem:HardyFunctionsAppendix}{Analytic properties of Hardy space functions (strip)}
    Let $\beta>0$.
    \begin{enumerate}
        \item\label{item:HardyPointWiseBound} Any $F\in\Hardy^2(\Strip_\beta)$ satisfies the pointwise bound
        \begin{align}
            |F(\zeta)|
            \leq
            \sqrt{\frac{2}{\pi}}\frac{\|F\|_{\Hardy^2(\Strip_\beta)}}{d(\zeta,\partial\Strip_\beta)^{1/2}},
            \qquad
            \zeta\in\Strip_\beta.
        \end{align}

        \item\label{item:AppHardyBoundaryValues1}\index{boundary values on a strip}\label{item:HardyBoundaryValues1} For any $F\in\Hardy^2(\Strip_\beta)$, the limits
        \begin{align}
            F_0:=\lim_{\la\downarrow0}F_\la,\qquad
            F_\beta:=\lim_{\la\uparrow\beta}F_\la
        \end{align}
        exist as norm limits in $L^2(\Rl)$. In particular, $F_0,F_\beta\in L^2(\Rl)$.

        \item\label{item:AppHardyBoundaryValues2}\label{item:HardyBoundaryValues2} The maps $\Hardy^2(\Strip_\beta)\ni F\mapsto F_0\in L^2(\Rl)$ and $\Hardy^2(\Strip_\beta)\ni F\mapsto F_\beta\in L^2(\Rl)$ are injective.

        \item\label{item:AppHardyFourierTransform} Given any $g\in\widetilde\CD_\beta$ \eqref{eq:FourierCbeta}, the function
        \begin{align}\label{eq:AppGtoF}
            F:\Strip_\beta\to\Cl,\qquad
            F(\te+i\la):=
            F_\la(\te):=
            \frac{1}{\sqrt{2\pi}}\int_\Rl e^{-i\te\xi}e^{\la\xi}g(\xi)\,d\xi
        \end{align}
        lies in $\Hardy^2(\Strip_\beta)$. Conversely, any $F\in\Hardy^2(\Strip_\beta)$ is of the form \eqref{eq:AppGtoF} for some $g\in\widetilde\CD_\beta$. The inverse Fourier-transformed boundary values of $F$ are then given by $(\EuF^{-1}F_0)(\xi)=g(\xi)$ and $(\EuF^{-1}F_\beta)(\xi)=e^{\beta\xi}g(\xi)$.
    \end{enumerate}
\end{theorem}

\begin{proof}
 \ref{item:HardyPointWiseBound} Let $\zeta\in\Strip_\beta$. For $r'<d(\zeta,\partial\Strip_\beta)$, the mean value property for analytic functions yields $F(\zeta)=\frac{1}{2\pi}\int_0^{2\pi}F(\zeta+r'e^{i\alpha})d\alpha$. Integrating $r'F(\zeta)$ in $r'$ over $[0,r]$, with $r<d(\zeta,\partial\Strip_\beta)$, and changing from polar to Cartesian coordinates gives
 \begin{align}\label{eq:MVP}
        F(\zeta)
        =
        \frac{1}{\pi r^2}\int_{D_r(\zeta)}F(\te+i\la)\,d\te\,d\la,
    \end{align}
 where $D_r(\zeta)\subset\Strip_\beta$ is the disc with radius $r$ and centre $\zeta$. Reading \eqref{eq:MVP} as an $L^2$-scalar product of $1$ and $F$, we can use the Cauchy-Schwarz inequality and the Hardy property of $F$ to estimate
 \begin{align*}
        |F(\zeta)|
        &\leq
        \frac{1}{\sqrt{\pi}r}\left(\int_{\Strip_{(\Im(\zeta)-r,\Im(\zeta)+r)}}|F(\te+i\la)|^2\,d\te\,d\la\right)^{1/2}
        \leq
        \sqrt{\frac{2}{\pi}}\frac{\|F\|_{\Hardy^2(\Strip_\beta)}}{\sqrt{r}}.
    \end{align*}
    The claimed inequality follows in the limit $r\to d(\zeta,\partial\Strip_\beta)$.

 \ref{item:AppHardyFourierTransform} Let $g\in\widetilde\CD_\beta$ and $\la\in[0,\beta]$. The measurable function $g_{(\la)}(\xi):=e^{\la\xi}g(\xi)$ satisfies
 \begin{align}\label{eq:HardyBound1}
        \int_\Rl|g_{(\la)}(\xi)|^2\,d\xi
        \leq
        \int_0^\infty e^{2\beta\xi}|g(\xi)|^2d\xi+\int_{-\infty}^0|g(\xi)|^2d\xi
        <
        \infty,
    \end{align}
 so $g_{(\la)}\in L^2(\Rl)$ and $\sup_{0\leq\la\leq\beta}\|g_{(\la)}\|_{L^2(\Rl)}<\infty$. 

For $R>0$, the truncated integral
    \begin{align*}
        F^{(R)}(\zeta):=\frac{1}{\sqrt{2\pi}}\int_{-R}^R e^{-i\zeta\xi}g(\xi)\,d\xi
    \end{align*}
    defines an entire function. By Plancherel's theorem and \eqref{eq:HardyBound1}, $F^{(R)}\in\Hardy^2(\Strip_\beta)$. For $R>S>0$, the same estimate applied to the difference gives
    \begin{align*}
        \|F^{(R)}-F^{(S)}\|_{\Hardy^2(\Strip_\beta)}^2
        \leq\int_{|\xi|>S}(1+e^{2\beta\xi})|g(\xi)|^2\,d\xi
        \longrightarrow 0 \qquad (S\to\infty).
    \end{align*}
    By \ref{item:HardyPointWiseBound}, the truncations therefore converge uniformly on compact subsets of $\Strip_\beta$ to an analytic function $F$. For $0<\la<\beta$, the function $\xi\mapsto e^{\la\xi}/(1+e^{2\beta\xi})^{1/2}$ is square integrable. Cauchy--Schwarz thus gives $e^{\la\xi}g(\xi)\in L^1(\Rl,d\xi)$, so the limit is precisely the integral in \eqref{eq:AppGtoF}. Finally, Plancherel's theorem gives $\|F_\la\|_2=\|g_{(\la)}\|_2$, uniformly bounded in $\la$ by \eqref{eq:HardyBound1}. Hence $F\in\Hardy^2(\Strip_\beta)$.

    We now consider $F\in\Hardy^2(\Strip_\beta)$. To prove that it is of the form \eqref{eq:AppGtoF}, it is sufficient to show that
 \begin{align}\label{eq:Hardyalmostall}
        \tilde F_\la(-\xi)e^{-\la\xi}=\tilde F_{\la'}(-\xi)e^{-\la'\xi} \quad\text{ for all } \la,\la'\in(0,\beta),\, \text{almost all }\xi\in\Rl.
    \end{align}
 Indeed, fix $\la_0\in(0,\beta)$ and choose a measurable representative of $g(\xi):=e^{-\la_0\xi}(\EuF^{-1}F_{\la_0})(\xi)$. By \eqref{eq:Hardyalmostall}, $e^{\la\xi}g(\xi)=(\EuF^{-1}F_\la)(\xi)$ almost everywhere for each fixed $\la\in(0,\beta)$. Thus
 \begin{align}
        \int_\Rl e^{2\la\xi}|g(\xi)|^2d\xi
        =
        \|F_\la\|_{L^2(\Rl)}^2
        \leq
        \|F\|^2_{\Hardy^2(\Strip_\beta)}
        <\infty.
    \end{align}
 Fatou's lemma, first along a sequence $\la\downarrow0$ and then along a sequence $\la\uparrow\beta$, gives $\|g\|_2^2,\|e^{\beta\xi}g\|_2^2\leq\|F\|_{\Hardy^2(\Strip_\beta)}^2$. Thus $g\in\widetilde\CD_\beta$. Taking the Fourier transform on each interior line gives \eqref{eq:AppGtoF}. Both sides are continuous in $\te$, so their $L^2$ equality holds pointwise.

    To establish \eqref{eq:Hardyalmostall}, we fix arbitrary $\la,\la'\in(0,\beta)$ and restrict $F$ to a substrip of the form $\Strip':=\Rl+iI$, with $I\subset(0,\beta)$ a compact interval, such that the two lines $\Rl+i\la$ and $\Rl+i\la'$ are contained in $\Strip'$. According to \ref{item:HardyPointWiseBound}, $F$ is bounded by a constant $C$ on~$\Strip'$ (with~$C$ depending on $\la,\la'$). To control the decay at infinity, we pick $\eps>0$ and consider $F^{(\eps)}(\zeta):=e^{-\eps\zeta^2}F(\zeta)$, which satisfies $|F^{(\eps)}(\zeta)|\leq C\,e^{-\eps(\Re\zeta)^2}e^{\eps(\Im\zeta)^2}$ for $\zeta\in\Strip'$. Hence the functions $\zeta\mapsto F^{(\eps)}(\zeta)e^{i\xi\zeta}$, with $\xi\in\Rl$, are analytic and of rapid decay at infinity in $\Strip'$. We may therefore use Cauchy's Theorem and a contour shift to conclude
    \begin{align*}
        \tilde F_\la^{(\eps)}(-\xi)e^{-\xi\la}
        &=
        \frac{1}{\sqrt{2\pi}}
        \int_\Rl F^{(\eps)}(\te+i\la)e^{i\xi(\te+i\la)}\,d\te
        \\
        &=
        \frac{1}{\sqrt{2\pi}}
        \int_\Rl F^{(\eps)}(\te+i\la')e^{i\xi(\te+i\la')}\,d\te
        =
        \tilde F_{\la'}^{(\eps)}(-\xi)e^{-\xi\la'}.
    \end{align*}
 For each fixed $\la\in(0,\beta)$, dominated convergence gives $F_\la^{(\eps)}\to F_\la$ in $L^2(\Rl)$ as $\eps\downarrow0$, since the Gaussian factor on that line is bounded by $e^{\eps\la^2}$. The inverse Fourier transforms therefore converge in $L^2(\Rl)$ as well. On $[-R,R]$ both multipliers $e^{-\la\xi}$ and $e^{-\la'\xi}$ are bounded, so we may pass to the limit in the contour-shift identity in $L^2([-R,R])$. Taking $R\to\infty$ proves \eqref{eq:Hardyalmostall} for almost all $\xi\in\Rl$ almost everywhere on $\Rl$ \eqref{eq:Hardyalmostall}.

    With this characterization of $\Hardy^2(\Strip_\beta)$ in place, the remaining parts \ref{item:AppHardyBoundaryValues1} and \ref{item:AppHardyBoundaryValues2} can be established easily:

    \ref{item:AppHardyBoundaryValues1} Let $g\in\widetilde{\CD}_\beta$ and $F\in\Hardy^2(\Strip_\beta)$ given by \eqref{eq:AppGtoF}. For the $L^2$-functions $g_{(\la)}(\xi):=e^{\la\xi}g(\xi)$, $0\leq\la\leq\beta$, an application of dominated convergence shows
 \begin{align}
        \lim_{\la\downarrow0}g_{(\la)}=g,\qquad
        \lim_{\la\uparrow\beta}g_{(\la)}=g_{(\beta)}
    \end{align}
 in the norm topology of $L^2(\Rl)$, where the integrable majorant is derived with the same technique as in \eqref{eq:HardyBound1}. By unitarity of the Fourier transform, this implies the $L^2$-limits $\lim_{\la\downarrow0}F_\la=F_0$ and $\lim_{\la\uparrow\beta}F_\la=F_\beta$. Now we also see that the unique representing $g\in L^2(\Rl)$ in \eqref{eq:AppGtoF}, which was defined by $g(\xi):=\tilde F_\la(-\xi)e^{-\la\xi}$ in the previous part, satisfies $(\EuF^{-1}F_0)(\xi)=g(\xi)$ and $(\EuF^{-1}F_\beta)(\xi)=e^{\beta\xi}g(\xi)$.

    \ref{item:AppHardyBoundaryValues2} Suppose that $F\in\Hardy^2(\Strip_\beta)$ satisfies $F_0=0$. Then its representing function $g=\EuF^{-1}F_0$ vanishes, hence $F=0$ by \eqref{eq:AppGtoF}. In case $F_\beta=0$, the function $\xi\mapsto e^{\beta\xi}g(\xi)$ vanishes, so again $g=0$ and $F=0$ follows.
 \end{proof}

\begin{remark}[$\Hardy^2(\Strip_\beta)$ as a subspace of $L^2(\Rl)$]\label{remark:Hardy}
    \leavevmode
    \begin{enumerate}
 \item\label{item:HardyDIdentification}\glsadd{boundaryDomain}
        It is clear from the representation \eqref{eq:AppGtoF} that the complex subspace
 \begin{align}
            \CD_\beta
            :=
            \{f\in L^2(\Rl) \col \exists F\in\Hardy^2(\Strip_\beta)\,\text{with}\, F_0=f\}
            \subset
            L^2(\Rl)
        \end{align}
        is dense (for instance by choosing $g$ to be compactly supported), and the boundary value map
        \begin{align}\label{eq:BoundaryValueMap}
            B_\beta:\Hardy^2(\Strip_\beta)\to\CD_\beta,\qquad F\mapsto F_0,
        \end{align}
 is a bijection. We also note that as a consequence of Theorem~\ref{theorem:HardyFunctionsAppendix}~\ref{item:AppHardyFourierTransform},
 \begin{align}
            \EuF:\widetilde{\CD}_\beta\to\CD_\beta
        \end{align}
        is a bijection.

        Using the boundary value map $B_\beta$ \eqref{eq:BoundaryValueMap}, we may, and often will, view $\Hardy^2(\Strip_\beta)$ as a dense subspace of $L^2(\Rl)$ by identifying it with $\CD_\beta$ and suppressing $B_\beta$. Viewing a function $f\in\CD_\beta$ as an element of $\Hardy^2(\Strip_\beta)$ (actually its analytic continuation $F=B_\beta^{-1}(f)$ lies in $\Hardy^2(\Strip_\beta)$), we write $f(\te+i\la)$ (for $0\leq\la\leq\beta$) instead of $(B_\beta^{-1}f)(\te+i\la)$.

        Of course, the same procedure applies to the upper boundary values at $\Rl+i\beta$, but we will not need this.

 \item\label{item:AppHardyNegativeStrip} We have so far only considered strips in the upper half-plane, i.e. $\beta>0$. Analogously, we may consider Hardy spaces $\Hardy^2(\Strip_{-\beta})$ (recall the notation $\Strip_{-\beta}=\Strip_{(-\beta,0)}$ for $\beta>0$) on strips in the lower half-plane (or on general strips). The complex conjugation map
 \begin{align}
            \Hardy^2(\Strip_{\beta})\to\Hardy^2(\Strip_{-\beta}),\qquad F\mapsto F^\flat,\qquad F^\flat(\zeta)=\overline{F(\overline{\zeta})},
        \end{align}
 provides a bijection between $\Hardy^2(\Strip_{\beta})$ and $\Hardy^2(\Strip_{-\beta})$, and all results in Theorem~\ref{theorem:HardyFunctionsAppendix} can be easily transferred to the negative strips.
 \end{enumerate}
\end{remark}

Using this identification of $\Hardy^2(\Stripb)\cong\CD_\beta$, we now consider the densely defined {\em boundary-to-boundary operator}\glsadd{boundaryTransfer}\index{boundary-to-boundary operator}
 \begin{align}\label{eq:BoundaryToBoundaryOperator}
    D_\beta:\CD_\beta\to L^2(\Rl),\qquad F_0\mapsto F_\beta,
\end{align}
mapping lower to upper boundary values, and the pointwise complex conjugation
\begin{align}
    J:L^2(\Rl)\to L^2(\Rl),\qquad f\mapsto\bar f.
\end{align}

 \begin{proposition}[The boundary-to-boundary operator]\label{proposition:AppHardyBoundaryOperator}\index{boundary-to-boundary operator}\glsadd{boundaryTransfer}
    Let $\beta>0$.
    \begin{enumerate}
        \item The operator $D_\beta$ is positive, self-adjoint and injective. Its domain is $\CD_\beta$, and in the Fourier representation it acts by
        \begin{align}\label{eq:AppHardyBoundaryMultiplier}
            D_\beta=\EuF M_{\beta}\EuF^{-1},
            \qquad \dom M_{\beta}=\widetilde\CD_\beta\subset L^2(\Rl,d\xi),
        \end{align}
        where $M_{\beta}$ denotes multiplication by $e^{\beta\xi}$.

        \item One has $JD_\beta=D_\beta^{-1}J$, and $S:=JD_\beta$ is a closed densely defined antilinear involution on $L^2(\Rl)$, with domain $\CD_\beta\cong\Hardy^2(\Stripb)$ and
        \begin{align}
            \Fix S
            =
            \{F\in\Hardy^2(\Stripb) \col F_\beta=\overline{F_0}\}.
        \end{align}
    \end{enumerate}
\end{proposition}
 \begin{proof}
 a) Let $F\in\Hardy^2(\Stripb)\cong\CD_\beta$. By Theorem~\ref{theorem:HardyFunctionsAppendix}~\ref{item:AppHardyFourierTransform}, $F_0=\EuF g$ and $F_\beta=\EuF(M_{\beta}g)$ for a unique $g\in\widetilde\CD_\beta$. Conversely, each such $g$ gives a Hardy function. This proves \eqref{eq:AppHardyBoundaryMultiplier}. Multiplication by the strictly positive function $e^{\beta\xi}$ on its maximal domain is positive, self-adjoint and injective.

    b) The operator $J$ is a conjugation. By Fourier transformation, we see that $\tilde J:=\EuF J\EuF^{-1}$ acts according to $(\tilde J g)(\xi)=\overline{g(-\xi)}$ on the Fourier transformed space, which implies $\tilde JM_\beta\tilde J=M_{-\beta}=M_\beta^{-1}$. This implies $JD_\beta=D_{\beta}^{-1}J$ and that $S=JD_\beta$ is closed. As $JD_\beta J$ acts by multiplication with $e^{-\beta\xi}$ in the Fourier transformed picture, we also see that $S$ is an antilinear involution. The fixed point description follows directly from the definition of $D_\beta$ and $J$.
\end{proof}

The operator $S$ is a Tomita operator and $H=\Fix S$ is a standard subspace in $L^2(\Rl)$. These standard subspaces appear in Example~\ref{example:U1current} and Section~\ref{section:U1DifferentSpaces}. We next show that the Hardy space is precisely the graph-norm Hilbert space of $D_\beta$ (cf.~Proposition~\ref{proposition:GraphHilbertSpaceOfDelta}).

 \begin{theorem}[$\Hardy^2(\Strip_\beta)$ as a Hilbert and Banach space]\label{theorem:HardySpace}\booktheoremindex{theorem:HardySpace}{Hardy space on a strip as a Hilbert and Banach space}
    Let $\beta>0$.
    \begin{enumerate}
        \item\label{item:AppHardyHilbert} $\Hardy^2(\Strip_\beta)$ is a Hilbert space with the scalar product
        \begin{align}
            \langle F,G\rangle_{\Hardy^2(\Strip_\beta)}
            :=
            \langle F_0,G_0\rangle_{L^2(\Rl)}
            +
            \langle F_\beta,G_\beta\rangle_{L^2(\Rl)},
        \end{align}
        and can be identified via $F\mapsto F_0$ with the graph norm Hilbert space of $D_\beta$.

        \item\label{item:AppHardyTomita} On this Hilbert space, the map
        \begin{align}\label{eq:AppHardyTomita}
            S:\Hardy^2(\Strip_\beta)\to\Hardy^2(\Strip_\beta),\qquad
            (SF)(\zeta)
            :=
            \overline{F(i\beta+\overline\zeta)}
        \end{align}
        is a conjugation satisfying $(SF)_0=\overline{F_\beta}$ and $(SF)_\beta=\overline{F_0}$.

        \item\label{item:HardyBanach} $\Hardy^2(\Strip_\beta)$ is a Banach space with the norm \eqref{eq:AppBanachNormHardy}. The Banach space norm \eqref{eq:AppBanachNormHardy} is different from the Hilbert norm from \ref{item:AppHardyHilbert}, but equivalent, namely
        \begin{align}\label{eq:EquivalentHardyNorms}
            \|F\|_{\Hardy^2(\Strip_\beta)}\leq
            \langle F,F\rangle_{\Hardy^2(\Strip_\beta)}^{1/2}
            \leq\sqrt{2}\|F\|_{\Hardy^2(\Strip_\beta)}.
        \end{align}
    \end{enumerate}
\end{theorem}
 \begin{proof}
 \ref{item:AppHardyHilbert} Identifying $\Hardy^2(\Stripb)\cong\CD_\beta$, the specified scalar product is precisely the graph scalar product of the boundary-to-boundary operator $D_\beta$. This implies the claim.

    \ref{item:AppHardyTomita} In terms of boundary values, the operator $S$ maps $F_0$ to $\overline F_\beta =JD_\beta F_0$. Hence $S=JD_\beta$ is a densely defined closed antilinear involution by the preceding proposition. Proposition~\ref{proposition:GraphHilbertSpaceOfDelta} shows that it is a conjugation in the graph Hilbert space.

 \ref{item:HardyBanach} It is clear that $\Hardy^2(\Strip_\beta)$ is a linear space and $\|\cdot\|_{\Hardy^2(\Strip_\beta)}$ \eqref{eq:AppBanachNormHardy} is a norm on it. Completeness follows once we have established the equivalence of this norm with the Hilbert norm $\TripNorm{F}^2=\langle F,F\rangle_{\Hardy^2(\Strip_\beta)}$.

For $\beta>0$, write $F_0=\EuF g$ as in Theorem~\ref{theorem:HardyFunctionsAppendix}~\ref{item:AppHardyFourierTransform}. Let $0<t<1$ and $\la:=t\beta$. Then Plancherel's theorem and H\"older's inequality give
    \begin{align*}
        \|F_{t\beta}\|_2^2
        &=\int_\Rl (|g(\xi)|^2)^{1-t}(e^{2\beta\xi}|g(\xi)|^2)^t\,d\xi\\
        &\leq \|F_0\|_2^{2(1-t)}\|F_\beta\|_2^{2t}
        \leq \max\{\|F_0\|_2^2,\|F_\beta\|_2^2\}.
    \end{align*}
    The $L^2$ boundary limits give the opposite inequality for the supremum. Hence
    \begin{align*}
        \|F\|_{\Hardy^2(\Strip_\beta)}^2=\max\{\|F_0\|_2^2,\|F_\beta\|_2^2\}.
    \end{align*}
    Comparing this maximum with the sum of the two squared norms gives \eqref{eq:EquivalentHardyNorms}.

    To see that the two norms are not equal, let $F$ be represented by $g\in\widetilde\CD_\beta$ as in \eqref{eq:AppGtoF}. Then
 \begin{align*}
        \TripNorm{F}^2
        =
        \int_\Rl(1+e^{2\beta\xi})|g(\xi)|^2\,d\xi,
        \qquad
        \|F\|_{\Hardy^2(\Strip_\beta)}^2
        =
        \sup_{0<\la<\beta}\int_\Rl e^{2\la\xi}|g(\xi)|^2\,d\xi.
    \end{align*}
 From here it is easy to see that the two norms are different ($\|\cdot\|_{\Hardy^2(\Strip_\beta)}$ is not a Hilbert norm).
 \end{proof}

\subsection{The radial Hardy space on the upper half-plane}\label{section:Hardy2Halfplane}

 Also this subsection is currently only a small snapshot, to be extended in a future release.

\medskip

Two Hardy spaces on $\Cl_+$ enter the comparison in Section~\ref{section:U1DifferentSpaces}. The classical space $\Hardy^2(\Cl_+)$\glsadd{H2C+}\glsadd{Cplus} uses horizontal lines and the measure $dx$ on $\Rl$ \cite[Thm.~19.2]{RudinRCA:1987}. In the radial space $\Hardy^2_\rad(\Cl_+)$ one integrates along rays with measure $dr/r$. For a holomorphic function $f$ the respective squared norms are
\begin{align*}
    \|f\|_{\Hardy^2(\Cl_+)}^2
        &=\sup_{q>0}\int_\Rl |f(x+iq)|^2\,dx,\\
    \|f\|_{\Hardy^2_\rad(\Cl_+)}^2
        &=\sup_{0<\alpha<\pi}\int_0^\infty |f(re^{i\alpha})|^2\,\frac{dr}{r}.
\end{align*}
Neither space contains the other. For instance, $(p+i)^{-1}$ belongs only to the classical space, whereas $(1-i\log p)^{-1}$ belongs only to the radial space. Here $0<\Im\log p<\pi$.

A change of variables by the logarithm gives an isometric identification with the strip space of Section~\ref{section:Hardy2Strip},
\begin{align}\label{eq:UStripToHalfplane}
    U:\Hardy^2(\Strip_\pi)\to\Hardy^2_\rad(\Cl_+),\qquad (UF)(p)=F(\log p).
\end{align}
The two radial boundary values of $UF$ are $F_0(\log r)$ and $F_\pi(\log r)$ in $L^2(\Rl_+,dr/r)$. For the sector formulation, see \cite[Sect.~6.1, Def.~4 and the discussion following it]{BardaroEtAl:2022} with $c=0$, $p=2$ and half-angle $\pi/2$. Their sector is centred on the positive real axis and their norm averages the integrals on two symmetric rays. Proposition~\ref{prop:Hardy2UpperHalfPlane} uses the supremum over individual rays displayed above.

\section{Notes and Remarks}

\subsubsection*{Section~\ref{section:HardySpaces} (\nameref{section:HardySpaces})}

Some general references on Hardy spaces are  \cite{Duren:1970,Koosis:1998,Garnett:2007}, see also \cite{Krantz:1951,SteinWeiss:1971}.

A dedicated analysis of strip Hardy spaces was carried out by Bakan and Kaijser \cite{BakanKaijser:2007}. The particular graph Hilbert space and the standard subspace used in this book can be found in Lechner et. al. \cite{LechnerLiQueffelecPiazza:2018}.

\newgeometry{left=2.7cm,right=2.7cm,top=2.7cm,bottom=2.9cm}
\begingroup
\newcommand{\notationtitle}{Notation for a standard subspace}
\chapter*{\notationtitle}
\addcontentsline{toc}{chapter}{Notation for a standard subspace}
\markboth{Notation for a standard subspace}{Notation for a standard subspace}
\label{back:StandardSubspaceNotation}

\noindent Throughout this table, $H$ is a standard subspace of the complex Hilbert space $\Hil$.

\normalsize
\newcommand{\notationgroup}[1]{\multicolumn{3}{@{}l}{\strut\bfseries #1}\\*}
\newcommand{\notationentry}[3]{#1 & #2 & \pageref{#3}\\}
\setlength{\LTleft}{0pt}
\setlength{\LTright}{0pt}
\setlength{\tabcolsep}{0pt}
\renewcommand{\arraystretch}{1.2}
\begin{longtable}{@{}p{.20\linewidth}@{\hspace{.02\linewidth}}>{\raggedright\arraybackslash}p{.69\linewidth}@{\hspace{.02\linewidth}}>{\raggedleft\arraybackslash}p{.07\linewidth}@{}}
  \symboltableheading
  \endhead
  \notationgroup{Subspaces and projections}
  \notationentry{\glsentryname{HlowC}, \glsentryname{HupC}}{$H_\Cl=H\cap iH=\{0\}$ and $H^\Cl=\overline{H+iH}=\Hil$.}{eq:TwoComplexSpaces}
  \notationentry{\glsentryname{Hprime}, \glsentryname{Hrperp}}{Symplectic complement $H'=(iH)^\rperp$ and real orthogonal complement $H^\rperp=iH'$.}{definition:SymplecticComplement}
  \notationentry{\glsentryname{ZH}}{Centre $H\cap H'$. The subspace $H$ is factorial if $Z(H)=\{0\}$.}{definition:HAbelianCenterEtc}
  \notationentry{\glsentryname{FH}}{Factorial part $H\cap Z(H)^\rperp$.}{proposition:DecomposeHIntoMaximallyAbelianAndFactorPart}
  \notationentry{\glsentryname{EH}}{Real orthogonal projection of $\Hil$ onto $H$.}{proposition:orthogonalprojectionmodulardata}
  \notationentry{\glsentryname{cutting}}{For factorial $H$, the cutting projection $h+h'\mapsto h$ on $H+H'$.}{definition:cuttingprojection}
  \notationentry{$E^{\Delta_H}_{\bullet}$}{Spectral projection for $(0,1)$, $(0,1]$, $\{1\}$, $[1,\infty)$ or $(1,\infty)$, respectively, for $\bullet$ equal to $<$, $\leq$, $1$, $\geq$ or $>$.}{eq:SpectralProjectionsDelta}
  \noalign{\smallskip}
  \notationgroup{Tomita operator and modular data}
  \notationentry{\glsentryname{SH}}{Tomita operator on $H+iH$, $h+ik\mapsto h-ik$ for $h,k\in H$.}{def:SH}
  \notationentry{\glsentryname{JH}, \glsentryname{DeltaH}}{Modular conjugation and modular operator in $S_H=J_H\Delta_H^{1/2}$.}{eq:SPolarDecomposition}
  \notationentry{$L_H$, $\Delta_H^{it}$}{Modular Hamiltonian $L_H=\log\Delta_H$ and modular group $\Delta_H^{it}=e^{itL_H}$, $t\in\Rl$.}{def:ModularHamiltonianAndGroup}
  \notationentry{\glsentryname{BJH}}{Real orthogonal isomorphism $\Fix J_H\to H$ given by $\sqrt{2}(1+\Delta_H)^{-1/2}$.}{item:FixJvsFixS}
  \noalign{\smallskip}
  \notationgroup{The pre-symplectic form and its polarizer}
  \notationentry{\glsentryname{sH}}{Pre-symplectic form $s_H(h,k)=\Im\langle h,k\rangle$.}{eq:sH}
  \notationentry{\glsentryname{polarizer}}{Polarizer on $H$, defined by $\Re\langle h,D_Hk\rangle=s_H(h,k)$.}{def:PolarizerOfStandardSubspace}
  \notationentry{\glsentryname{IH}}{Phase in $D_H=I_H|D_H|$. Its restriction to $F(H)$ is the internal complex structure.}{item:DHPolar}
  \notationentry{$\scpr_H$ on $H$}{For factorial $H$, the internal complex scalar product on the real space $H$ with complex structure $I_H$.}{item:InternalSP}
  \notationentry{$A_H$}{Skew-adjoint generator of $\Delta_H^{it}|_H$ on the real Hilbert space $H$, given by $iL_H$ on $H\cap\dom L_H$.}{lemma:PolarizationsAndOrthogonalDynamics}
  \noalign{\smallskip}
  \notationgroup{Vector entropy}
  \notationentry{$\E_H$, \glsentryname{entropyvector}}{Entropy operator and entropy of a vector $\psi\in\Hil$ relative to $H$.}{def:entropyoperator}
\end{longtable}
\endgroup
\restoregeometry

\printnoidxglossary[style=booksymbols,sort=standard,title={List of Symbols}]

\begin{theindex}
{\textbf{A}}\nopagebreak

  \item anti-locality, \hyperpage{100}
  \item antiunitary (Def.~\ref{def:AntiunitariesAndConjugations}), 
		\hyperpage{12}

  \indexspace
{\textbf{B}}\nopagebreak

  \item boundary values on a strip, \hyperpage{112, 113}
  \item boundary-to-boundary operator, \hyperpage{116}

  \indexspace
{\textbf{C}}\nopagebreak

  \item $C^*$-dynamical system, \hyperpage{97}
  \item centre of a subspace, \hyperpage{19}
  \item classification of standard subspaces
    \subitem by dual pairs (factorial case), \hyperpage{80}
    \subitem by graphs of positive operators, \hyperpage{43}
    \subitem by modular data, \hyperpage{36}
    \subitem by polarized real Hilbert spaces, \hyperpage{75}
    \subitem by positive contractions, \hyperpage{40}
    \subitem by spectral multiplicity (separable case), \hyperpage{96}
    \subitem by symmetric measure classes, \hyperpage{94}
    \subitem by symmetric representations, \hyperpage{93}
    \subitem by Tomita operators, \hyperpage{23}
  \item complement
    \subitem orthogonal, \hyperpage{6}
    \subitem symplectic (Def.~\ref{definition:SymplecticComplement}), 
		\hyperpage{13}
  \item complete passivity, \hyperpage{92}
  \item complex scalar product
    \subitem internal to $H$ (Cor.~\ref{corollary:FactorialSpacesAreComplexHilbert}), 
		\hyperpage{72}
  \item complex structure
    \subitem general (Def.~\ref{def:ComplexStructures}), \hyperpage{68}
    \subitem internal to $H$ (Cor.~\ref{corollary:FactorialSpacesAreComplexHilbert}), 
		\hyperpage{72}
    \subitem orthogonal (Def.~\ref{def:ComplexStructures}), 
		\hyperpage{68}
  \item complexification
    \subitem canonical, \hyperpage{70}
  \item conjugation (Def.~\ref{def:AntiunitariesAndConjugations}), 
		\hyperpage{12}

  \indexspace
{\textbf{D}}\nopagebreak

  \item direct
    \subitem integral, \hyperpage{87}
    \subitem sum, \hyperpage{84}
  \item direct summand, \hyperpage{86}
  \item dual pair
    \subitem defined by positive operator (Def.~\ref{def:PositiveOperatorDualPair}), 
		\hyperpage{79}
    \subitem of real spaces (Def.~\ref{def:DualPairGeneral}), 
		\hyperpage{78}

  \indexspace
{\textbf{E}}\nopagebreak

  \item entropy of a vector, \hyperpage{102}
  \item entropy operator of $H$, \hyperpage{102}

  \indexspace
{\textbf{F}}\nopagebreak

  \item factorial part (Prop.~\ref{proposition:DecomposeHIntoMaximallyAbelianAndFactorPart}), 
		\hyperpage{21}

  \indexspace
{\textbf{G}}\nopagebreak

  \item Gibbs density matrix, \hyperpage{98}
  \item graph Hilbert space of the modular operator, \hyperpage{44}
  \item graph standard subspace (Example~\ref{example:StandardSubspacesFromGraphs}), 
		\hyperpage{43}
  \item Grassmannian, \hyperpage{5}
    \subitem real, \hyperpage{10}

  \indexspace
{\textbf{H}}\nopagebreak

  \item Hadamard product, \see{Schur product theorem}{105}
  \item Hardy space, \hyperpage{45}
    \subitem on strip, \hyperpage{62}, \hyperpage{112}
    \subitem radial on $\Cl_+$, \hyperpage{64}

  \indexspace
{\textbf{I}}\nopagebreak

  \item involution (Def.~\ref{def:Involutions}), \hyperpage{9}

  \indexspace
{\textbf{K}}\nopagebreak

  \item Klein--Gordon field
    \subitem local standard subspaces, \hyperpage{100}
  \item KMS
    \subitem condition (Def.~\ref{definition:KMSCondition}), 
		\hyperpage{46}, 
		\seealso{KMS characterization of the modular group}{47}
    \subitem state, \hyperpage{97}
    \subitem state, normal, \hyperpage{98}
  \item KMS characterization of the modular group, \hyperpage{47}

  \indexspace
{\textbf{L}}\nopagebreak

  \item localization by distributional support, \hyperpage{67}

  \indexspace
{\textbf{M}}\nopagebreak

  \item measurable
    \subitem field, \hyperpage{88}
  \item modular
    \subitem conjugation, \hyperpage{24}
    \subitem function, \hyperpage{38}
    \subitem group, \hyperpage{31}
    \subitem Hamiltonian, \hyperpage{31}
    \subitem operator, \hyperpage{24}
    \subitem unitaries, \hyperpage{31}
  \item multiplicity-free standard subspace, \hyperpage{95, 96}

  \indexspace
{\textbf{O}}\nopagebreak

  \item operator
    \subitem decomposable, \hyperpage{85}
    \subitem direct integral, \hyperpage{89}
    \subitem direct sum, \hyperpage{85}
  \item orthogonal dynamics (Def.~\ref{def:RealOrthogonalDynamicalSystem}), 
		\hyperpage{74}

  \indexspace
{\textbf{P}}\nopagebreak

  \item passivity, \hyperpage{31}
  \item passivity of $\Delta_H$ (Prop.~\ref{prop:PassivityOfDelta}), 
		\hyperpage{31}
  \item polarization of a real Hilbert space (Def.~\ref{def:PolarizedHilbertSpace}), 
		\hyperpage{73}
  \item polarized real Hilbert space (Def.~\ref{def:PolarizedHilbertSpace}), 
		\hyperpage{73}
  \item polarizer
    \subitem general (Def.~\ref{def:ComplexStructures}), \hyperpage{68}
    \subitem of $H$ (Def.~\ref{def:PolarizerOfStandardSubspace}), 
		\hyperpage{71}
  \item positive contraction
    \subitem classification of standard subspaces, \hyperpage{40}
  \item pre-symplectic form (Def.~\ref{def:ComplexStructures}), 
		\hyperpage{68}
  \item projection
    \subitem cutting (Def.~\ref{definition:cuttingprojection}), 
		\hyperpage{35}
    \subitem general, \hyperpage{7}
    \subitem real orthogonal (Def.~\ref{def:RealOrthogonalProjection}), 
		\hyperpage{11}

  \indexspace
{\textbf{R}}\nopagebreak

  \item real orthogonal dynamical system (Def.~\ref{def:RealOrthogonalDynamicalSystem}), 
		\hyperpage{74}
  \item reflection positivity, \hyperpage{48}, \hyperpage{58}

  \indexspace
{\textbf{S}}\nopagebreak

  \item Schur product theorem, \hyperpage{105}
  \item spectral gap (Def.~\ref{def:SpectralGap}), \hyperpage{29}
  \item standard subspace, \see{subspace, standard}{14}
  \item standard subspace realization of a polarized real Hilbert space (Def.~\ref{def:PolarizedHilbertSpace}), 
		\hyperpage{73}
  \item strip reflection, \hyperpage{46}
  \item subspace, \hyperpage{5}
    \subitem abelian (Def.~\ref{definition:HAbelianCenterEtc}), 
		\hyperpage{19}
    \subitem coisotropic, \hyperpage{20}
    \subitem cyclic (Def.~\ref{definition:CyclicSeparating}), 
		\hyperpage{13}
    \subitem factorial (Def.~\ref{definition:HAbelianCenterEtc}), 
		\hyperpage{19}
    \subitem independent, \hyperpage{7}
    \subitem isotropic, \hyperpage{19}
    \subitem Lagrangian, \hyperpage{20}
    \subitem maximal abelian (Prop.~\ref{proposition:AbelianSpaces}), 
		\hyperpage{20}
    \subitem separating (Def.~\ref{definition:CyclicSeparating}), 
		\hyperpage{13}
    \subitem standard (Def.~\ref{definition:standardsubspace}), 
		\hyperpage{14}
    \subitem symplectic, \hyperpage{19}
  \item symmetric function on a strip, \hyperpage{46}
  \item symplectic form (Def.~\ref{def:ComplexStructures}), 
		\hyperpage{68}

  \indexspace
{\textbf{T}}\nopagebreak

  \item tensor category of standard subspaces, \hyperpage{93}
  \item tensor product
    \subitem real Hilbert spaces, \hyperpage{90}
    \subitem standard subspaces, \hyperpage{90}
  \item theorem
    \subitem Analytic properties of Hardy space functions (current net) (\ref *{theorem:HardyFunctionsInText}), 
		\hyperpage{62}
    \subitem Analytic properties of Hardy space functions (strip) (\ref *{theorem:HardyFunctionsAppendix}), 
		\hyperpage{113}
    \subitem Boundary values and the three-lines theorem (\ref *{theorem:HinftyBoundaryValues}), 
		\hyperpage{112}
    \subitem Closed subspaces and orthogonal projections (\ref *{theorem:ClosedSpacesAndOrthogonalProjections}), 
		\hyperpage{7}
    \subitem Comparison of real and ambient complex tensor products (\ref *{theorem:TensorProductComparison}), 
		\hyperpage{90}
    \subitem Complete passivity and the modular group (\ref *{theorem:competepassivity<=>modulargroup}), 
		\hyperpage{92}
    \subitem Graph model of a standard subspace (\ref *{theorem:GraphStandardSubspaces}), 
		\hyperpage{43}
    \subitem Hardy space on a strip as a Hilbert and Banach space (\ref *{theorem:HardySpace}), 
		\hyperpage{116}
    \subitem KMS characterization of the modular group (\ref *{theorem:KMS}), 
		\hyperpage{47}
    \subitem Lattice of closed subspaces (\ref *{theorem:Grassmannian}), 
		\hyperpage{6}
    \subitem Net of standard subspaces of the U(1) current (\ref *{theorem:U1C+}), 
		\hyperpage{54}
    \subitem Schur product theorem (\ref *{thmSchur}), \hyperpage{105}
    \subitem Spectral multiplicity classification of standard subspaces (\ref *{thm:mf}), 
		\hyperpage{96}
    \subitem Standard subspace realization of polarized real Hilbert spaces (\ref *{thm:PolarizedRealHilbertSpacesAreStandardSubspaces}), 
		\hyperpage{75}
    \subitem Standard subspaces and dual pairs given by a positive operator (\ref *{theorem:PositiveOperatorStandardSubspaces}), 
		\hyperpage{79}
    \subitem Standard subspaces and positive contractions up to unitary equivalence (\ref *{theorem:StdAndContractions}), 
		\hyperpage{40}
    \subitem Tomita--Takesaki theorem for standard subspaces (\ref *{theorem:Tomita}), 
		\hyperpage{26}
  \item three-lines theorem, \hyperpage{112}
  \item Tomita operator (Def.~\ref{definition:Tomitaoperator}), 
		\hyperpage{22}
  \item Tomita--Takesaki theorem
    \subitem for standard subspaces, \hyperpage{26}

  \indexspace
{\textbf{U}}\nopagebreak

  \item $U(1)$ current net, \hyperpage{53}
    \subitem affine covariance, \hyperpage{53}
    \subitem duality, \hyperpage{54}
    \subitem half-line localization, \hyperpage{63}, \hyperpage{66, 67}
    \subitem isotony, \hyperpage{53}
    \subitem locality, \hyperpage{54}

  \indexspace
{\textbf{W}}\nopagebreak

  \item $W^*$-dynamical system, \hyperpage{97}

\end{theindex}

\AtNextBibliography{\normalsize}
\newrefcontext[sorting=anyt]
\printbibliography
\end{document}